\documentclass[journal]{IEEEtran}
\usepackage{cite}
\usepackage{amsmath,amssymb,amsfonts}
\usepackage{amsmath}
\usepackage{algorithmic}
\usepackage{graphicx}
\usepackage{subfigure}
\usepackage{textcomp}
\usepackage{balance}

\begin{document}
\title{A model-guided deep network for limited-angle computed tomography}
\author{Wei Wang, Xiang-Gen Xia, Chuanjiang He, Zemin Ren, Jian Lu,  Tianfu Wang and  Baiying Lei
	\thanks{This work was supported partly by National Natural Science Foundation of China (Nos.61871274 and 61801305), China Postdoctoral Science Foundation
		(2018M64081), Peacock Plan (No. KQTD2016053112051497), Shenzhen Key Basic Research Project (Nos. JCYJ20180507184647636, JCYJ20170818094109846, and JCYJ20190808155618806).}
	\thanks{Wei Wang, Tianfu Wang, and  Baiying Lei are with the School of Biomedical Engineering, Shenzhen University,
		National-Regional Key Technology Engineering Laboratory for Medical Ultrasound, Guangdong Key Laboratory for Biomedical Measurements and Ultrasound Imaging, School of Biomedical Engineering, Health Science Center, Shenzhen University, Shenzhen, China. (e-mail: wangwei@szu.edu.cn, leiby@szu.edu.cn, tfwang@szu.edu.cn). }
	\thanks{Xiang-Gen Xia is  with the Department of Electrical and Computer Engineering, University of Delaware, Newark, DE 19716, USA.  (e-mail: xxia@ee.udel.edu).}
	\thanks{Chuanjiang He is with the
		College of Mathematics and Statistics, Chongqing University, Chongqing, China (e-mail: cjhe@cqu.edu.cn).}
	\thanks{Zemin Ren is with the
		College of Mathematics and Physics, Chongqing University of Science and Technology, Chongqing, China (e-mail: zeminren@cqu.edu.cn).}
	\thanks{Jian Lu is with the
		Shenzhen Key Laboratory of Advanced Machine Learning and Applications, Shenzhen University, Shenzhen, China (e-mail: jianlu@szu.edu.cn).}
}
\markboth{Journal of \LaTeX\ Class Files,~Vol.~14, No.~8, August~2015}%
{Shell \MakeLowercase{\textit{et al.}}: Bare Demo of IEEEtran.cls for IEEE Journals}
\maketitle

\begin{abstract}
In this paper, we first propose a variational model for the  limited-angle computed tomography (CT) image reconstruction and then convert the model into an end-to-end deep network. We use the penalty method to solve the model and divide it into three iterative subproblems, where  the first subproblem completes  the sinograms by utilizing the prior information of  sinograms in the frequency domain and the second refines the CT images by using the prior information of  CT images in the spatial domain,  and the last merges the outputs of the first two subproblems. In each iteration, we use the convolutional neural networks (CNNs) to approxiamte the solutions of the first two subproblems and, thus,   obtain an  end-to-end deep network for the  limited-angle CT image reconstruction. Our network tackles both the sinograms and the CT images, and can simultaneously suppress the artifacts caused by the incomplete data and recover fine structural information in the CT images. Experimental results show that our method outperforms the existing algorithms for the limited-angle CT image reconstruction.
\end{abstract}


\section{Introduction}
\label{sec:introduction}
\IEEEPARstart{C}{omputed} Tomography (CT) is a fundamental imaging tool widely used in many areas, including industrial non-destructive tests, medical diagnoses and security checks. To stably and exactly reconstruct a CT image, the test object is required to be scanned under consecutive $180^\circ$ or $180^\circ$+fan angles for parallel-beam or fan-beam geometries, respectively. The measured data obtained by scanning objects are collectively called as sinograms. In some cases, the ranges of the scanning angles are less than $180^\circ$ or $180^\circ$+fan, which are known as the limited-angle CT. The reasons for limited-angle scanning might be to reduce the scanning time and X-ray dose exposed to patients or physical constraints (e.g., the object is too large and the scanner can’t image all of the object). The limited-angle CT image reconstruction is a severely ill-posed problem \cite{doi:10.1137/1.9780898719284.fm}. Standard analytic algorithms such as filtered back-projection (FBP) for limited-angle CT will generally produce images with heavy directional artifacts and intensity inhomogeneities.

So far, many algorithms have been proposed to suppress the artifacts and improve the qualities of the limited-angle CT images. Among these algorithms, completing the missing data of sinograms by interpolation is the most straightforward way for limited-angle CT image reconstructions \cite{ISI:000288875602078}. However, due to the complexity of real data and the difficulty in interpolating sinograms, the performance improvements of such interpolation methods are very small.

To deeply understand the limited-angle CT image reconstruction problem, the ill-conditioned nature of the problem \cite{ISI:A1983QM05500014} and the characterization of artifacts in limited-angle CT images \cite{ISI:000327794100007}\cite{ISI:A1993LV86200005} were studied. Based on the characterization of artifacts in limited-angle CT images, anisotropic total variation (TV) models were proposed to suppress the artifacts \cite{ISI:000327794100007}\cite{ISI:000316181300010}\cite{ ISI:000413319200020}. Using microlocal analysis, Quinto proved that edges tangent to available X-rays can be stably reconstructed while those whose singularities are not tangent to any X-ray lines cannot be reconstructed easily \cite{ISI:A1993LV86200005}. 

Inspired by the compressed sensing (CS) theory, the variational models with sparse regularizations have been widely researched and also used to reconstruct limited-angle CT images. Under some conditions, these variational methods can reconstruct high quality CT images from sampled sinograms of rates far fewer than the Nyquist sampling rate. Typical regularizations used in the variational models involve the TV regularization \cite{ISI:000258537000022}\cite{ ISI:000287848600003}\cite{ISI:000316181300010}\cite{ISI:000454918800001}, the  higher-order derivative regularization \cite{ISI:000274931100013}\cite{ISI:000332599500012}, the wavelet sparse regularization \cite{ISI:000235321000008}\cite{ISI:000385277200007}, the curvelet sparse regularization \cite{INSPEC:11719708}, the shearlet sparse regularization \cite{ ISI:000325827200018} and the dictionary sparse regularization \cite{ISI:000310148600002}\cite{ISI:000298147700012}. Optimal iterative algorithms are usually employed to solve these variational models and so their computational costs are usually very high. Meanwhile, there exist many parameters in a variational model and its associated algorithm that need to be set by hand, which greatly influences the quality of the reconstructed CT images.

Recently, machine-learning techniques especially the deep learning using convolutional neural networks (CNNs) have achieved great success in a wide range of image processing areas including the limited-angle CT image reconstruction. In \cite{ISI:000456950500003}, the traditional machine-learning technique  based on handcrafted features was used to reduce the artifacts in the limited-angle CT images. In \cite{ISI:000434302700015}, W\"urfl et al. presented a new deep learning
framework for the limited-angle CT image reconstruction, where the filtered back-projection-type algorithms were mapped to the neural networks. In \cite{ISI:000519725000026}, Huang et al. used a U-Net network to reduce artifacts in the CT images reconstructed by the FBP algorithm for transmission X-ray microscopy systems. In \cite{ ISI:000457843606052}, Anirudh et al. proposed a CTNet to reconstruct limited-angle CT images, where the CTNet is a system of 1D and 2D convolutional neural networks that operates directly on the limited-angle sinograms. In \cite{ISI:000470704700001}, Bubba et al. proposed a shearlet regularization model to reconstruct the visible parts of  CT images and a U-Net network with dense blocks to predict the invisible parts. In \cite{ ISI:000535354300071}, Li et al. proposed a GAN-based inpainting method to restore the missing sinogram data for the limited-angle scannings. In \cite{INSPEC:19432610}, Ghani utilized conditional generative adversarial networks (cGANs) in both the data and the image domain for the limited-angle CT image reconstruction, where the cGANs are combined through a consensus process.

In this paper, we first propose a variational model with two regularizations for the limited-angle CT image reconstruction, where one regularization utilizes the prior information of  sinograms in the frequency domain and another utilizes the prior information of  CT images in the spatial domain. Then we use the penalty method to solve our model and divide it into three iterative subproblems, where the first subproblem completes  the sinograms by utilizing the prior information of  sinograms in the frequency domain and the second refines the CT images by using the prior information of  CT images in the spatial domain,  and the last merges the outputs of the first two subproblems. Then we  unroll the iterative scheme to an end-to-end deep network, where the solution of the first two subproblems are approximated by two  residual subnetworks and the last subproblem corresponds to a layer that merges the outputs of the first two subnetworks. Therefore, our network utilizes the information of  both the sinograms and the CT images, and can simultaneously suppress the artifacts caused by the incomplete data and reconstruct good images from limited-angle sinograms.

The rest of the paper is organized as follows. Section II introduces the preliminaries and related works. Section III describes the proposed model and network. Section IV performs the simulated experiments and Section IV gives the  conclusion.

\section{Preliminaries and Related works}
In this section, we briefly introduce some mathematical notations and related works.
\subsection{CT Reconstruction Model}
The process of measuring the sinograms can be mathematically expressed by the following formula:
\begin{equation}\label{e1}
g(\theta,\gamma)=\int_{0}^{\infty}u(\Gamma(\gamma)+s\theta)ds,
\end{equation}
where $\Gamma(\gamma)$ is a continuous curve representing the position  of the X-ray source, $S^{n-1}=\{\theta\in R^n:|\theta|=1\}$ is the unit  sphere in $R^n$,  and $\theta\in S^{n-1}$ represents the diverging direction of the X-ray beam.

Discretizing equation (\ref{e1}), we can formulate the CT reconstruction model as a linear equation set:
\begin{equation}\label{e2}
Wu=g,
\end{equation}
where $u\in R^{M_1M_2\times1}$ is the CT image to be reconstruct, $M_1$ and $M_2$ are the width and height of the CT image (for a 3D object, it can be reshape into a 2D image), respectively, $g\in R^{MN\times1}$ is the measured sinogram, $M$ and $N$ 
are the sampling numbers of $\gamma$ and $\theta$, respectively, and  $W\in R^{MN\times M_1M_2}$ is a sparse matrix representing the discrete line integral $\mathcal{R}$ in equation (\ref{e1}), which is feasible for different scanning geometries, including parallel-beam, fan-beam, and 3-D cone-beam.  Each element $w_{i,j}$ in matrix $W$ can be computed by discretizing equation (\ref{e1}) via interpolation or calculating the intersection length of the $i$ th ray through the $j$ th pixel \cite{ISI:A1985AFZ6900022}.

Instead of solving the linear equation set (\ref{e2}) directly, one may solve   the following optimal problem:
\begin{equation}\label{e3}
\mathop{\arg\min}\limits_{u\in R^{M_1M_2\times 1}}\|Wu-g\|_2^2.
\end{equation}
The Euler-equation of problem (\ref{e3}) is
\begin{equation}\label{e4}
W^{T}Wu=W^{T}g.
\end{equation}
For a CT system with a high enough sampling rate, the inverse matrix $(W^{T}W)^{-1}$ may exist and the equation set can be solved by the  conjugate gradient method. However, due to the singularity of $(W^{T}W)^{-1}$, the solution of equation (\ref{e4}) is usually not robust (ie, adding a little noise  to the sinogram $g$ will result in the solution $\bar u$ far way from the original $u$). For a undersampled CT system, the inverse matrix $(W^{T}W)^{-1}$ does not exist and so the conjugate gradient algorithm used to find the inverse matrix  $(W^{T}W)^{-1}$ usually doesn't converge. To address these issues, extra regularizations  were usually proposed to add to the energy functional   (\ref{e3}).

\subsection{Related Works}
In \cite{ISI:000454918800001}, Tovey et. al  proposed a directional TV model for the sinogram inpainting and limited-angle CT image reconstruction:
\begin{equation}\label{e5}
\begin{aligned}
(u, v)=&\underset{u \ge 0}{\operatorname{argmin}} E(u, v)\\
=&\underset{u \ge 0}{\operatorname{argmin}} \frac{1}{2}\|\mathcal{R} u-v\|_{\alpha_{1}}^{2}+\frac{\alpha_{2}}{2}\|S \mathcal{R} u-g\|_{2}^{2} \\
&+\frac{\alpha_{3}}{2}\|S v-g\|_{2}^{2}+\beta_{1} \operatorname{TV}(u)+\beta_{2} \operatorname{DTV}_{\mathcal{R} u}(v),
\end{aligned}
\end{equation}
where $u\ge0$ is the reconstructed CT image,  $v$ is the completed sinogram, $g$ is the measured sinogram, $\alpha_i,\beta_i$ are weighting parameters, $\alpha_1$ is embedded in the norm,
$\mathcal{R}$ is the Radon transform, $S$ represents the subsampling on limited angles, $\operatorname{TV}(u)$ is the TV regularization, $\operatorname{DTV}_{\mathcal{R} u}(v)$ is the  directional TV regularization,
\begin{equation}\label{e6}
\operatorname{DTV}_{\mathcal{R} u}(v)=\int \sqrt{c_{1}^{2}\left|\left\langle\mathbf{e}_{1}, \nabla v\right\rangle\right|^{2}+c_{2}^{2}\left|\left\langle\mathbf{e}_{2}, \nabla v\right\rangle\right|^{2}} \mathrm{d} x,
\end{equation}
where $c_{1}=\frac{1}{\sqrt{1+\Sigma^{2}}}$ and $c_2=1$, or $c_1=\epsilon$ and $c_{2}=\varepsilon+\exp \left(-1 / \Delta^{2}\right)$ for some $\epsilon>0$, $\Sigma=\lambda_1+\lambda_2$, $\Delta=\lambda_1-\lambda_2$, $\lambda_1$, $\lambda_2$ and $\mathbf{e}_{1}$, $\mathbf{e}_{2}$ are, respectively, the eigenvalues and eigenvectors of the structure tensor of $\mathcal{R} u$.

In \cite{ISI:000413319200020}, Wang et. al proposed a reweighted anisotropic TV  model for the limited-angle CT image reconstruction:
\begin{equation}\label{e7}
\begin{aligned}
&\min _{u}\|u\|_{\mathrm{RwATV}} \\
&\text { s.t. } W u=g, ~u \geq 0,
\end{aligned}
\end{equation}
where 
\begin{equation}\label{e8}
\begin{aligned}
\|u\|_{\mathrm{RwATV}}=&\left\|R \nabla_{A, B} u\right\|_{1}=\sum_{i, j} r_{i, j}\left|\nabla_{A, B} u_{i, j}\right|,\\
r_{i, j}^{(l+1)}=&\frac{1}{\left|\nabla_{A, B} u_{i, j}^{(l)}\right|+\xi},\\
\left|\nabla_{A, B} u_{i, j}\right|=&\sqrt{A\left(u_{i, j}-u_{i-1, j}\right)^{2}+B\left(u_{i, j}-u_{i, j-1}\right)^{2}},
\end{aligned}
\end{equation}
and $A>0, B>0, \xi>0$ are three parameters.

In \cite{ISI:000470704700001}, Bubba et. al proposed a  hybrid deep learning-shearlet framework for the
limited-angle CT image reconstruction, where they divided the shearlet coefficients into visible and invisible. First, they obtained the visible shearlet coefficients via the following model:
\begin{equation}\label{e9}
u^{*}:=\underset{u}{\operatorname{argmin}} \frac{1}{2}\left\|\mathcal{R}_{\phi} u-g\right\|_{2}^{2}+\lambda \cdot\left\|\mathbf{SH}_{\psi}(u)\right\|_{1},
\end{equation}
where $\mathcal{R}_{\phi}$ represents the limited-angle Radon transform, $\mathbf{SH}_{\psi}$ is the shearlet transform and $g$ is the measured sinogram.
Then they trained a CNN to  estimate the invisible shearlet coefficients
from the visible ones:
\begin{equation}\label{e10}
{\xi}=\mathrm{CNN}_{\boldsymbol{\theta}}\left(\mathbf{SH} \left({u}^{*}\right)\right) \quad\left(\approx  \mathbf{SH} ({u})_{\mathcal{I}_{\mathrm{inv}}}\right),
\end{equation}
At last, the reconstructed CT image was obtained by
\begin{equation}\label{e11}
u_{\mathrm{LtI}}=\mathbf{SH}^{-1}\left(\mathbf{SH}\left(u^{*}\right)_{\mathcal{I}_{\mathrm{vis}}}+\xi\right).
\end{equation}

In \cite{ISI:000457604700005}, He et. al adopted the
following dual-domain general penalized weighted least-squares (PWLS) model for the CT image reconstruction:
\begin{equation}\label{e12}
\min _{u,v} \frac{1}{2}\|v-\hat{v}\|_{\Sigma_{v}^{-1}}^{2}+\frac{1}{2}\|W u-v\|_{\Sigma_{u}^{-1}}^{2}+\lambda R_{v}(v)+\lambda R_{u}(u) ,
\end{equation}
where $\hat v$ represents the measured sinogram, $v$ is the desired
sinogram,  $u$ represents the desired CT image to be
reconstructed,  $W$ denotes the system matrix, $\Sigma_{u}$ and $\Sigma_{v}$ are two diagonal weighted matrices, $R_{u}(u)$ is a regularization of $u$ and $R_{v}(v)=\frac{\gamma}{2} \sum_{i} \sum_{m \in N_{i}} r_{j m}\left(v_{j}-v_{m}\right)^{2}$ is a regularization of $v$, $N_i$ represents the neighbor of pixel $i$, $r_{j m}>0$ is a weight value. Then, they used the parameterized plug-and-play alternating direction method of multipliers(3pADMM) algorithm to solve the model:
\begin{equation}\label{e13}
\left\{\begin{aligned}
v^{(n)}=&\left(I-\tilde{\Sigma}_{v}^{(n)}-\tilde{\Sigma}_{u}^{(n)}-\lambda^{(n)} I\right) v^{(n-1)}\\
&+\tilde{\Sigma}_{v}^{(n)} \hat{v}+\tilde{\Sigma}_{u}^{(n)} A u^{(n-1)}+\tilde{\lambda}^{(n)} D v^{(n-1)} \\
u^{(n)}=&\left(1-\theta^{(n)}\right) u^{(n-1)}+\theta^{(n)}\left(z^{(n-1)}-\beta^{(n-1)}\right)\\
&-A^{T} \tilde{\Sigma}^{(n)}\left(A u^{(n-1)}-v^{(n)}\right) \\
z^{(n)}=&\left(1-\tilde{\theta}^{(n)}\right) z^{(n-1)}+\tilde{\theta}^{(n)}\left(u^{(n)}+\beta^{(n-1)}\right)\\ &-\tilde{\gamma}^{(n)} \operatorname{Res}^{(n)}\left(z^{(n-1)}\right) \\
\beta^{(n)}=& \beta^{(n-1)}+\tilde{\eta}^{(n)}\left(u^{(n)}-z^{(n)}\right)
\end{aligned}\right.,
\end{equation}
where $\tilde{\Sigma}_{v}^{(n)}=l_{r v} \Sigma_{v}^{-1}$, $\tilde{\Sigma}_{u}^{(n)}=l_{r v} \Sigma_{u}^{-1}$, $\tilde{\lambda}^{(n)}=\lambda$, $\theta^{(n)}=l_{r u} \rho$, $\tilde{\Sigma}^{(n)}=l_{r u} \Sigma_{u}^{-1}$, $\tilde{\theta}^{(n)}=l_{r z} \rho$, $\tilde{\gamma}^{(n)}=\gamma$, $\tilde{\eta}^{(n)}=\eta$,  $D$ denotes a filtering operation corresponding to the weight value $r_{jm}$ and $\operatorname{Res}^{(n)}$ is a residual CNN used to approximate the gradient $\nabla R_{u}$. Instead of setting these parameters $\left\{\tilde{\Sigma}_{v}^{(n)}, \tilde{\Sigma}_{u}^{(n)}, \tilde{\lambda}^{(n)}, \theta^{(n)}, \tilde{\Sigma}^{(n)}, \tilde{\theta}^{(n)}, \tilde{\gamma}^{(n)}, \tilde{\eta}^{(n)}\right\}$ by hand, they trained them by supervised leaning and, thus, obtained an end-to-end deep network for CT image reconstruction.

\section{Proposed model and Network}
\subsection{Proposed Model}
In this paper, we  propose the following model for the limited-angle CT image reconstruction:
\begin{equation}\label{e14}
\mathop{\arg\min} \limits_{u} E(u)=\|S(Wu)-g\|_2^2+\lambda_1R_1(u)+\lambda_2R_2(\mathcal{F}(Wu)),
\end{equation}
where $u\in R^{M_1M_2\times1}$ is the desired CT image,  $W\in R^{MN\times M_1M_2}$ is the system matrix, $g\in R^{M_3N\times1}$ is the measured sinogram, $M_3<M$ is the sampling number of the limited scanning angles $\gamma$.  $R_1(u)$ and $R_2(\mathcal{F}(Wu))$ are two regularizations, $\mathcal{F}$ is the Fast Fourier transform and $S:D\to D_1$  represents downsampling the full-angle sinogram $g_1$ on the limited-angle scanning domain $D_1$:
\begin{equation}\label{e15}
S(g_1(x))=
g_1(x), ~~\text{for}~~ x\in D_1,
\end{equation} 
where $D$ is the full-angle scanning domain. For example, for 2D parallel-beam scanning geometry, $D=[0,\pi]\times [0,s]$ and $D_1=[0,\phi]\times [0,s]$, where $\phi<\pi$ is the maximal scanning angle of the limited scanning geometry and $s$ is the width of detector array. Note that the definition domain of $g_1(x)\in R^{MN\times1}$ is $D$ while that of $S(g_1(x))\in R^{M_3N\times1}$ is $D_1$.

In theory, the regularization $R_1(u)$ utilizes the prior information of CT images in the spatial domain (such as sparsities of gradients) while $R_2(\mathcal{F}(Wu))$ utilizes the prior information of sinograms in the frequency domain (such as sparsities). There are two reasons that we use the Fourier transform in the regularization $R_2$. On one hand,  when we complete the missing data of  sinograms in $D\backslash D_1$, there is no local  information in $D\backslash D_1$ that we can use (since the values are all zeros). What we can utilize is the  global information extracted from $D_1$. The information in frequency domain is naturally global. Therefore, to better utilize the global  information, we adopt $R_2(\mathcal{F}(Wu))$ as our regularization of sinograms. On the other hand,  when training a CNN network to predict the missing data of sinograms in $D\backslash D_1$, the input sinograms $g_1$  are always zeros-padded, i.e. 
\begin{equation}\label{e16}
g_1(x)=\left\{ \begin{array}{l}
g(x), \text{~if~~} x\in D_1\\
0,    \text{~~~~~if~~} x\in D\backslash D_1
\end{array} \right.,
\end{equation}
where $g(x)$ is the limited-angle sinogram. Therefore, the CNN is required to map the patch in  $D\backslash D_1$ of zero values to the true values of the missing data. Because of the local receptive field of CNN, it's infeasible and will result in the training process  not convergent. After converting our model to a deep network, by experiments, we find that using $R_2((Wu))$ as a regularization will also lead to the training process of our network not convergent and so we use $R_2(\mathcal{F}(Wu))$ instead.

We use the penalty method to solve  model (\ref{e14}). Let
  $z $ and $\varsigma$ be two auxiliary variables, corresponding to $u$ and $\mathcal{F}(Wu)$, respectively. Then problem (\ref{e14}) can be rewritten as:
\begin{equation}\label{e17}
\begin{aligned}
\mathop{\arg\min} \limits_{\varsigma,z,u}& \|S(Wu)-g\|_2^2+\lambda_1R_1(z)+\lambda_1R_2(\varsigma)\\
&+ \beta_1\|u-z\|_2^2+ \beta_2\|\mathcal{F}(Wu)-\varsigma \|_2^2.
\end{aligned}
\end{equation}
Problem (\ref{e17}) can be solved by alternatively minimizing $\varsigma$, $z$ and $u$, respectively. Let $u^0$ be the initial CT image reconstructed by conventional algorithms such as FBP. Then problem (\ref{e17}) is equivalent to iteratively solving the following three subproblems.

\textbf{Subproblem 1: updating $\varsigma$} ~The objective function for iterating  $\varsigma$ is
\begin{equation}\label{e18}
\varsigma^n=\mathop{\arg\min} \limits_{\varsigma} \lambda_2R_2(\varsigma)+ \beta_2\|(\mathcal{F}Wu^n)-\varsigma \|_2^2.
\end{equation}

\textbf{Subproblem 2: updating $z$} ~The objective function for iterating  $z$ is
\begin{equation}\label{e19}
z^n=\mathop{\arg\min} \limits_{z} \lambda_1R_1(z)+ \beta_1\|u^n-z\|_2^2.
\end{equation}

\textbf{Subproblem 3: updating $u$} ~The objective function for iterating  $u$ is
\begin{equation}\label{e20}
\begin{aligned}
u^{n+1}=&\mathop{\arg\min} \limits_{u} \|S(Wu)-g\|_2^2+ \beta_1\|u-z^n\|_2^2+\\ &\beta_2\|\mathcal{F}(Wu)-\varsigma^n \|_2^2.
\end{aligned}
\end{equation}

\subsection{Proposed Network}
\begin{figure*}[!t]
	\centerline{\includegraphics[scale=0.9]{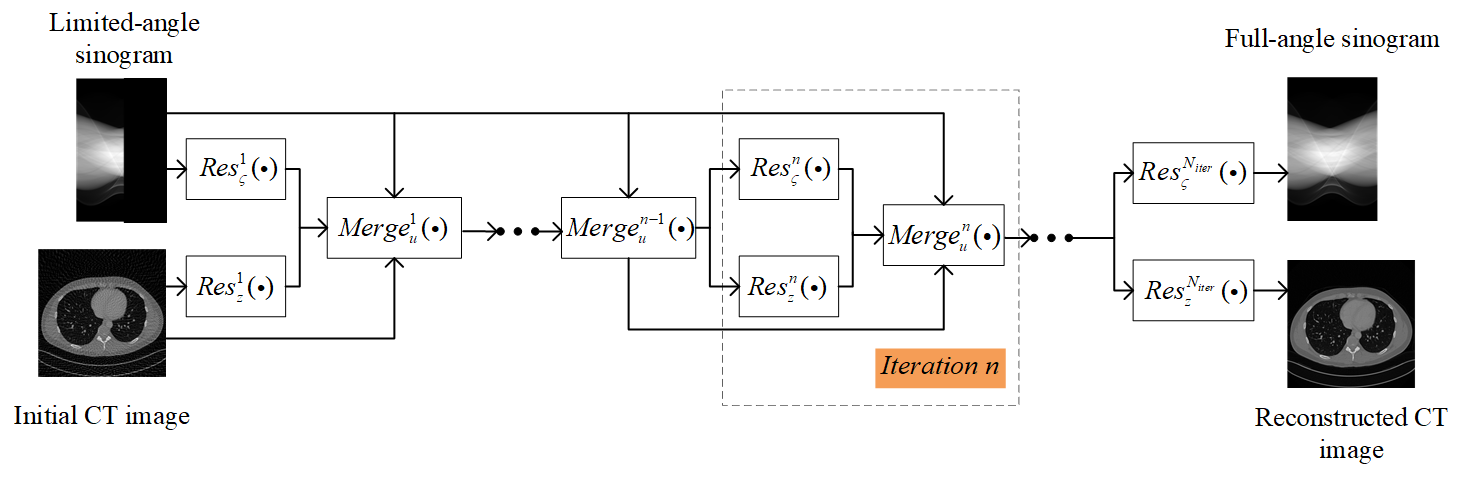}}
	\caption{The overall structure of our network.}
	\label{F1}
\end{figure*}

Instead of giving the explicit form of regularizations $R_1$ and $R_2$ and using optimal algorithm to solve subproblems (\ref{e18}) and (\ref{e19}), we use two  subnetworks to automatically extract the prior information and approximate the solutions of subproblems (\ref{e18}) and (\ref{e19}), respectively. By this way, we convert our model to an end-to-end deep network for the limited-angle CT image reconstruction.
The overall architecture of our network is shown in Fig. \ref{F1}. It is composed of $N_{iter}$ iteration blocks, and each block has three components that correspond to  subproblems (\ref{e18}), (\ref{e19}) and (\ref{e20}), respectively. A main difference between  our method and 3pADMM is that we use subnetworks to approximate the solutions of the regulariztion-based models while 3pADMM uses  subnetworks to approximate the gradients of the regularizations. Since our subnetworks don't need to approximate the gradients of the fidelity term  and regularization term, the structures of our subnetworks are looser than those of 3pADMM. 

\subsubsection{Structure of Sub-network $Res_\varsigma^n(\cdot)$ for Updating $\varsigma$} 
\begin{figure*}[!t]
	\centerline{\includegraphics[scale=0.8]{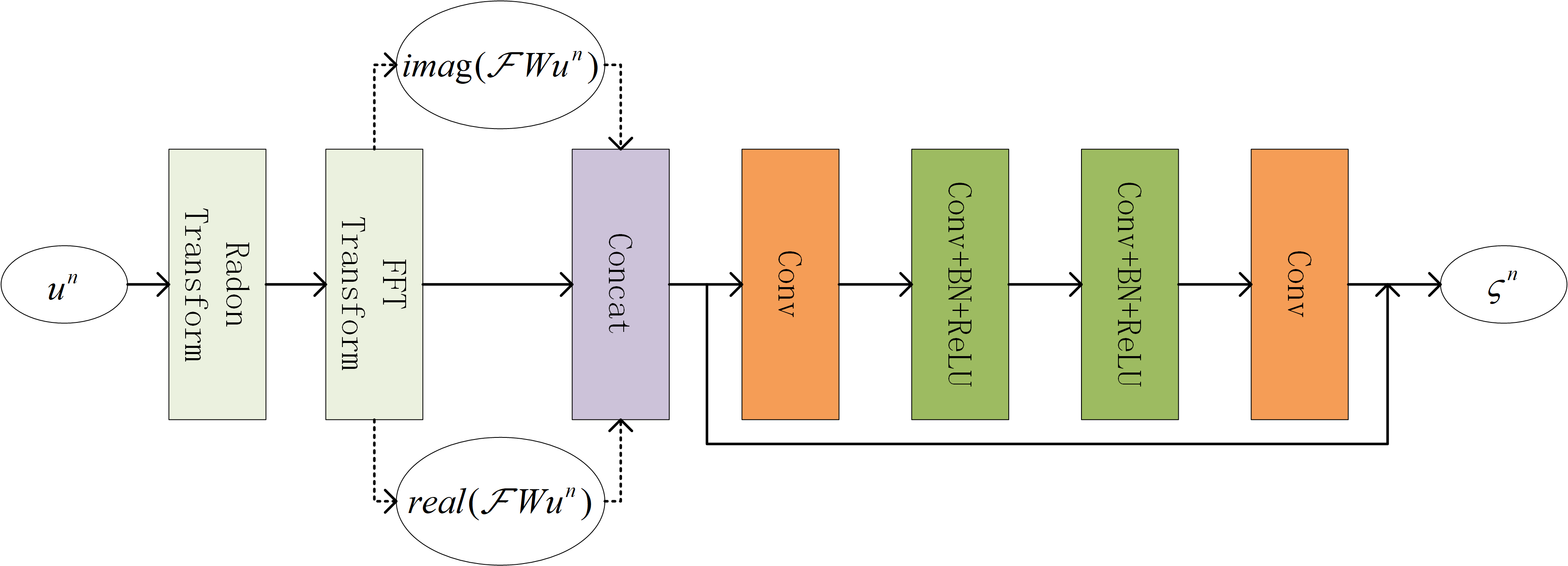}}
	\caption{The structure of $Res_\varsigma^n(\cdot)$ for updating $\varsigma$.}
	\label{F2}
\end{figure*}

We use a residual CNN of four layers to approximate the solution of subproblem (\ref{e18}). As can be seen from subproblem (\ref{e18}), the input of this sub-network is  $\mathcal{F}(Wu^n)$, which can't be directly tackled by CNNs since the input value is  complex. To remedy this issue, one approach is to use two subnetworks to process the real part and image part of $\mathcal{F}(Wu^n)$ separately. To save the GPU memory, we make the two subnetworks share the same weights.  The detailed structure of sub-network $Res_\varsigma^n(\cdot)$  is shown in Fig. \ref{F2}.

\begin{figure}[!t]
	\centerline{\includegraphics[scale=0.8]{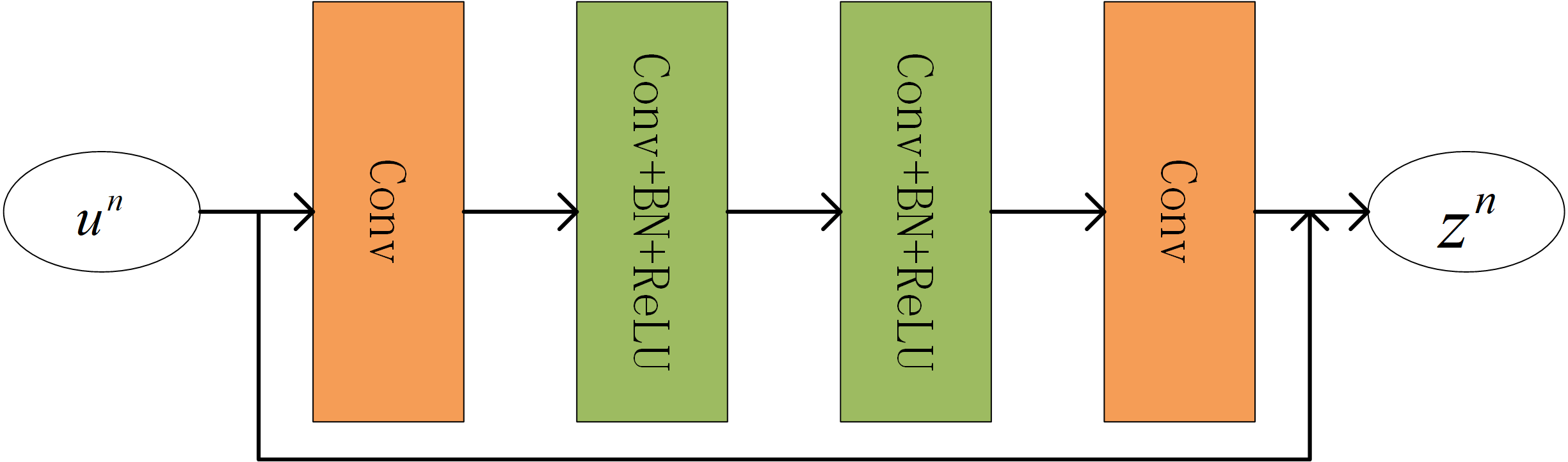}}
	\caption{The  structure of $Res_z^n(\cdot)$ for updating $z$.}
	\label{F3}
\end{figure}

\begin{figure}[!t]
	\centerline{\includegraphics[scale=0.8]{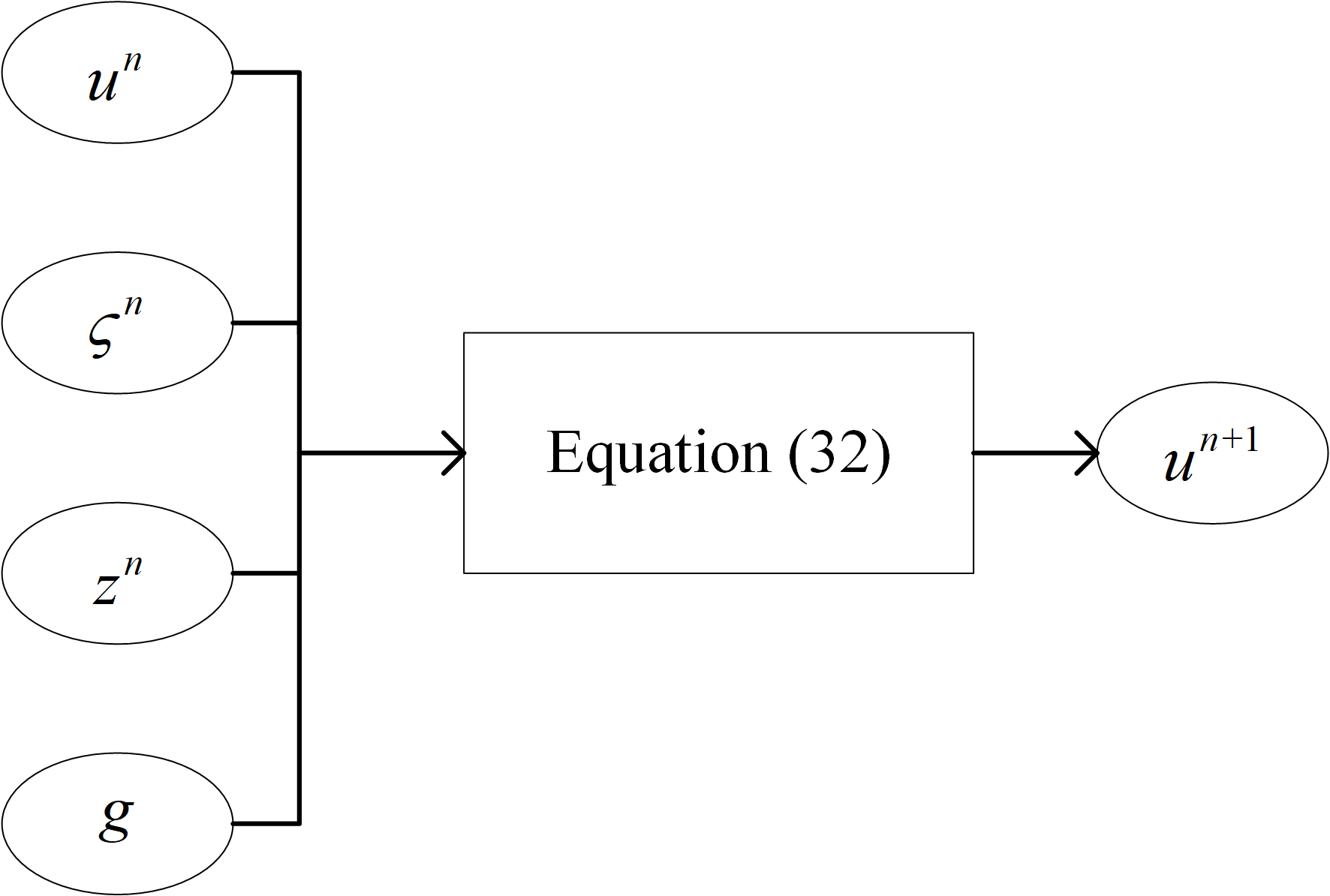}}
	\caption{The structure of $\mathop{Merge}_u^n(\cdot)$ for updating $u$.}
	\label{F4}
\end{figure}
Let $\mathop{real}(\mathcal{F}(Wu^n))\in R^{batch\times M\times N\times 1}$ and  $\mathop{image}(\mathcal{F}(Wu^n))\in R^{batch\times M\times N\times 1}$ be, respectively, the real and image parts of $\mathcal{F}(Wu^n)$ input to $Res_\varsigma(\cdot)$, where $batch$ is the number of input samples.  For the first layer, the convolution unit with filters of size $5\times5$ is used to generate 64 features, and  the activation function Rectified Linear Unit (ReLU) is used
for nonlinearity:
\begin{equation}\label{e21}
\begin{aligned}
input_\varsigma^n&=\mathop{concat}[\mathop{real}(\mathcal{F}(Wu^n)),\mathop{image}(\mathcal{F}(Wu^n))],\\
L_{\varsigma,1}^n&=input_\varsigma^n\otimes F_{\varsigma,1}^n+B_{\varsigma,1}^n,\\
L_{\varsigma,1}^n&=\max(L_{\varsigma,1}^n,0),\\
\end{aligned}
\end{equation}
where $\mathop{concat}$ merges the real part and image part of $\mathcal{F}(Wu^n)$
into one tensor, $ F_{\varsigma,1}^n\in R^{5\times5\times1\times64}$ and $B_{\varsigma,1}^n\in R^{1\times1\times1\times64}$ are the weights and biases of the convolution units, $\otimes$ is the convolution operator:
\begin{equation}
\begin{aligned}
&x\otimes F_{\varsigma,1}^n(m0,m1,m2,m3)=\\ 
&\sum_{d_1,d_2,d_3}x(m0,m1+d_1,m2+d_2,d_3)F_{\varsigma,1}^n(d_1,d_2,d_3,m_3),
\end{aligned}
\end{equation}
where $d1, d2$ and $d3$ go through the height, width and channels
of the filter $F_{\varsigma,1}^n$ and so $L_{\varsigma,1}^n\in R^{batch\times M\times N\times 64}$.

For Layer2 and Layer 3, the convolution units, batch normalized
(BN) units and activation units ReLU are used: 
\begin{equation}\label{e22}
\begin{aligned}
L_{\varsigma,2}^n&=L_{\varsigma,1}^n\otimes F_{\varsigma,2}^n+B_{\varsigma,2}^n,\\
L_{\varsigma,2}^n&=\frac{\tau_{\varsigma,2}^n(L_{\varsigma,2}^n-\mu_{\varsigma,2}^n)}{\nu_{\varsigma,2}^n}+\kappa_{\varsigma,2}^n,\\
L_{\varsigma,2}^n&=\max(L_{\varsigma,2}^n,0),\\
L_{\varsigma,3}^n&=L_{\varsigma,2}^n\otimes F_{\varsigma,3}^n+B_{\varsigma,3}^n,\\
L_{\varsigma,3}^n&=\frac{\tau_{\varsigma,3}^n(L_{\varsigma,3}^n-\mu_{\varsigma,3}^n)}{\nu_{\varsigma,3}^n}+\kappa_{\varsigma,3}^n,\\
L_{\varsigma,3}^n&=\max(L_{\varsigma,3}^n,0),\\
\end{aligned}
\end{equation}
where $F_{\varsigma,2}^n,F_{\varsigma,3}^n\in R^{5\times5\times64\times64}$ and $B_{\varsigma,2}^n,B_{\varsigma,3}^n\in R^{1\times1\times1\times64}$ are the weights and biases of the convolution units, $\mu_{\varsigma,2}^n,\mu_{\varsigma,3}^n \in R^{1\times1\times1\times64}$ and $\nu_{\varsigma,2}^n,\nu_{\varsigma,3}^n\in R^{1\times1\times1\times64}$ are  the mean and variance of each batch data, $\tau_{\varsigma,2}^n,\tau_{\varsigma,3}^n\in R^{1\times1\times1\times64}$ and $\kappa_{\varsigma,2}^n,\kappa_{\varsigma,3}^n\in R^{1\times1\times1\times64}$ are the scale and offset parameters of the batch normalized units,  $L_{\varsigma,2}^n,L_{\varsigma,3}^n\in R^{batch\times M\times N\times 64}$ are the outputs of Layer 2 and Layer 3.

For Layers 4, the convolution units, batch normalized
(BN) units, activation units ReLU and skip connection are used: 
\begin{equation}\label{e23}
\begin{aligned}
L_{\varsigma,4}^n&=L_{\varsigma,3}^n\otimes F_{\varsigma,4}^n+B_{\varsigma,4}^n,\\
L_{\varsigma,4}^n&=\frac{\tau_{\varsigma,4}^n(L_{\varsigma,4}^n-\mu_{\varsigma,4}^n)}{\nu_{\varsigma,4}^n}+\kappa_{\varsigma,4}^n,\\
L_{\varsigma,4}^n&=\max(L_{\varsigma,4}^n,0),\\
{\varsigma}^n&=input_\varsigma^n+L_{\varsigma,4}^n,
\end{aligned}
\end{equation}
where $F_{\varsigma,4}^n\in R^{5\times5\times64\times1}$, and $B_{\varsigma,4}^n\in R^{1\times1\times1\times64}$ are the weights and biases of the convolution units, $\mu_{\varsigma,4}^n\in R^{1\times1\times1\times64}$ and $\nu_{\varsigma,4}^n\in R^{1\times1\times1\times64}$ are  the mean and variance of each batch data, $\tau_{\varsigma,4}^n\in R^{1\times1\times1\times64}$ and $\kappa_{\varsigma,4}^n\in R^{1\times1\times1\times64}$ are the scale and offset parameters of the batch normalized units, respectively, $L_{\varsigma,4}^n\in R^{batch\times M\times N\times 1}$ and ${\varsigma}^n\in R^{batch\times M\times N\times 1}$ is the output.

\subsubsection{Structure of Sub-network $Res_z^n(\cdot)$ for Updating $z$} We also use a residual CNN of four layers to approximate the solution of subproblem (\ref{e19}).  The detailed structure of sub-network $Res_z^n(\cdot)$  is given in Fig. \ref{F3}.

Let $u^n\in R^{batch\times M_1\times M_2\times 1}$ be the input of $Res_z^n(\cdot)$.  For the first layer of $Res_z^n(\cdot)$, the convolution unit with filters of size $5\times5$ is used to generate 64 features, and  the activation function Rectified Linear Unit (ReLU) is used
for nonlinearity:
\begin{equation}\label{e24}
\begin{aligned}
L_{z,1}^n&=u\otimes F_{z,1}^n+B_{z,1}^n,\\
L_{z,1}^n&=\max(L_{z,1}^n,0),\\
\end{aligned}
\end{equation}
where  $F_{z,1}^n\in R^{5\times5\times1\times64}$ and $B_{z,1}^n\in R^{1\times1\times1\times64}$ are the weights and biases of the convolution units, respectively.

For Layer 2 and Layer 3 of $Res_z^n(\cdot)$, the convolution units, batch normalized
(BN) units and activation units ReLU are used: 
\begin{equation}\label{e25}
\begin{aligned}
L_{z,2}^n&=L_{z,1}^n\otimes F_{z,2}^n+B_{z,2}^n,\\
L_{z,2}^n&=\frac{\tau_{z,2}^n(L_{z,2}^n-\mu_{z,2}^n)}{\nu_{z,2}^n}+\kappa_{z,2}^n,\\
L_{z,2}^n&=\max(L_{z,2}^n,0),\\
L_{z,3}^n&=L_{z,2}^n\otimes F_{z,3}^n+B_{z,3}^n,\\
L_{z,3}^n&=\frac{\tau_{z,3}^n(L_{z,3}^n-\mu_{z,3}^n)}{\nu_{z,3}^n}+\kappa_{z,3}^n,\\
L_{z,3}^n&=\max(L_{z,3}^n,0),\\
\end{aligned}
\end{equation}
where $F_{z,2}^n,F_{z,3}^n\in R^{5\times5\times64\times64}$ and $B_{z,2}^n,B_{z,3}^n\in R^{1\times1\times1\times64}$ are the weights and biases of the convolution units, $\mu_{z,2}^n,\mu_{z,3}^n \in R^{1\times1\times1\times64}$ and $\nu_{z,2}^n,\nu_{z,3}^n\in R^{1\times1\times1\times64}$ are  the mean and variance of each batch data, $\tau_{z,2}^n,\tau_{z,3}^n\in R^{1\times1\times1\times64}$ and $\kappa_{z,2}^n,\kappa_{z,3}^n\in R^{1\times1\times1\times64}$ are the scale and offset parameters of the batch normalized units,  $L_{z,2}^n,L_{z,3}^n\in R^{batch\times M\times N\times 64}$ are the outputs of Layer 2 and Layer 3, respectively.

For Layer 4 of $Res_{z}^n(\cdot)$, the convolution units, batch normalized
(BN) units, activation units ReLU and skip connection are used: 
\begin{equation}\label{e26}
\begin{aligned}
L_{z,4}^n&=L_{z,3}^n\otimes F_{z,4}^n+B_{z,4}^n,\\
L_{z,4}^n&=\frac{\tau_{z,4}^n(L_{z,4}^n-\mu_{z,4}^n)}{\nu_{z,4}^n}+\kappa_{z,4}^n,\\
L_{z,4}^n&=\max(L_{z,4}^n,0),\\
{z}^n&=u^n+L_{z,4}^n,
\end{aligned}
\end{equation}
where $F_{z,4}^n\in R^{5\times5\times64\times1}$, and $B_{z,4}^n\in R^{1\times1\times1\times64}$ are the weights and biases of the convolution units, $\mu_{z,4}^n\in R^{1\times1\times1\times64}$ and $\nu_{z,4}^n\in R^{1\times1\times1\times64}$ are  the mean and variance of each batch data, $\tau_{z,4}^n\in R^{1\times1\times1\times64}$ and $\kappa_{z,4}^n\in R^{1\times1\times1\times64}$ are the scale and offset parameters of the batch normalized units, respectively, $L_{z,4}^n\in R^{batch\times M_1\times M_2\times 1}$ and ${z}^n\in R^{batch\times M_1\times M_2  \times 1}$ is the output.

\subsubsection{Structure of Sub-network $\mathop{Merge}_u^n(\cdot)$ for Updating $u$}  The detailed structure of sub-network $\mathop{Merge}_u^n(\cdot)$ is shown in Fig. \ref{F4}. By the Plancherel's theorem, subproblem (\ref{e20}) is equivalent to 
\begin{equation}\label{e27}
\begin{aligned}
u^{n+1}=&\mathop{\arg\min} \limits_{u} \|S(Wu)-g\|_2^2+ \beta_1\|u-z^n\|_2^2+\\ &\beta_2\|Wu-\mathcal{F}^{-1}\varsigma^n \|_2^2.
\end{aligned}
\end{equation}

Problem (\ref{e27}) requires its solution $u^{n+1}$ to be consistent with the measured sinogram $g$, the output CT images $z^n$ from subnetwork $Res_z^n(\cdot)$ and the output sinogram $\mathcal{F}^{-1}\varsigma^n$ from subnetwork $Res_\varsigma^n(\cdot)$. Therefore, the subnetwork corresponding to problem (\ref{e27}) merges the information of the outputs of $Res_z^n(\cdot)$ and $Res_\varsigma^n(\cdot)$, and 
we call it as 'Merge'.

We use the semi-implicit gradient-descent algorithm to solve problem (\ref{e27}).
The gradient-descent equation of problem (\ref{e27}) is
\begin{equation}\label{e29}
\begin{aligned}
\frac{u^{n+1}-u^{n}}{t}=&-W^{*}S^*(SWu^{n}-g)-\beta_1(u^{n+1}-z^n)-\\
&\beta_2 W^*(Wu^{n}-\mathcal{F}^{-1}\varsigma^n),
\end{aligned}
\end{equation}
where $t$ is the time step size, $W^*$ and $S^*$ are the dual of $W$ and $S$, respectively. By equation (\ref{e15}), it's not hard to derive 
\begin{equation}\label{e30}
S^{*}(g(x))=\left\{\begin{array}{l}
g(x) ~~\text{if}~~x\in D_1\\
0    ~~~~~~\text{if}~~ x\in D\backslash D_1
\end{array}\right.,
\end{equation}
where  $g(x)\in R^{M_3N\times1}$ and $S^{*}(g(x))\in R^{MN\times1}$. As discussed in \cite{ISI:000385396100016}, using inverse Radon transform $\mathcal{R}^{-1}$ to replace $W^*$  may make the iteration (\ref{e29}) have a higher convergence rate (for 3D CT, $\mathcal{R}^{-1}$ can be replaced by the FDK \cite{ISI:A1994NZ27700005} algorithm). Therefore, in this work, we also replace $W^*$ by  $\mathcal{R}^{-1}$ and  equation (\ref{e29}) is reduced to:
\begin{equation}\label{ee1}
\begin{aligned}
u^{n+1}=&\frac{1}{1+t\beta_1}u^n-\frac{t}{1+t\beta_1}\mathcal{R}^{-1}S^*(SWu^{n}-g)\\
&-\frac{t\beta_2}{1+t\beta_1}\mathcal{R}^{-1}(Wu^{n}-\mathcal{F}^{-1}\varsigma^n)+\frac{t\beta_1}{1+t\beta_1}z^n
\end{aligned}
\end{equation}

Let $\varsigma^n\in R^{batch\times M\times N\times1}$, $z^n\in R^{batch\times M_1\times M_2\times1}$ and $u^n\in R^{batch\times M_1\times M_2\times1}$ be the inputs to sub-network $\mathop{Merge}_u^n(\cdot)$, where $u^n$ is the output of $\mathop{Merge}_u^{n-1}(\cdot)$ in  iteration $\mathop{n-1}$, and $\varsigma^n$ and $z^n$ are the outputs of $Res_\varsigma^n(\cdot)$  and $Res_z^n(\cdot)$ in  iteration $n$, respectively, and $t_1=\frac{1}{1+t\beta_1}$, $t_2=-\frac{t}{1+t\beta_1}$, $t_3=-\frac{t\beta_2}{1+t\beta_1}$ and $t_4=\frac{t\beta_1}{1+t\beta_1}$ be four parameters to be learned in $\mathop{Merge}_u^n(\cdot)$. Then, according to equation (\ref{ee1}), the forward propagation formula of sub-network $\mathop{Merge}_u^n(\cdot)$ can be represented as:
\begin{equation}\label{e31}
\begin{aligned}
u^{n+1}=&t_1u^{n}+t_2\mathcal{R}^{-1}S^*(SWu^{n}-g)+\\
&t_3\mathcal{R}^{-1}(Wu^{n}-\mathcal{F}^{-1}\varsigma^n)+t_4z^n.
\end{aligned}
\end{equation}

\subsection {Interpretability and Consistency}
In the subsection, we discuss the interpretability and consistency of our network.
	\subsubsection{Interpretability} Our network iteratively approximates the solution of problem (\ref{e17}). In each iteration, the subnetwork $Res_\varsigma^n(\cdot)$ automatically extracts the information of sinograms in frequency domain  to complete the limited-angle sinograms, the subnetwork  $Res_z^n(\cdot)$ extracts local information of CT images in spatial domain to
	 refine the reconstructed  CT images  and the merge layer $\mathop{Merge}_u^n(\cdot)$ merges the outputs of $Res_\varsigma^n(\cdot)$ and $Res_z^n(\cdot)$. The inputs of $Res_\varsigma^n(\cdot)$ and $Res_z^n(\cdot)$ are the output of $\mathop{Merge}_u^{n-1}(\cdot)$ in the former iteration, which is closer to the true solution than the CT image reconstructed from the measured sinogram. Therefore, Our network is superior to those none-iterative post-processing networks whose inputs are the CT images directly reconstructed  from the sinograms. 
	\subsubsection{Consistency}  For an inverse problem, its solution is desired  to
	be consistent	with its  measurements. For most post-processing CT reconstruction networks \cite{ISI:000417913600012}\cite{ISI:000405701500004}\cite{ISI:000434302700011}, this requirement is not satisfied. To address this issue, Gupta et al. \cite{ISI:000434302700014} trained a CNN network to replace the projector in a projected gradient, where the CNN is used to  recursively project the solution closer to the desired reconstruction  images. 
	
	From model (\ref{e14}) and equation (\ref{e31}), we can see that our network is partially consistent with the measured sinograms $g$. The last term $-t_3\mathcal{R}^{-1}S^*(SWu^{n}-g)$ in equation (\ref{e31}) is responsible for correcting the reconstructed CT image $u^n$ such that the limited-angle sinogram $SWu^{n}$ does not deviate from the measured sinogram $g$ too much.

\subsection{Loss Function}
In our network, the parameters need to be learned are $\Theta=(\Theta_\varsigma^1\cup...\cup\Theta_\varsigma^{N_{iter}})\cup(\Theta^n_1\cup...\cup\Theta^{N_{iter}}_z)\cup\Theta_u$, where
\begin{align*}
&\Theta_\varsigma^n=\{F_{\varsigma,1}^n, F_{\varsigma,2}^n, F_{\varsigma,3}^n, B_{\varsigma,1}^n, B_{\varsigma,2}^n, B_{\varsigma,3}^n, \tau_{\varsigma,2}^n, \tau_{\varsigma,3}^n, \kappa_{\varsigma,2}^n, \kappa_{\varsigma,3}^n\} \\
&\text{in}~~ Res_\varsigma^n(\cdot) ~~\text{for}~~ n=1,2,...,N_{iter},\\
 &\Theta_z^n=\{F_{z,1}^n, F_{z,2}^n, F_{z,3}^n, B_{z,1}^n, B_{z,2}^n, B_{z,3}^n, \tau_{z,2}^n, \tau_{z,3}^n, \kappa_{z,2}^n, \kappa_{z,3}^n\} \\
&\text{in}~~ Res_z^n(\cdot)~~ \text{for}~~ n=1,2,...,N_{iter},\\
& \Theta_u=\{t_1, t_2, t_3\} ~~\text{in}~~ {\mathop{Merge}}_u^n(\cdot). 
\end{align*}

Note that the parameters $\Theta_u$ in different iteration blocks are the same, while the parameters $\Theta_\varsigma^n$ and $\Theta_z^n$ vary with $n$. We can also set the parameters $\Theta_\varsigma^n$ and $\Theta_z^n$ to be the same for all $n$ and so the forward propagation of the network is more like the iterative algorithm of problem (\ref{e17}). However, if setting them to be the same,  the capacity of the whole network will be reduced to about $\frac{1}{N_{iter}}$ of the original. To make  the  network capable of fitting all the training samples, the number of layers of $Res_{z}^n(\cdot)$ and $Res_{\varsigma}^n(\cdot)$ must be increased.  Therefore, to save the GPU memory, we set the parameters $\Theta_\varsigma^n$ and $\Theta_\varsigma^n$ to be different in each iteration block $n$.

Our network has three outputs, $u^{N_{iter}}$, $z^{N_{iter}}$ and $\varsigma^{N_{iter}}$, where $u^{N_{iter}}$ and  $z^{N_{iter}}$ correspond to the CT images, and $\varsigma^{N_{iter}}$ corresponds to the measured sinogram. We train our network by minimizing the following loss function:
\begin{equation}\label{e32}
\arg\min\limits_{\Theta}\frac{1}{N_{s}}\sum_{i=1}^{N_{s}}\{\lambda\|z_i^{N_{iter}}-z_i^{label}\|_2^2+(1-\lambda)\|\varsigma_i^{N_{iter}}-\varsigma_i^{label}\|_2^2\},
\end{equation}
where $z_i^{label}$ is the  sinogram label and $\varsigma_i^{label}$ is the  CT image label, $N_s$ is the number of training samples and $\lambda$ is a balancing parameter. In our work, if not specifically given, we set $\lambda=0.5$.

This minimization problem (\ref{e32}) can be solved by various algorithms;
in this work, we adopt the Adam 
%[ADAM: A METHOD FOR STOCHASTIC OPTIMIZATION] 
algorithm with the learning rate $lr=0.001$ to train our network.  

\section{Experimental Results}
In this section, we give some simulated experimental results and compare them
with those of the FBP algorithm, TV regularization algorithm, Red-CNN \cite{ISI:000417913600012}, FBP-Conv
\cite{ISI:000405701500004} and DD-Net \cite{ISI:000434302700011}. The codes for implementing our networks
can be downloaded from https://github.com/wangwei-cmd/limited-angle-CT-reconstruction.

For the compared TV regularization algorithm, we use the  alternating direction method of multipliers (ADMM) \cite{INSPEC:12637129}  to solve the following model:
\begin{equation}\label{e33}
u=\mathop{\arg\min}\limits_{u\in BV({D_1})}\lambda_3|\nabla u|_{1}+\int_{D_1}(W_1u-g)^2,
\end{equation}
where $D_1$ is the limited-angle definition domain of the sinograms $g$, $BV(D_1)$ is the space  of functions of bounded variation, $|\nabla u|_{1}$ is the total variation of $u$, $W_1$ is the system matrix of the limited-angle CT, and $\lambda_3>0$ is a balanced parameter.
let $v=\nabla u$ and $v^0=c^0=0$, then  problem (\ref{e33}) is equivalent to the following iterations
\begin{equation}\label{e34}
\begin{aligned}
u^{k+1}=&\mathop{\arg\min}\limits_{u}\int_{D_1}(W_1u-g)^2+\frac{\rho}{2}\int_{D_1}(v^{k}-\nabla u-c^{k})^2,\\
v^{k+1}=&\mathop{\arg\min}\limits_{v}\lambda{_3}|v|_{1}+\frac{\rho}{2}\int_{D_1}(v-\nabla u^{k+1}-c^{k})^2,\\
c^{k+1}=&c^{k}+\nabla u^{k+1}-v^{k+1}
\end{aligned}
\end{equation}
For the $u$-subproblem, we use the gradient descent method to solve:
\begin{equation}\label{e35}
u^{n+1}=u^{n}-t_5[\mathcal{R}^{-1}(W_1u^k-g)-\rho \mathop{div}(\nabla u^k-v^k+c^k)],
\end{equation}
where $t_5$ is the time step size.
The solution for $v$-subproblem is 
\begin{equation}\label{e36}
v^{k+1}=\mathop{sign}(\nabla u^{k+1}+c^k)*\max(|\nabla u^{k+1}+c^k|-\frac{\lambda_3}{\rho},0),
\end{equation} 

All the networks are trained by using the software Tensorflow 2 on a personal computer with a Ubuntu 18.04 operating system, an Intel Core i7-8650U central processing unit (CPU), 256GB random access memory (RAM) and a  Nvidia GTX Titan GPU card of 12000MB memory.  

%In our experiments, we only perform the 2D CT image reconstructions for  parallel-beam and fan-beam scanning geometries. In theory, our network can also reconstruct CT images from the  3D cone-beam geometry sinograms. However, to  implement and train our network for 3D cone-beam CT image reconstruction, a GPU card of larger memory ($>$12000MB) is needed. Therefore, in this paper, we only give the simulated experimental  results  for  2D CT image reconstructions.

In our experiments, we  perform the 2D CT image reconstructions for  parallel-beam and fan-beam scanning geometries and the 3D CT image reconstruction for circle  cone-beam scanning geometry.

\subsection{Parallel-beam}
\begin{figure*}[!t]
	\centerline{
		\includegraphics[width=0.28\columnwidth]{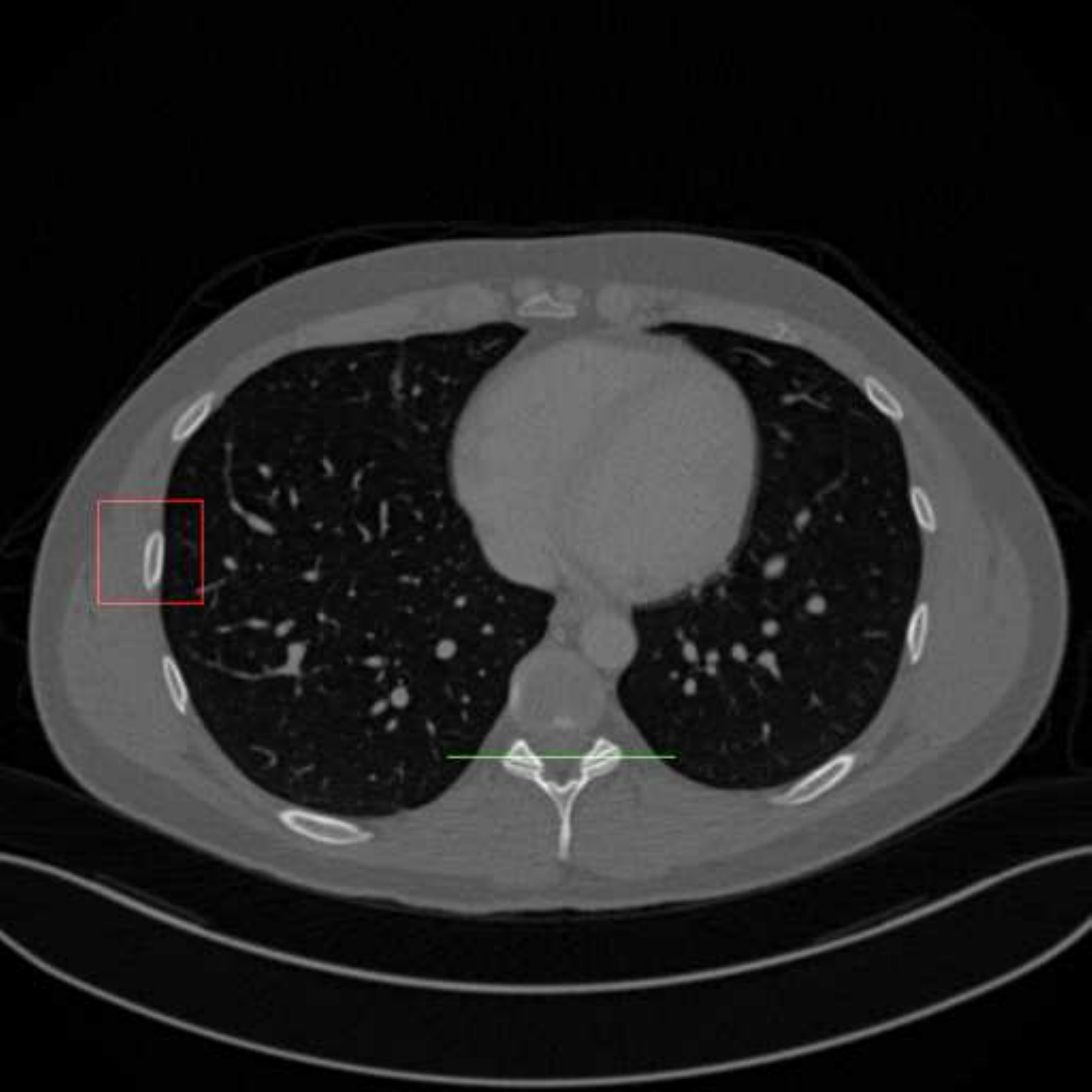}
		\includegraphics[width=0.28\columnwidth]{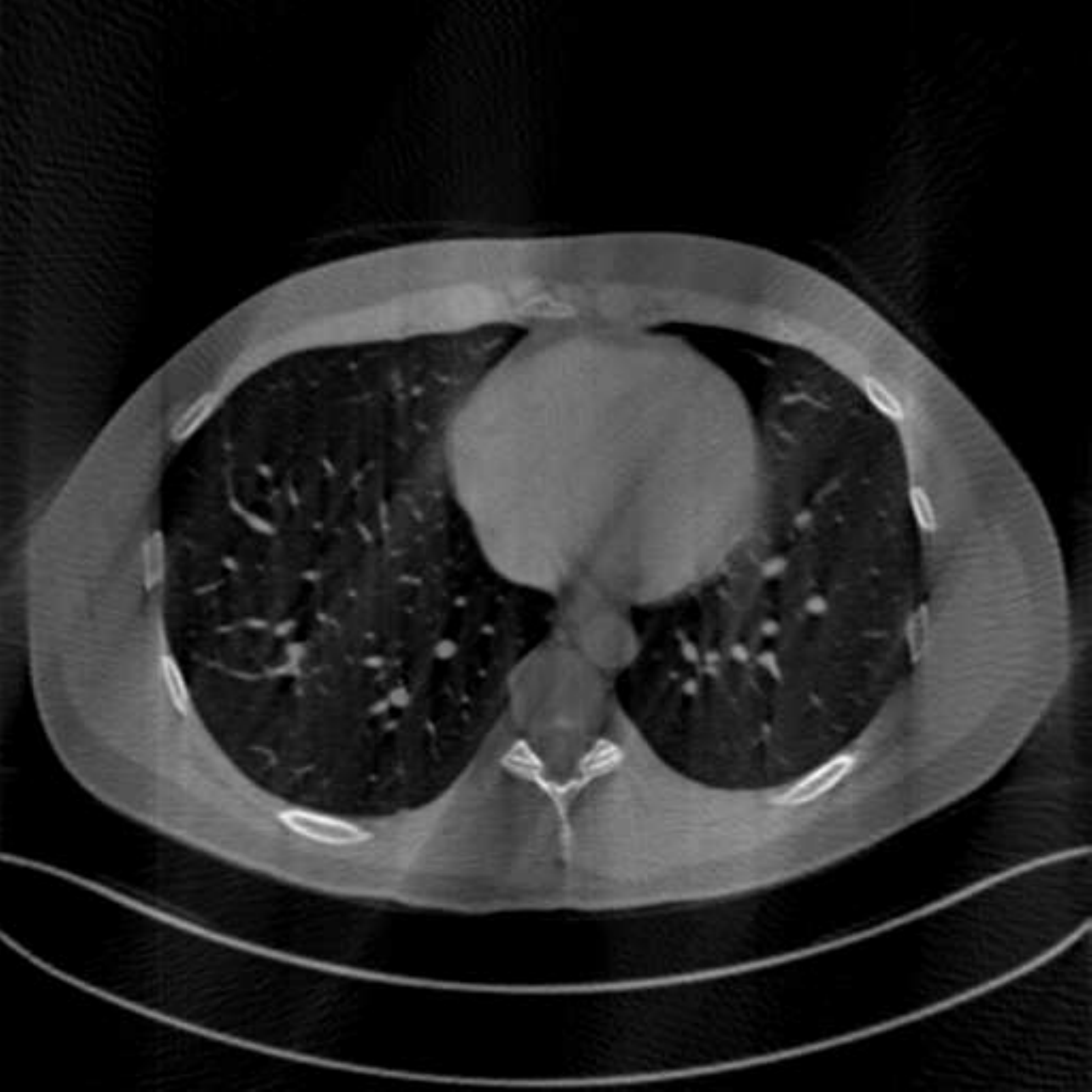}
		\includegraphics[width=0.28\columnwidth]{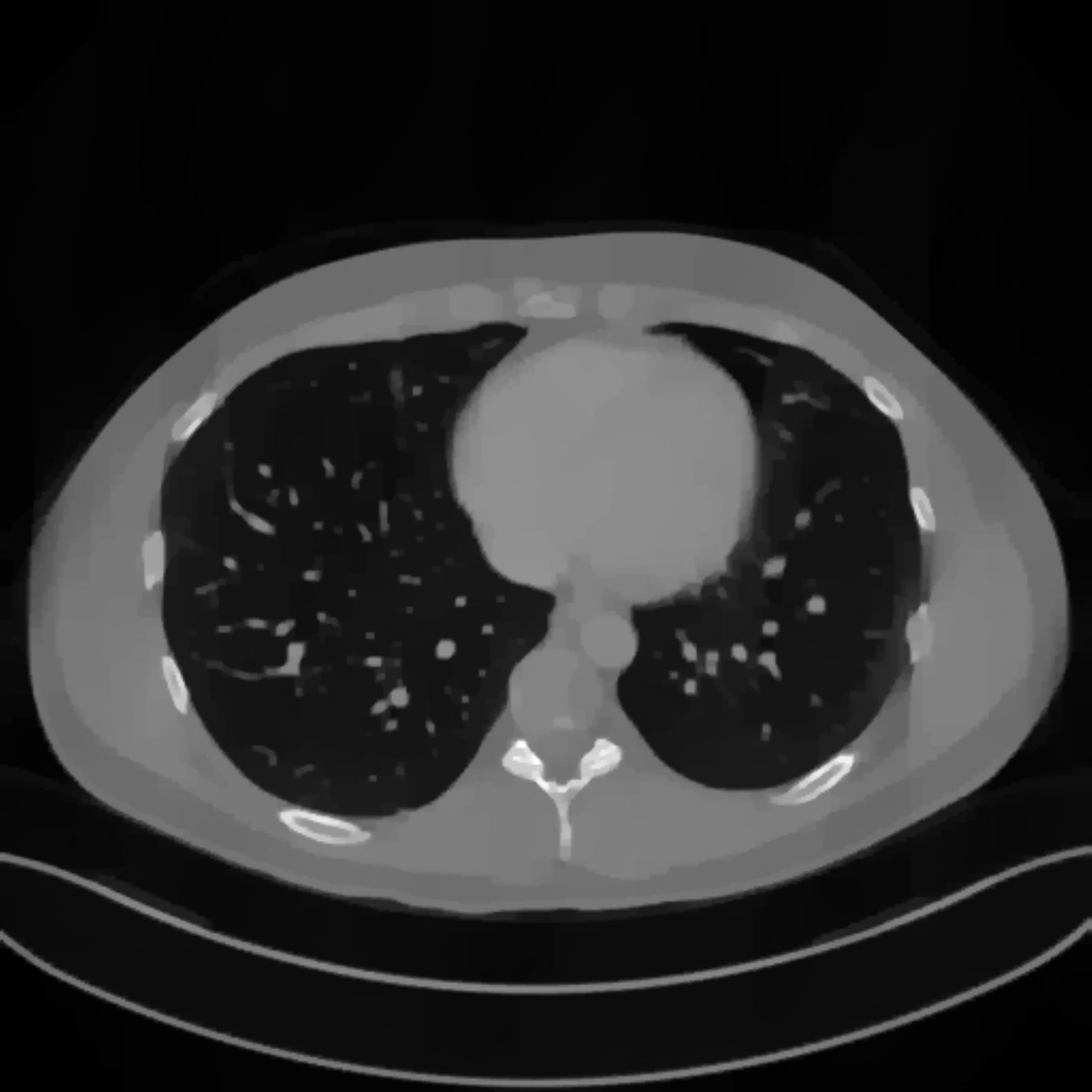}
		\includegraphics[width=0.28\columnwidth]{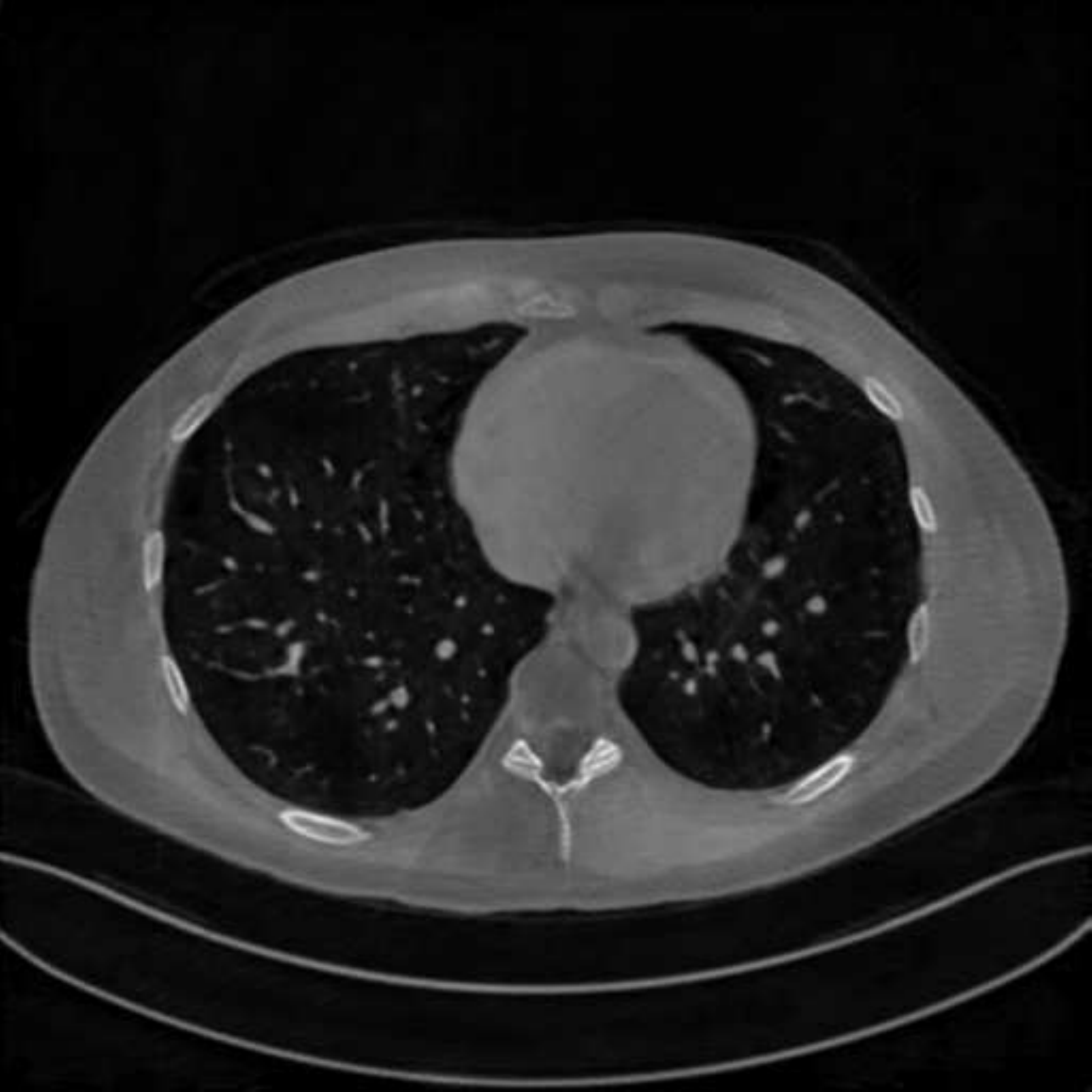}
		\includegraphics[width=0.28\columnwidth]{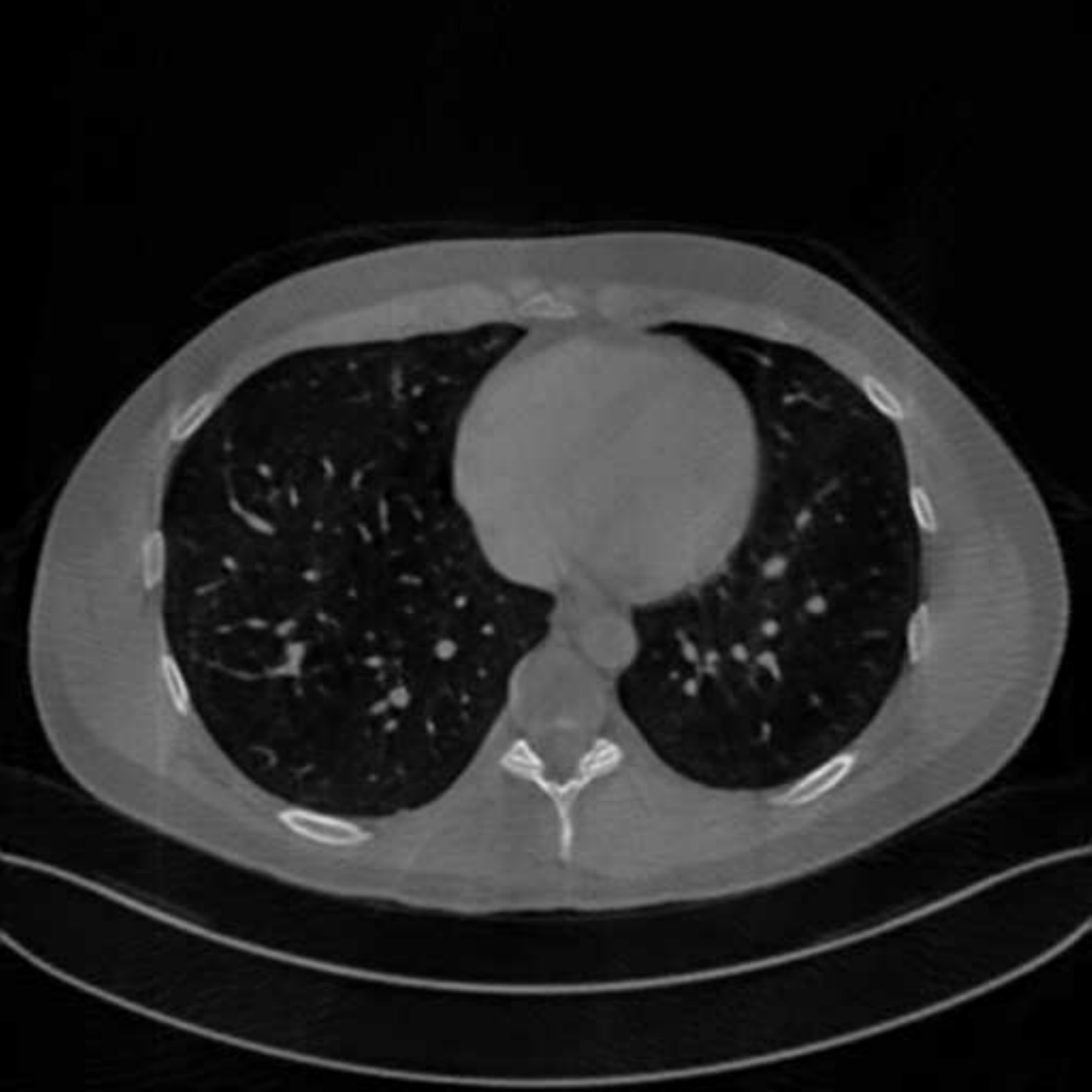}
		\includegraphics[width=0.28\columnwidth]{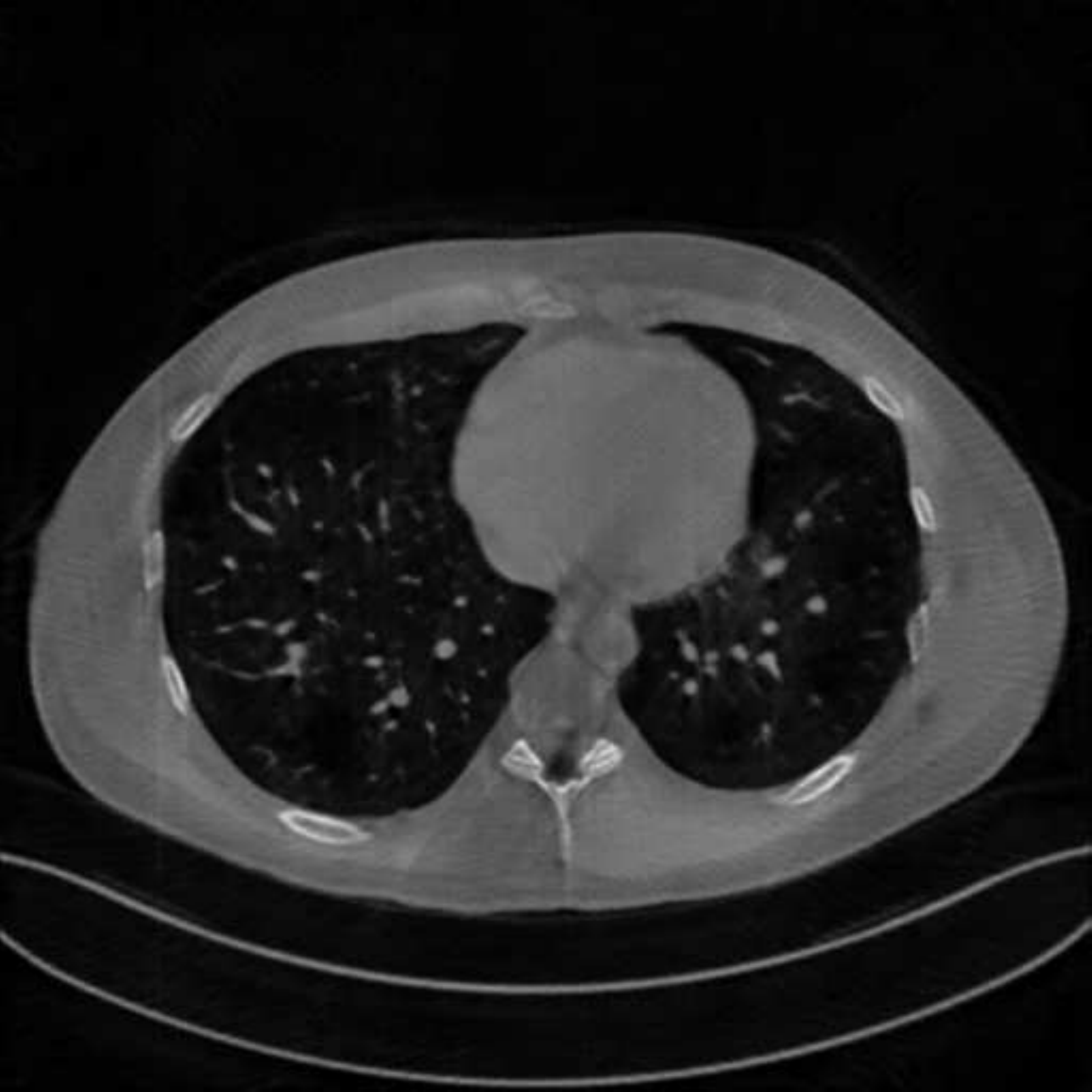}
		\includegraphics[width=0.28\columnwidth]{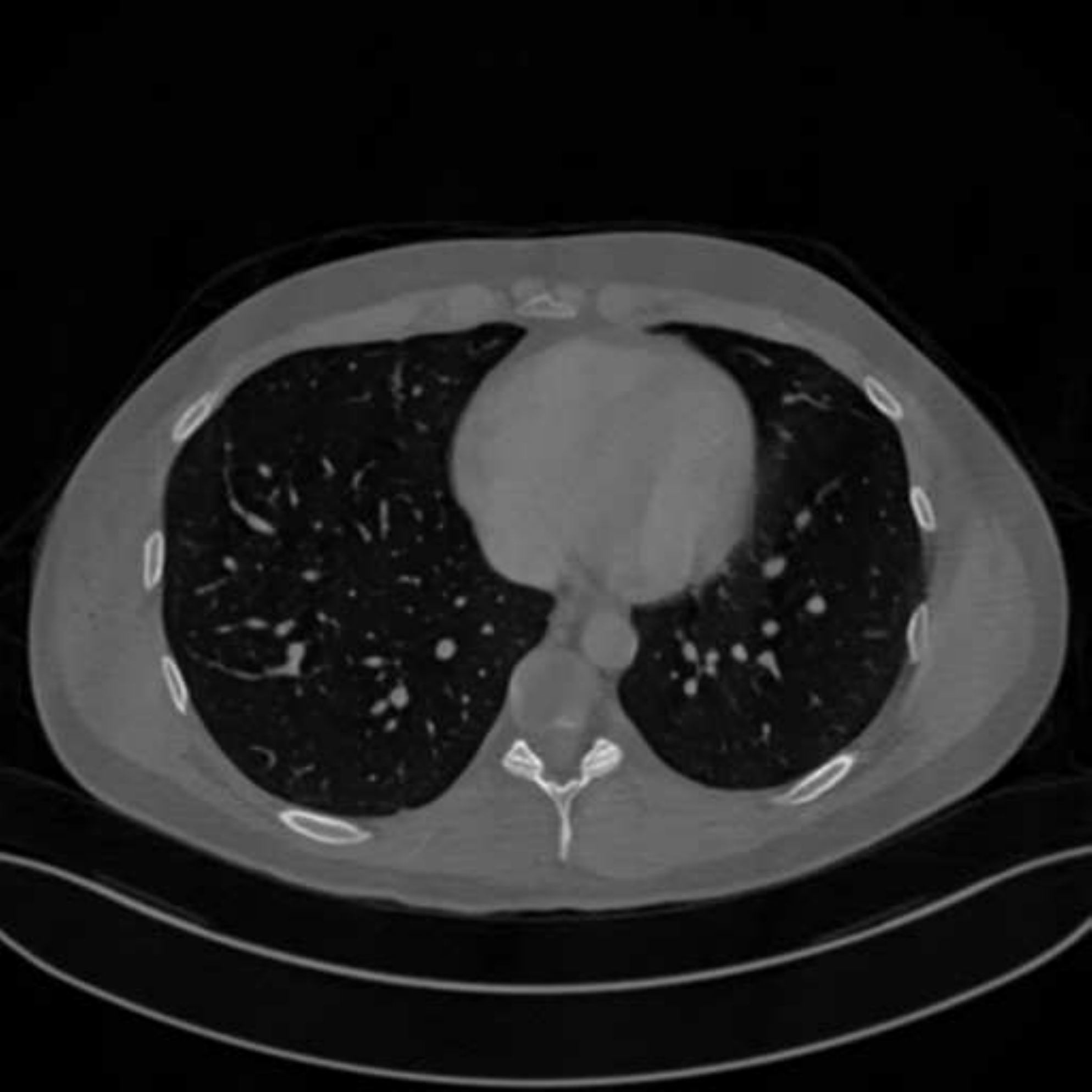}
	}
	\centerline{
		\includegraphics[width=0.28\columnwidth]{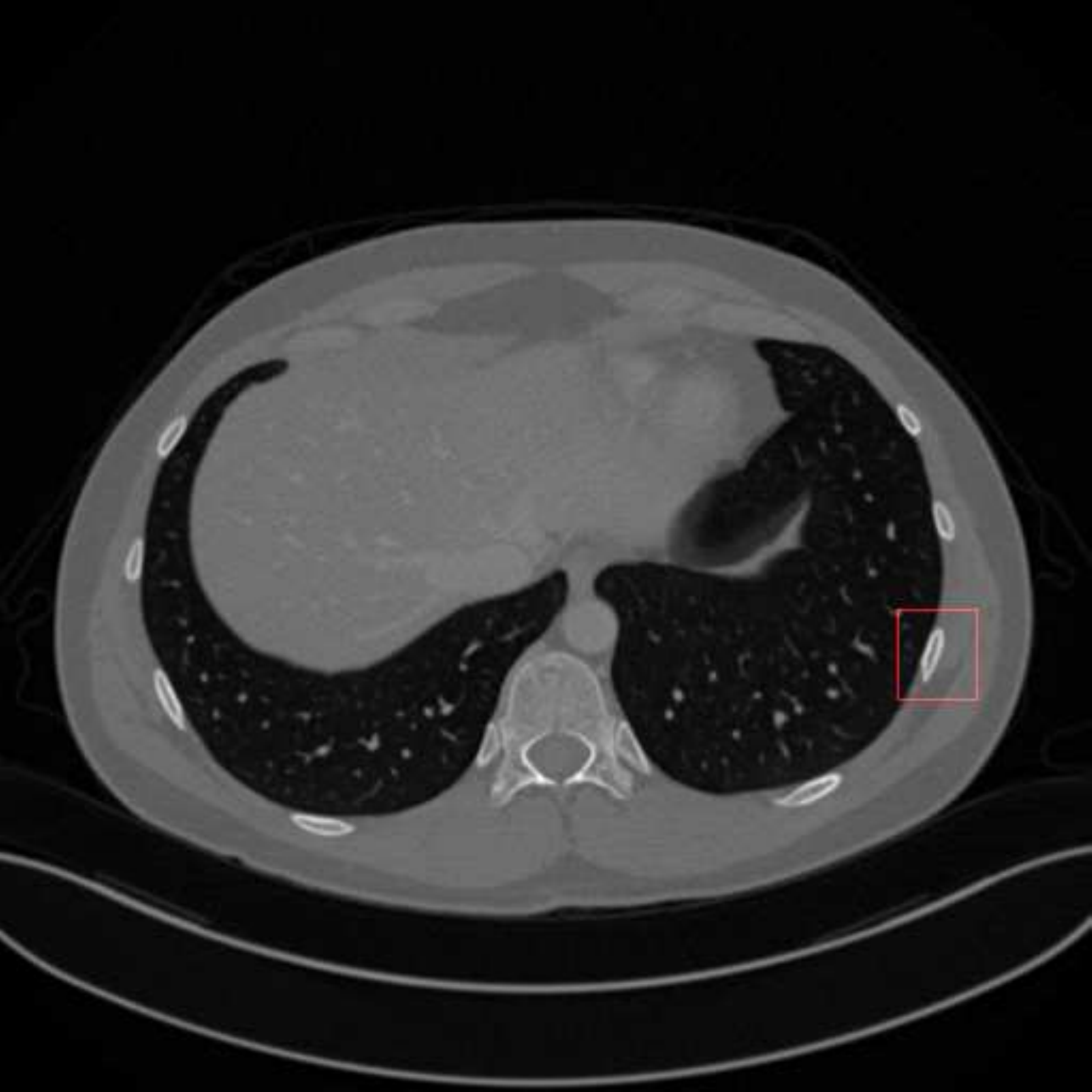}
		\includegraphics[width=0.28\columnwidth]{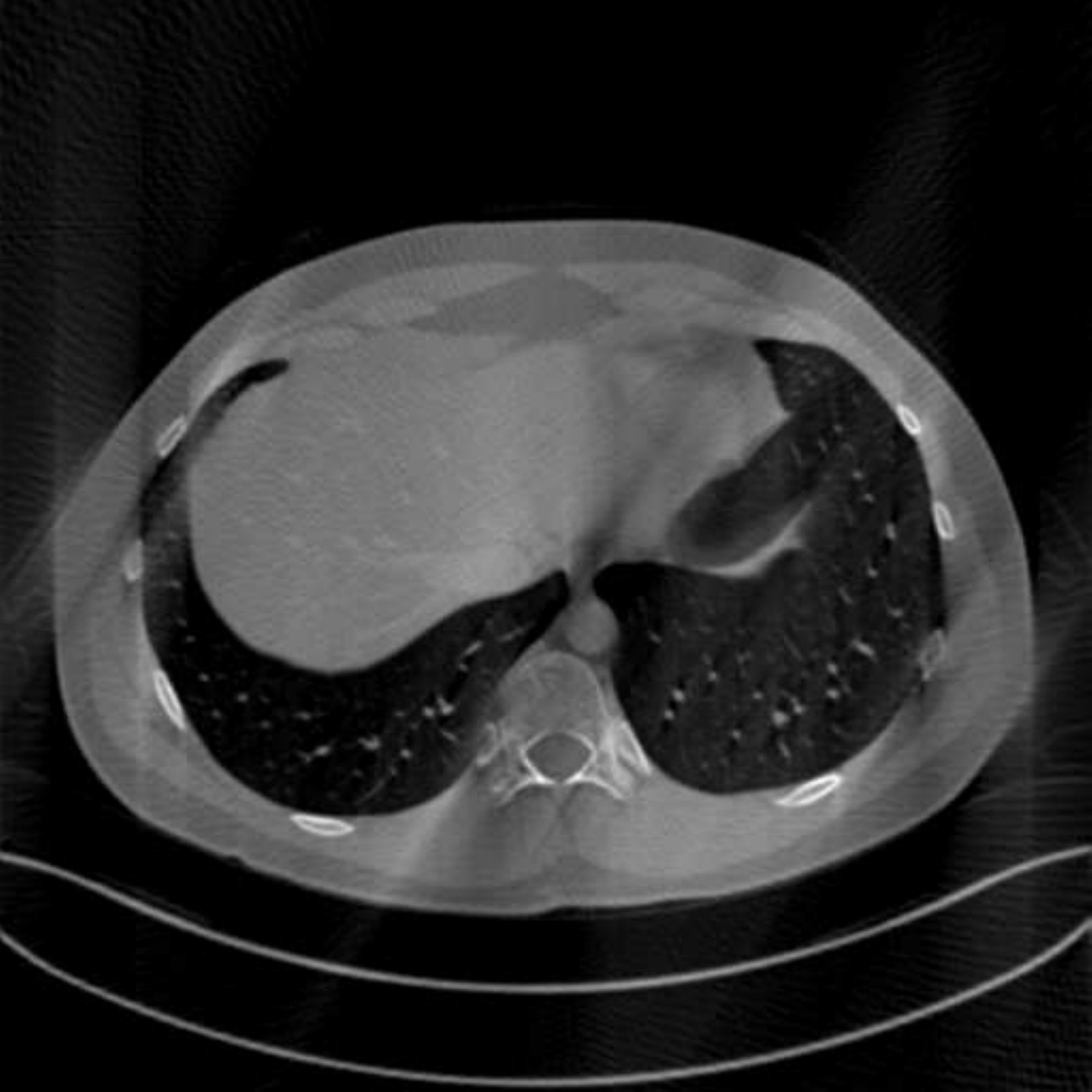}
		\includegraphics[width=0.28\columnwidth]{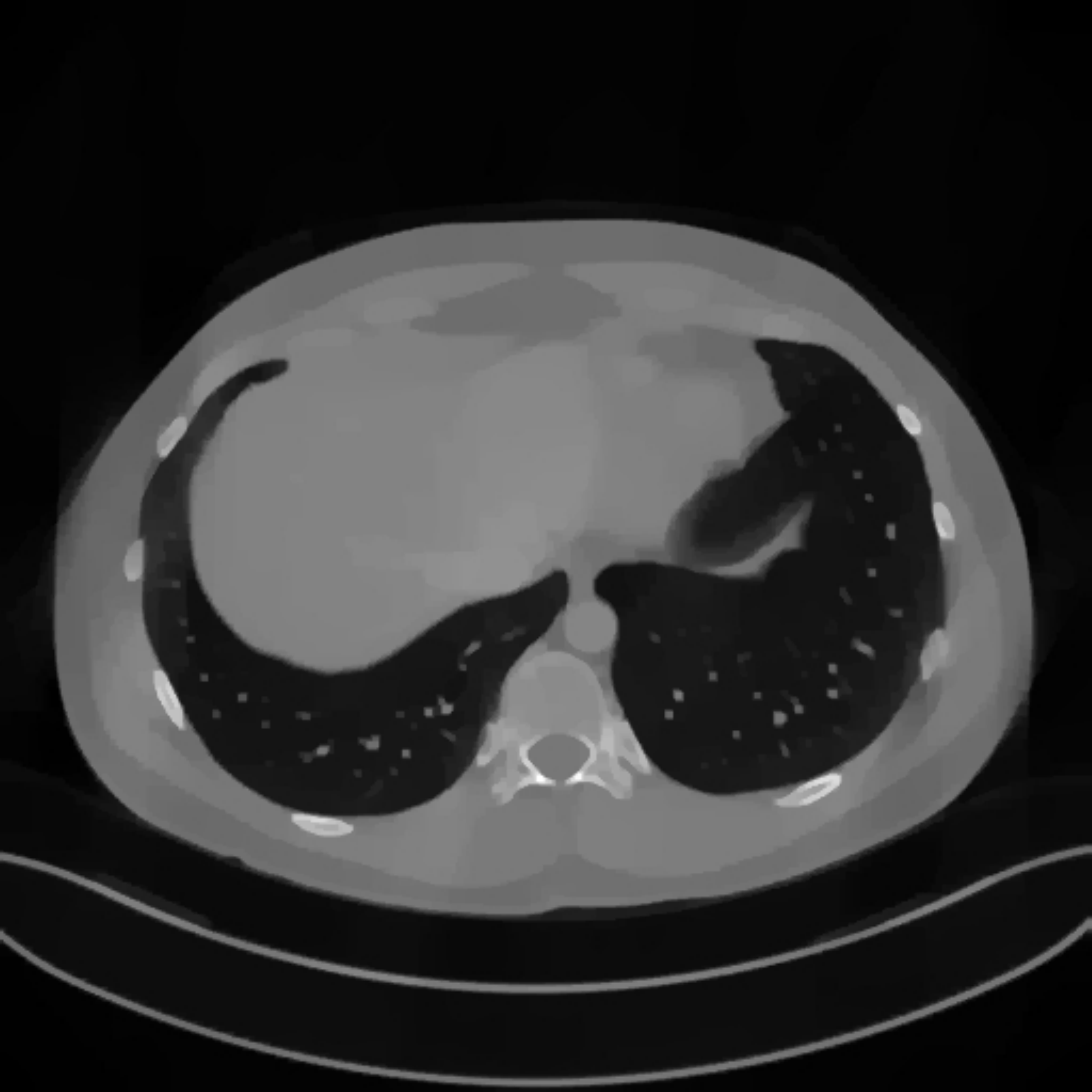}
		\includegraphics[width=0.28\columnwidth]{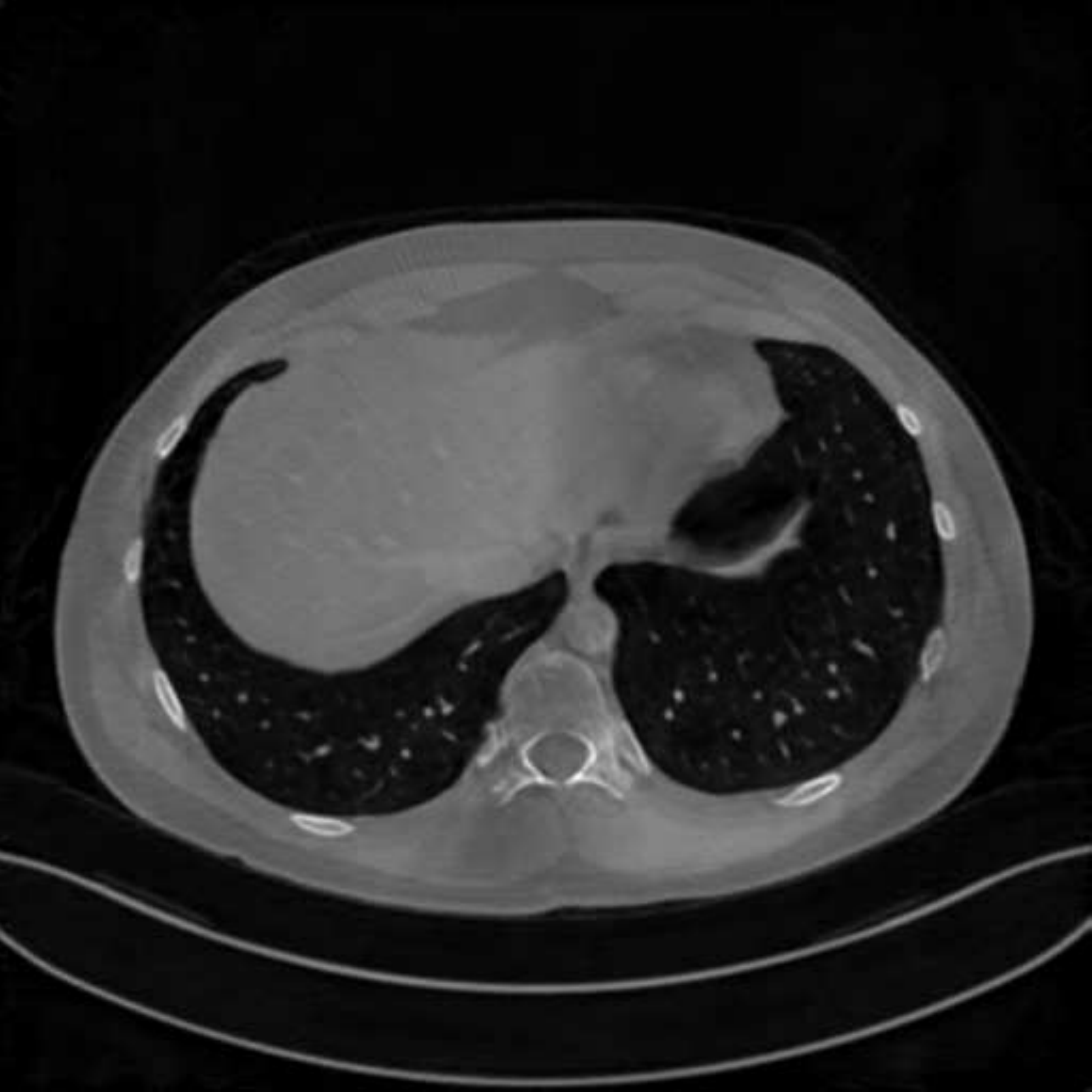}
		\includegraphics[width=0.28\columnwidth]{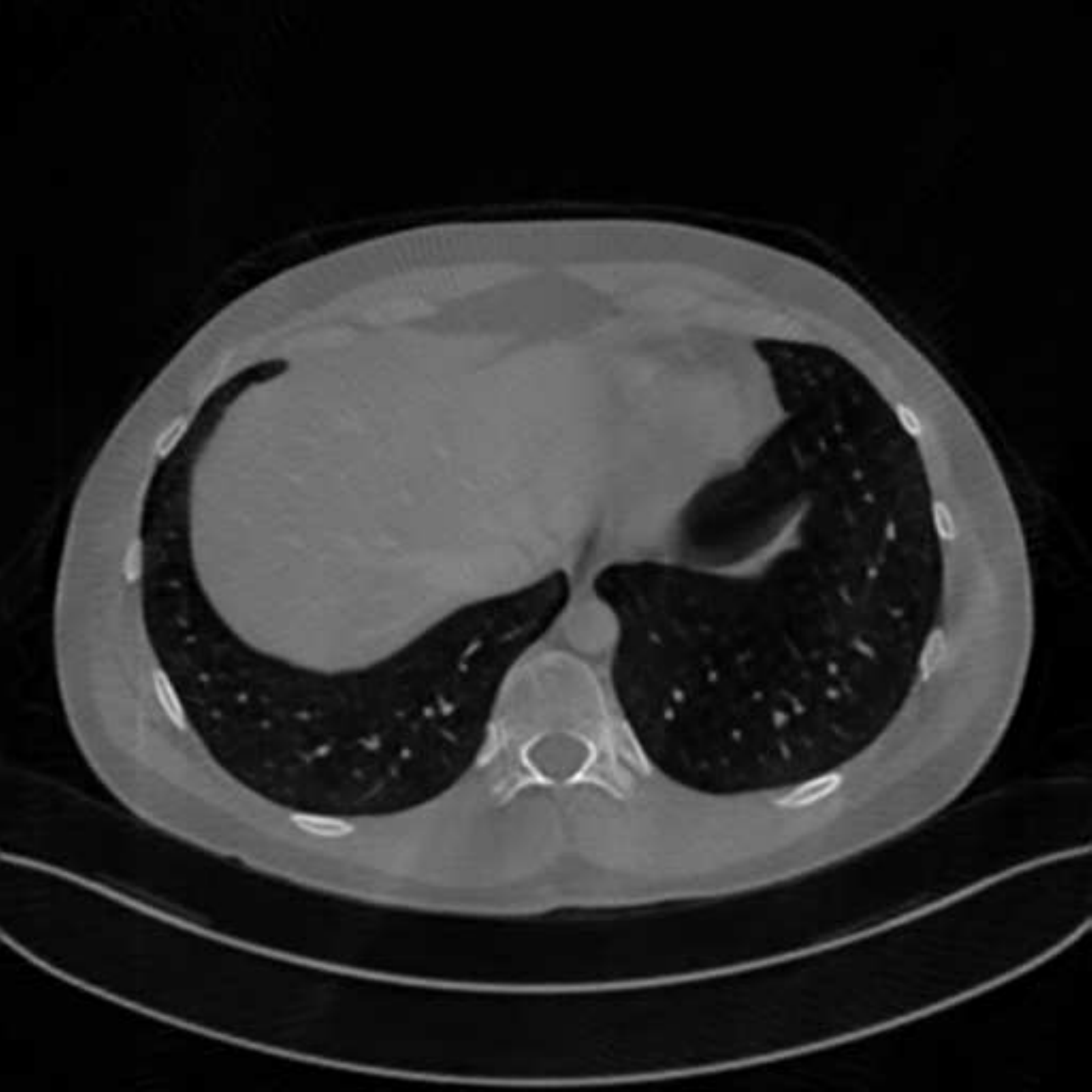}
		\includegraphics[width=0.28\columnwidth]{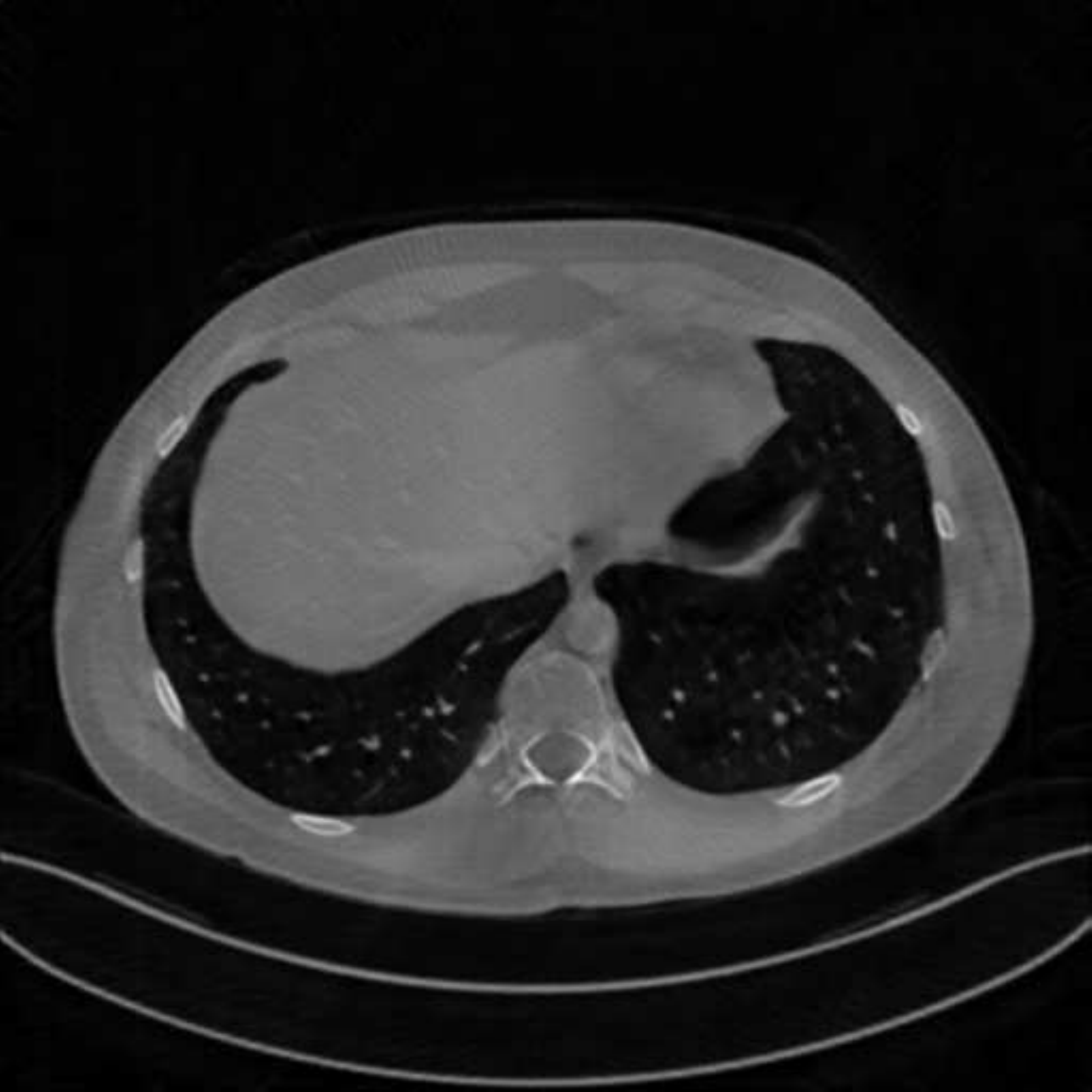}
		\includegraphics[width=0.28\columnwidth]{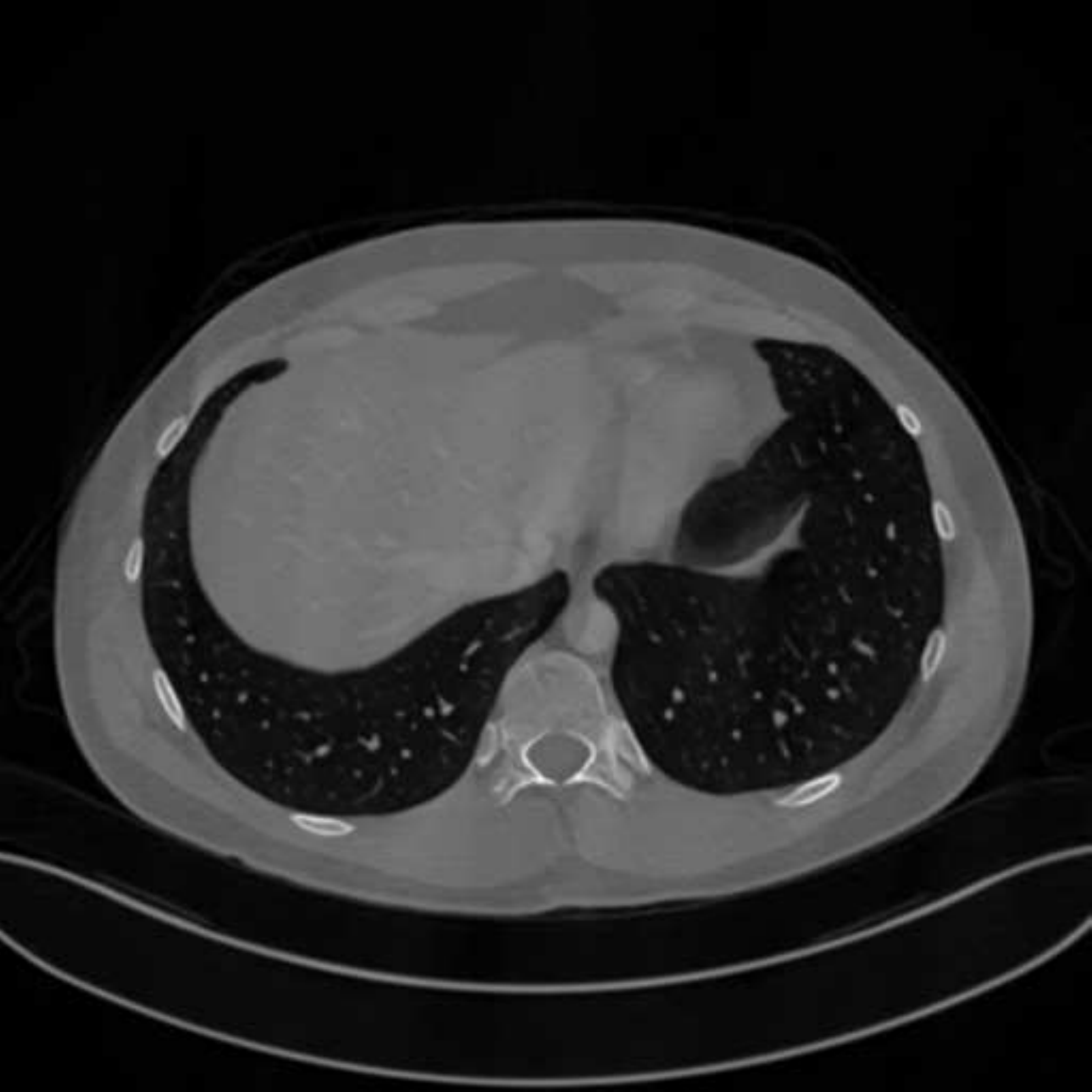}
	}
	\centerline{
		\includegraphics[width=0.28\columnwidth]{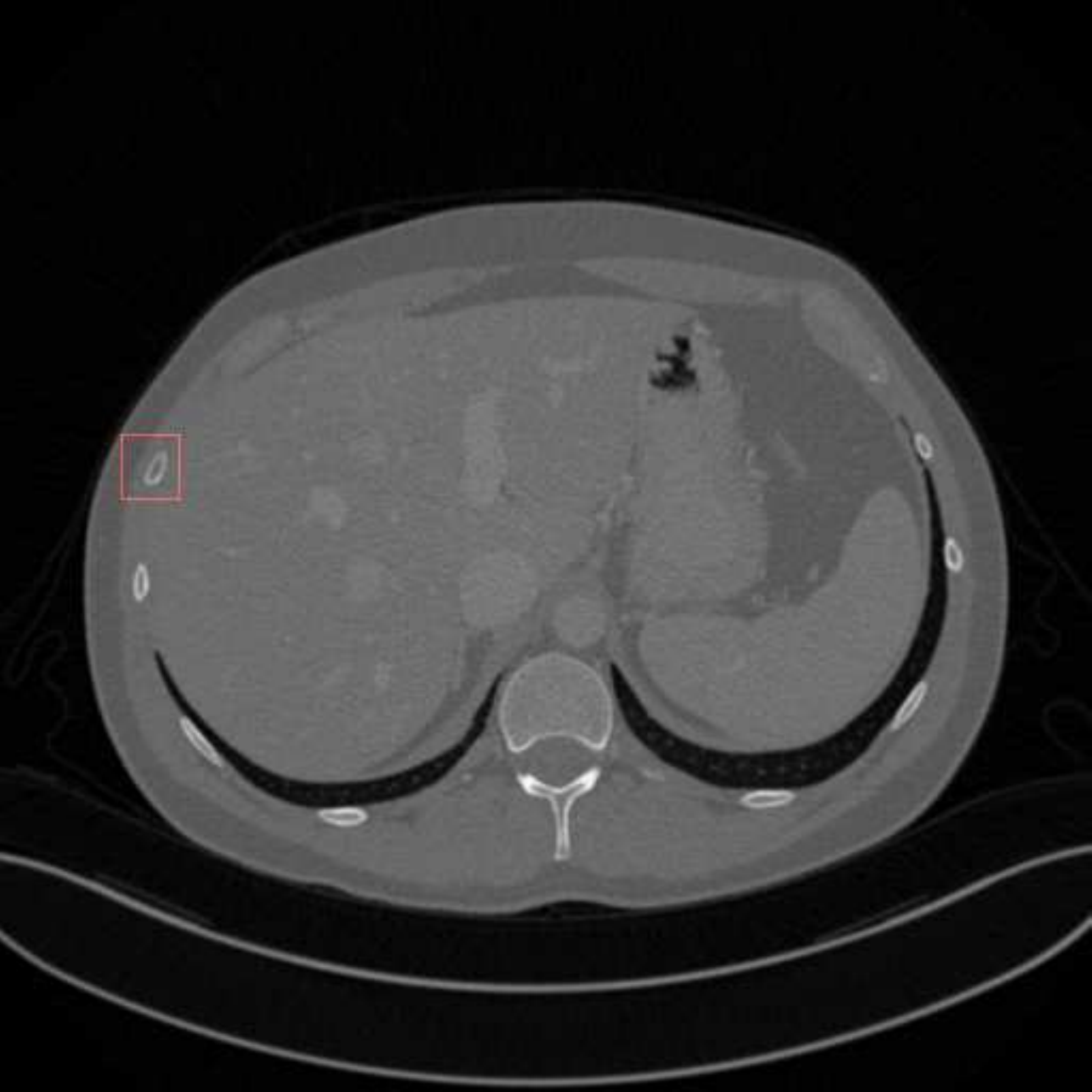}
		\includegraphics[width=0.28\columnwidth]{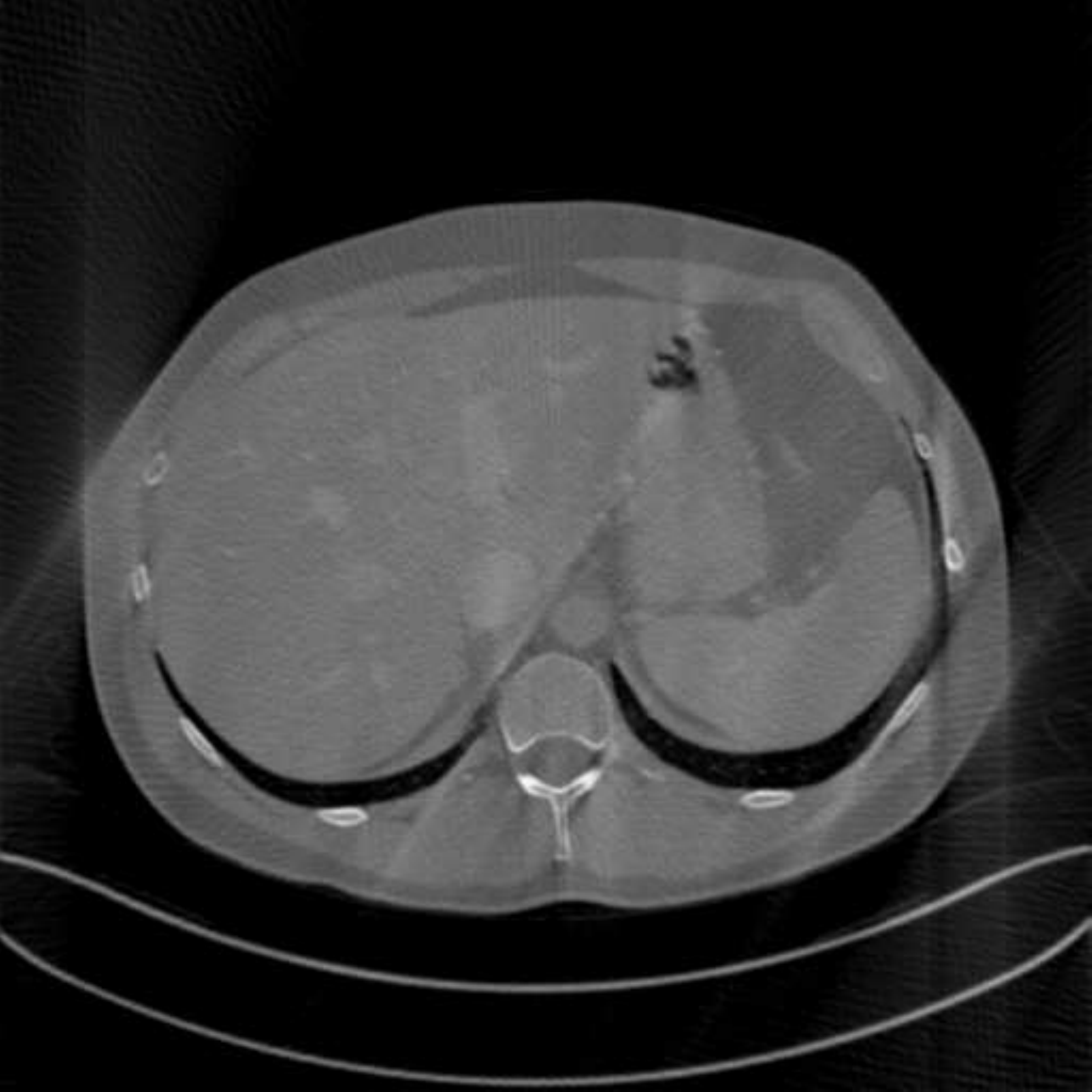}
		\includegraphics[width=0.28\columnwidth]{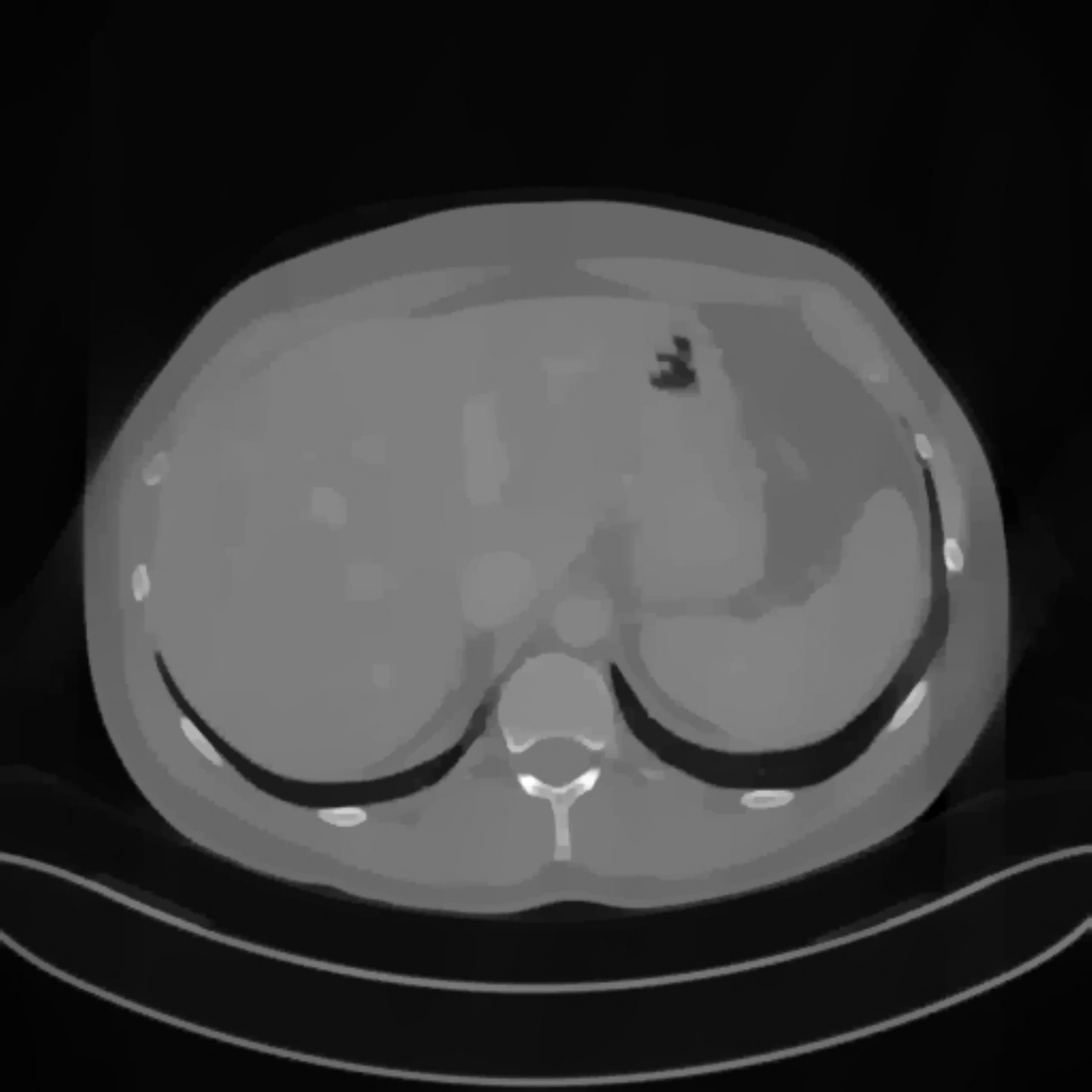}
		\includegraphics[width=0.28\columnwidth]{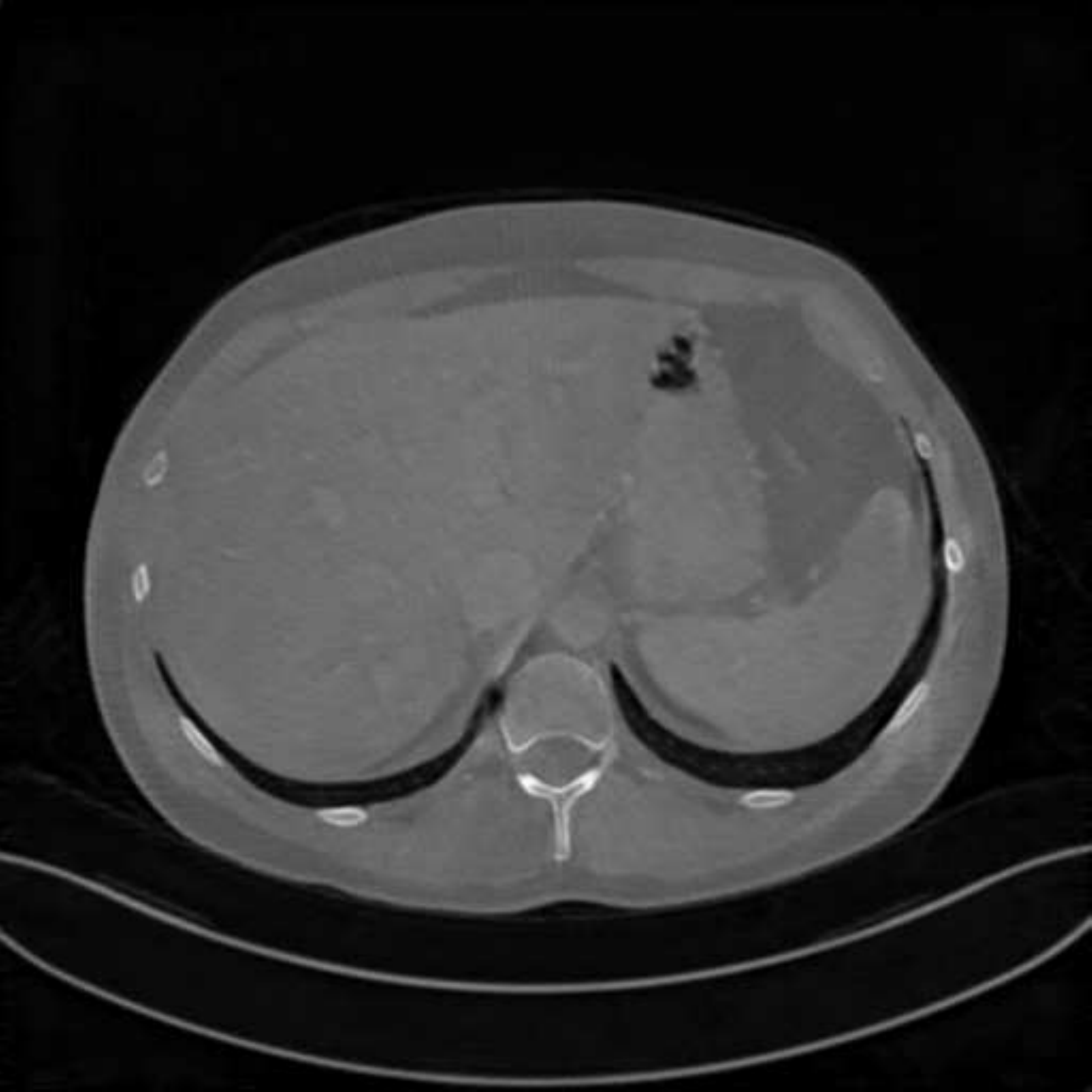}
		\includegraphics[width=0.28\columnwidth]{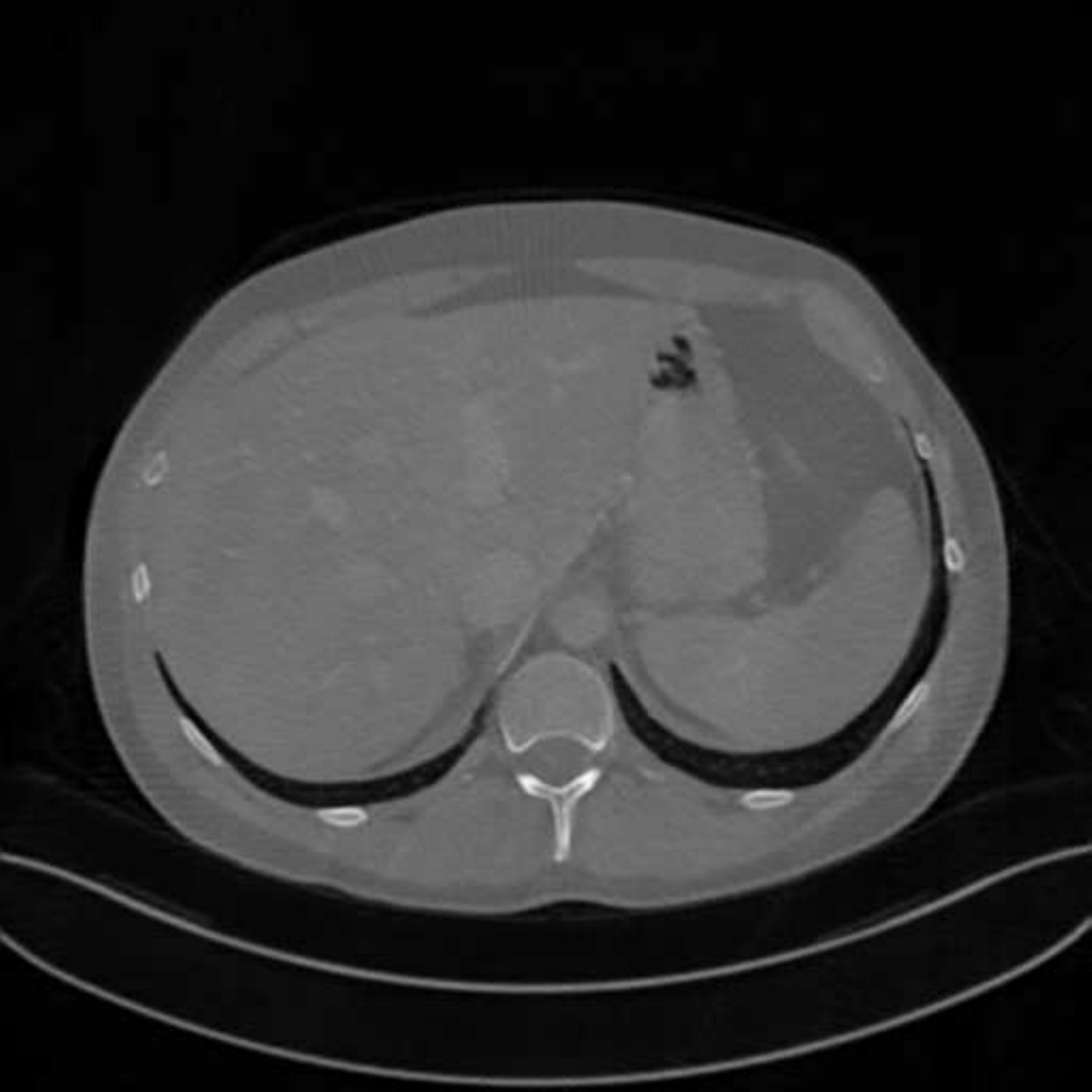}
		\includegraphics[width=0.28\columnwidth]{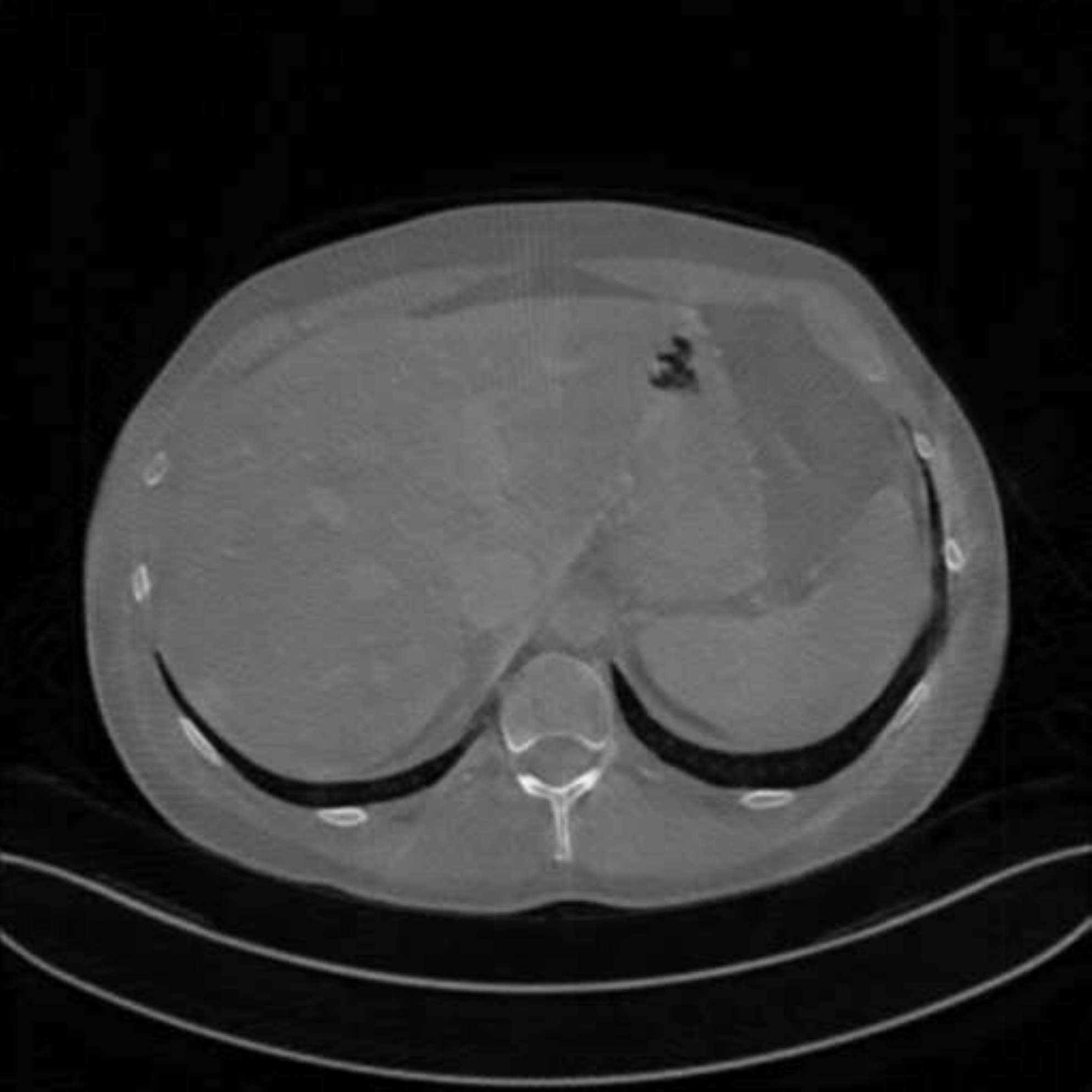}
		\includegraphics[width=0.28\columnwidth]{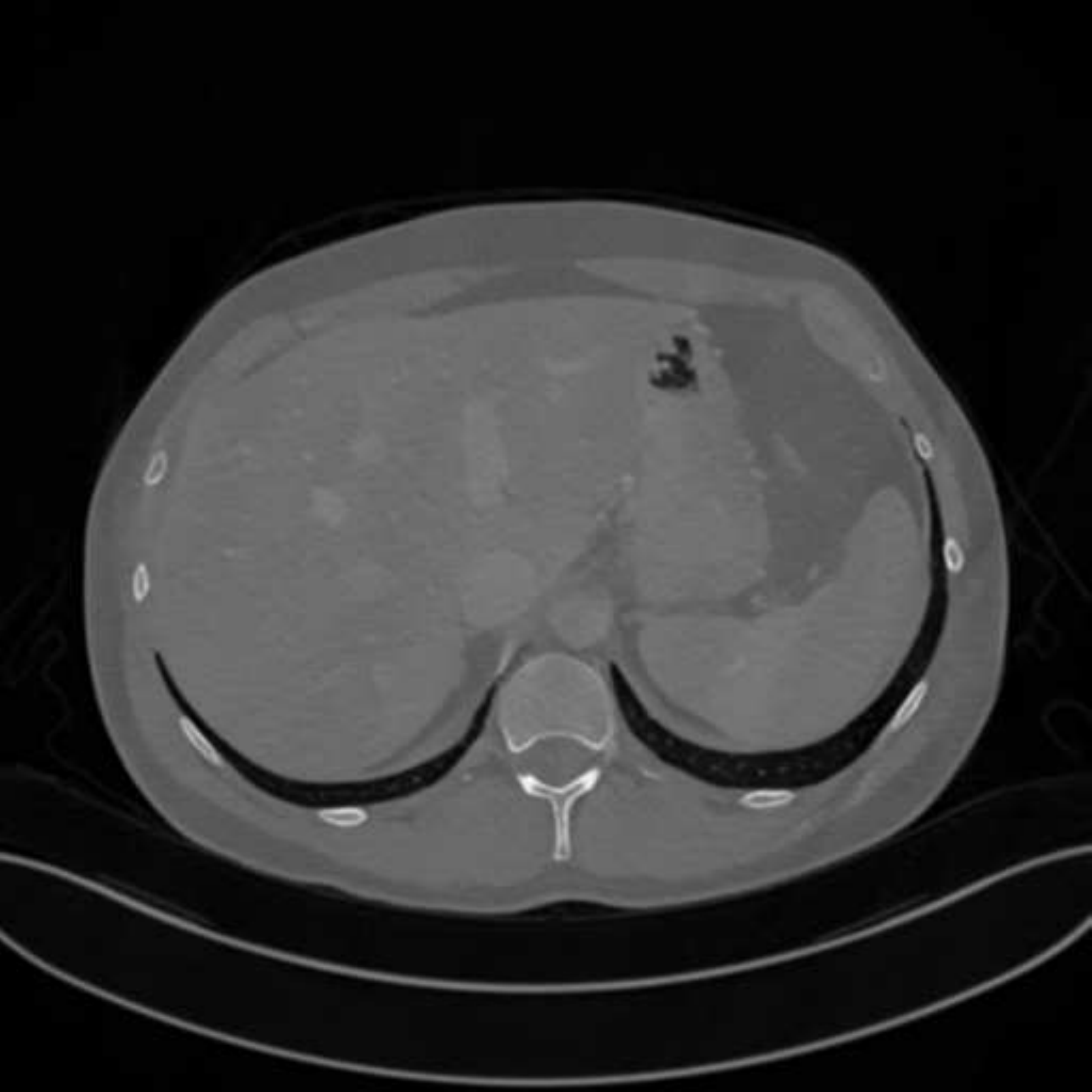}
	}
	\centerline{
		\includegraphics[width=0.28\columnwidth]{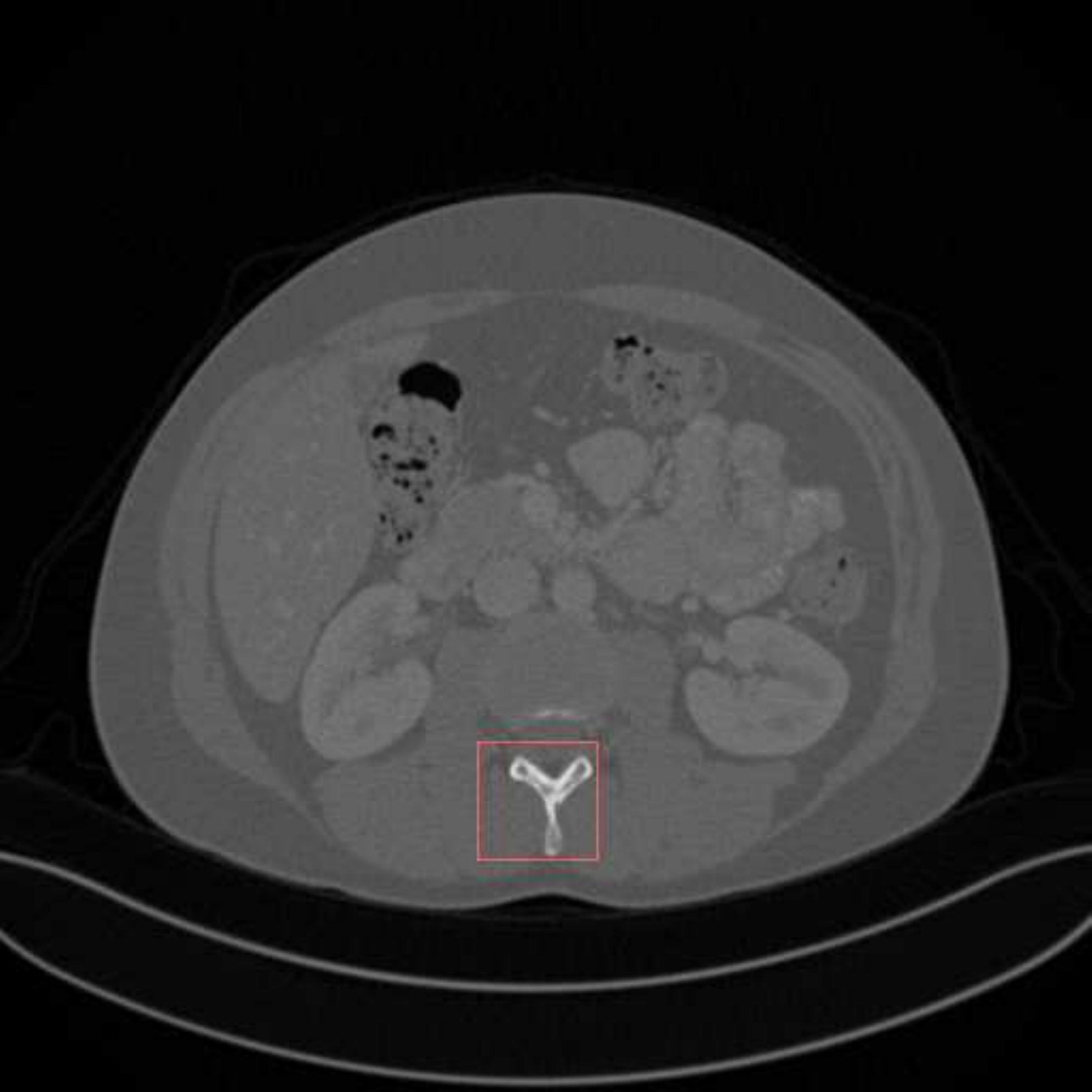}
		\includegraphics[width=0.28\columnwidth]{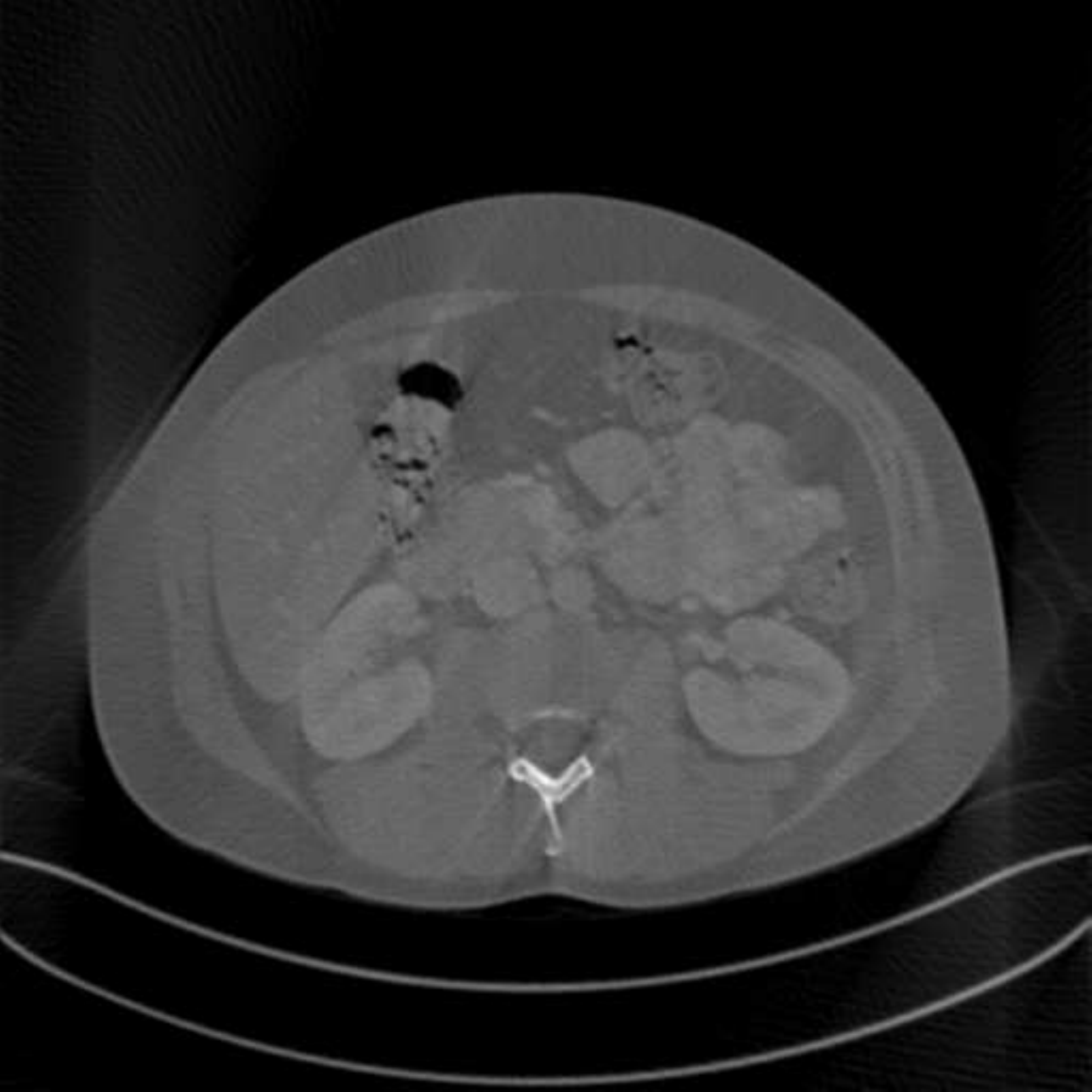}
		\includegraphics[width=0.28\columnwidth]{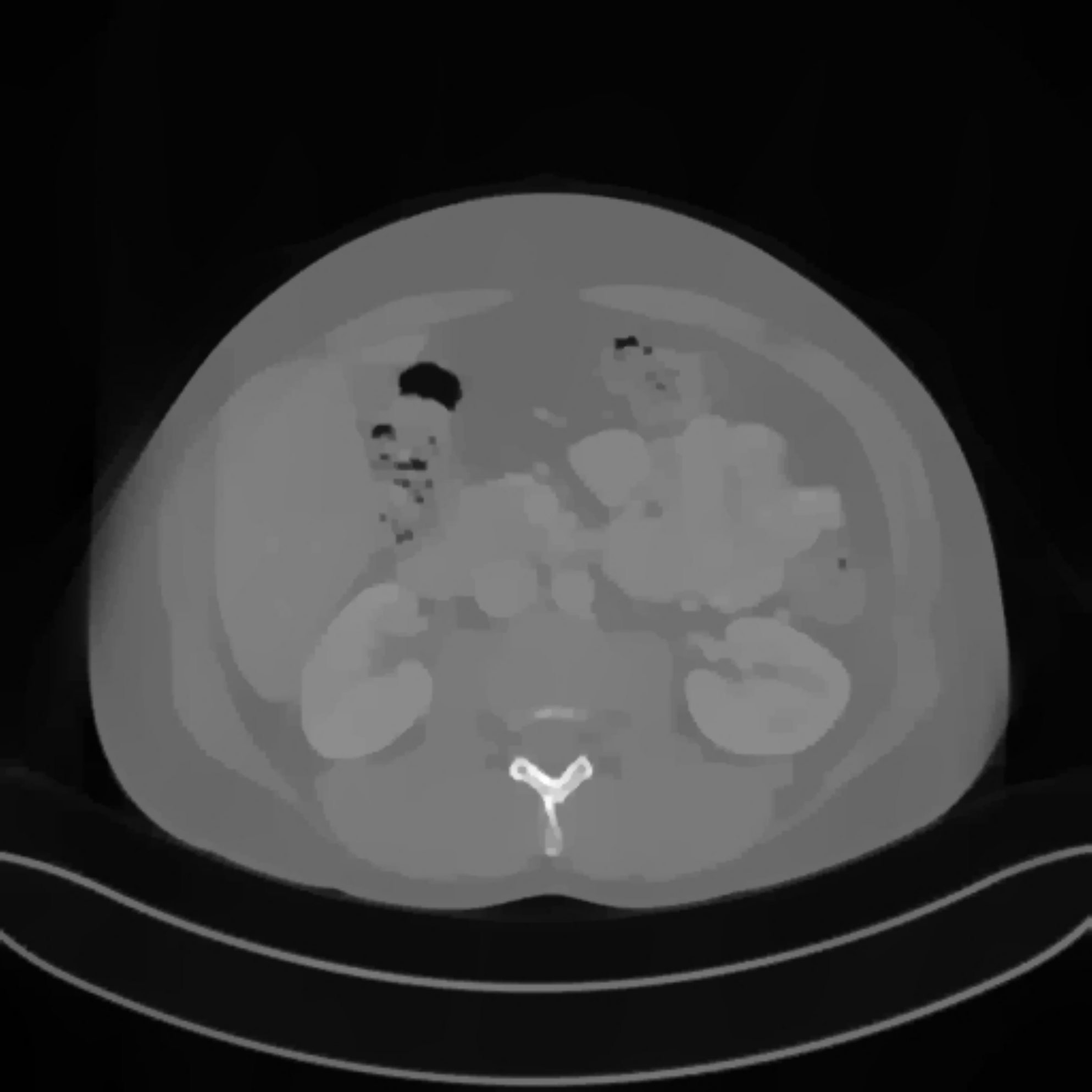}
		\includegraphics[width=0.28\columnwidth]{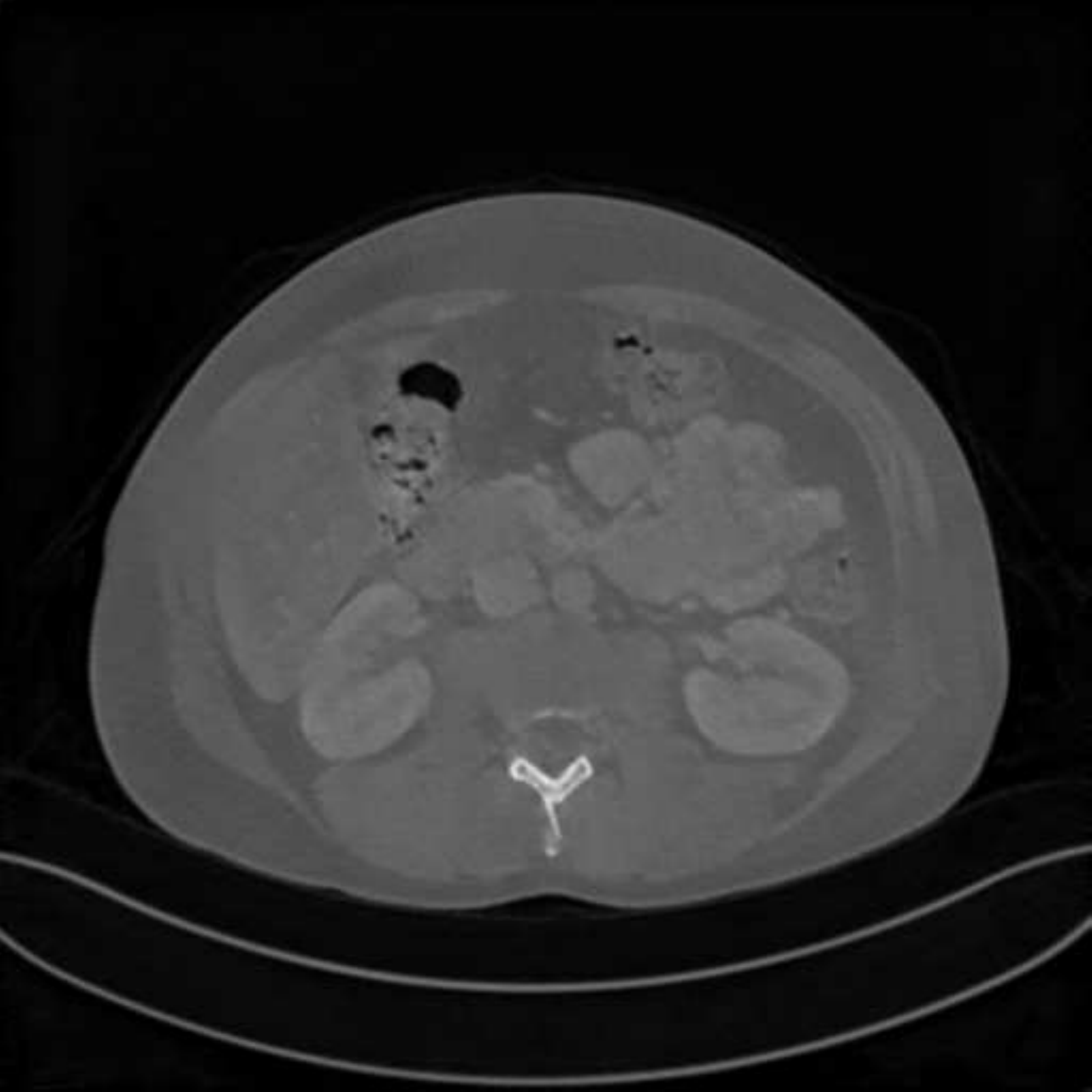}
		\includegraphics[width=0.28\columnwidth]{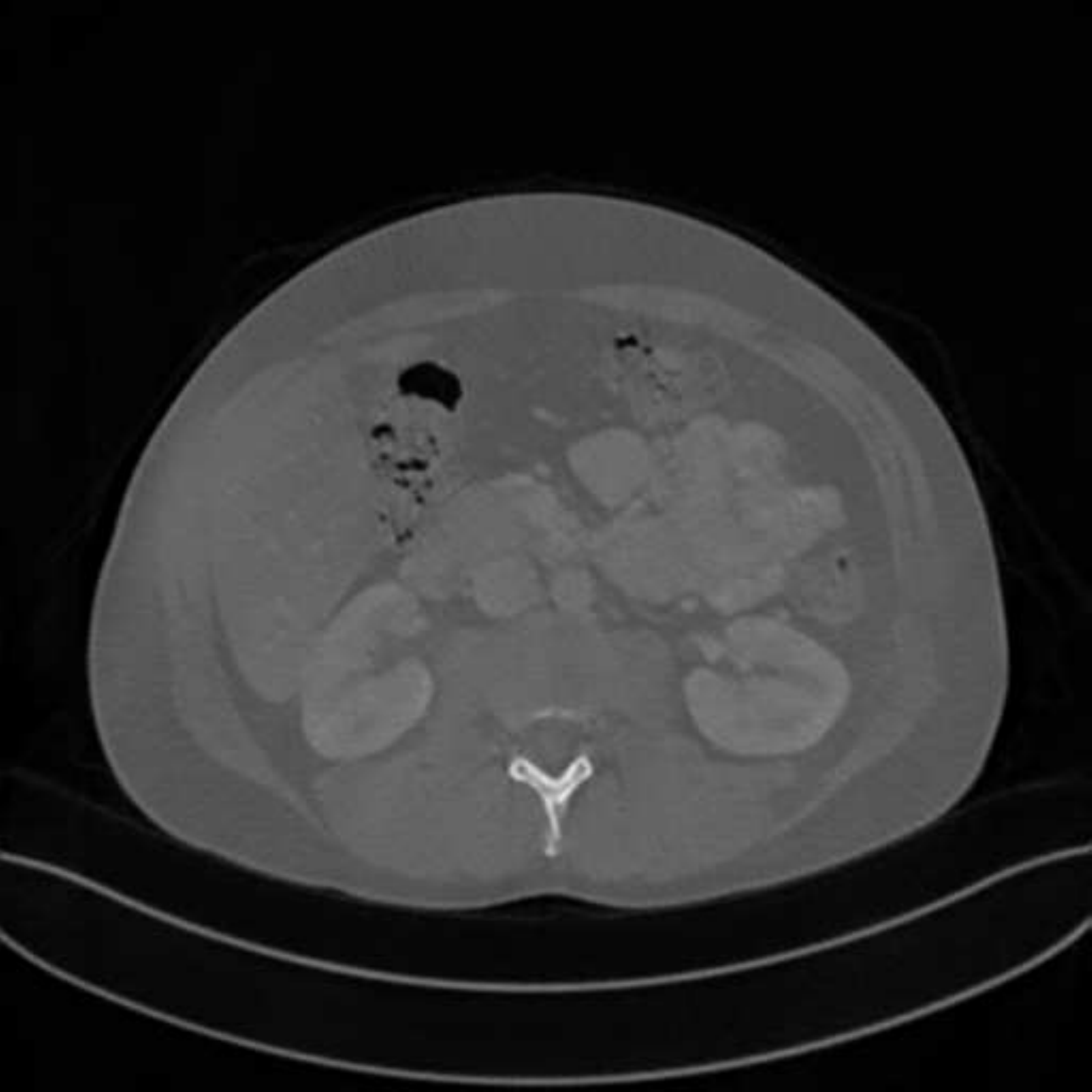}
		\includegraphics[width=0.28\columnwidth]{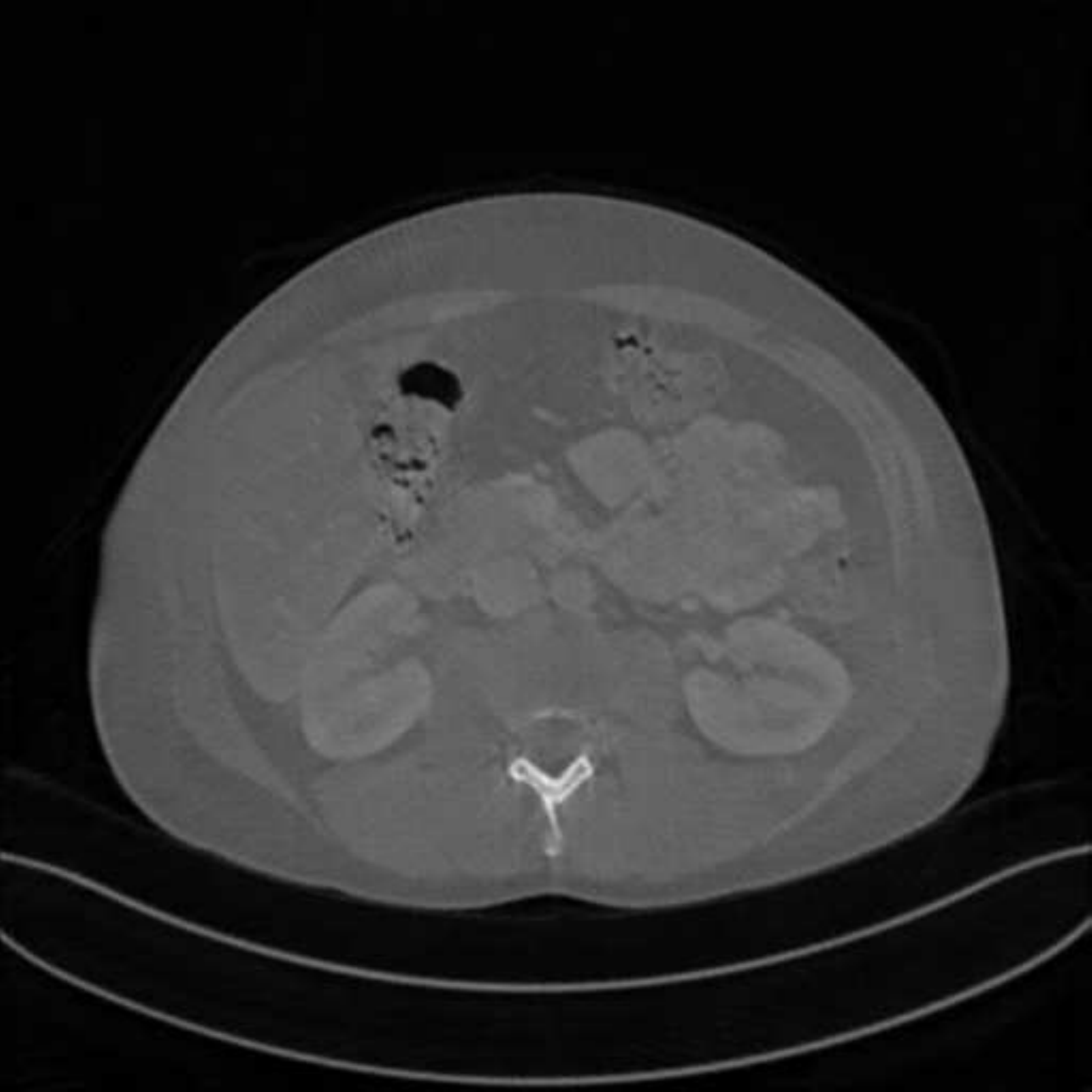}
		\includegraphics[width=0.28\columnwidth]{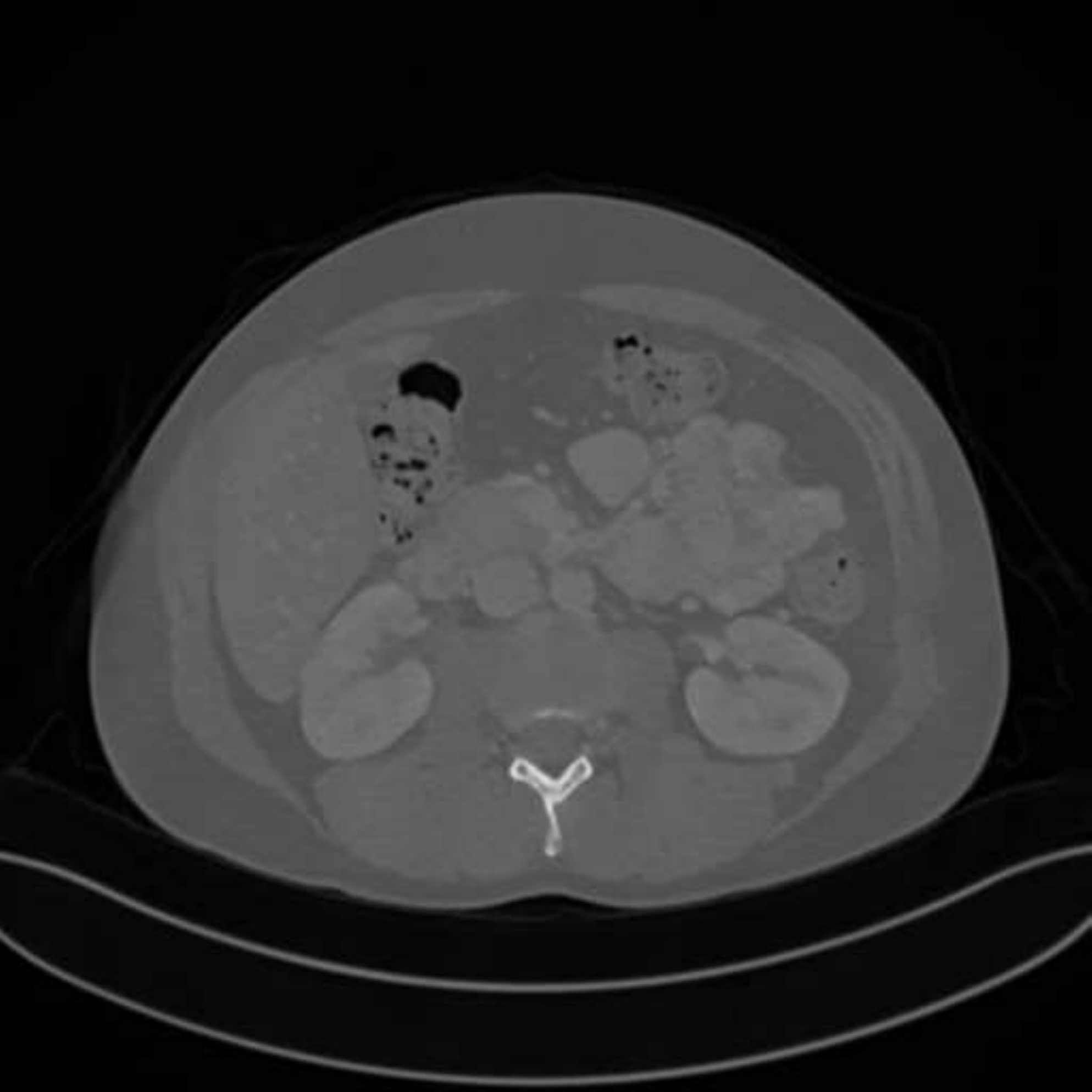}
	}
	\centerline{
		\subfigure[Label]{\includegraphics[width=0.28\columnwidth]{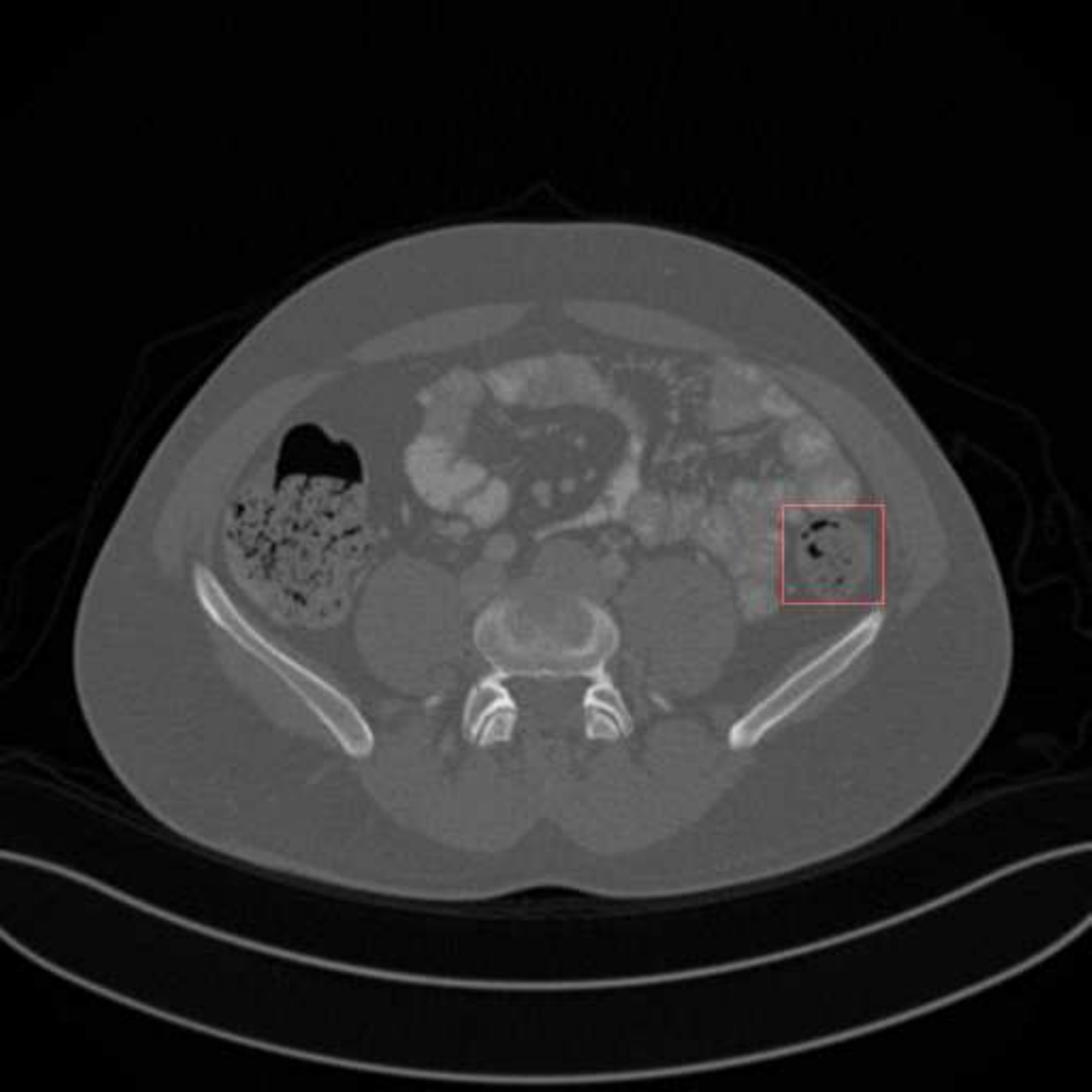}}
		\subfigure[FBP]{\includegraphics[width=0.28\columnwidth]{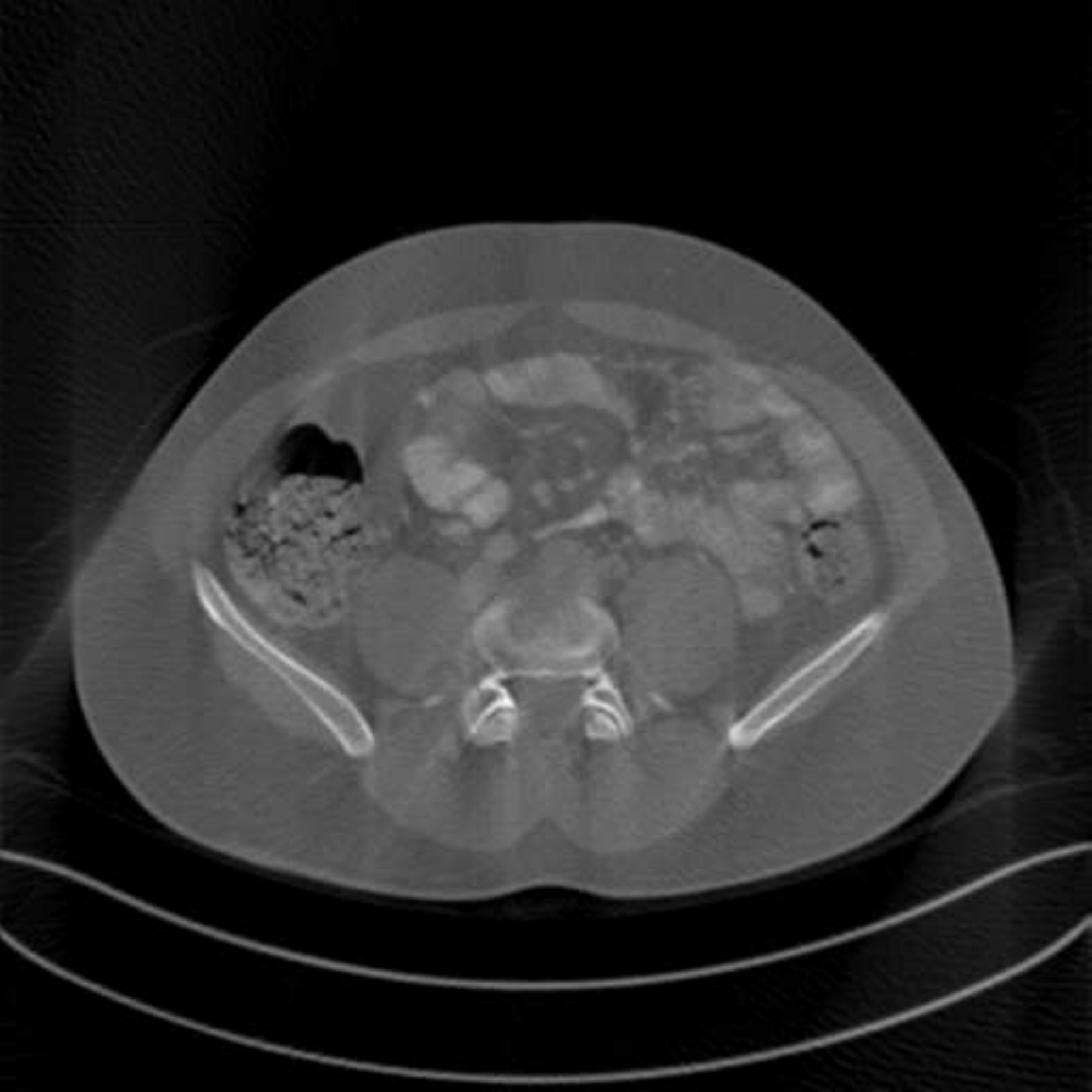}}
		\subfigure[TV-regularization]{\includegraphics[width=0.28\columnwidth]{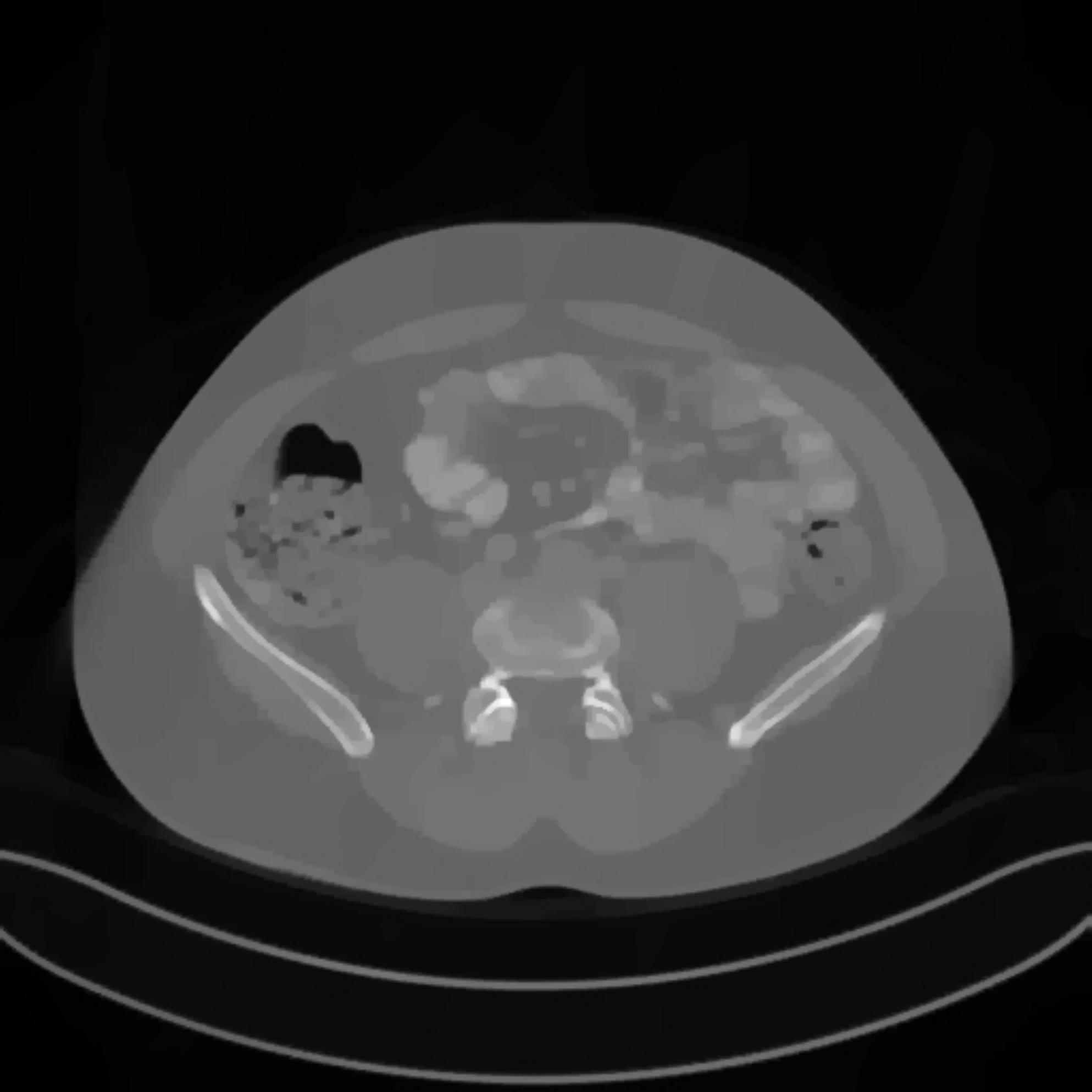}}
		\subfigure[Red-CNN]{\includegraphics[width=0.28\columnwidth]{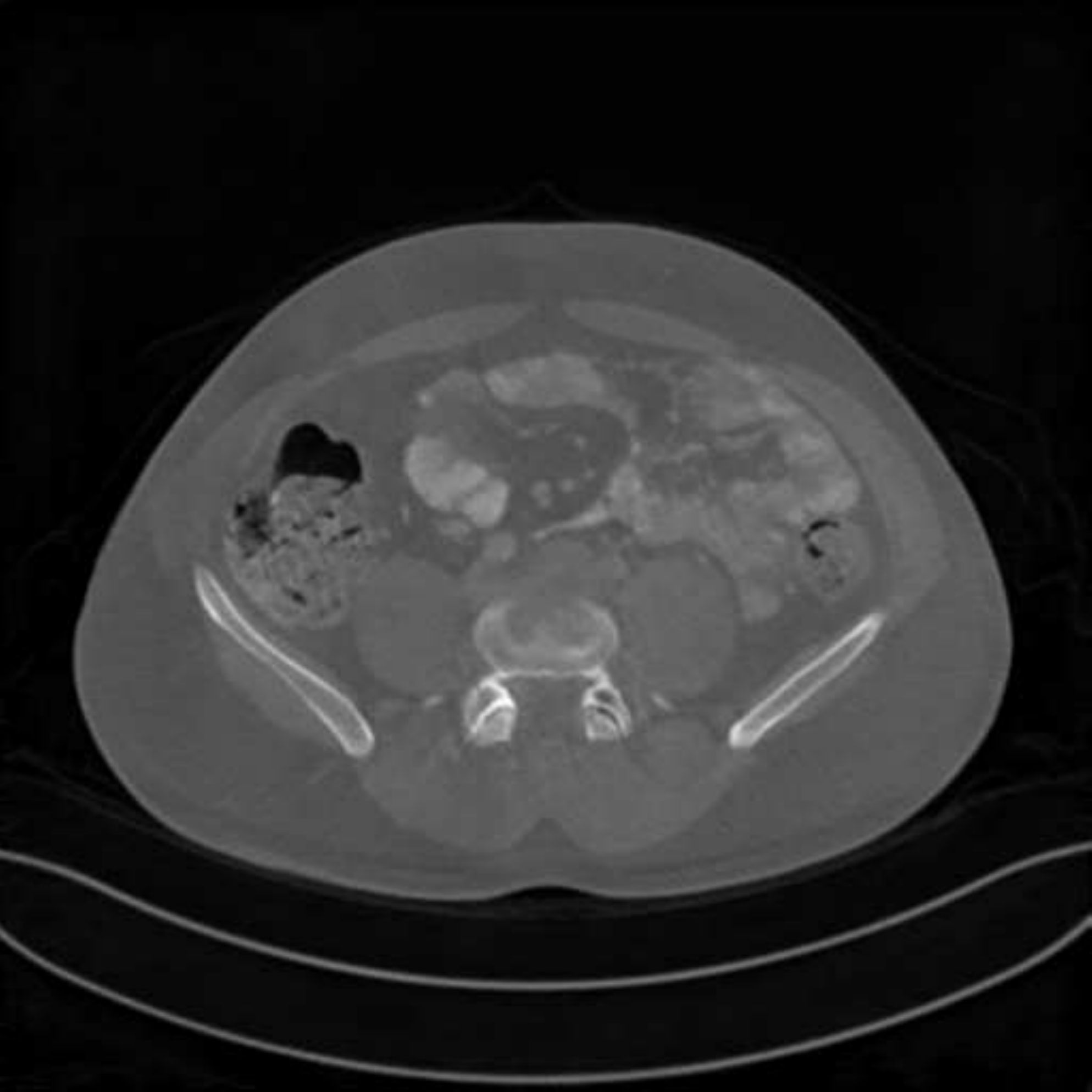}}
		\subfigure[FBP-Conv]{\includegraphics[width=0.28\columnwidth]{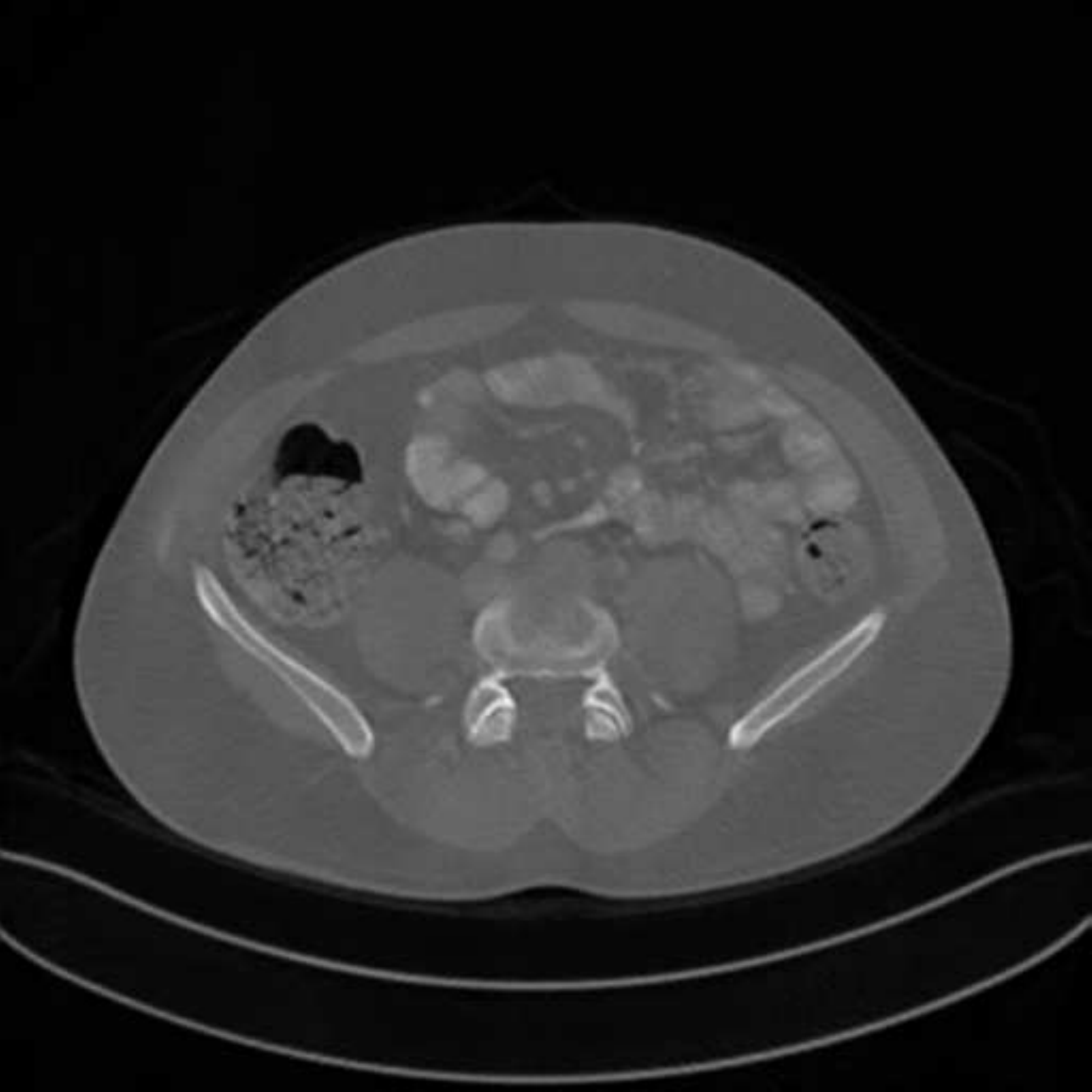}}
		\subfigure[DD-Net]{\includegraphics[width=0.28\columnwidth]{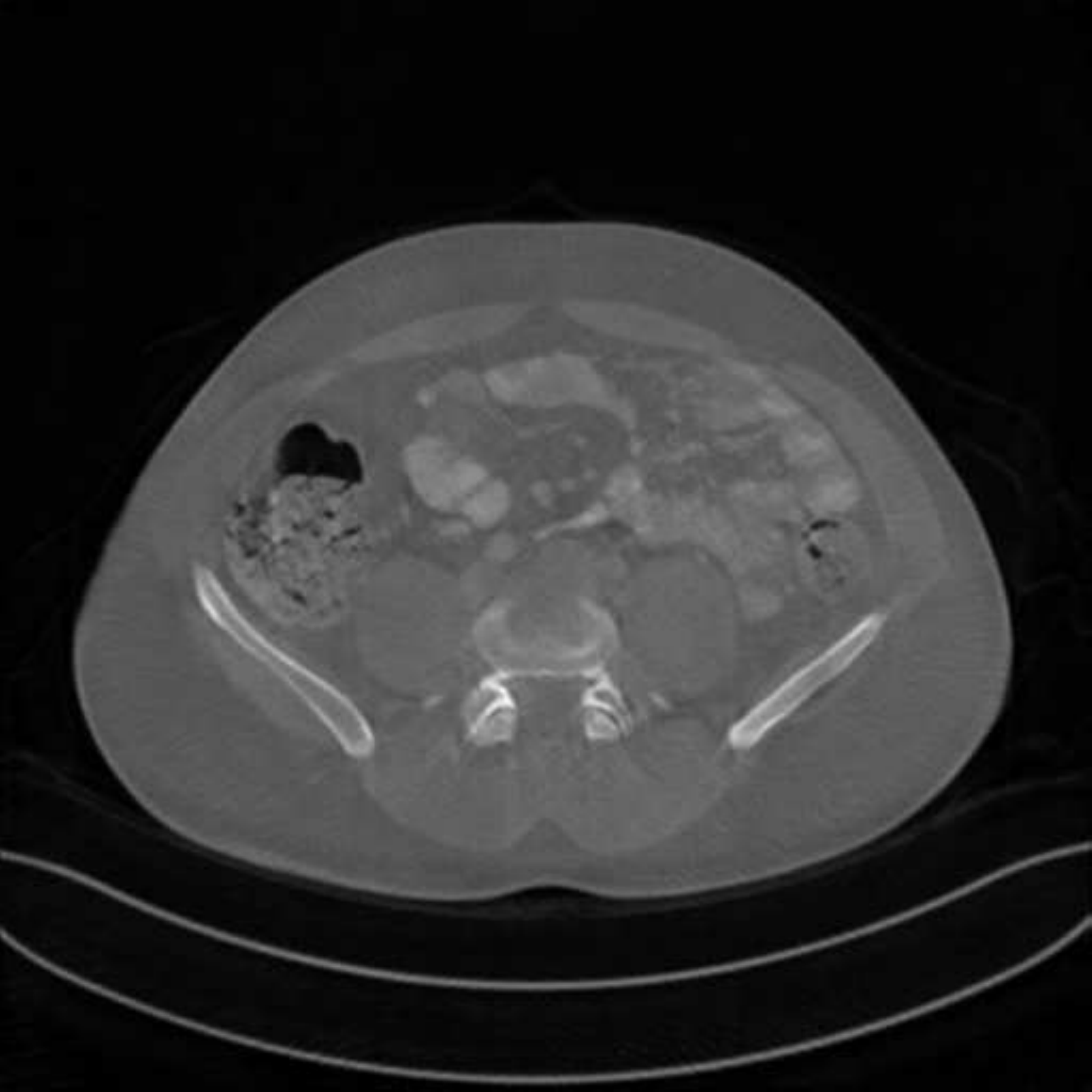}}
		\subfigure[Ours]{\includegraphics[width=0.28\columnwidth]{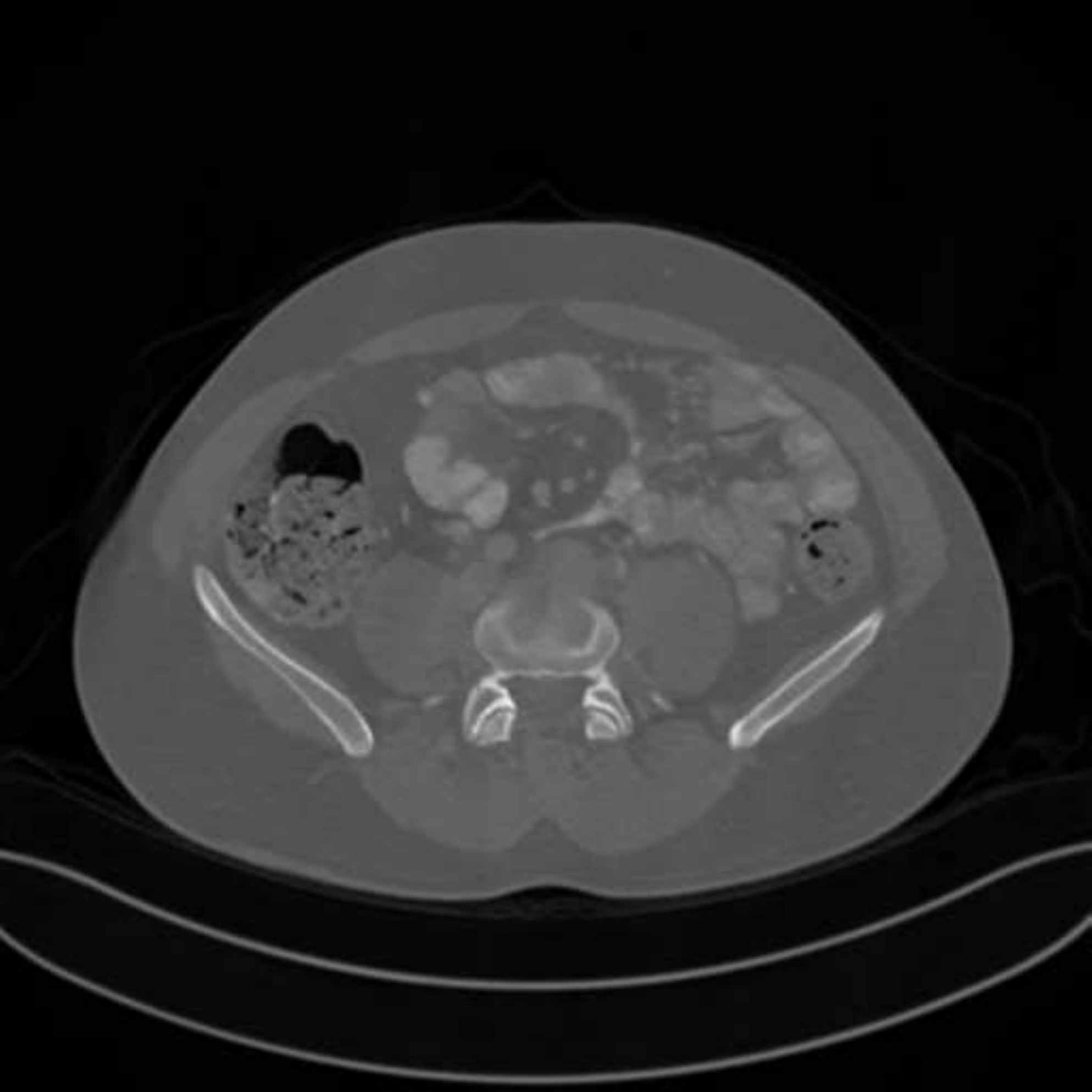}}
	}
	
	\caption{The reconstructed CT images by the six methods for parallel-beam geometry.}
	\label{F5}
\end{figure*}

\begin{figure*}[!t]
	\centerline{
		\includegraphics[width=0.28\columnwidth]{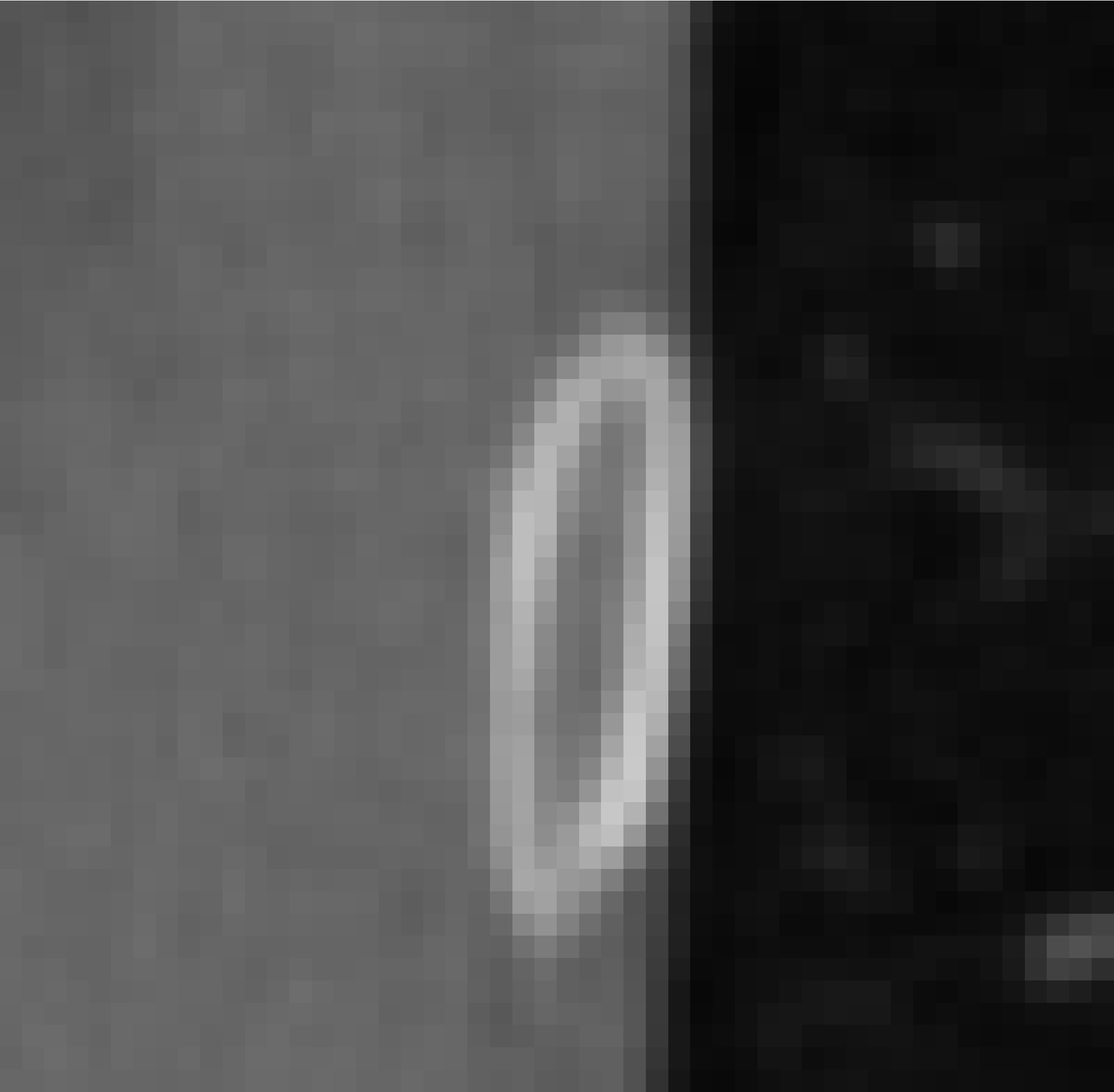}
		\includegraphics[width=0.28\columnwidth]{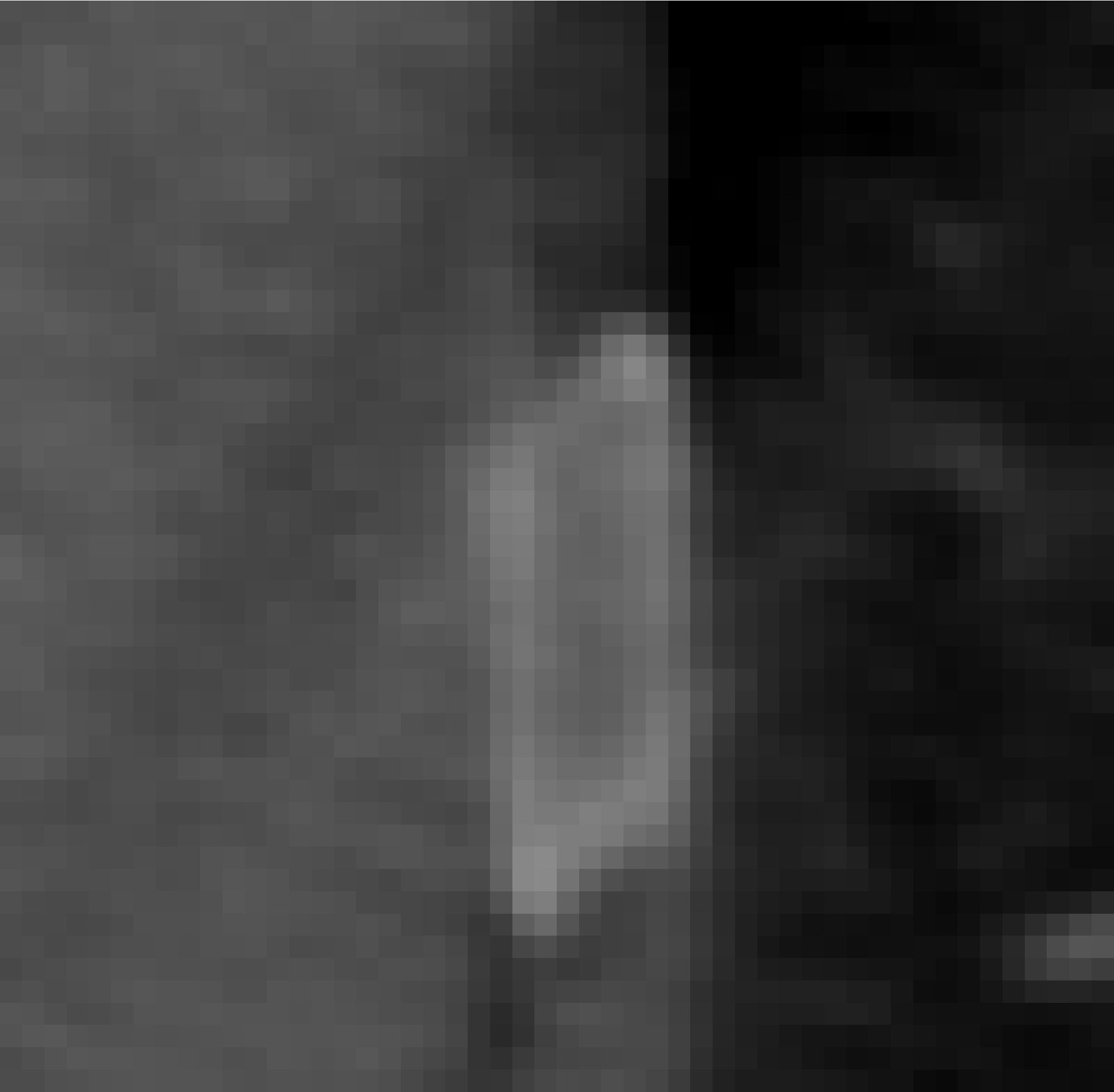}
		\includegraphics[width=0.28\columnwidth]{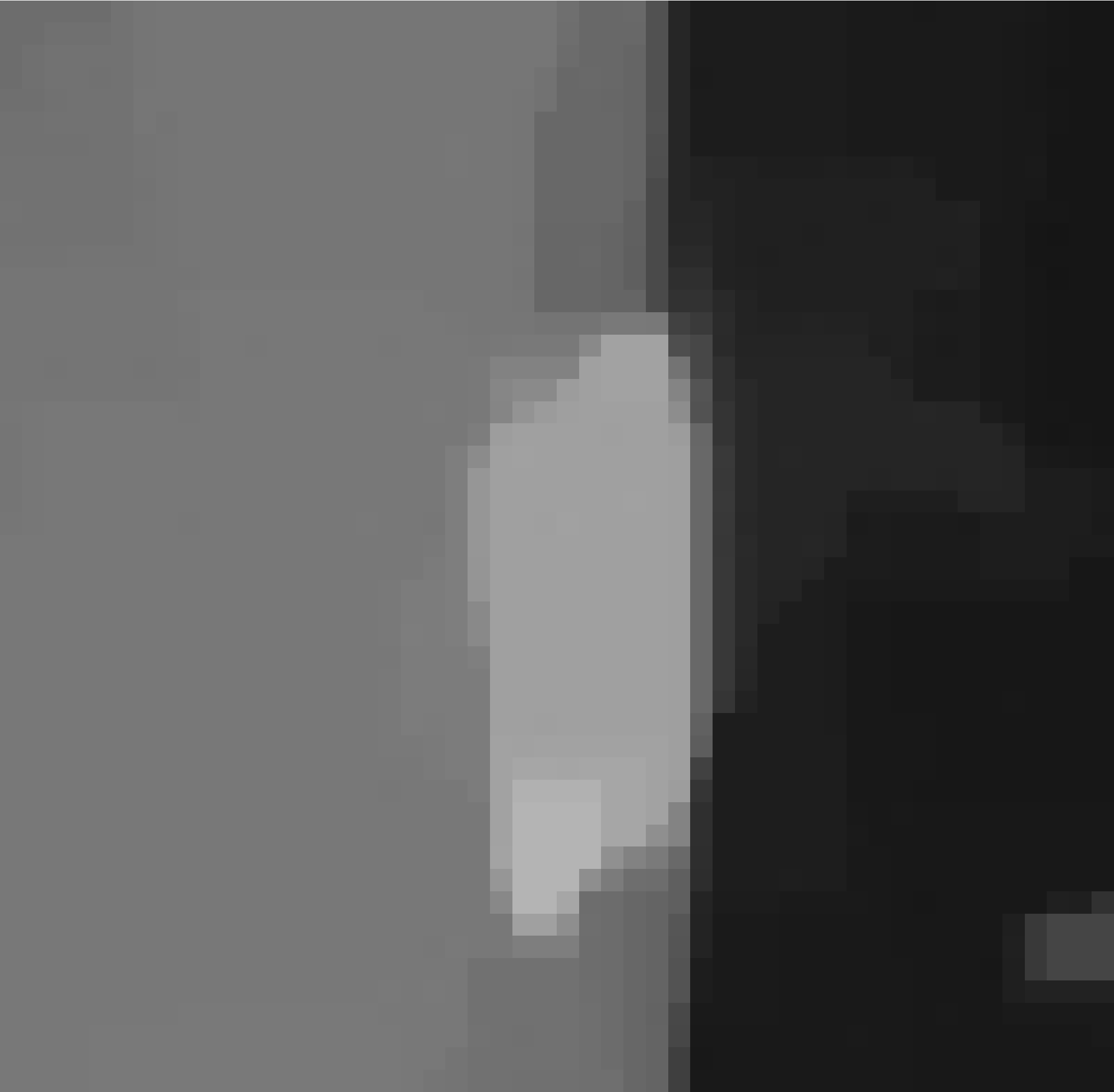}
		\includegraphics[width=0.28\columnwidth]{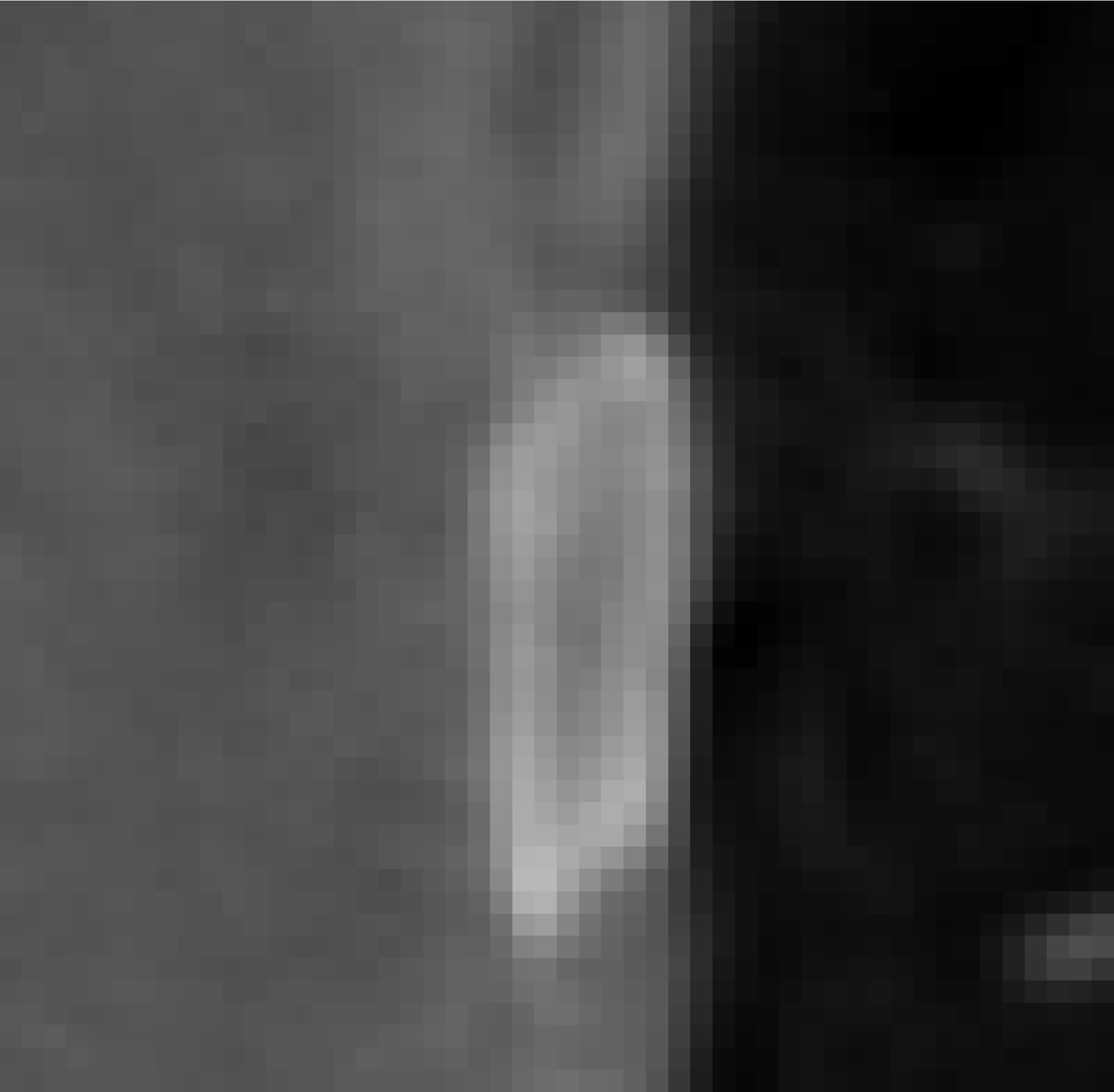}
		\includegraphics[width=0.28\columnwidth]{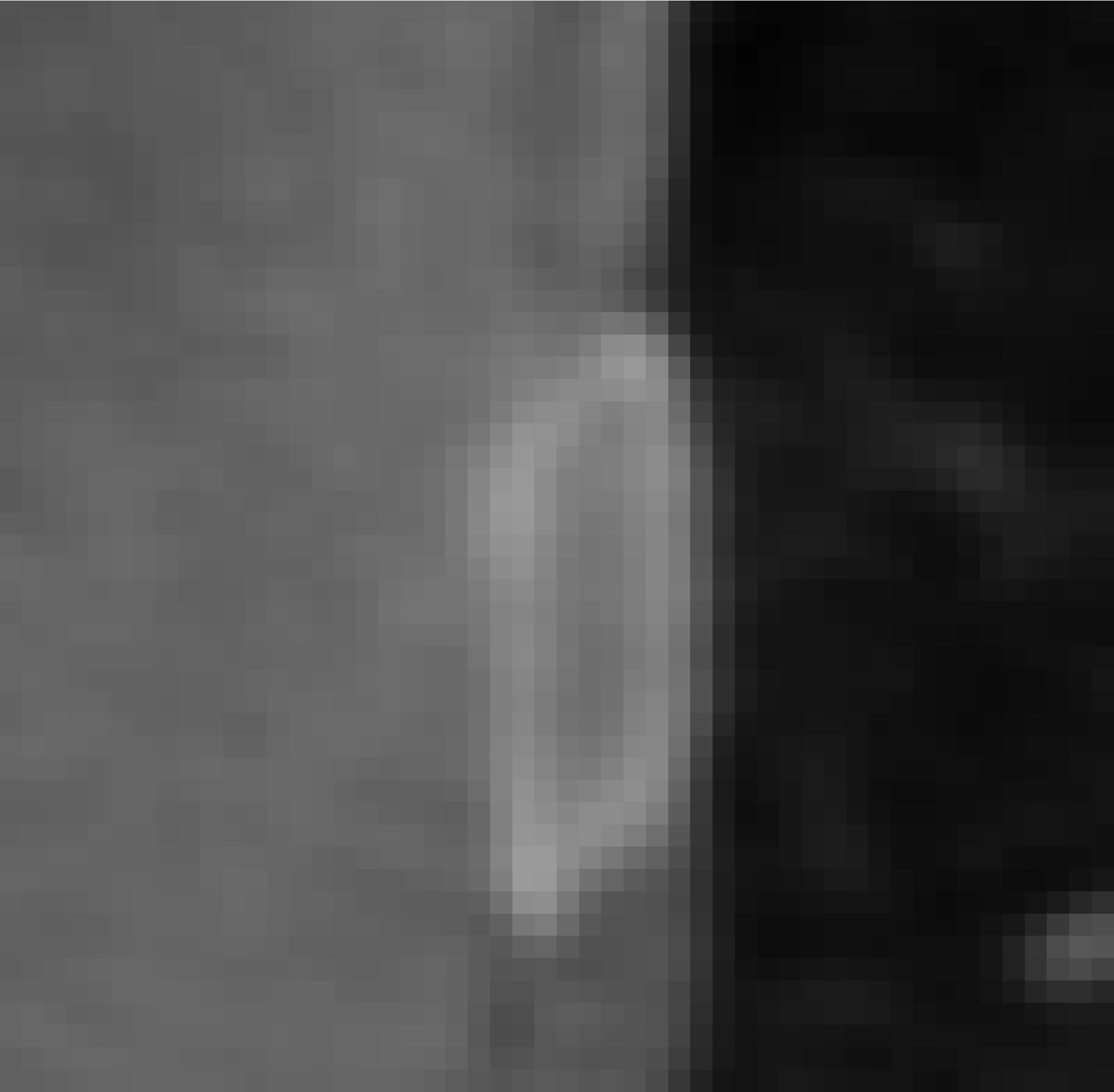}
		\includegraphics[width=0.28\columnwidth]{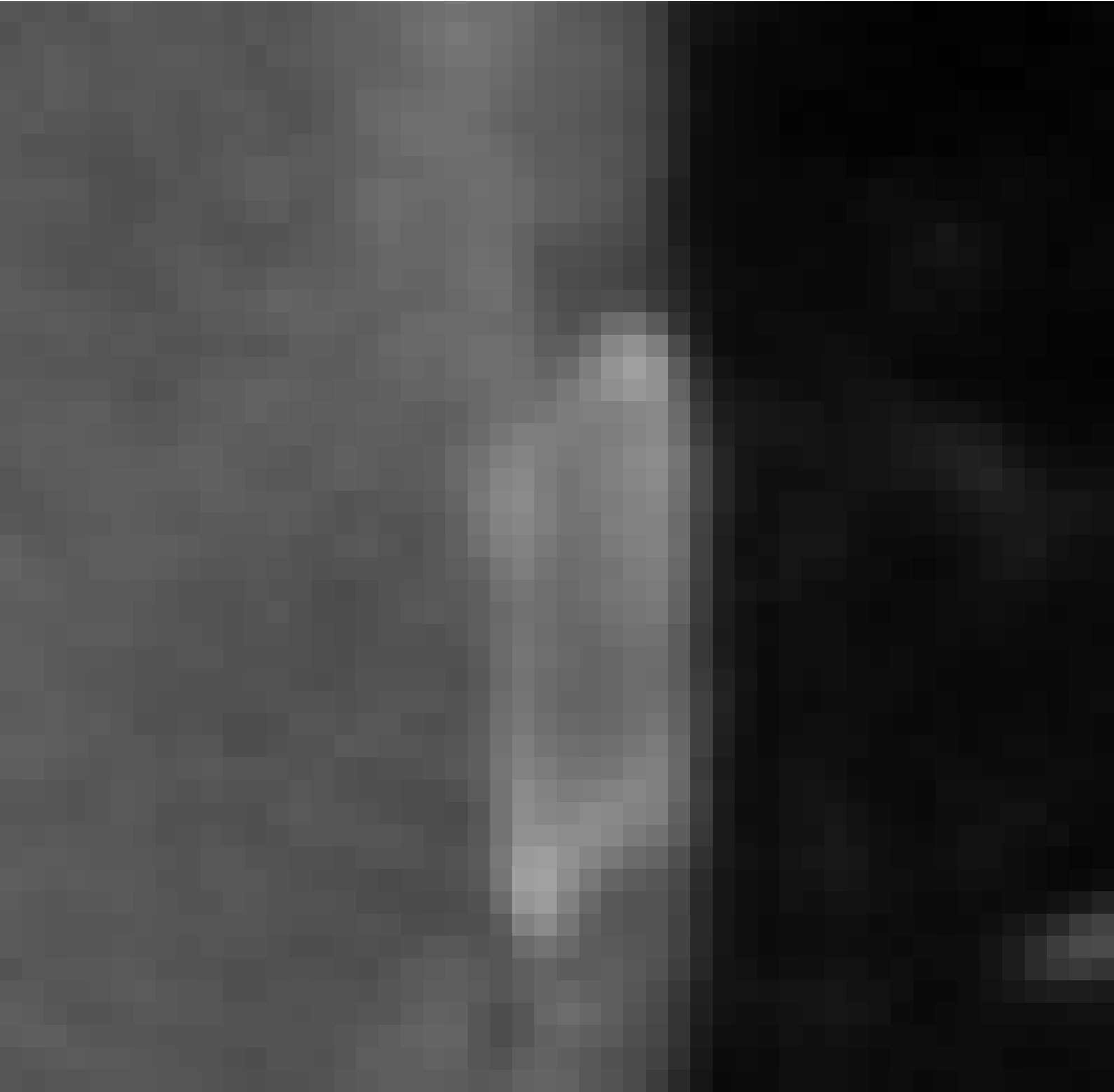}
		\includegraphics[width=0.28\columnwidth]{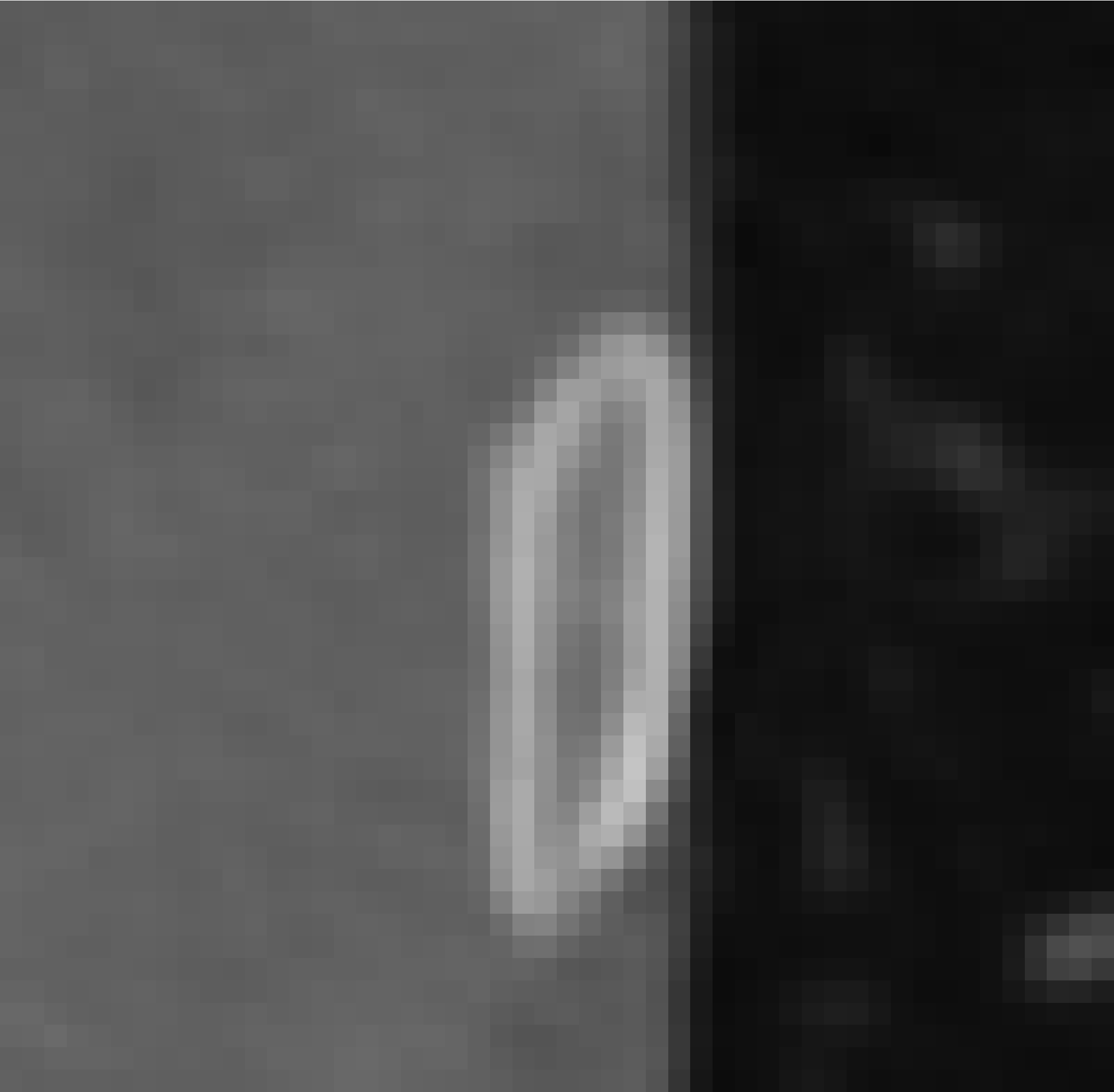}
	}
	\centerline{
		\includegraphics[width=0.28\columnwidth]{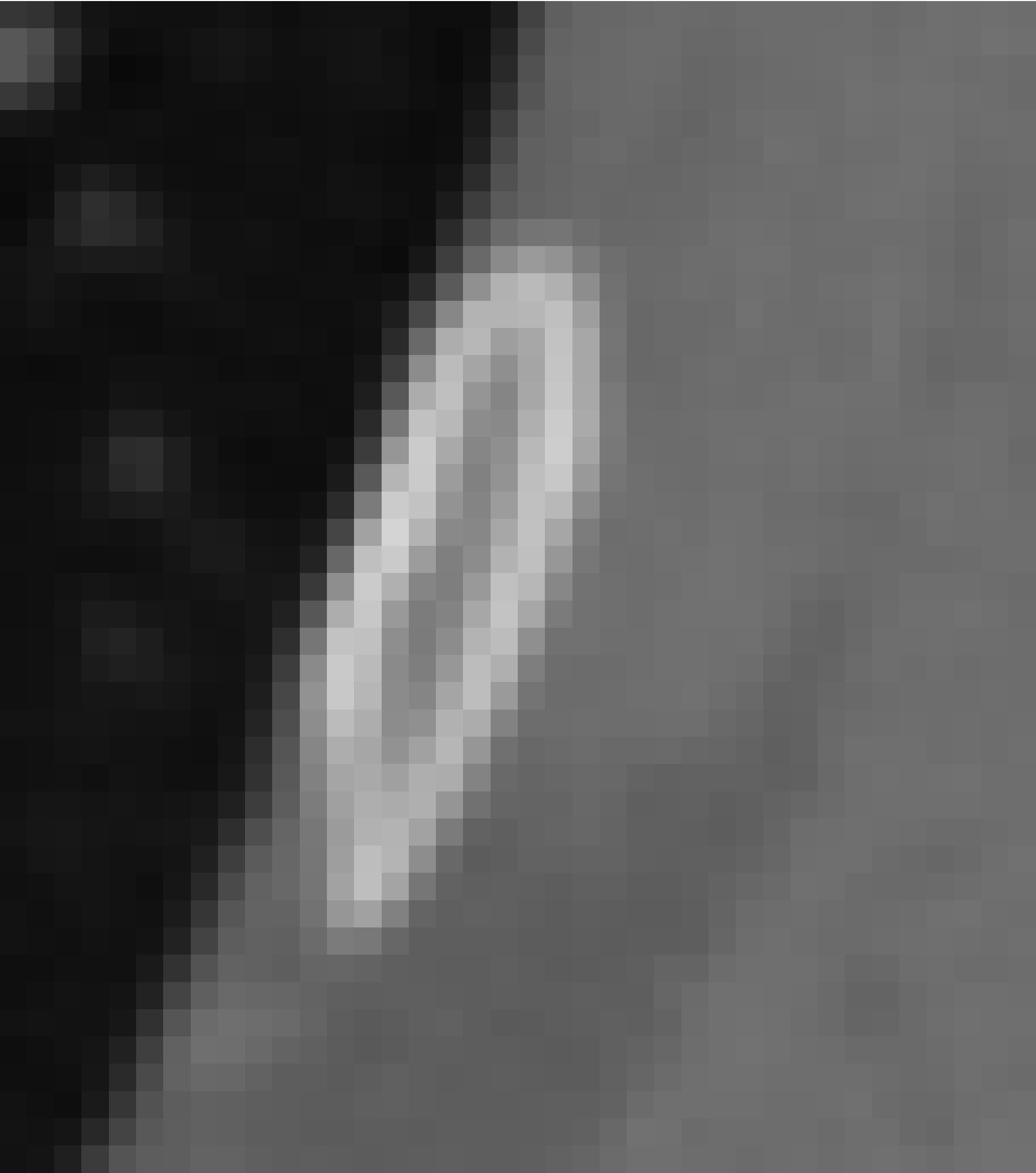}
		\includegraphics[width=0.28\columnwidth]{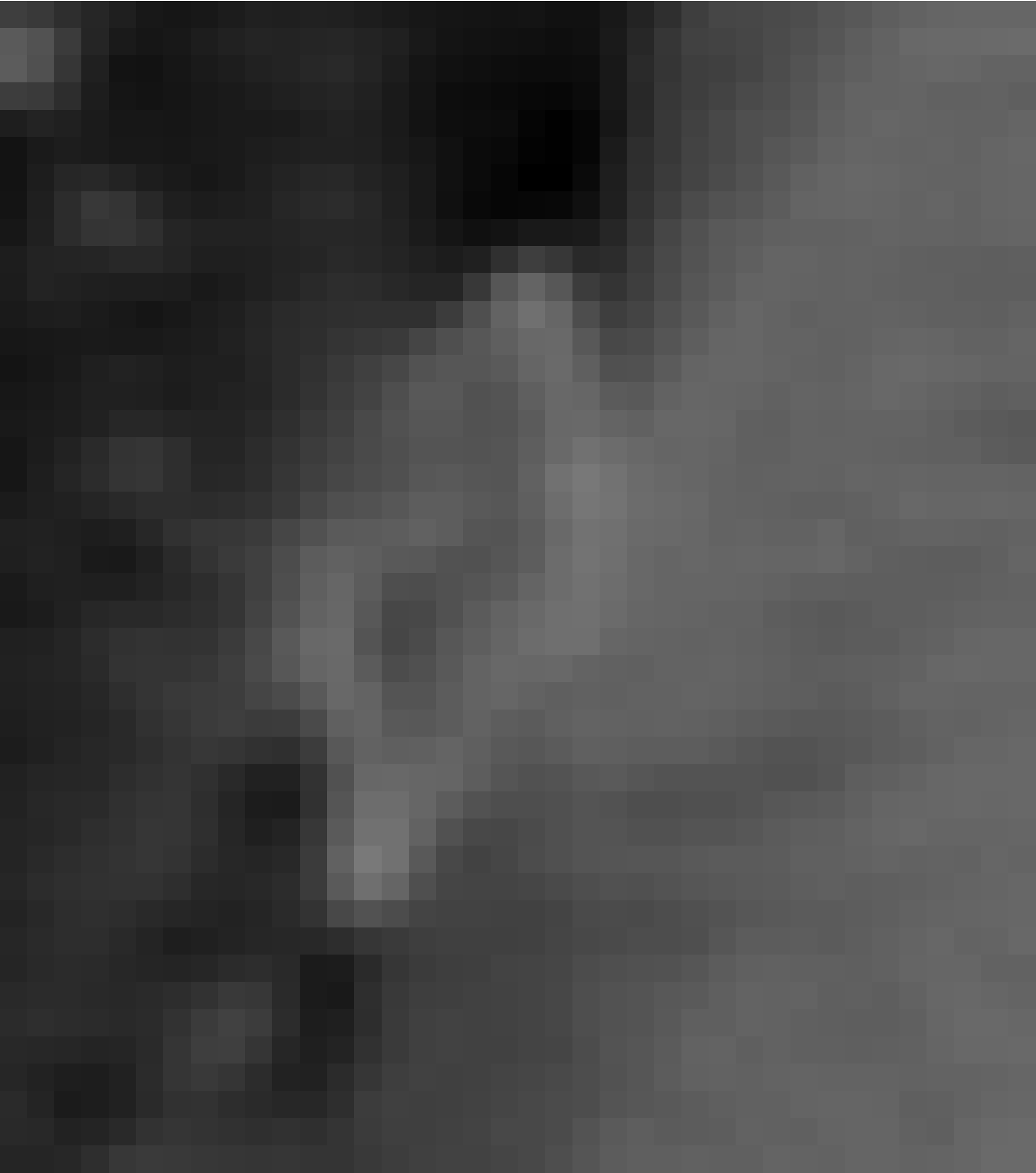}
		\includegraphics[width=0.28\columnwidth]{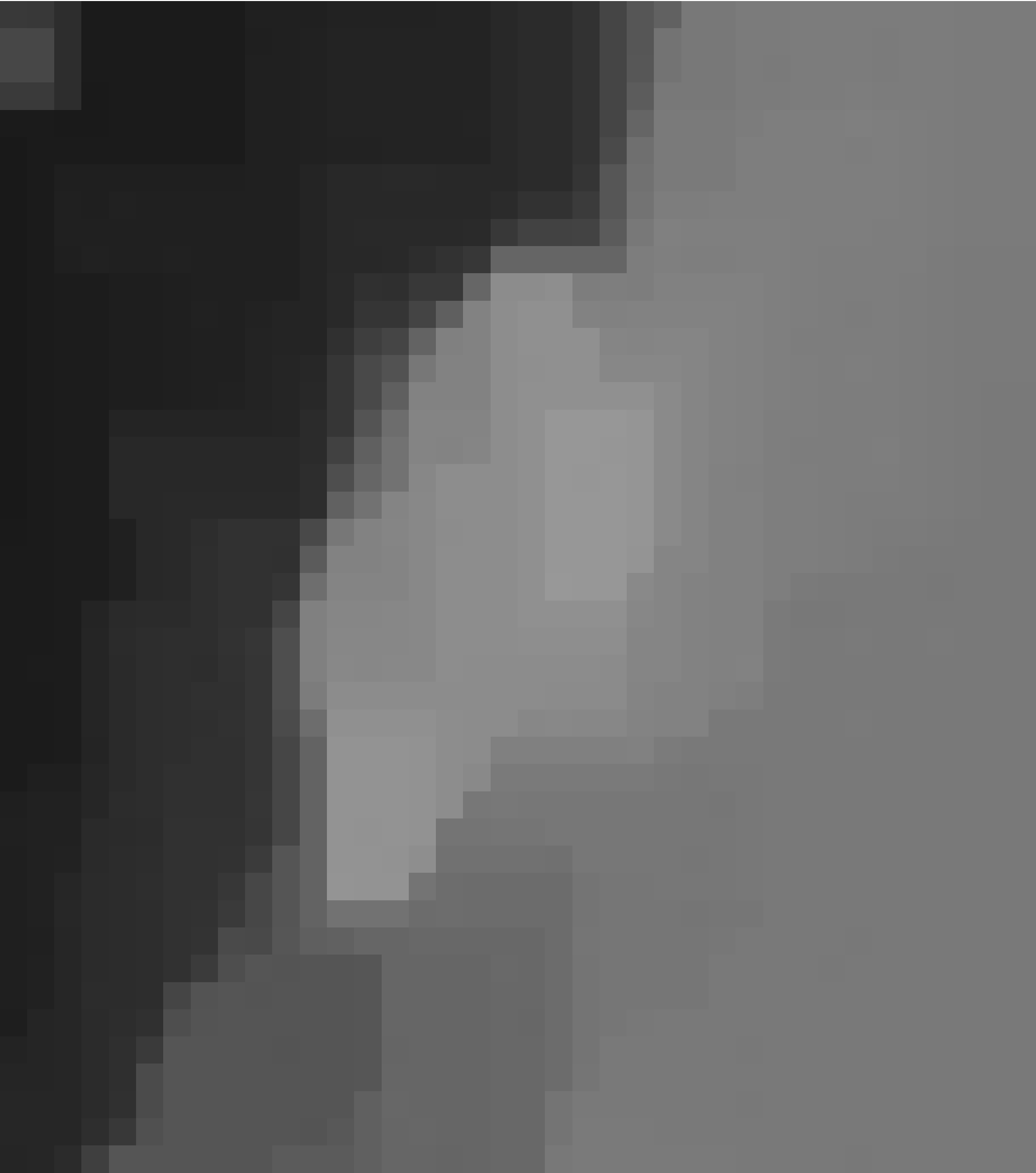}
		\includegraphics[width=0.28\columnwidth]{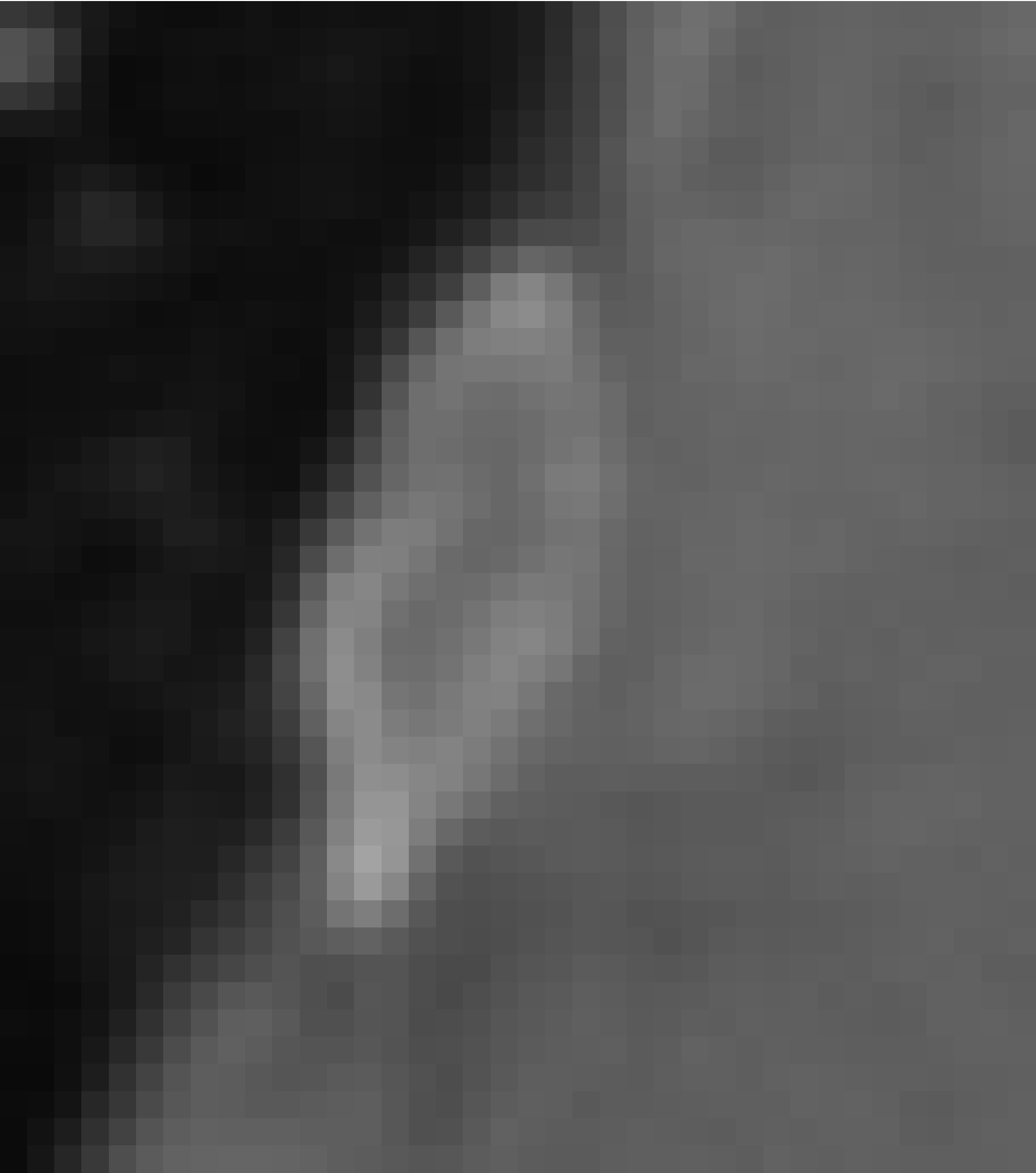}
		\includegraphics[width=0.28\columnwidth]{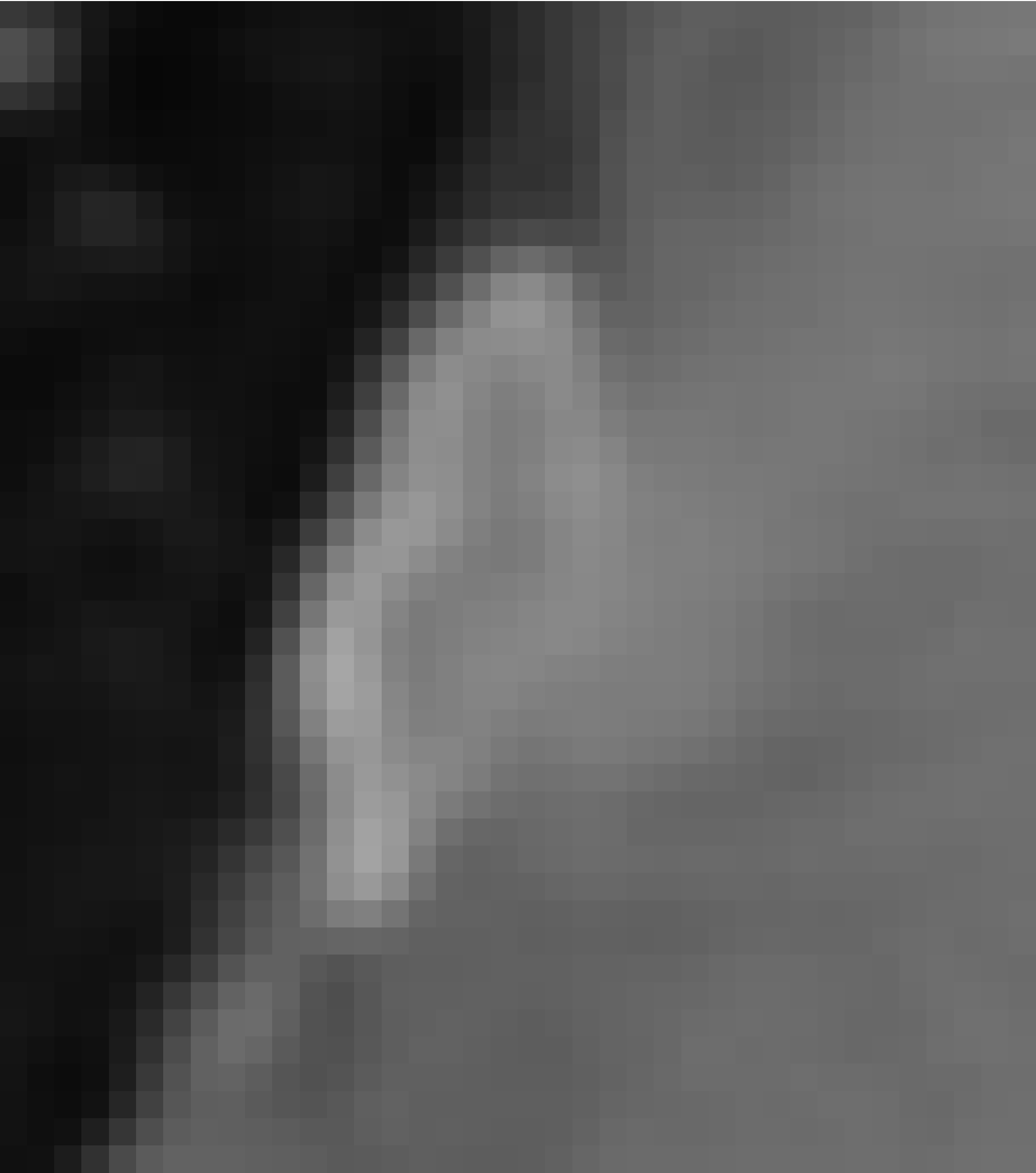}
		\includegraphics[width=0.28\columnwidth]{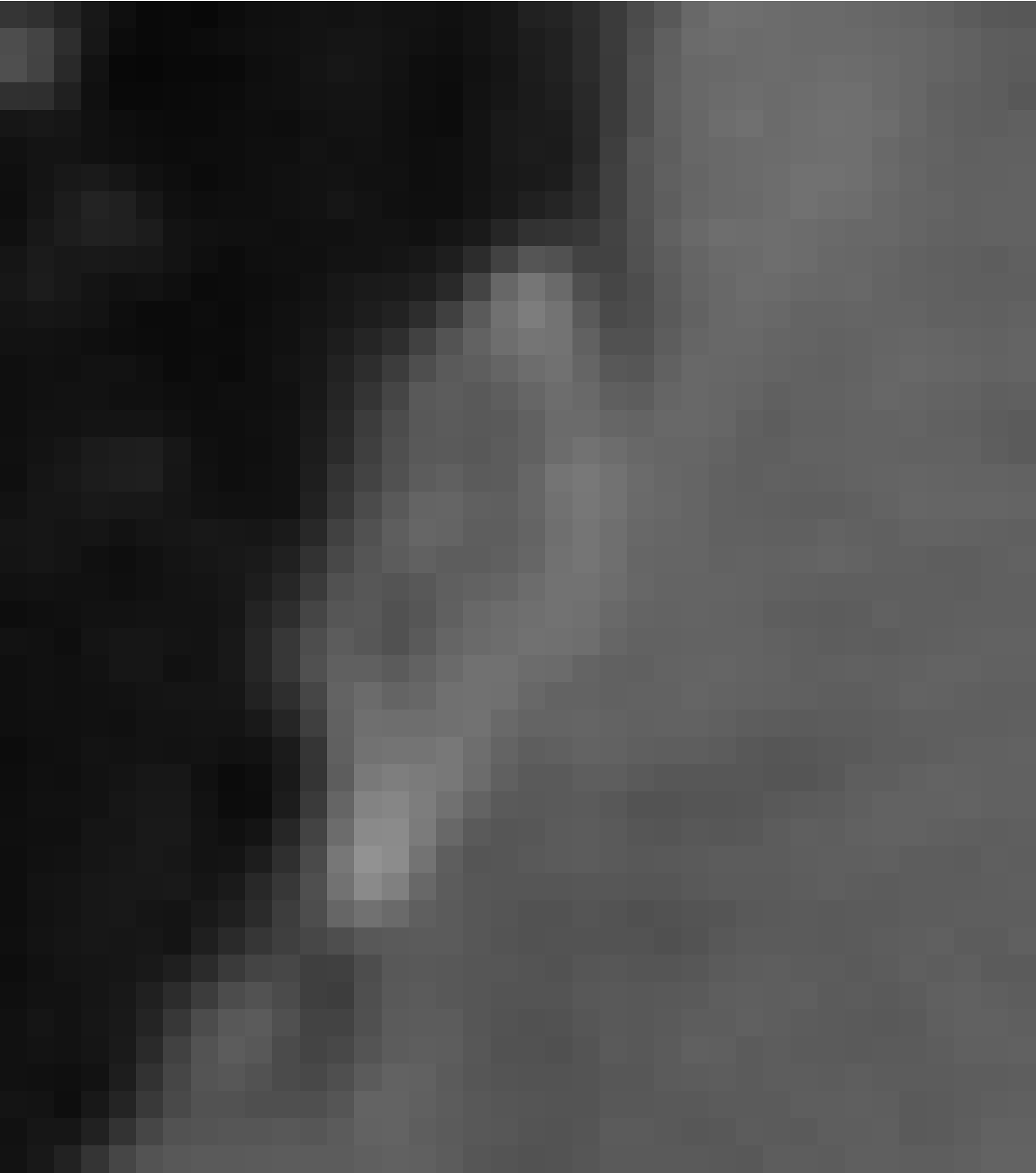}
		\includegraphics[width=0.28\columnwidth]{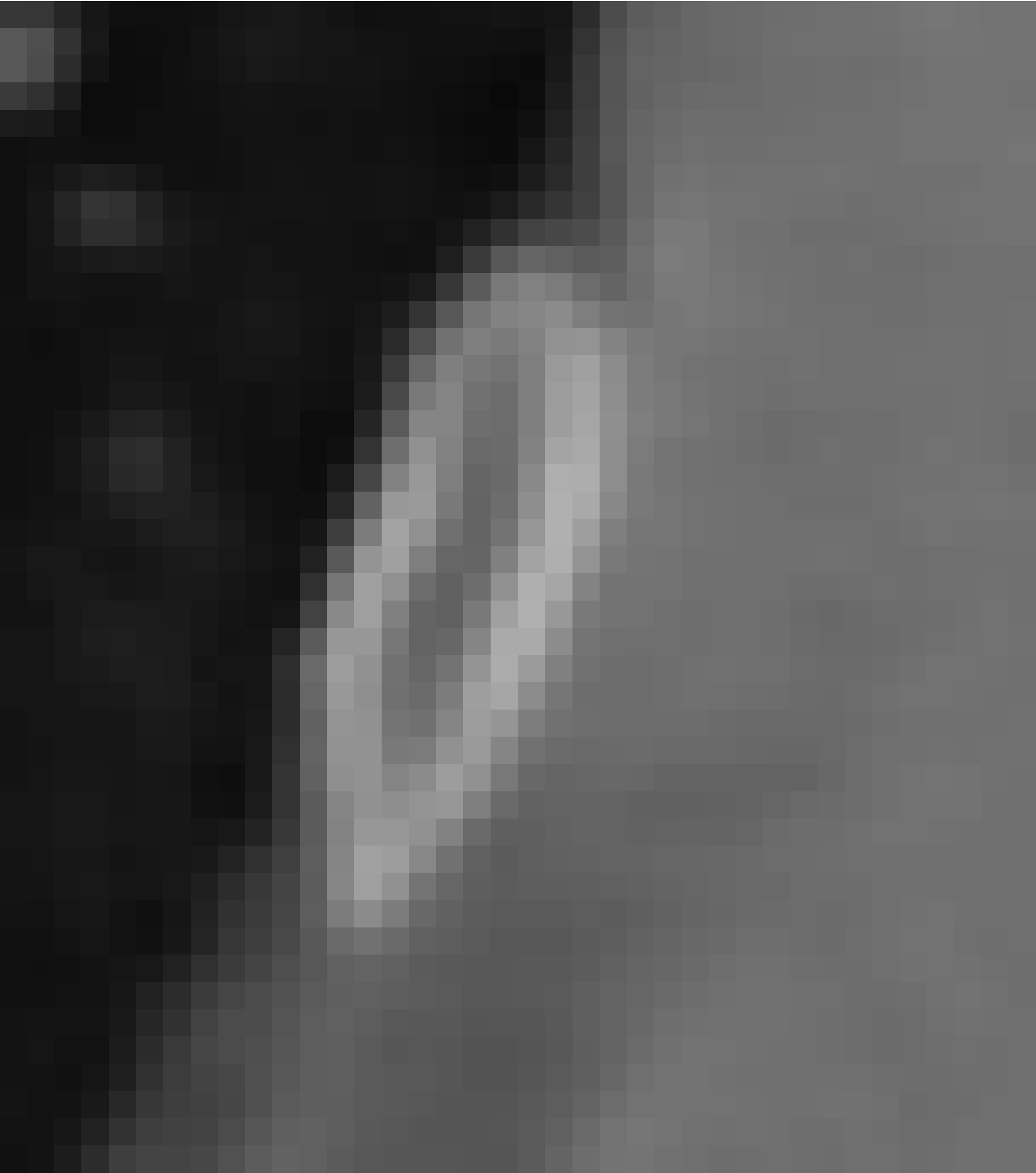}
	}
	\centerline{
		\includegraphics[width=0.28\columnwidth]{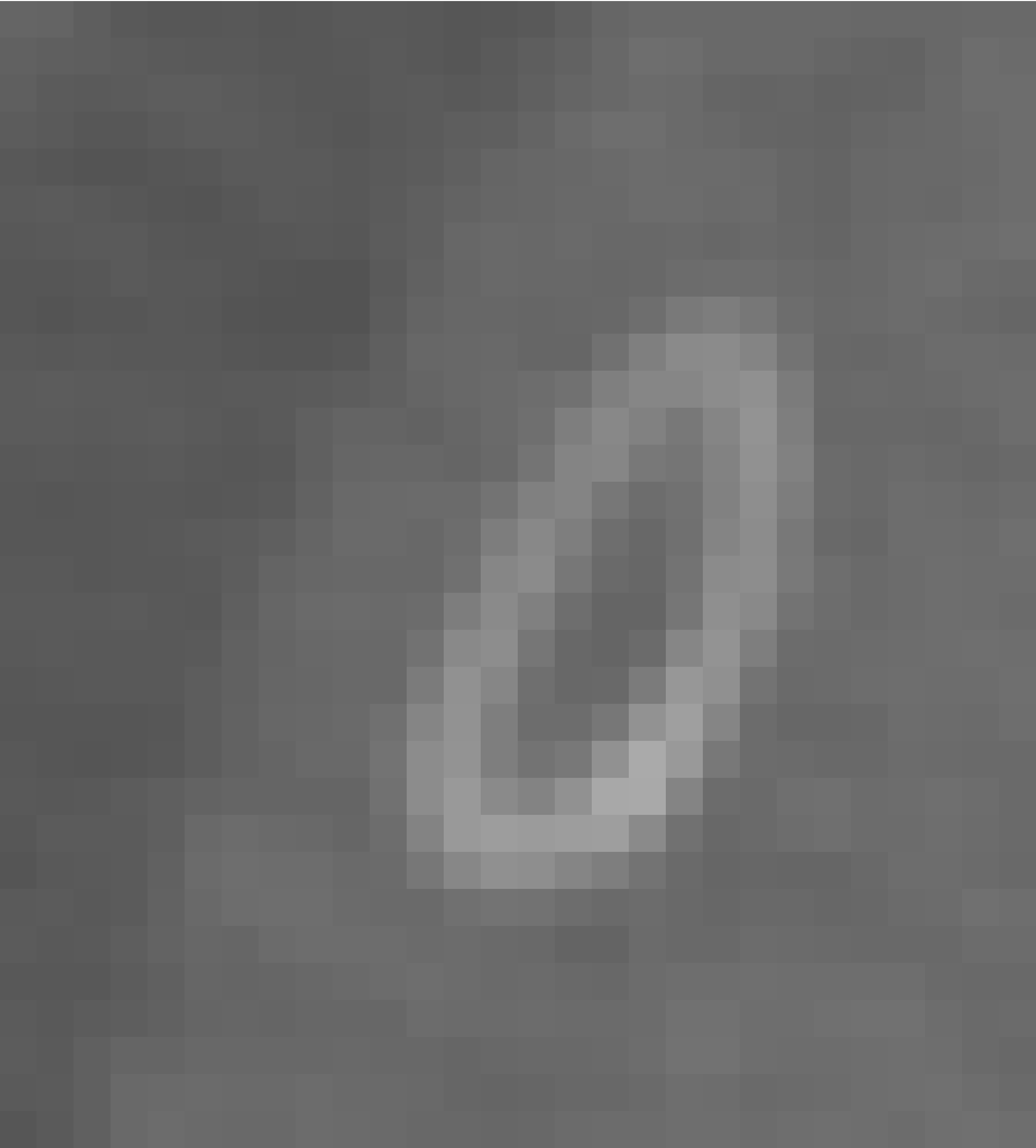}
		\includegraphics[width=0.28\columnwidth]{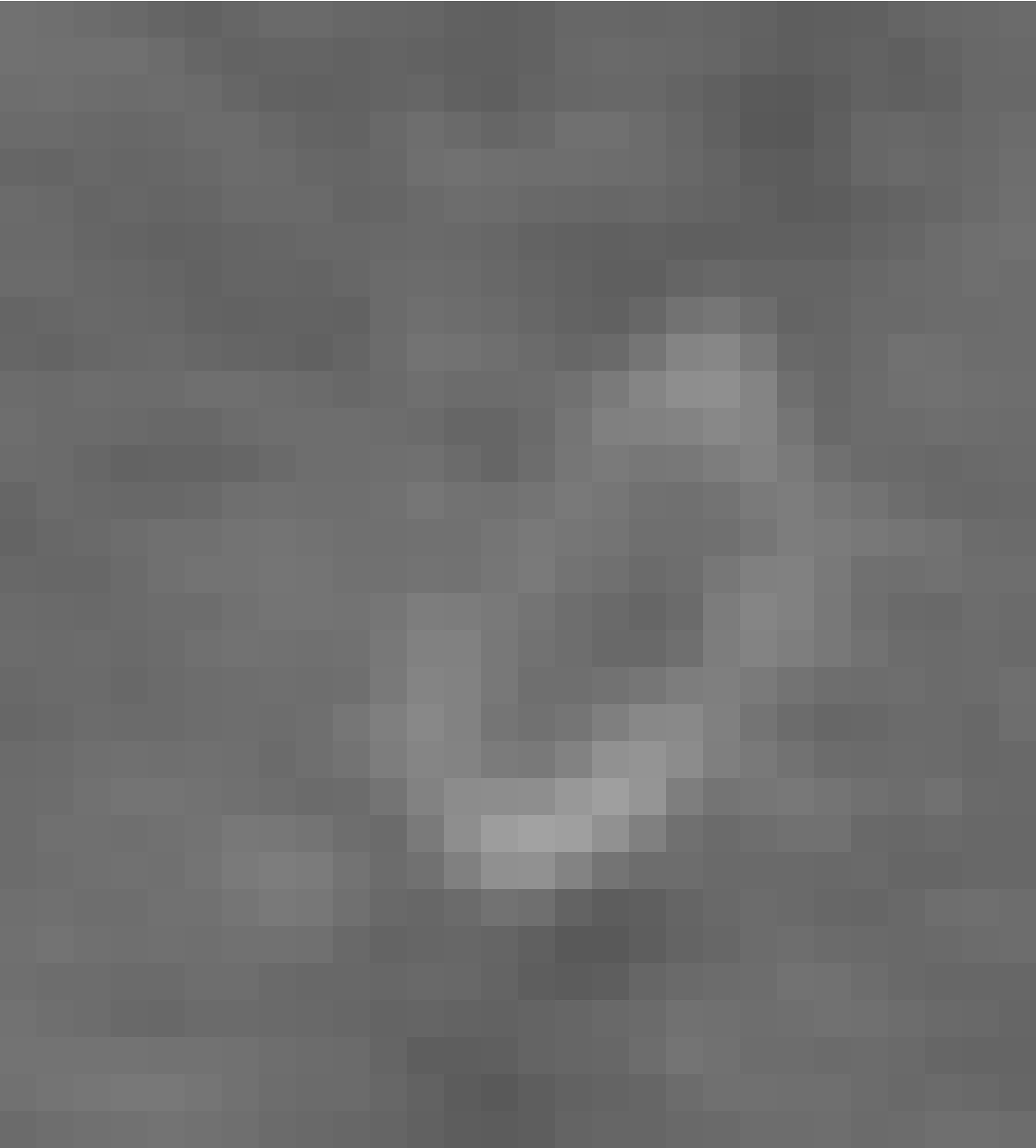}
		\includegraphics[width=0.28\columnwidth]{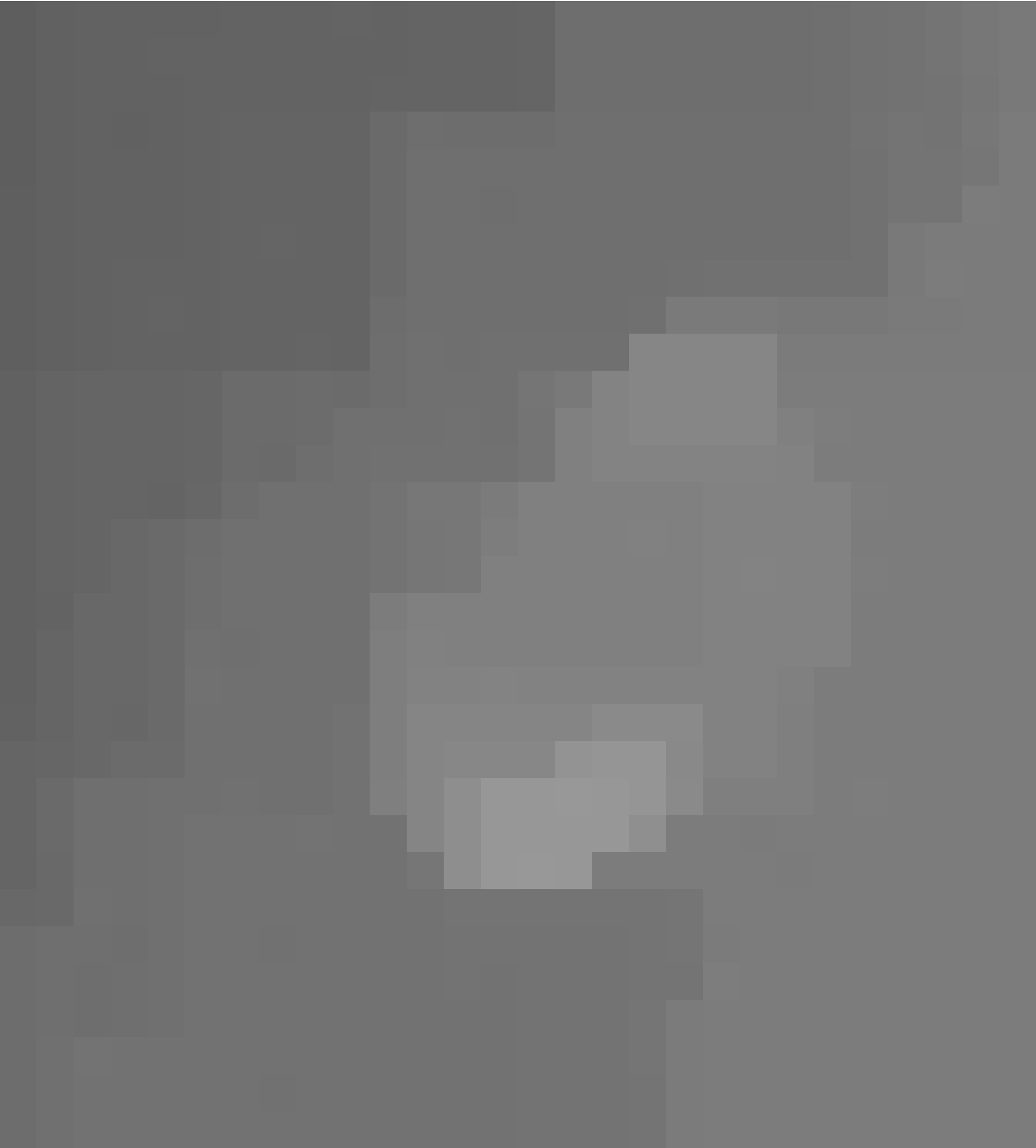}
		\includegraphics[width=0.28\columnwidth]{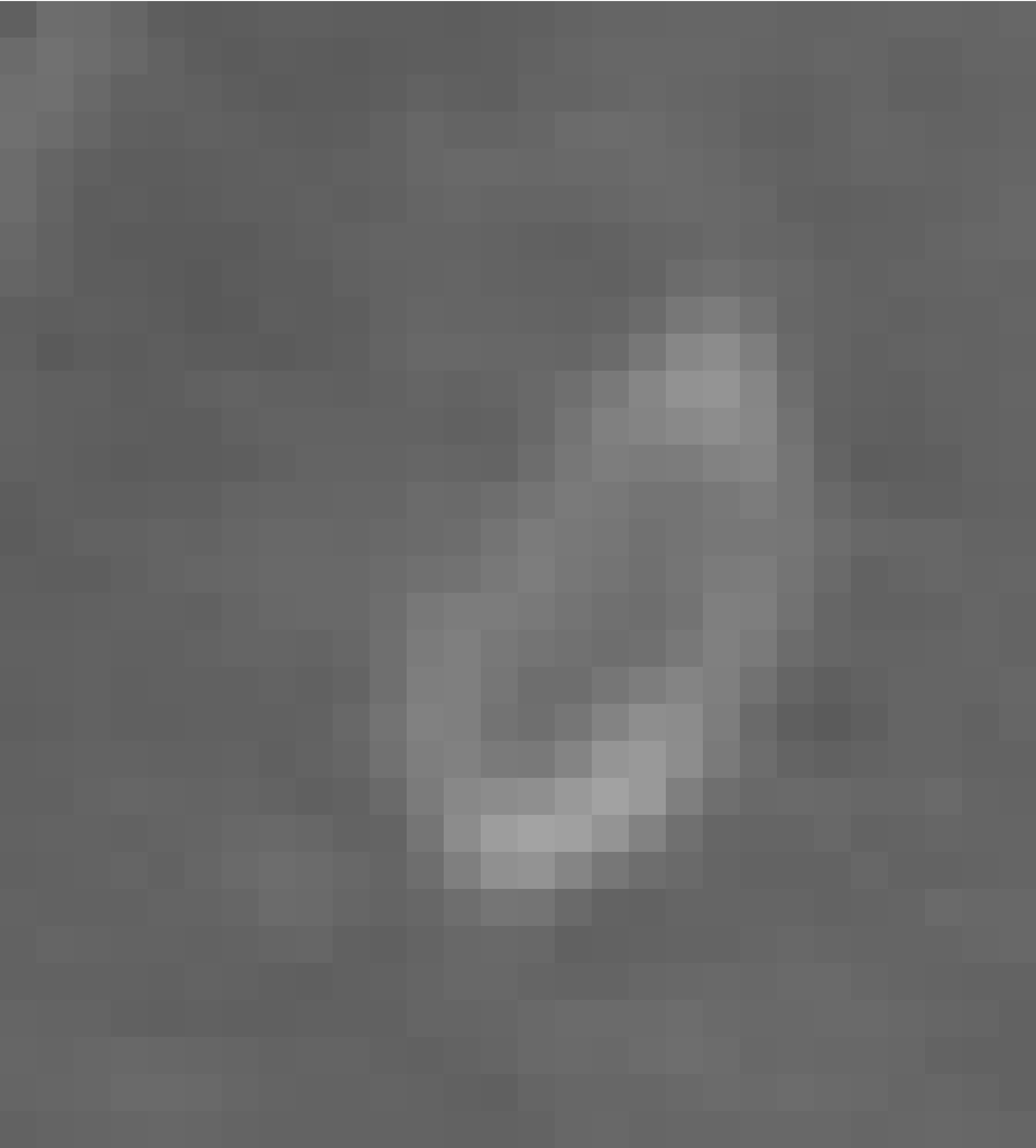}
		\includegraphics[width=0.28\columnwidth]{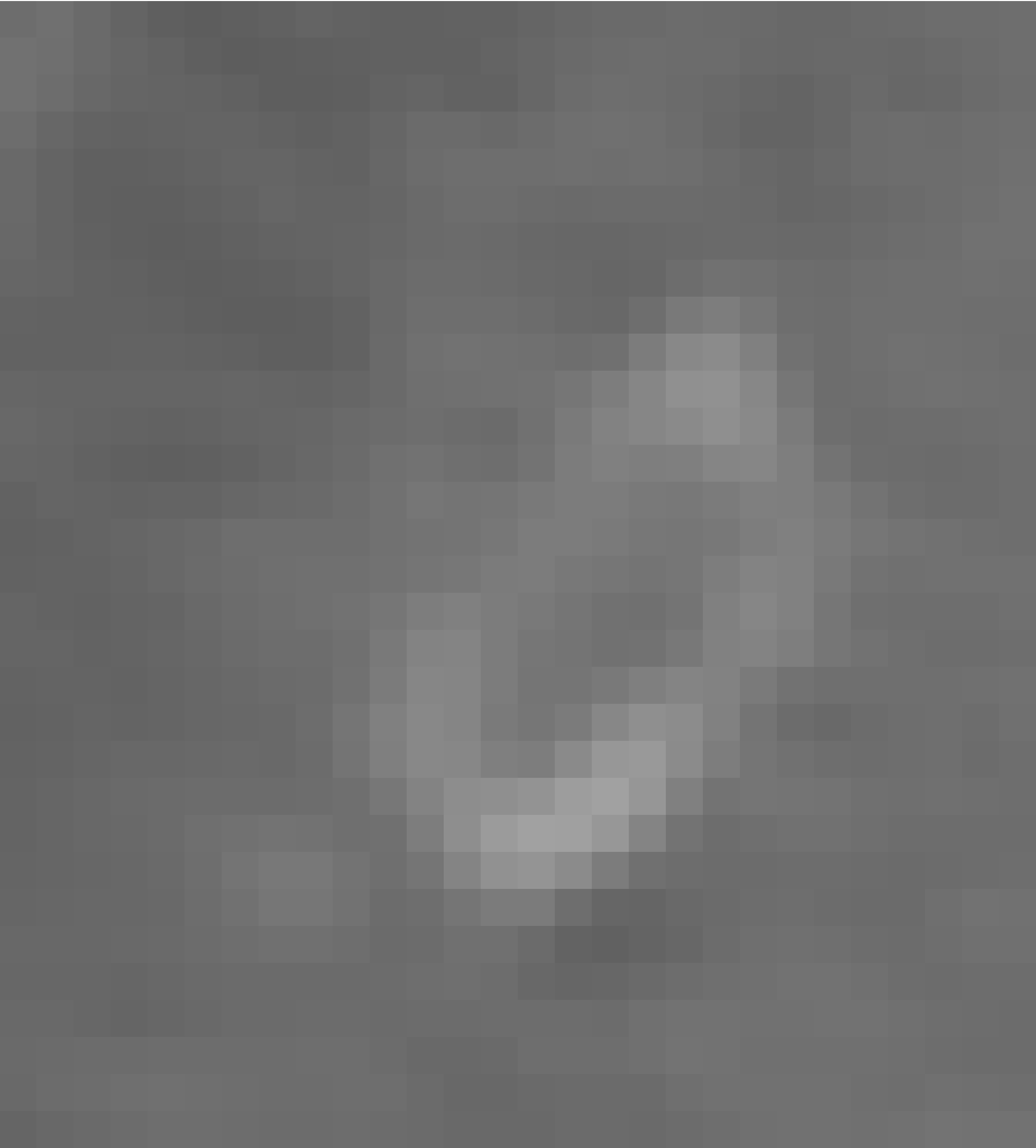}
		\includegraphics[width=0.28\columnwidth]{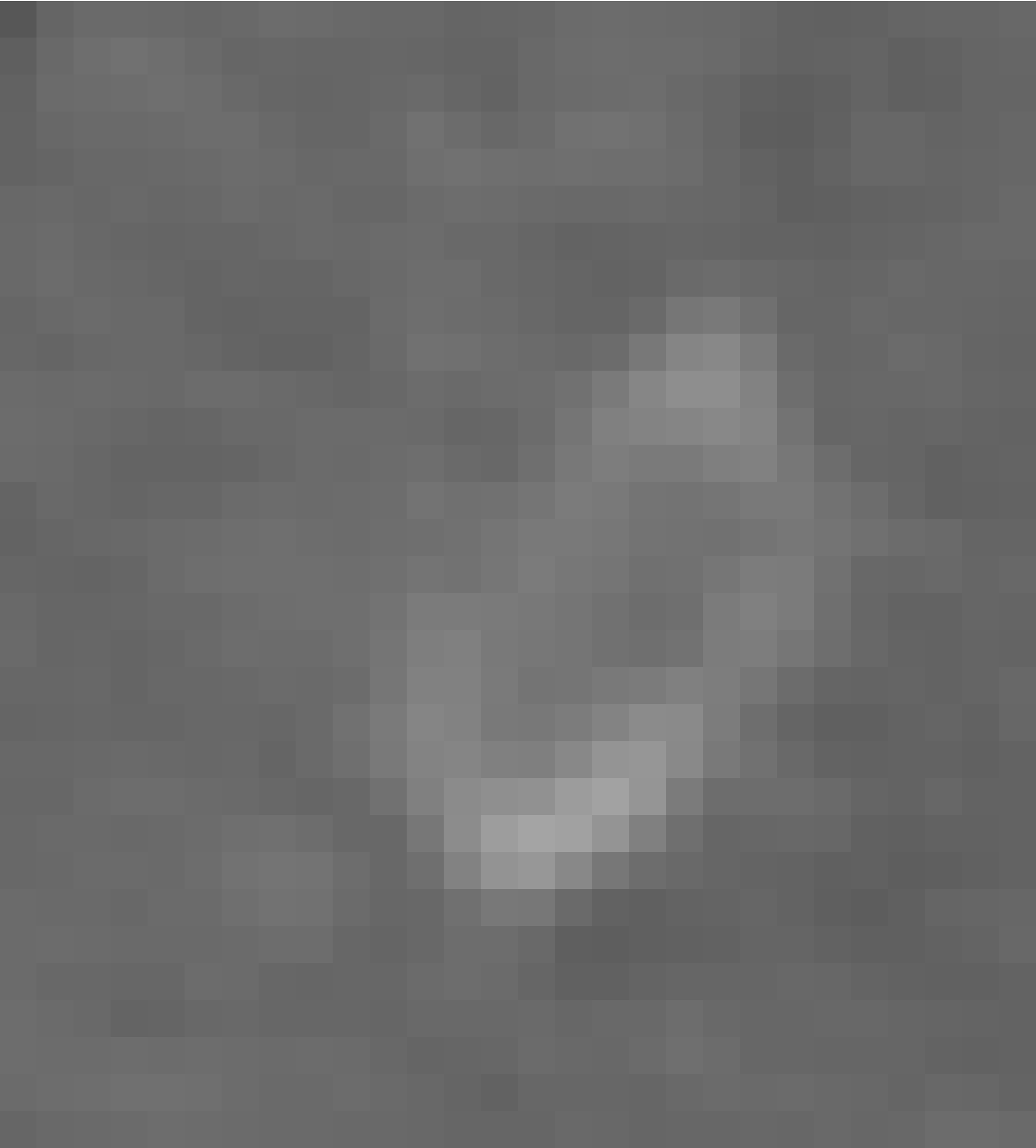}
		\includegraphics[width=0.28\columnwidth]{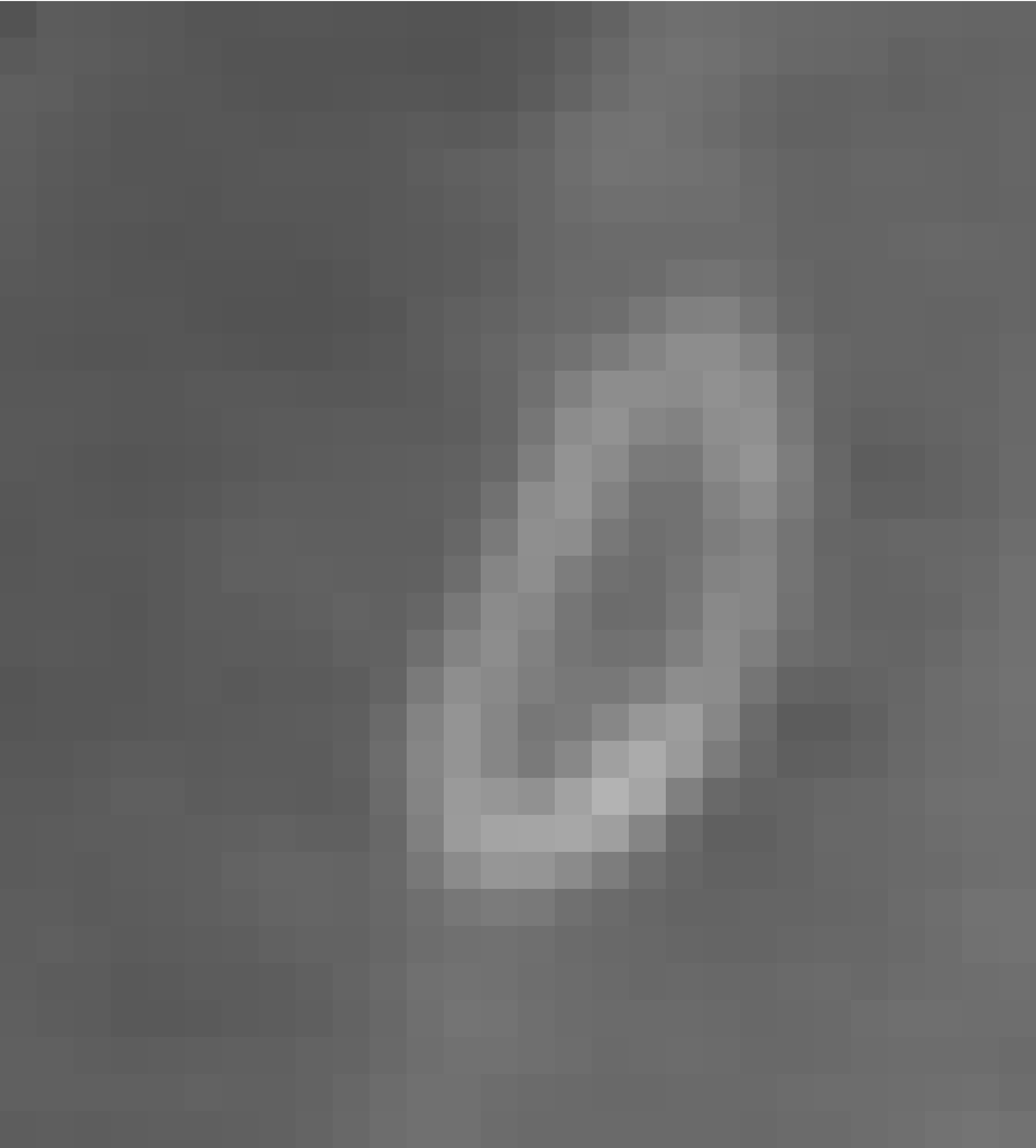}
	}
	\centerline{
		\includegraphics[width=0.28\columnwidth]{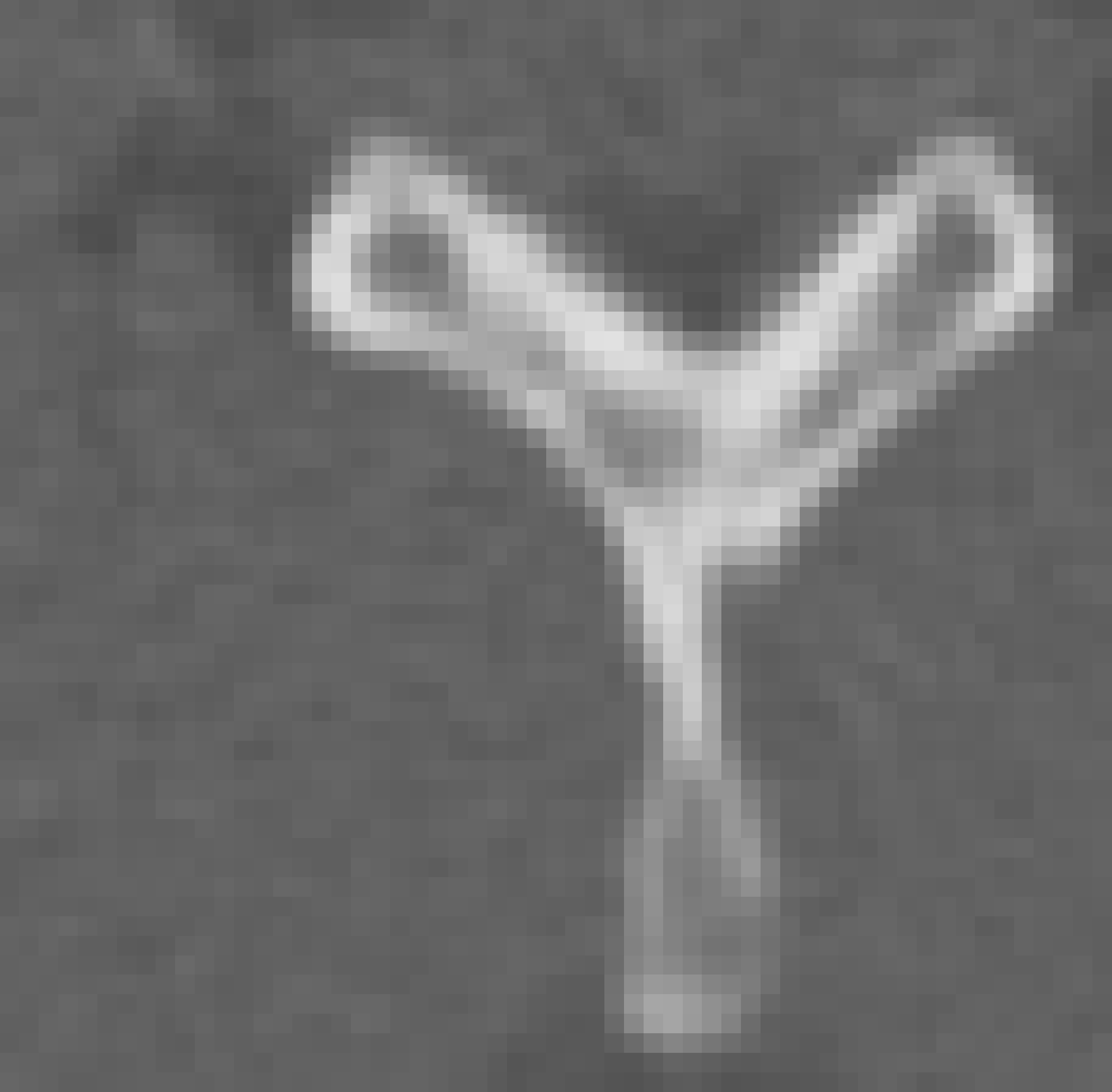}
		\includegraphics[width=0.28\columnwidth]{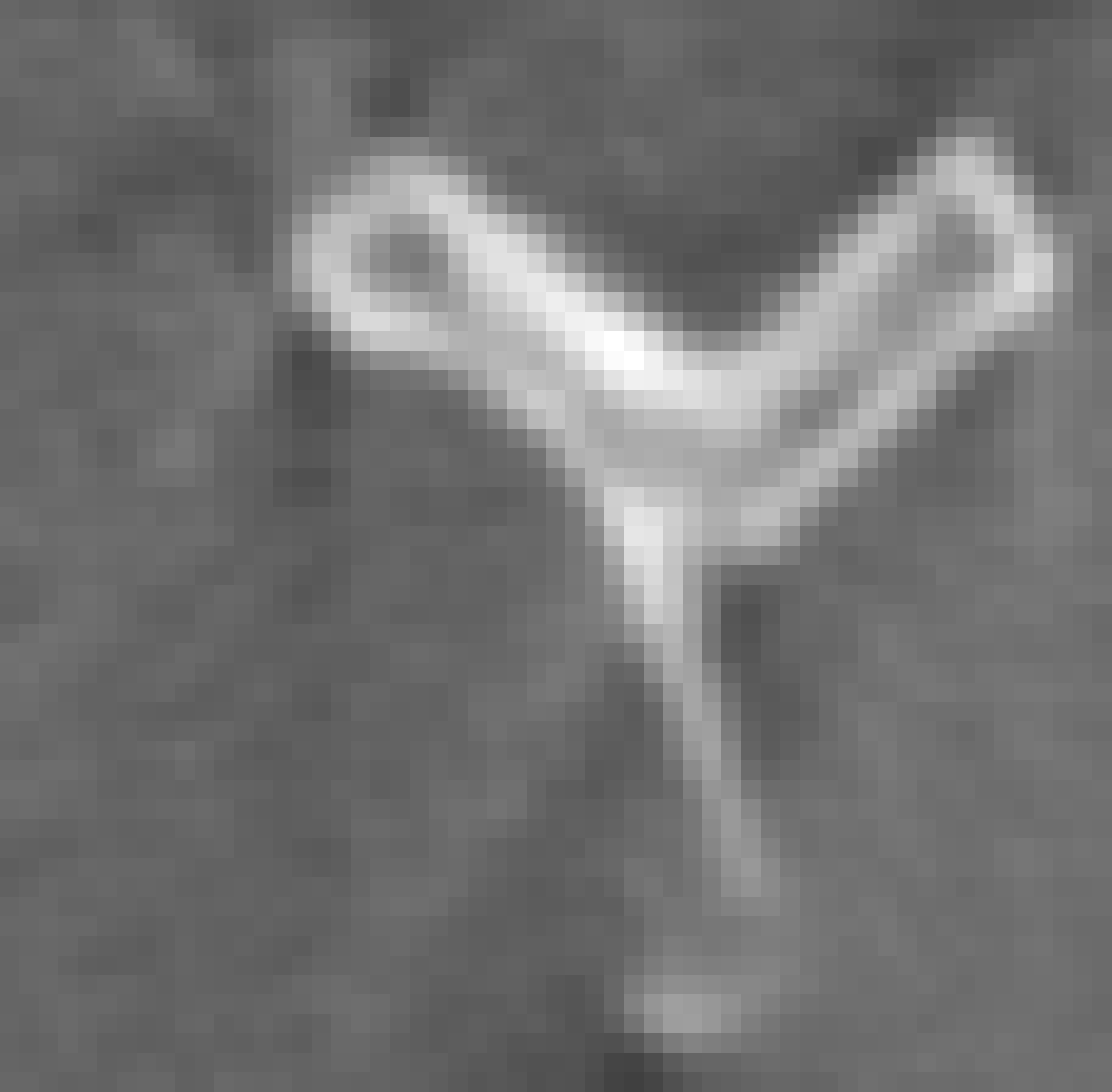}
		\includegraphics[width=0.28\columnwidth]{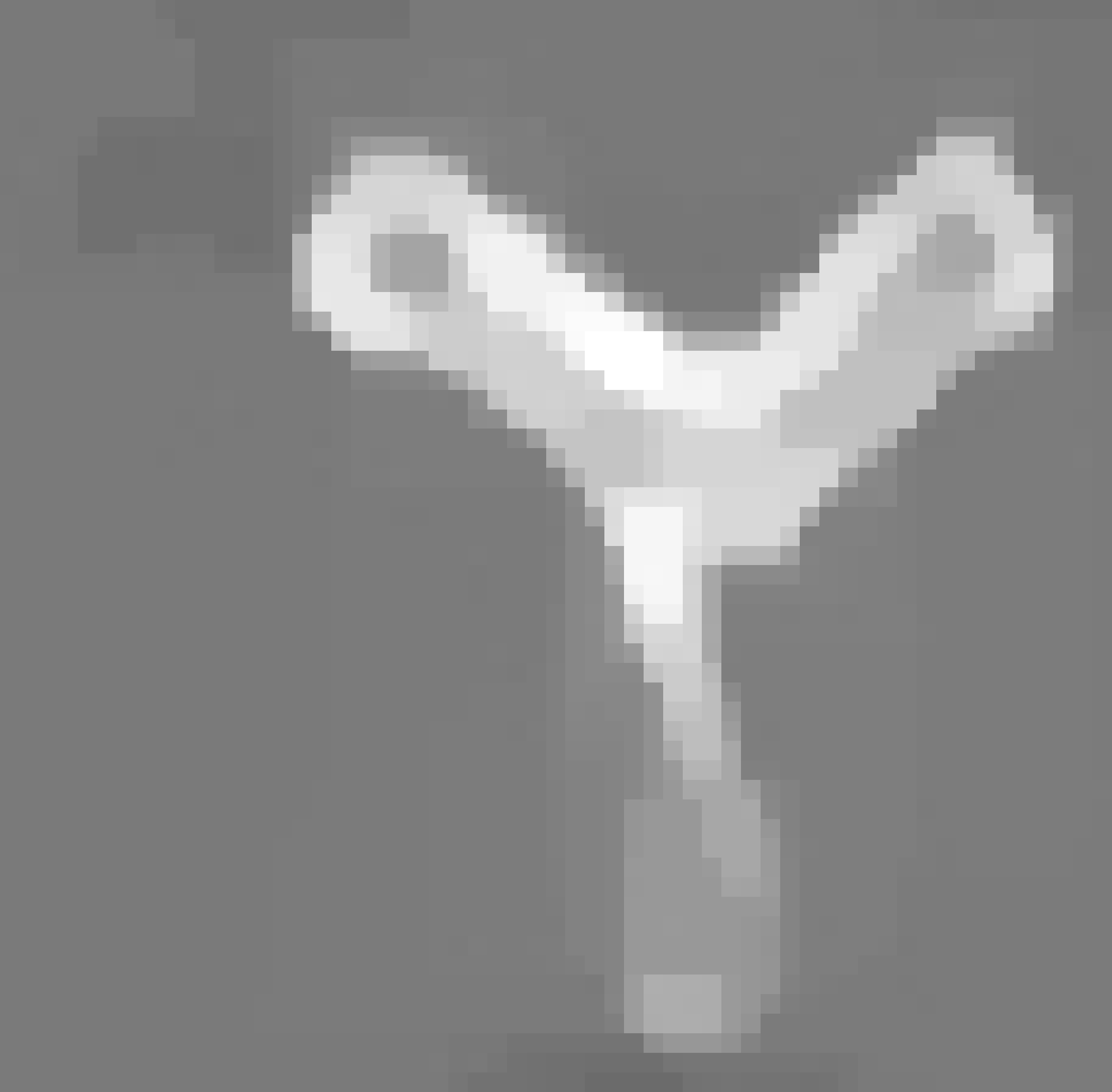}
		\includegraphics[width=0.28\columnwidth]{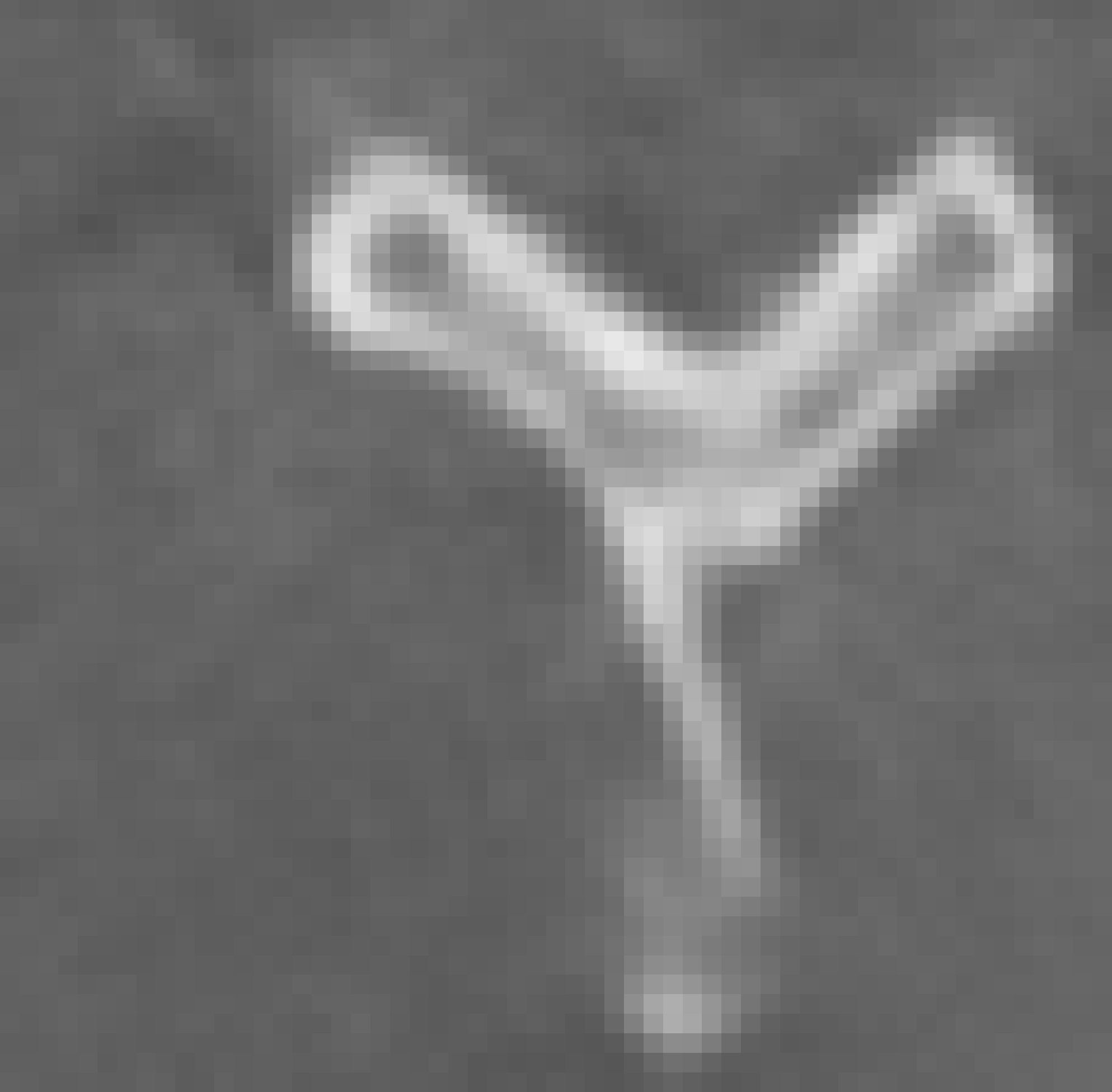}
		\includegraphics[width=0.28\columnwidth]{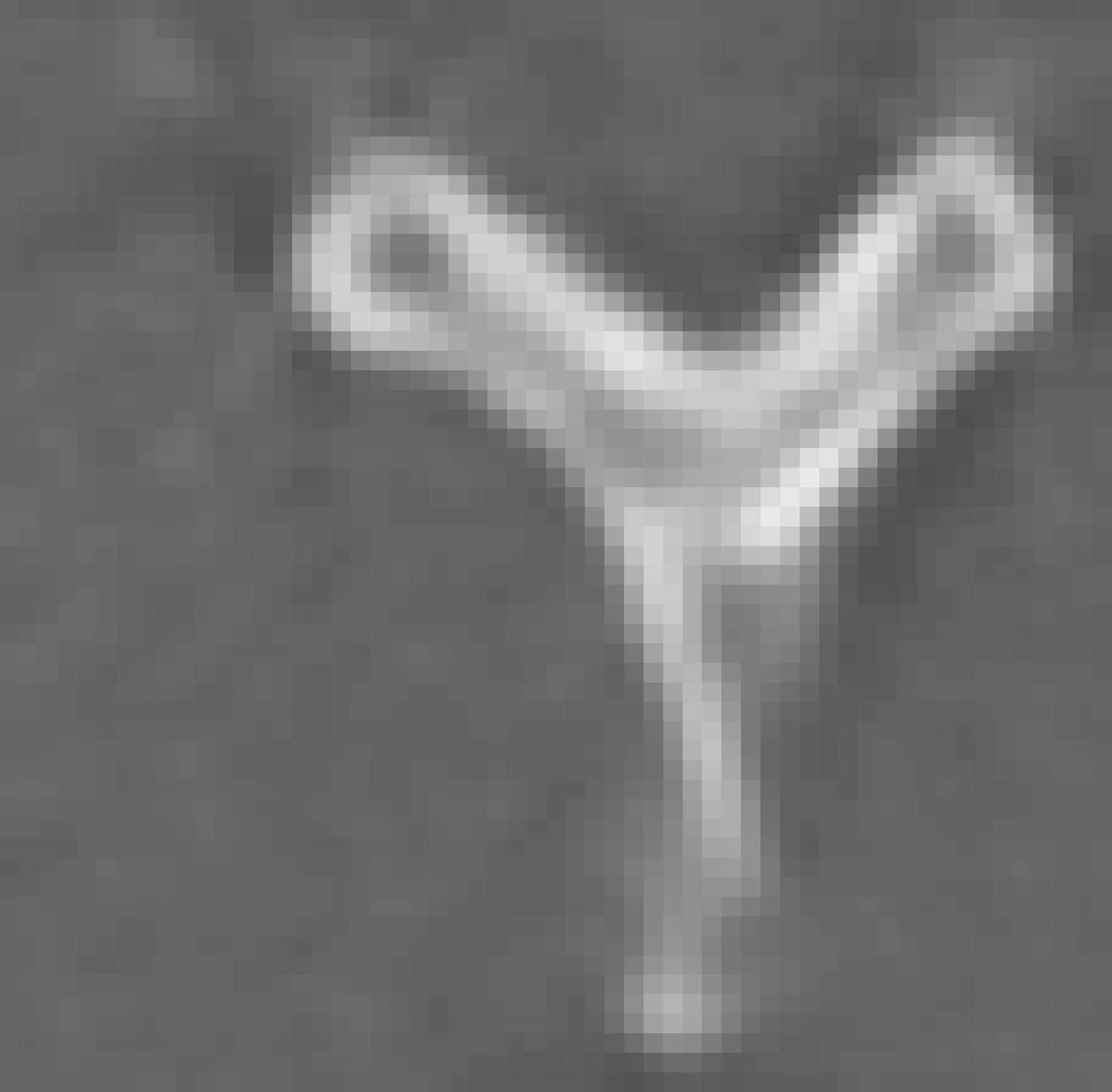}
		\includegraphics[width=0.28\columnwidth]{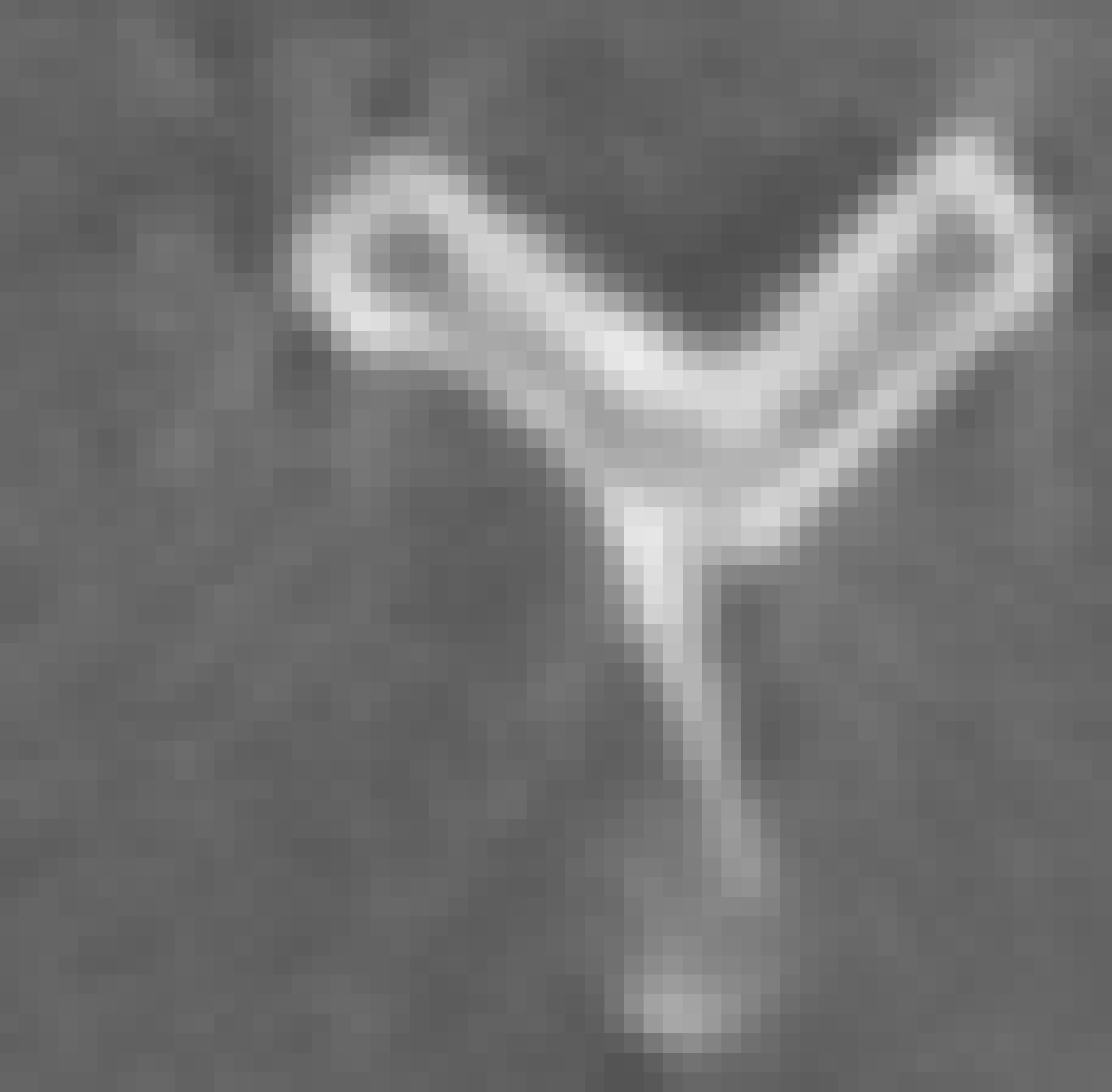}
		\includegraphics[width=0.28\columnwidth]{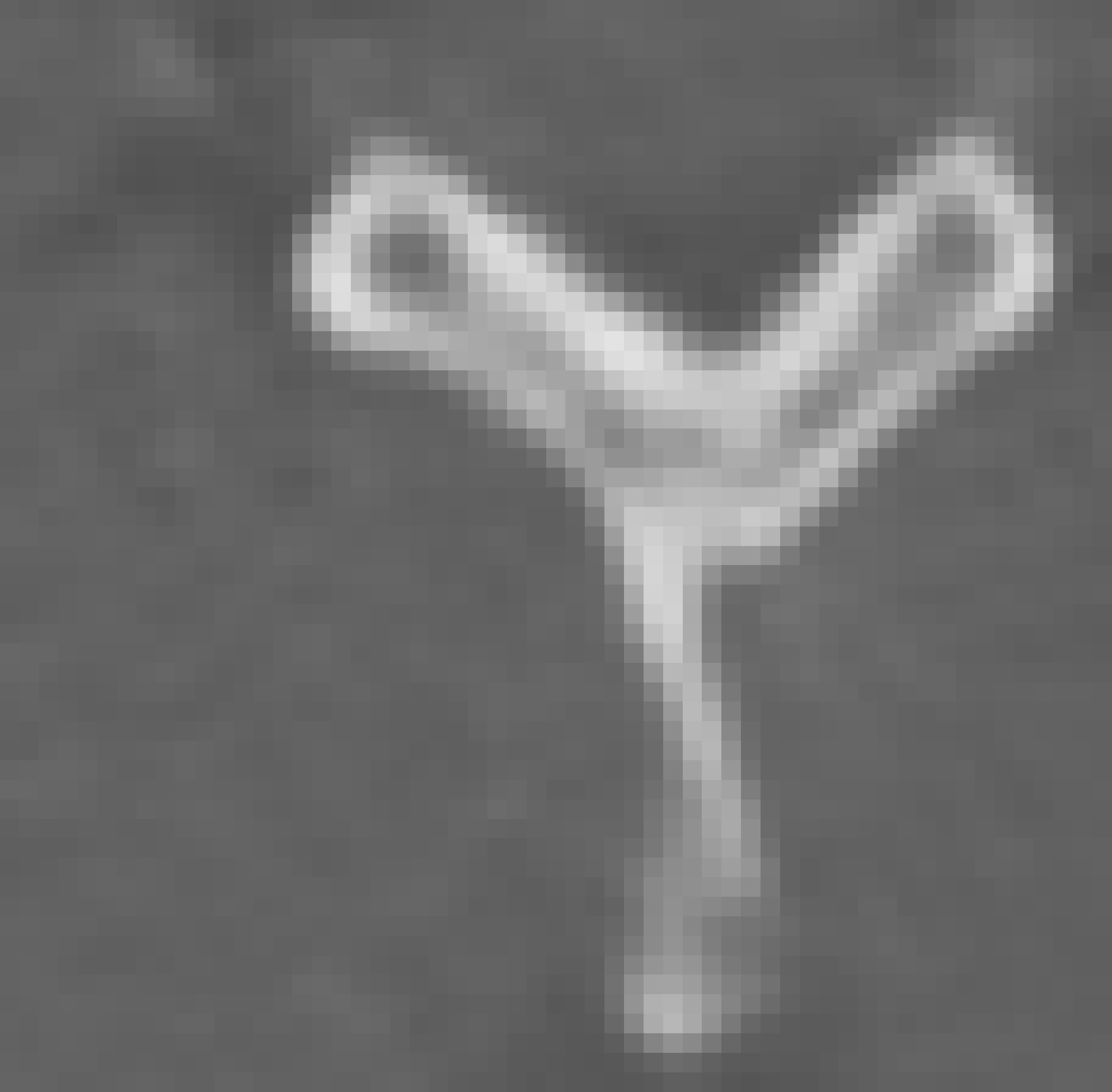}
	}
	\centerline{
		\subfigure[Label]{\includegraphics[width=0.28\columnwidth]{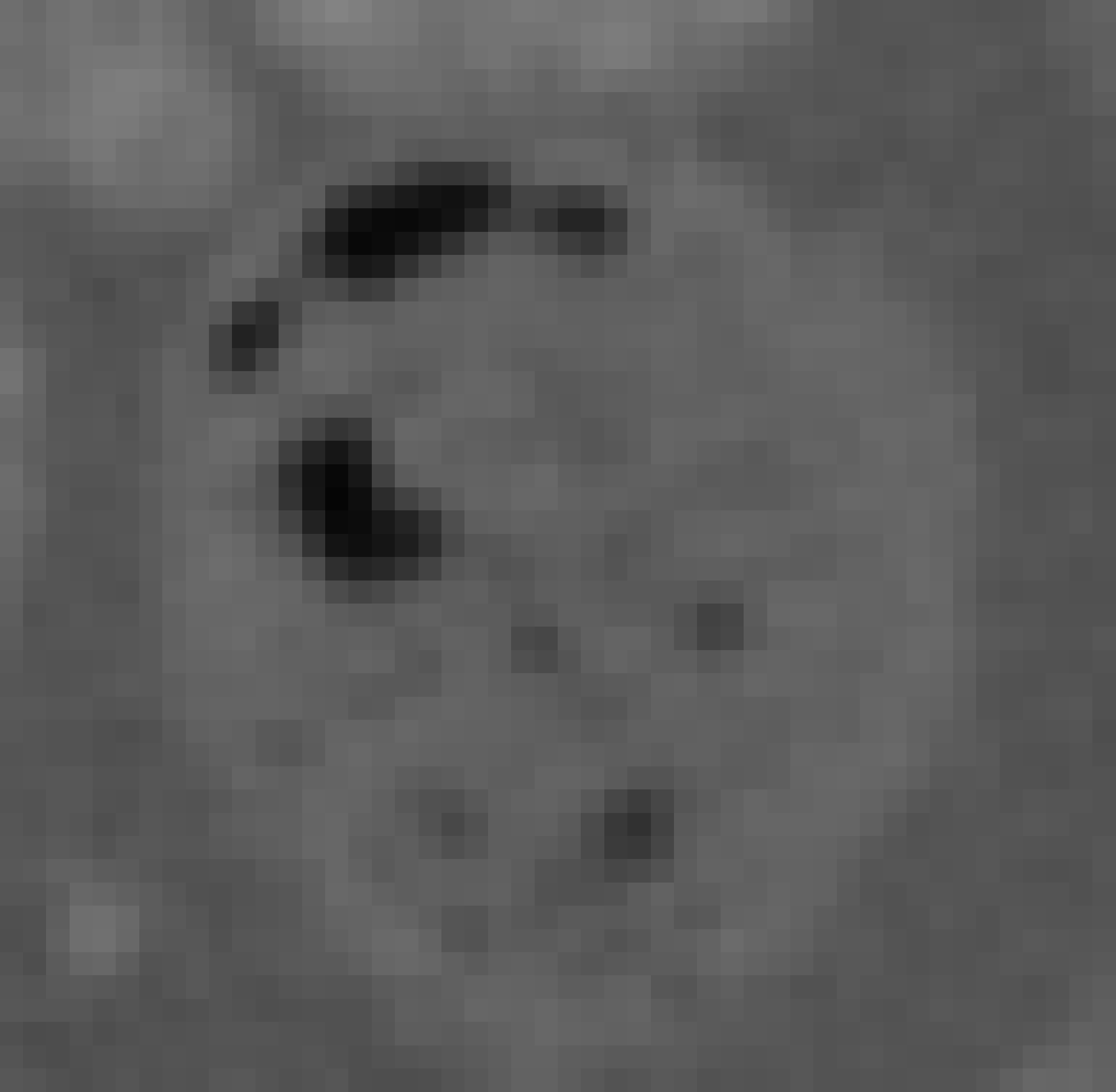}}
		\subfigure[FBP]{\includegraphics[width=0.28\columnwidth]{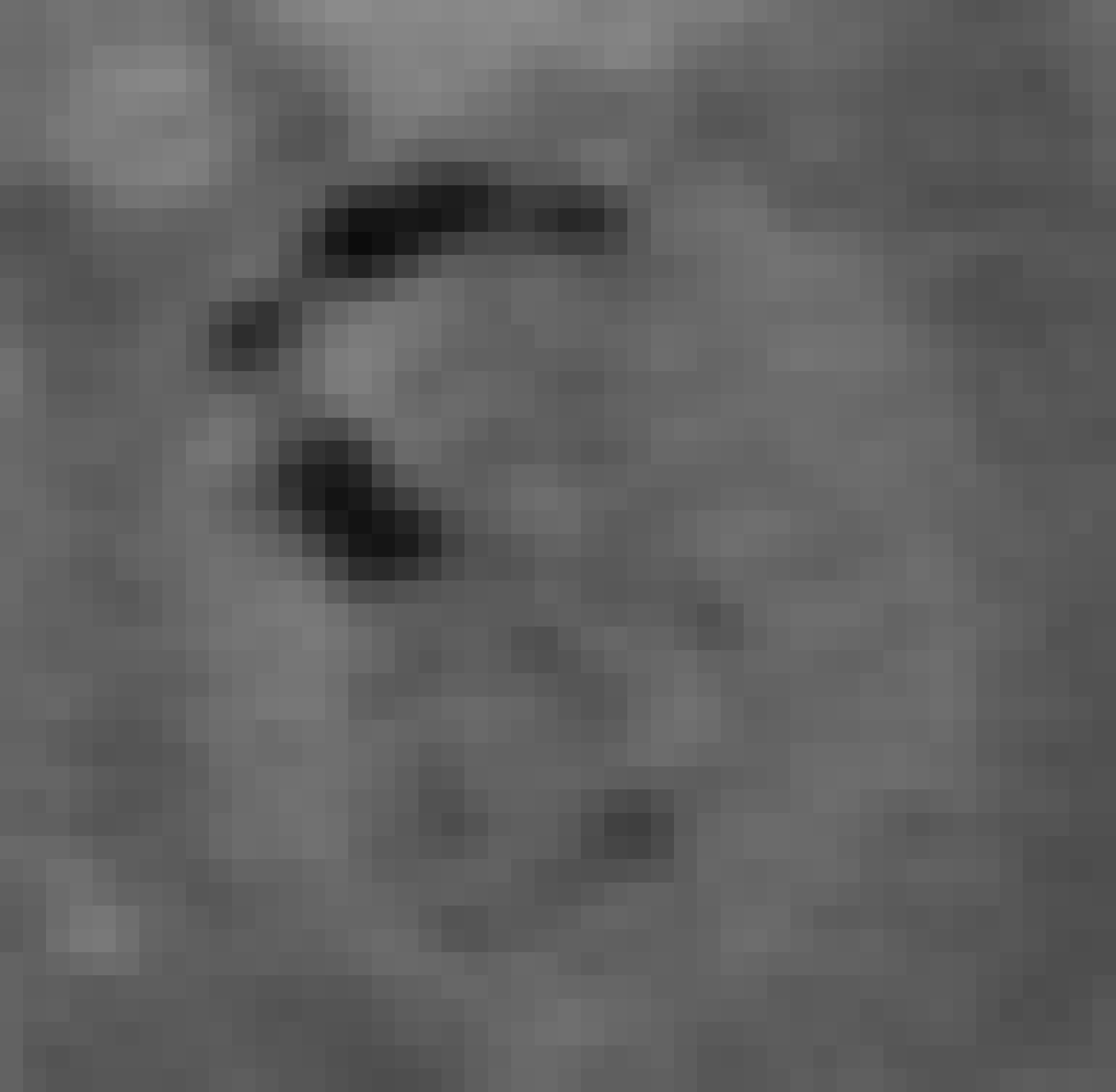}}
		\subfigure[TV-regularization]{\includegraphics[width=0.28\columnwidth]{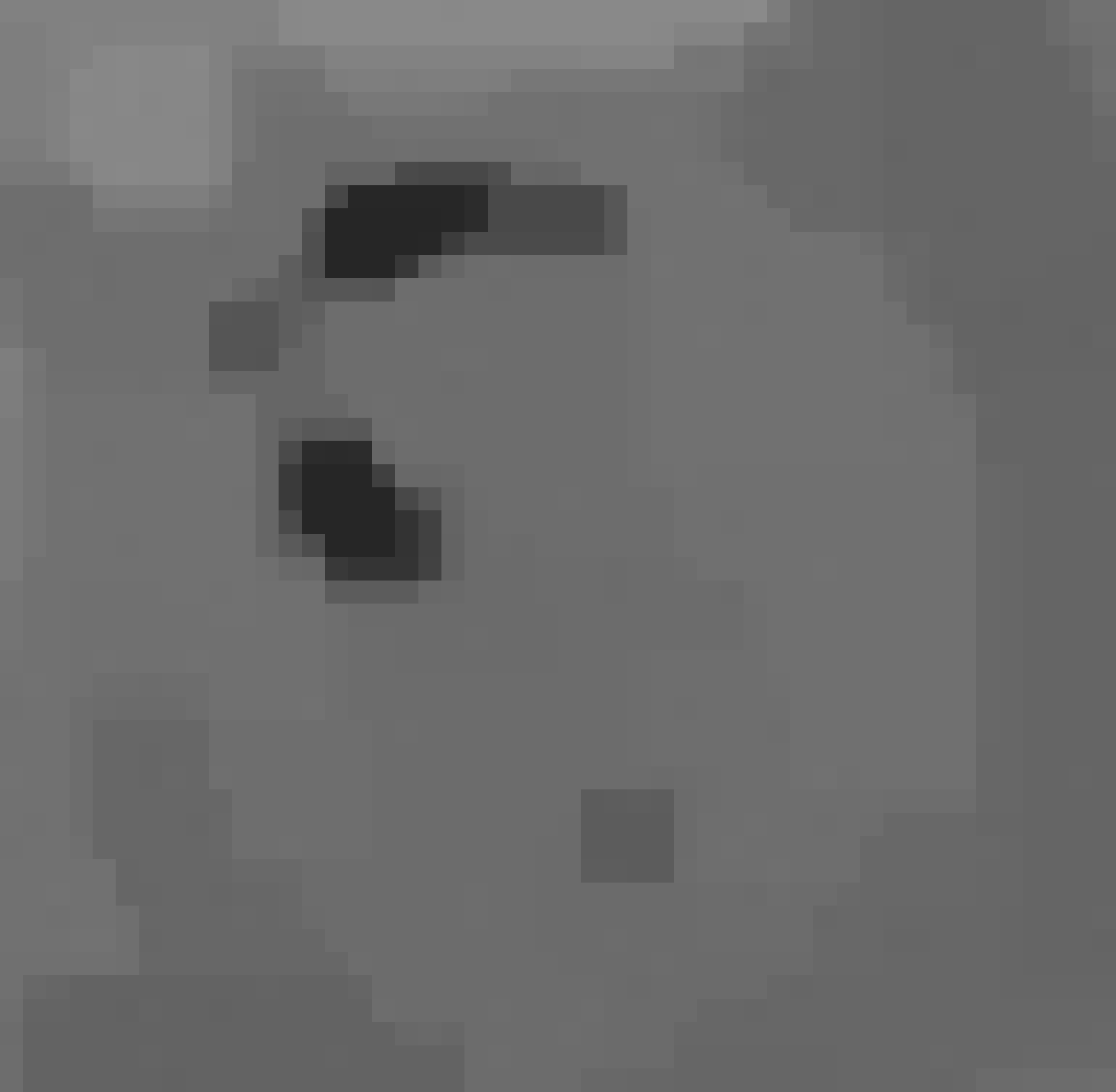}}
		\subfigure[Red-CNN]{\includegraphics[width=0.28\columnwidth]{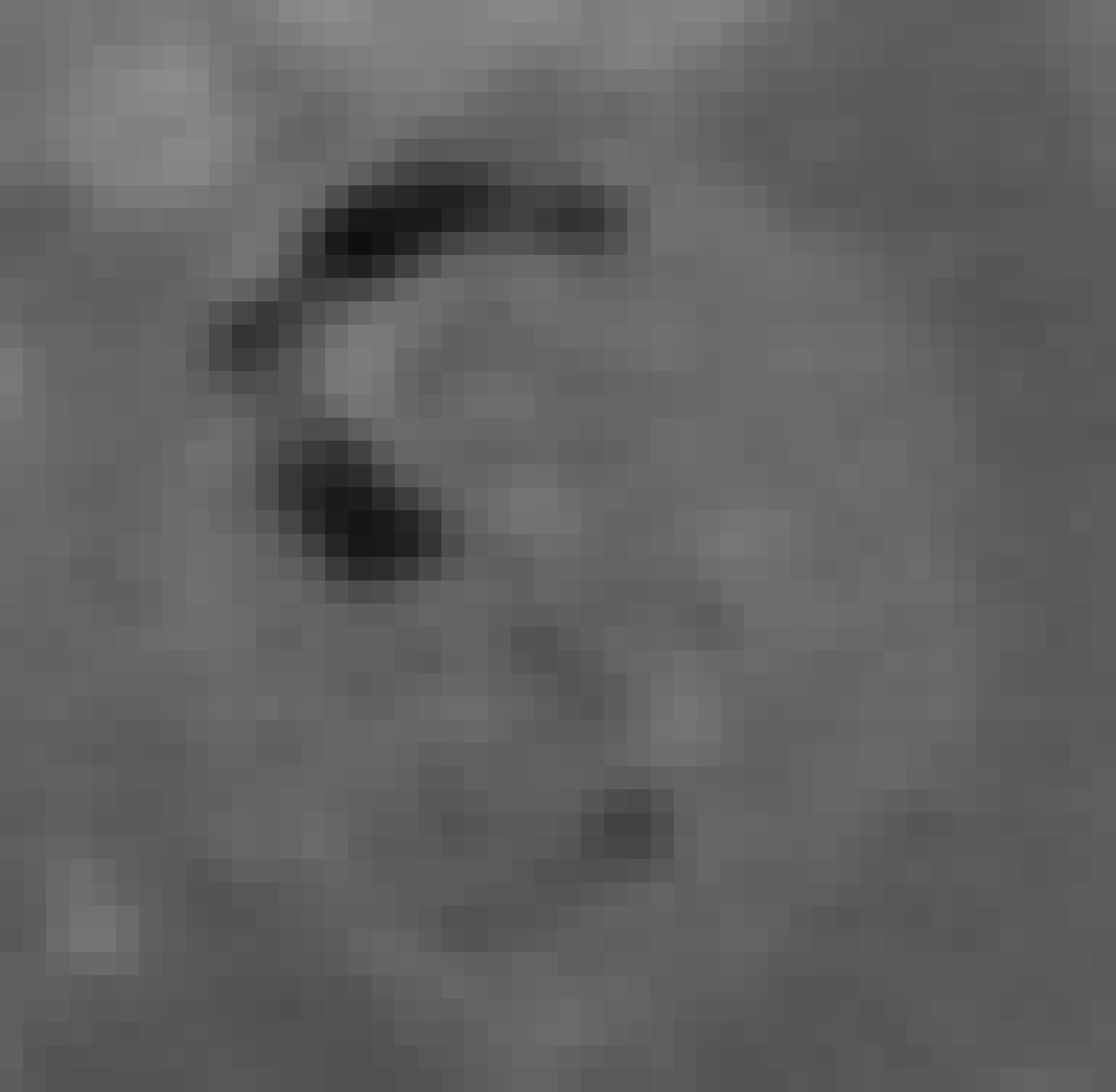}}
		\subfigure[FBP-Conv]{\includegraphics[width=0.28\columnwidth]{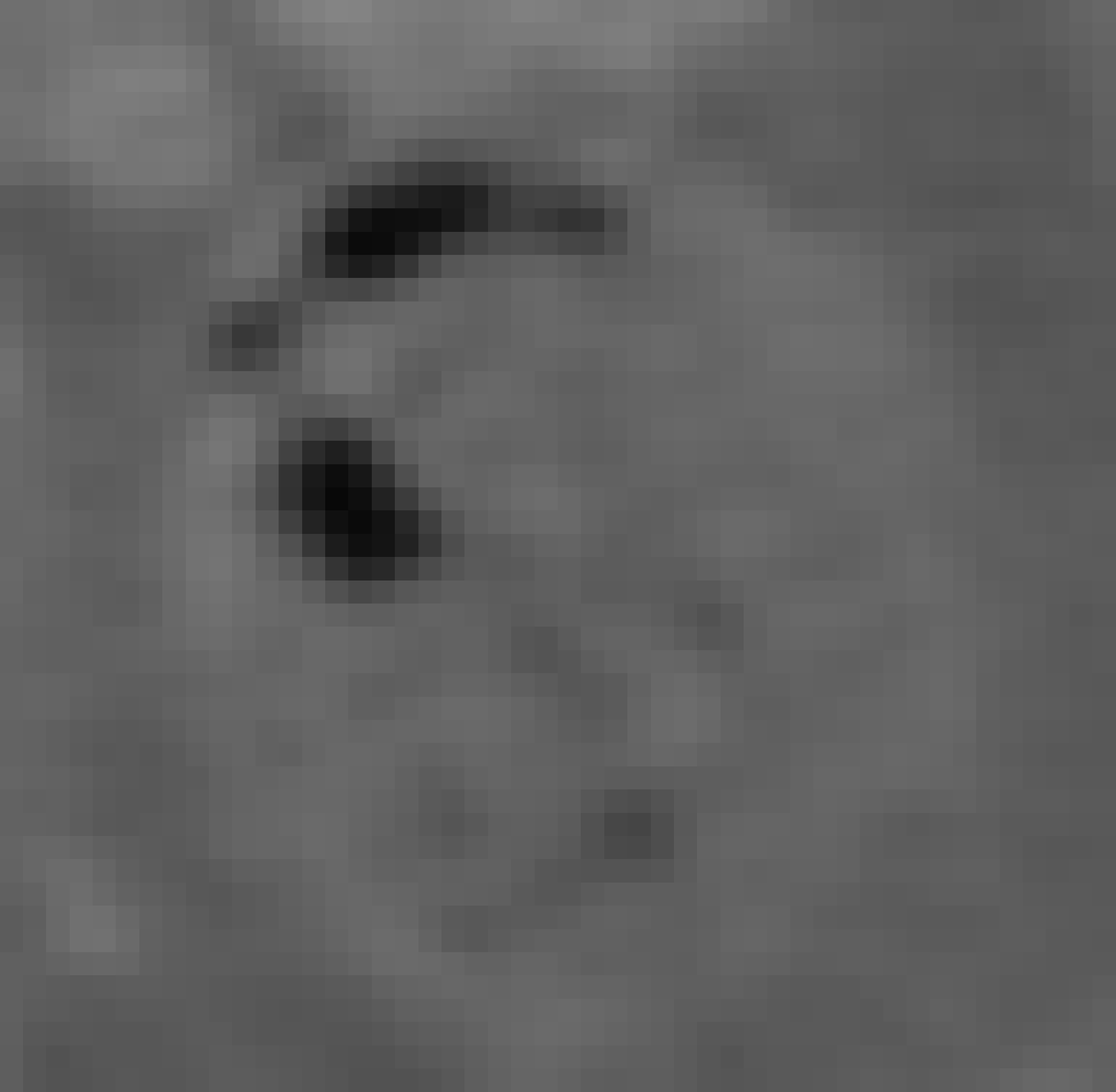}}
		\subfigure[DD-Net]{\includegraphics[width=0.28\columnwidth]{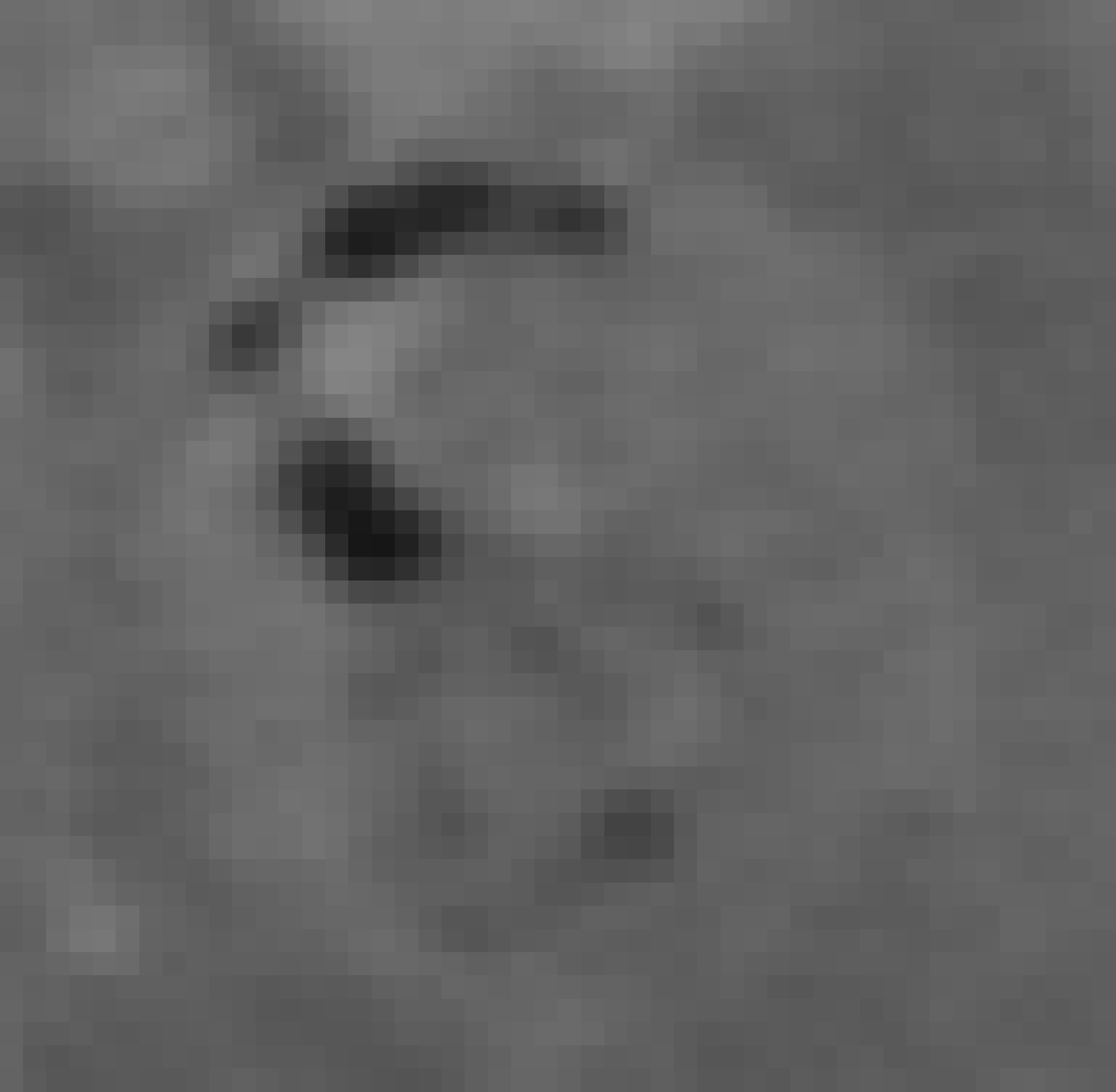}}
		\subfigure[Ours]{\includegraphics[width=0.28\columnwidth]{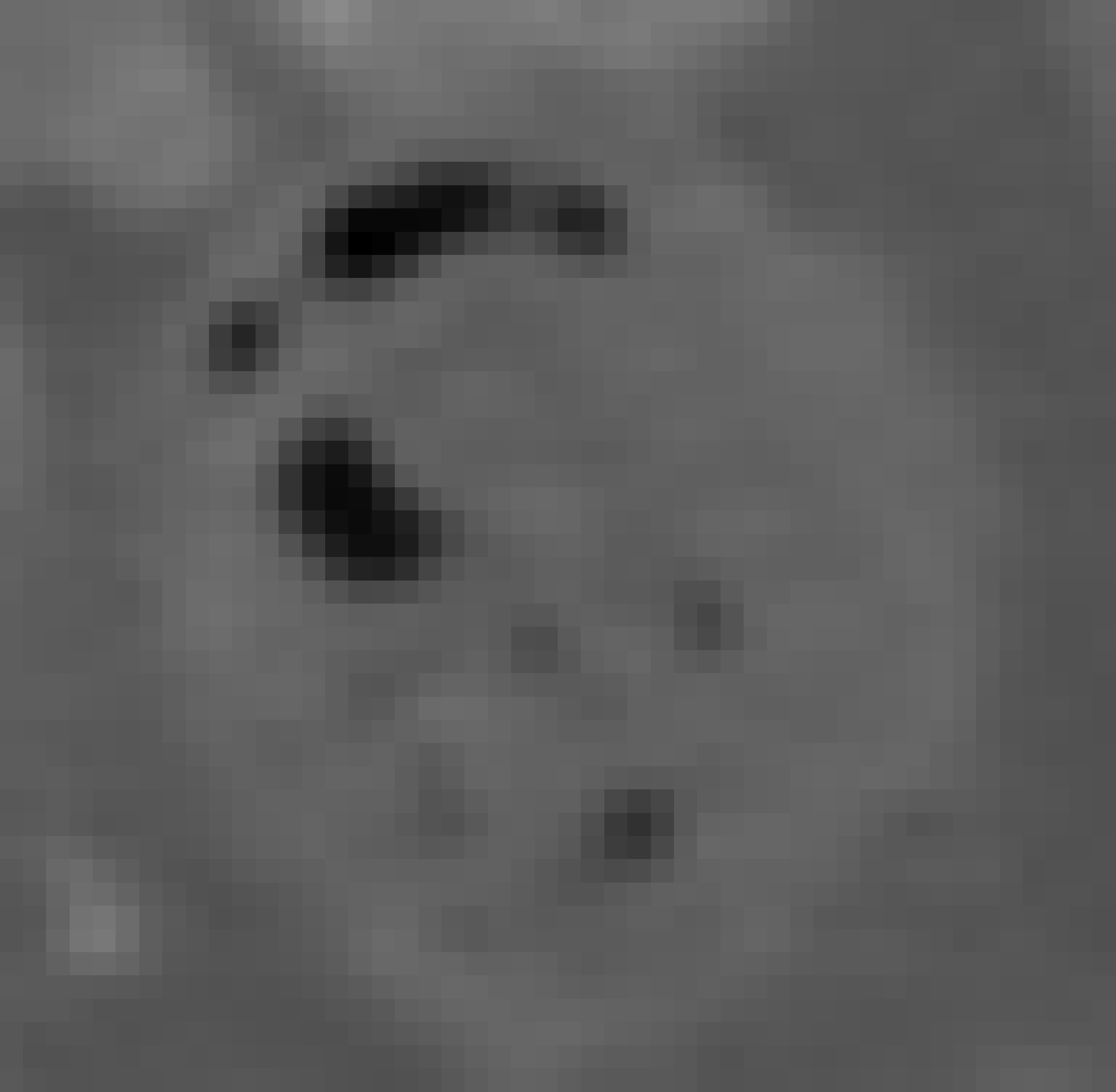}}
	}
	
	\caption{The zoomed regions marked by the red box in Fig. \ref{F5}a.}
	\label{F6}
\end{figure*}
\subsubsection{Data Preparation}
A set of clinical CT images from ``the 2016 NIH-AAPM-Mayo
Clinic Low Dose CT Grand Challenge" \cite{data} authorized by Mayo Clinics are used to be the training and test data. We randomly choose 1500 full dose CT images (of size 512$\times$512) from
this data-set as our CT image labels $\varsigma^{label}$, where 995 of them
are used to train the networks, another 5 of them are used as the
validation set and the other 500 as the test set. We use the Matlab function ``Radon" to  generate the sinogram labels $z^{label}$ of size $725\times180$, where 180 is the number of scanning angles corresponding to $\frac{\pi}{180}\times[0:1:179]$ and $725$ is the number of detectors corresponding to $[-362:1:362]$.  We downsample  $z^{label}$ at angles $\frac{\pi}{180}\times[0:1:149]$ to get the simulated limited-angle sinogram $g$ of size $150\times725$.  The limited-angle sinogram $g$ is the input of our network and  the reconstructed CT image from $g$ by the FBP algorithm is used as the initial guesse of the CT image $u_0$. For the  compared networks, Red-CNN, FBP-Conv and DD-Net, the reconstructed CT image from $g$ by the FBP algorithm is also used as their inputs.

\subsubsection{Parameter Setup}
The parameters $\Theta_\varsigma^n$ and $\Theta_z^n$ in $Res_\varsigma^n(\cdot)$ and $Res_z^n(\cdot)$ of our network are automatically initialized by Tensorflow using  the default values. The initial values of parameters $\Theta_u$ in $Merge_u^n(\cdot)$ are set as $t_1=1$ and $t_2=t_3=t_4=0.1$. The number of iterations is set as $N_{iter}=5$. The batch size is set as 1  and the number of training epochs is $100$.

The parameters for Red-CNN, FBP-Conv, and DD-NeT are set as
described in their corresponding papers and the initial values in their networks are initialized by Tensorflow automatically. The batch size is 5.
The training epochs
for Red-CNN is 100 (since it takes too much time to complete
one epoch iteration, we train Red-CNN by learning only 100
epochs) and those for FBP-Conv and DD-Net are  500. 

The parameters of the TV regularization algorithm for parallel-beam CT are set as
$\lambda_3=100$, $\rho=0.1$, $t_5=0.1$, the number of maximal iterations $M_{iter}=300$ and the convergent criteria $\frac{\|u^{n+1}-u^{n}\|^2_2}{\|u^{n+1}\|^2_2}\le\epsilon=10^{-4}$.

\subsubsection{Subjective Evaluation}
Fig. \ref{F5} shows some reconstructed CT images from the test set by the six methods. As can be observed, the results in Fig. \ref{F5}b by the FBP algorithm have some inhomogeneous intensities and  many artifacts. For the TV regularization algorithm,  the parameters need to be tuned carefully to make a tradeoff between the effect of artifact removal and details preservation. We try our best to tune the parameters such that the average Peak Signal to Noise Ratio (PSNR)  of the reconstructed results from the  validation set has the highest value.
From Fig. \ref{F5}c, we  can observe that the results of the TV regularization algorithm  are very smooth and have almost no  artifacts. However, it is at the cost of losing some fine details and structures. From Fig. \ref{F5}d to Fig. \ref{F5}g, it can be observed that
the artifacts are suppressed to different degrees by
the different networks. However, we can find that our method is best
for artifact removing and  small structures preserving through visual inspection.

To better demonstrate that our method can simultaneously  remove artifacts and preserve more fine details, Fig. \ref{F6} shows the zoomed regions marked by the  red box in Fig. \ref{F5}a. It can be easily observed that  the results of our method have more well-defined boundaries and
fine details compared to those of the other methods. 

\subsubsection{Objective Evaluation}

To objectively evaluate the performances of  these  methods, the PSNR and  Structural SIMilarity (SSIM) are used to measure the similarities between the reconstructed CT images and the label CT images. The PSNR and SSIM between the reconstructed CT image $u$ and the referenced image $u^\text{label}$ are, respectively, defined as
\begin{equation}
\begin{aligned}
\text{PSNR}&=\log_{10}\frac{(\max(u^\text{label})-\min(u^\text{label}))^2}{\|u-u^\text{label}\|_2^2},\\
\text{SSIM}&=\frac{(2\mu_1\mu_2+c_1)(2\sigma_{1,2}+c_2)}{(\mu_1^2+\mu_2^2+c_1)(\sigma_1^2+\sigma_2^2+c_2)},\\
\end{aligned}
\end{equation}
where $\mu_1$ and $\mu_2$ are the mean values of $u$ and $u^\text{label}$, $\sigma_1$ and $\sigma_2$ are the variances of  $u$ and $u^\text{label}$, respectively, $\sigma_{1,2}$ is the covariance between $u$ and $u^\text{label}$, 
$C_1=(0.01\times(\max(u^\text{label})-\min(u^\text{label})))^2$ and $C_2=(0.03\times(\max(u^\text{label})-\min(u^\text{label})))^2$.

In Table \ref{T1}, the averaged PSNR and SSIM of 500  CT images   reconstructed from the test set by the six methods are listed. From this table, we can see that our method clearly outperforms the other methods and  produces the highest PSNR and SSIM, which agrees with our visual observations. The PSNR of our method is about 1.3db higher than the second highest, DD-Net.

\begin{table}[htbp]
	\caption{{The Averaged PSNR and SSIM of  CT Images Reconstructed by the Six Methods for Parallel-Beam CT}}
	\label{T1}
	\setlength{\tabcolsep}{3pt}
	\centering
	\begin{tabular}{c|c|c}
		\hline
		& PSNR  & SSIM   \\ \hline
		FBP	     &25.15 	&0.43 \\
		TV regularization & 34.92  &0.91\\
		Red-CNN	 &35.58 	&0.92 \\
		FBP-Conv &34.42 	&0.92  \\
		DD-Net	 &36.73	&0.94  \\
		Ours     &\textbf{38.19 }    &\textbf{0.95}   \\
		\hline
	\end{tabular}%
\end{table}%

\begin{figure*}[!t]
	\centerline{
		\includegraphics[width=2\columnwidth]{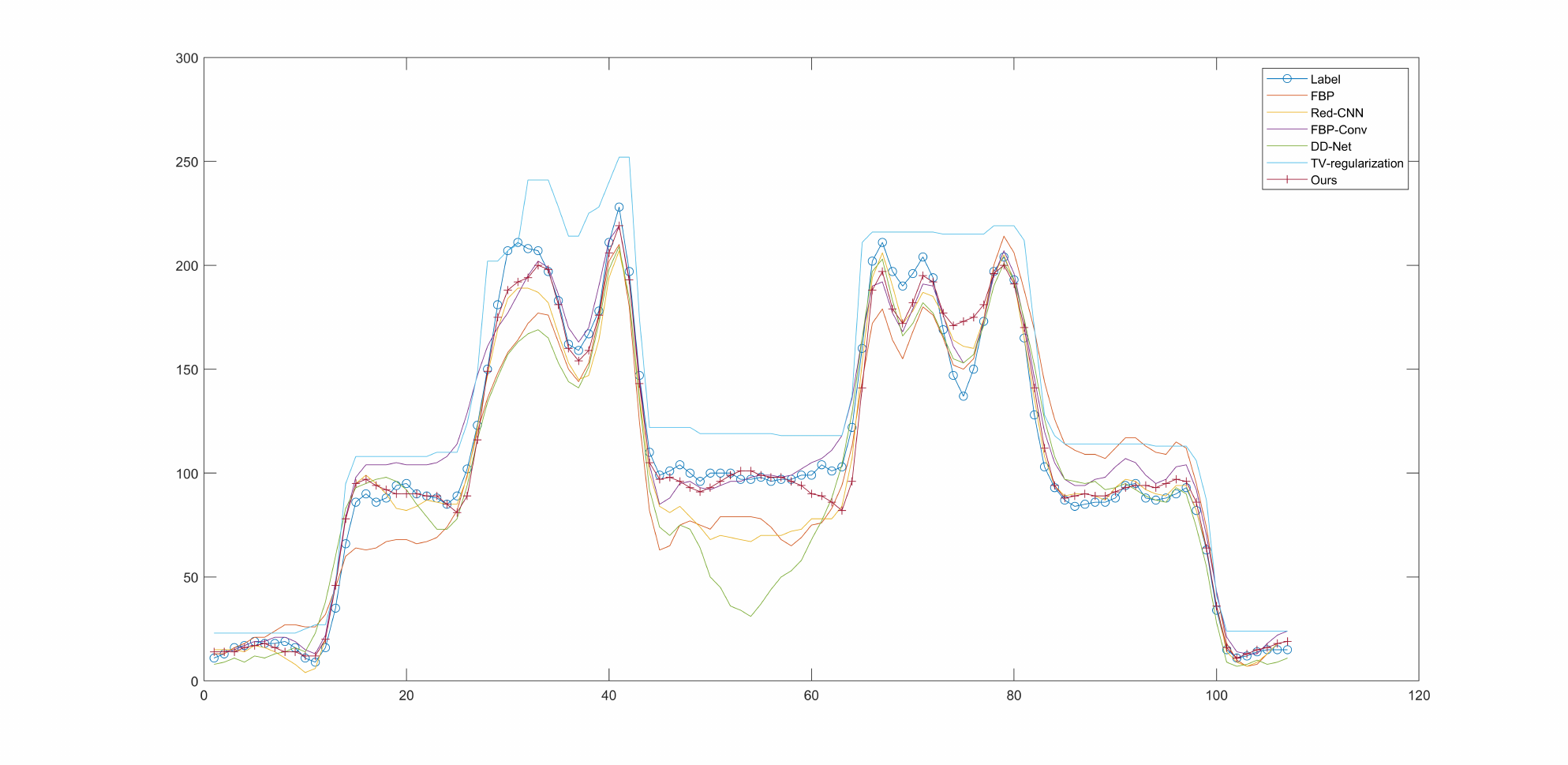}
	}
	
	\caption{The 1D intensity profiles passing through the green line in Row 1 of  Fig. \ref{F5}a.}
	\label{F7}
\end{figure*}
Fig. \ref{F7} depicts the intensity profiles at pixel
positions of i = 356 and j from 211 to 317, corresponding
to the green line in Fig. \ref{F5}a.  Through visual inspection, it can be  observed that the intensity profile of our method is the most consistent to the label image in most locations. This comparison  
demonstrates the advantage of our
method over the other post-processing CT reconstruction networks on edge and detail preservation.

\subsection{Fan-beam}
\begin{figure*}[!t]
	\centerline{
		\includegraphics[width=0.28\columnwidth]{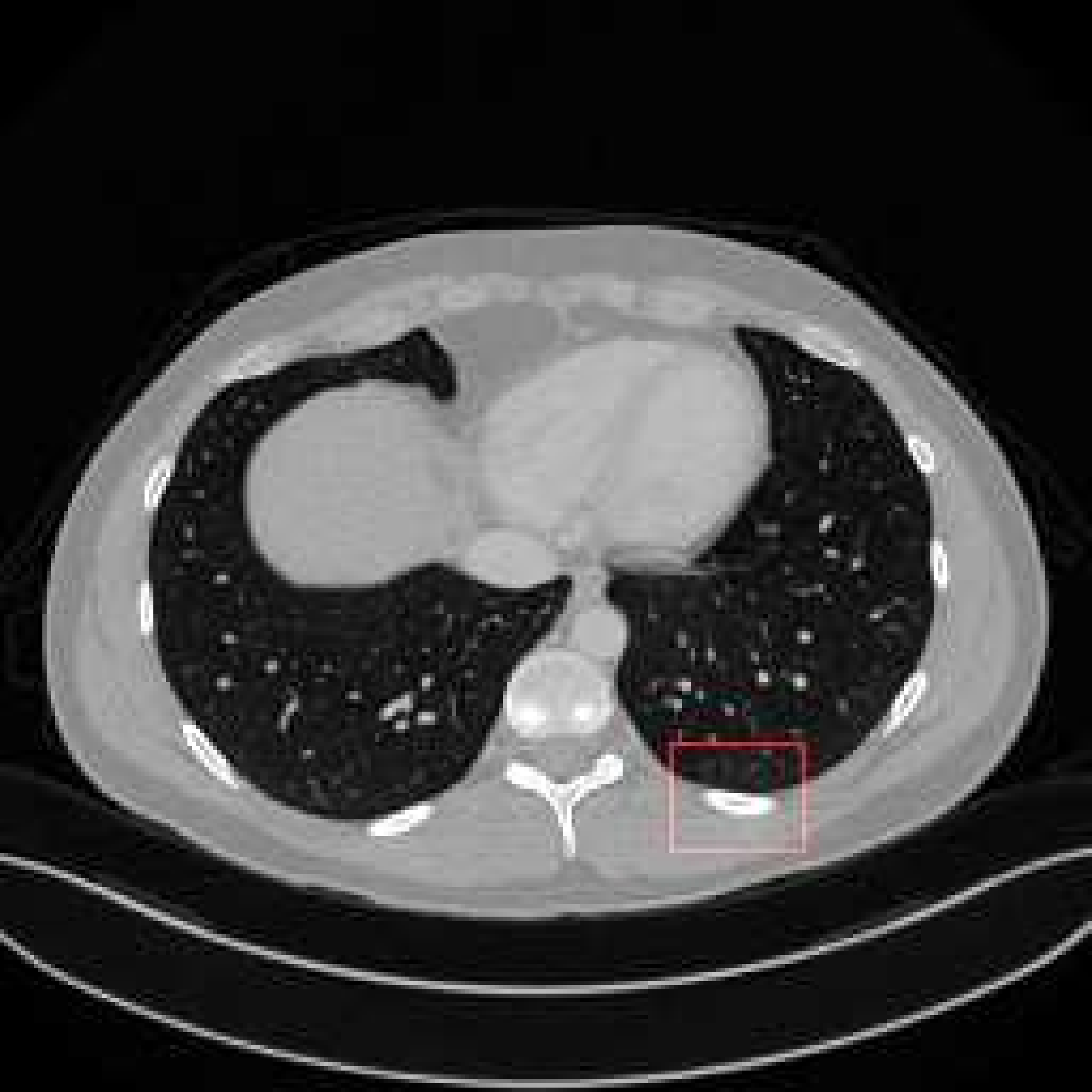}
		\includegraphics[width=0.28\columnwidth]{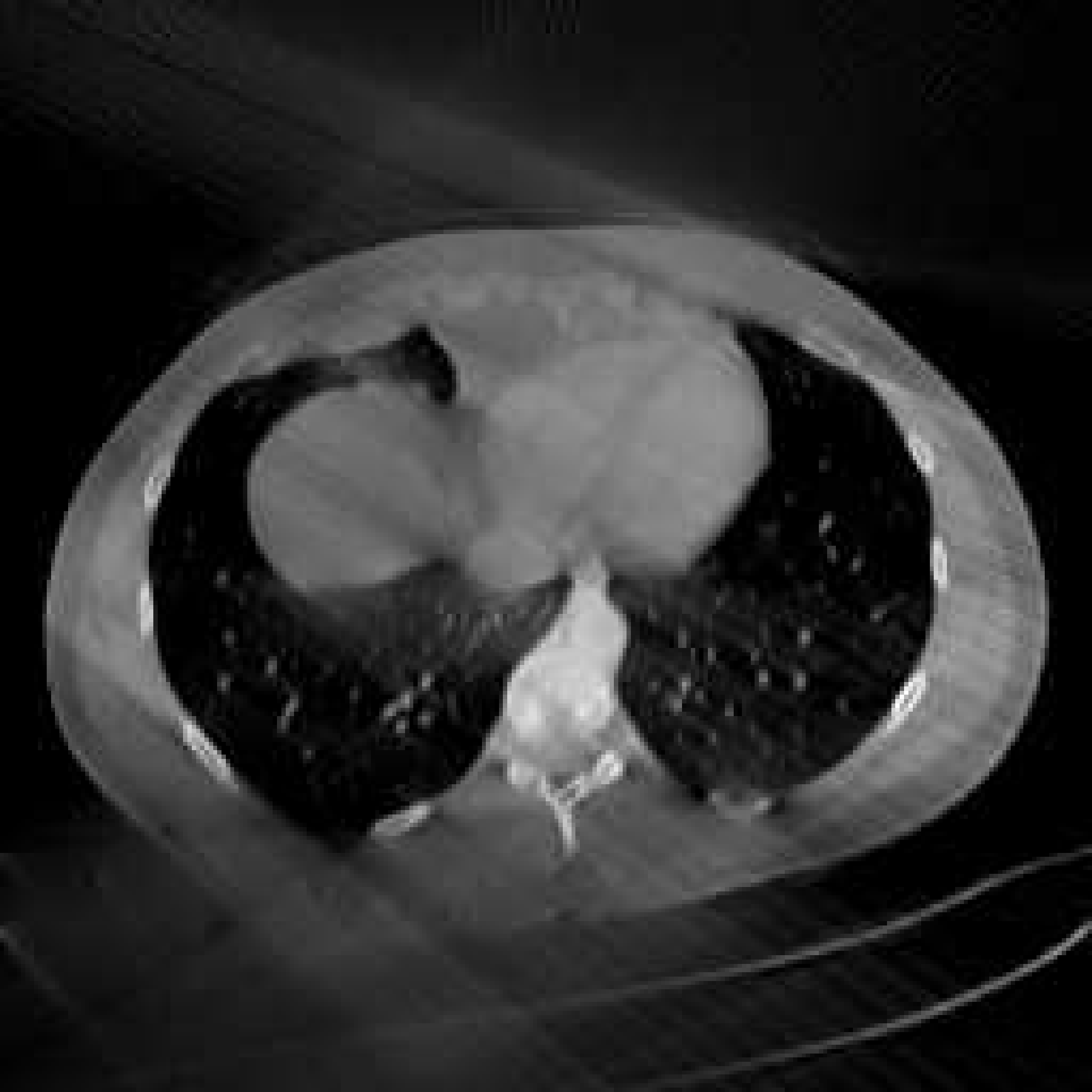}
		\includegraphics[width=0.28\columnwidth]{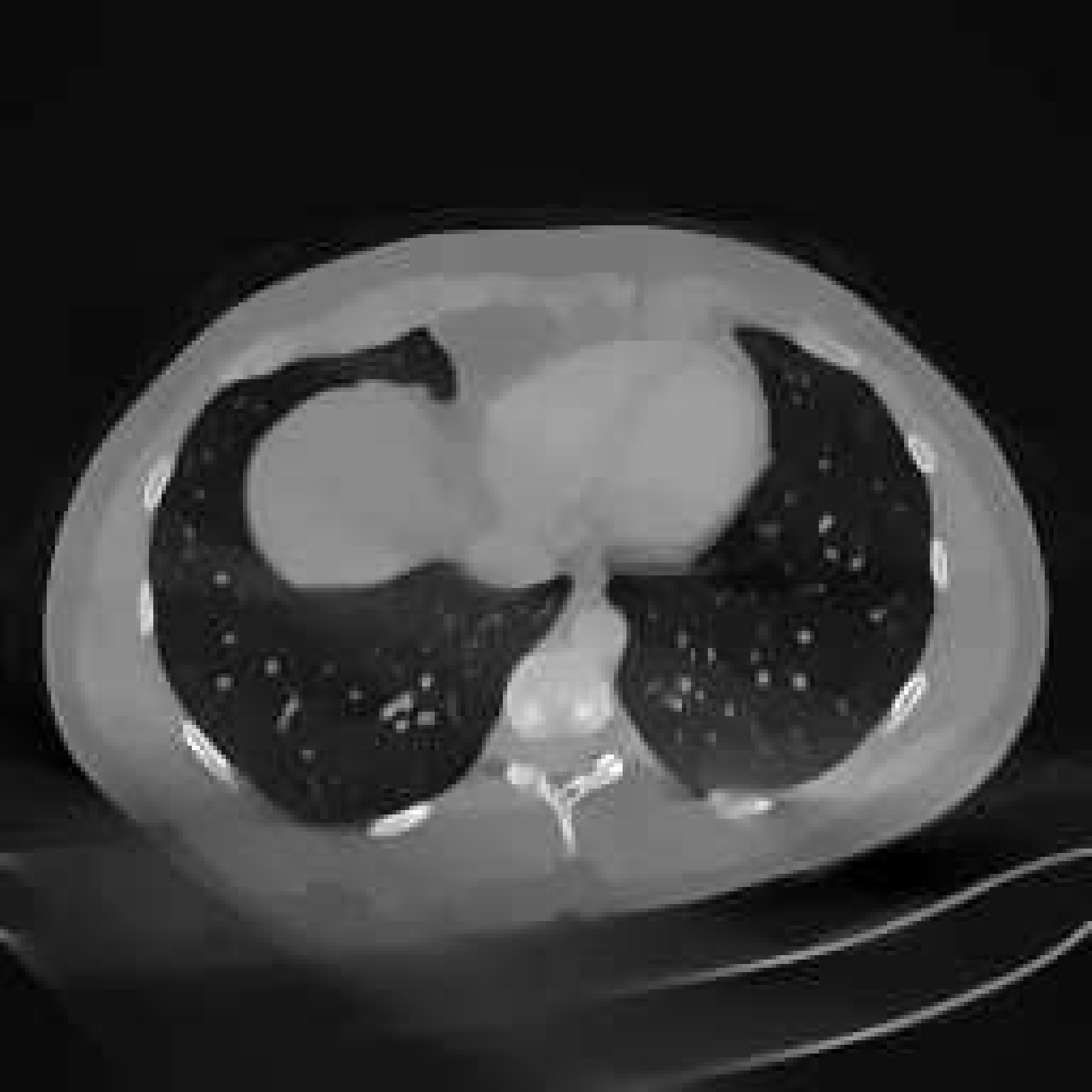}
		\includegraphics[width=0.28\columnwidth]{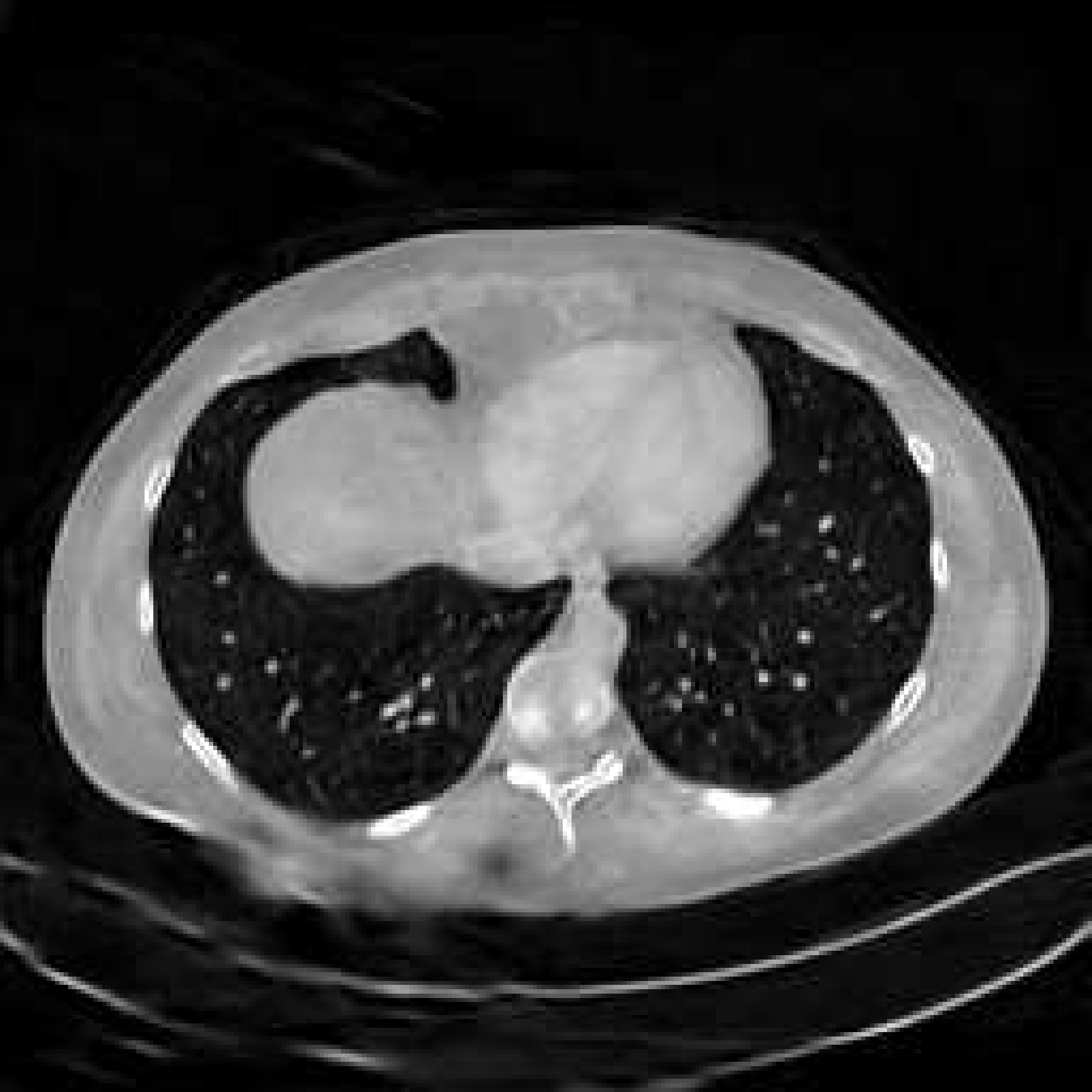}
		\includegraphics[width=0.28\columnwidth]{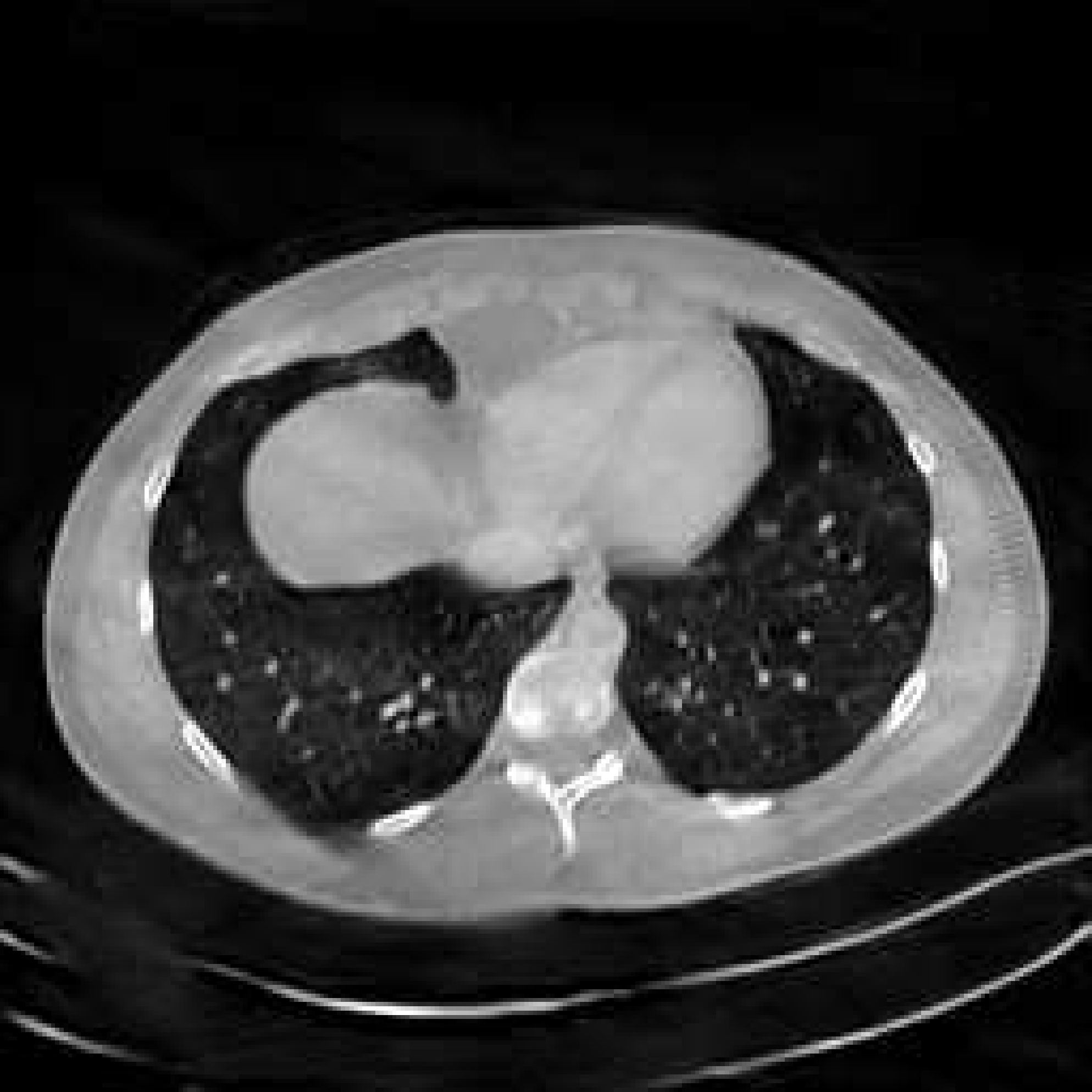}
		\includegraphics[width=0.28\columnwidth]{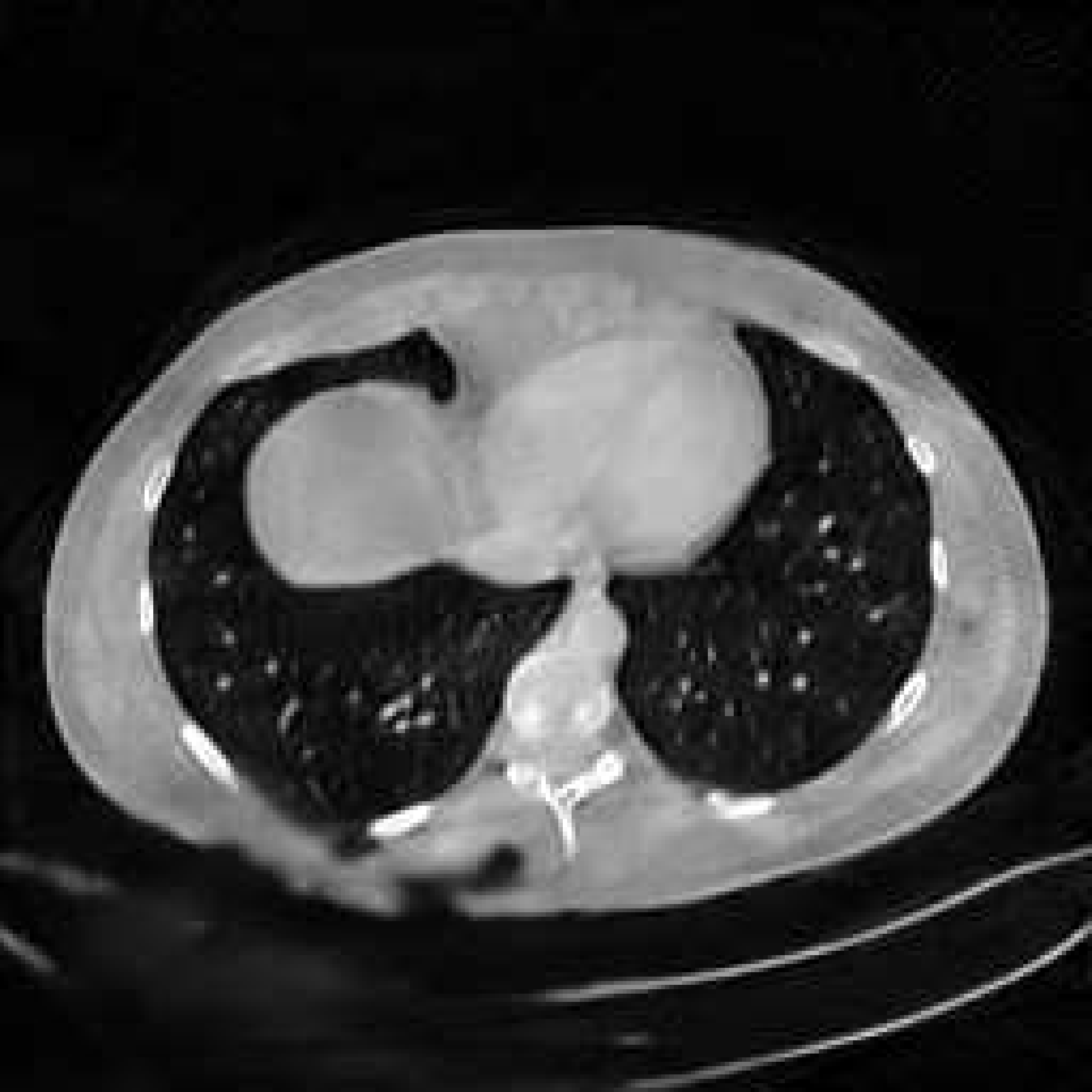}
		\includegraphics[width=0.28\columnwidth]{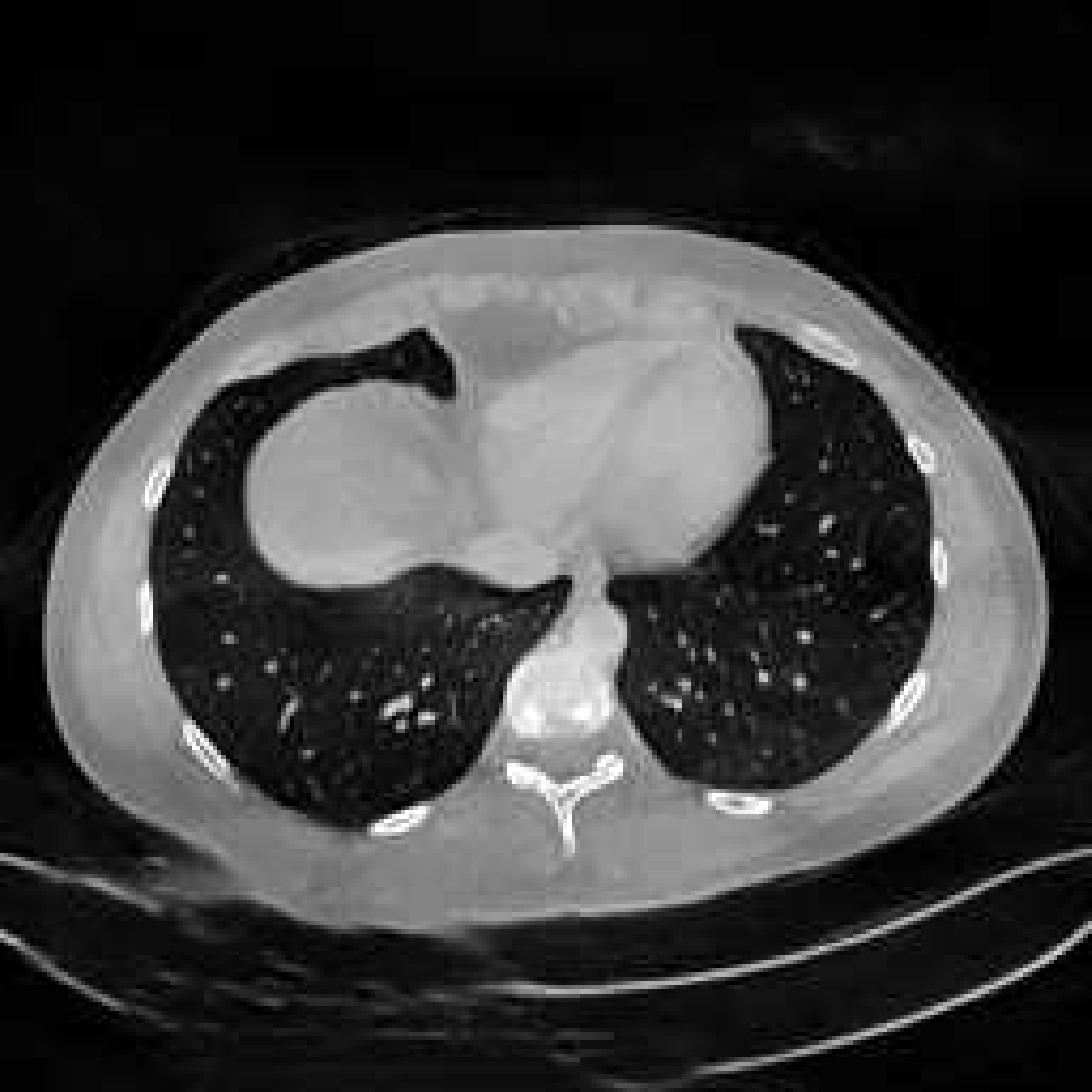}
	}
	
	\centerline{
		\includegraphics[width=0.28\columnwidth]{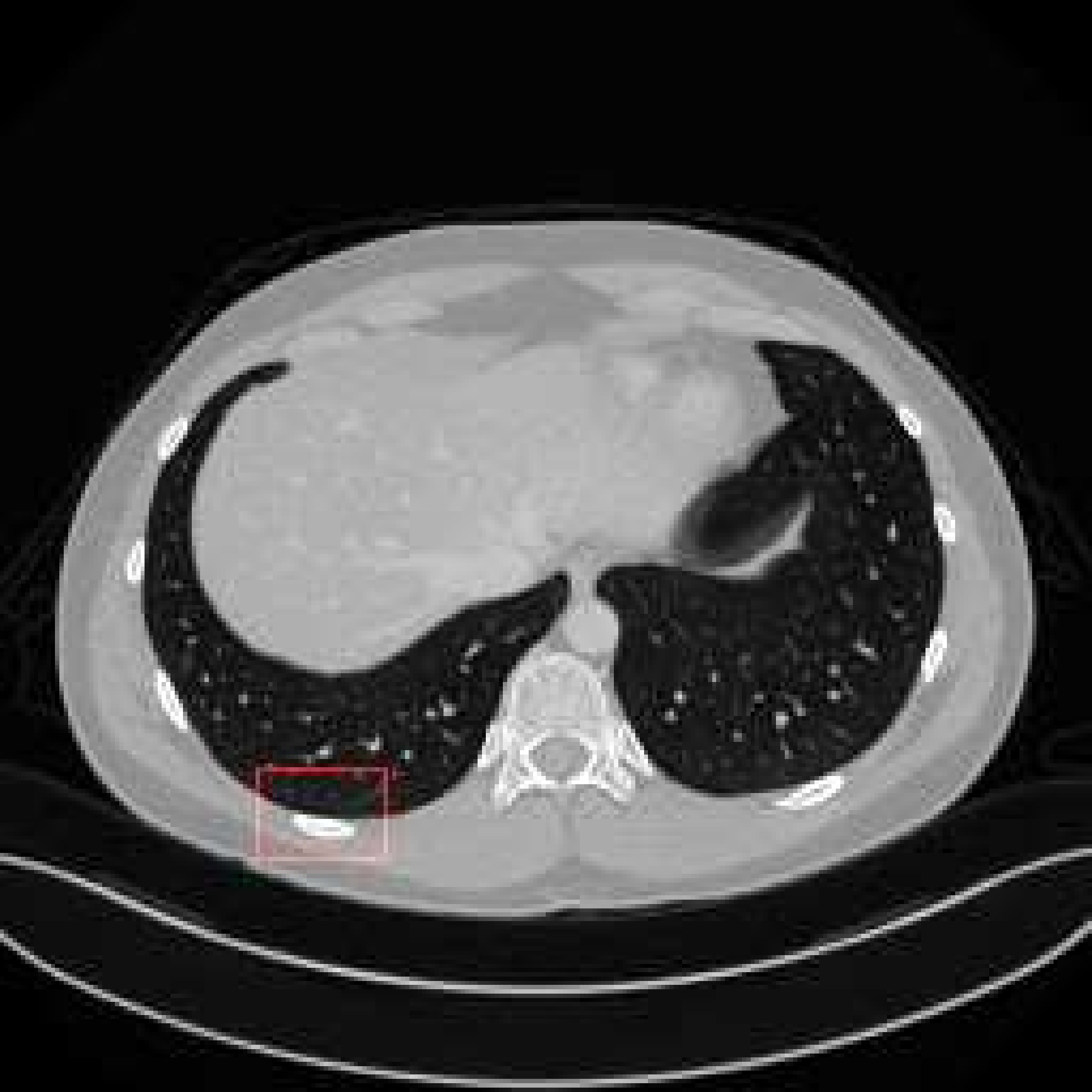}
		\includegraphics[width=0.28\columnwidth]{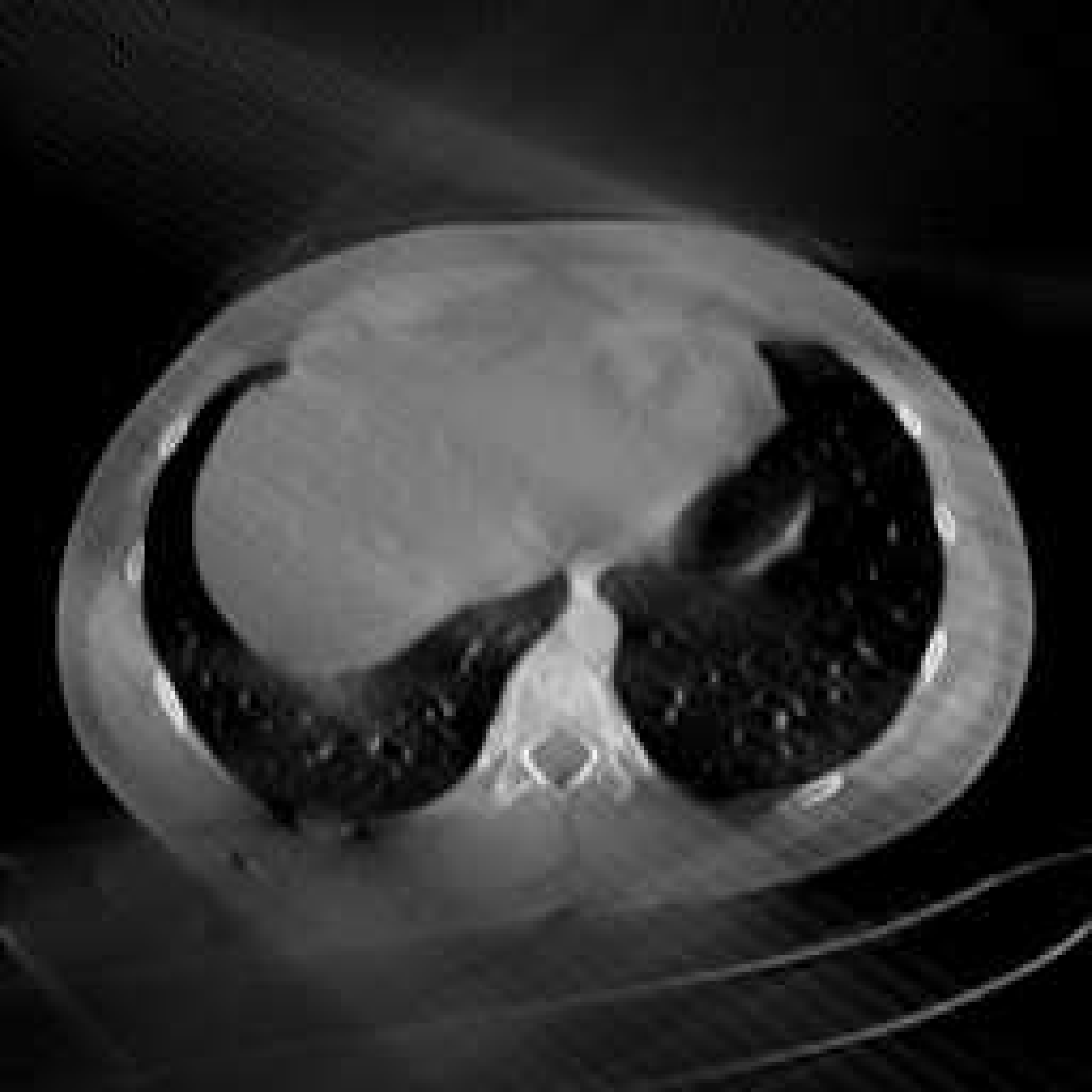}
		\includegraphics[width=0.28\columnwidth]{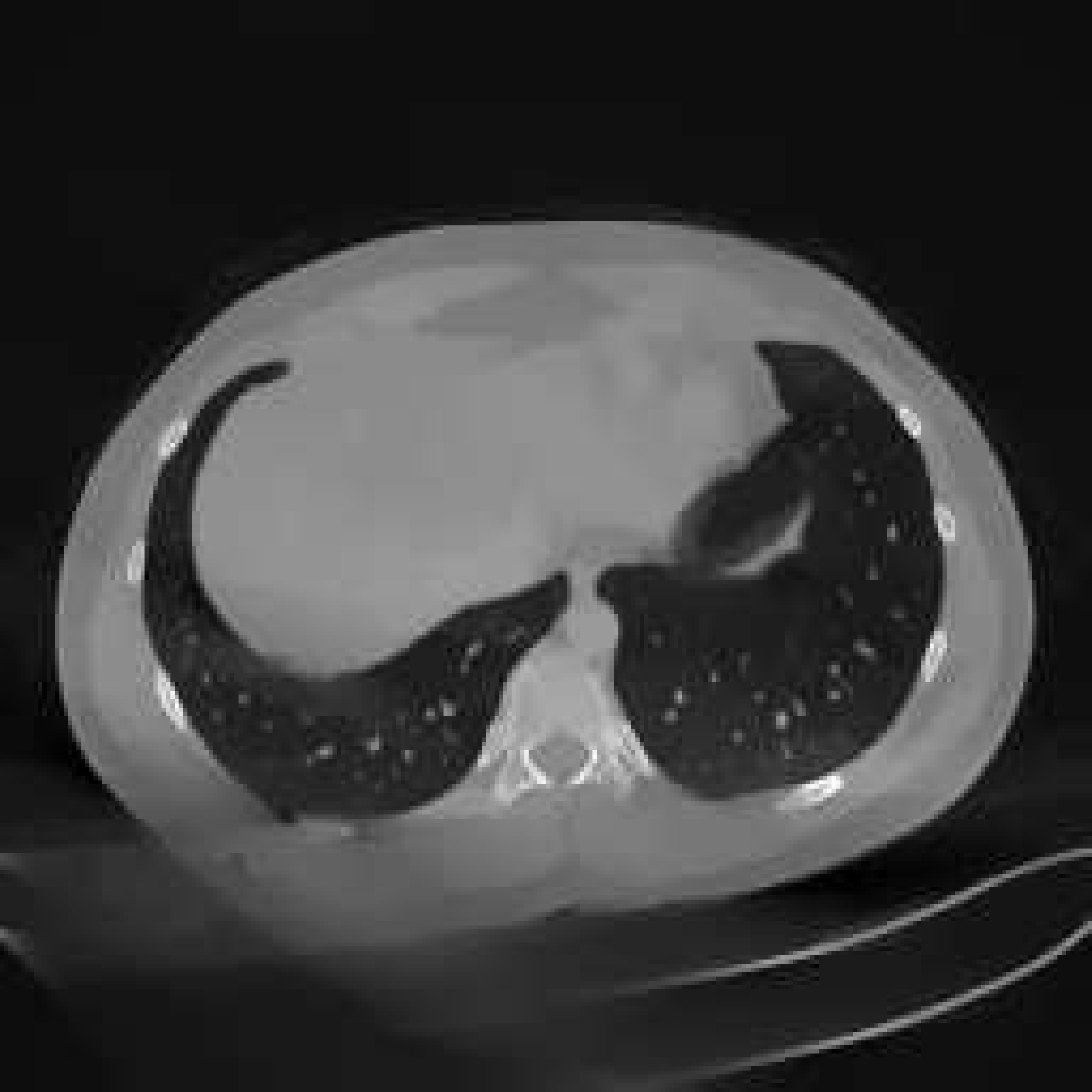}
		\includegraphics[width=0.28\columnwidth]{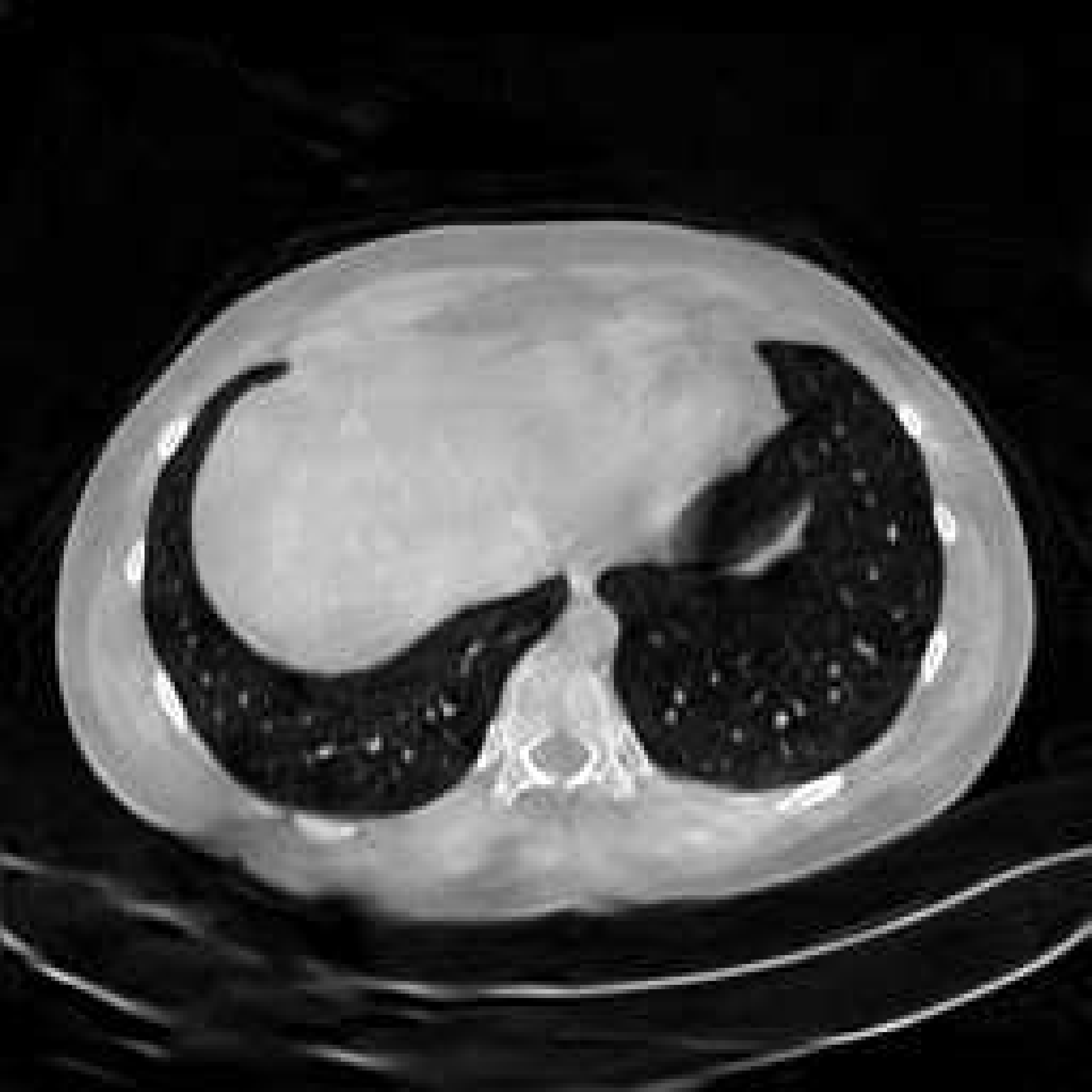}
		\includegraphics[width=0.28\columnwidth]{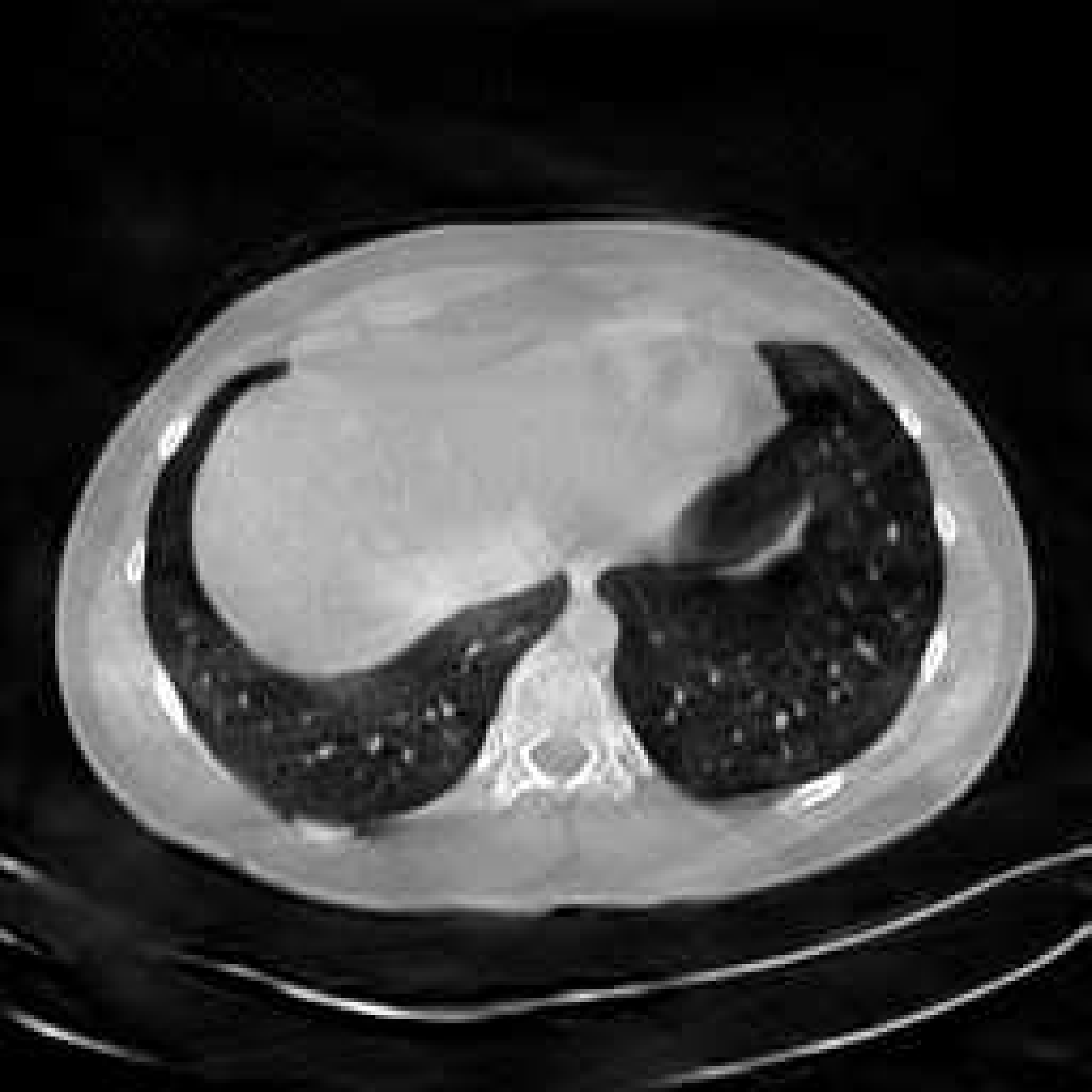}
		\includegraphics[width=0.28\columnwidth]{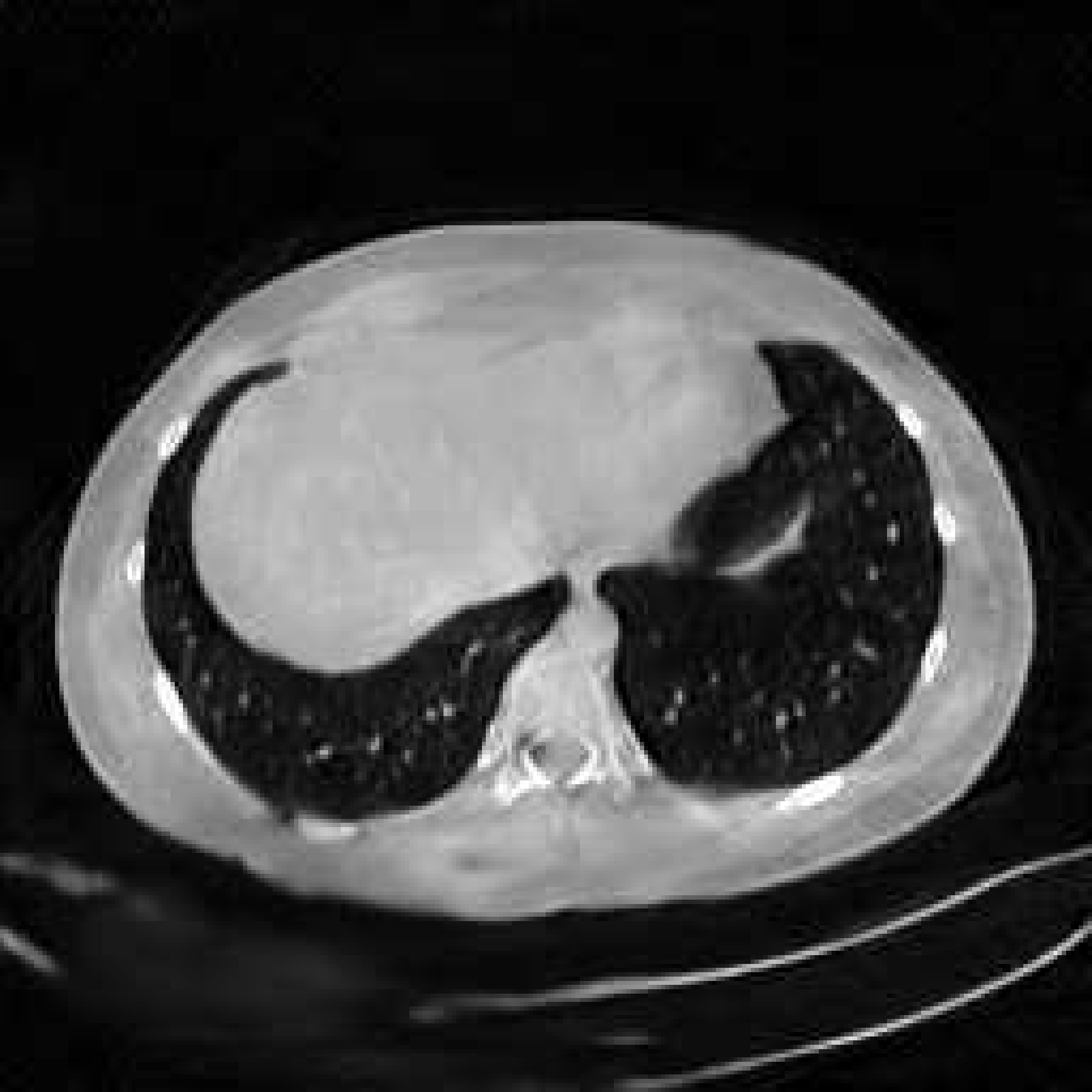}
		\includegraphics[width=0.28\columnwidth]{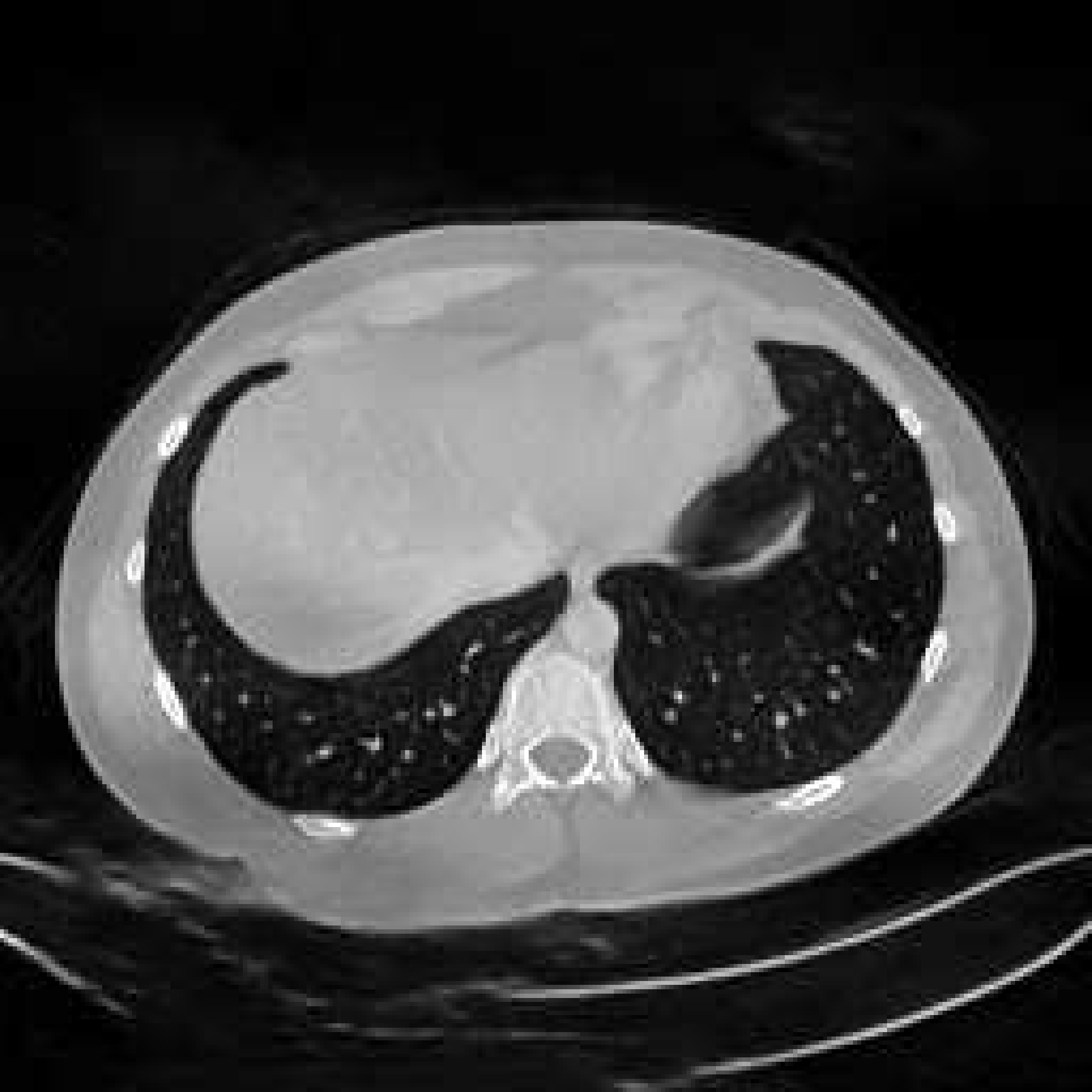}
	}
	\centerline{
		\includegraphics[width=0.28\columnwidth]{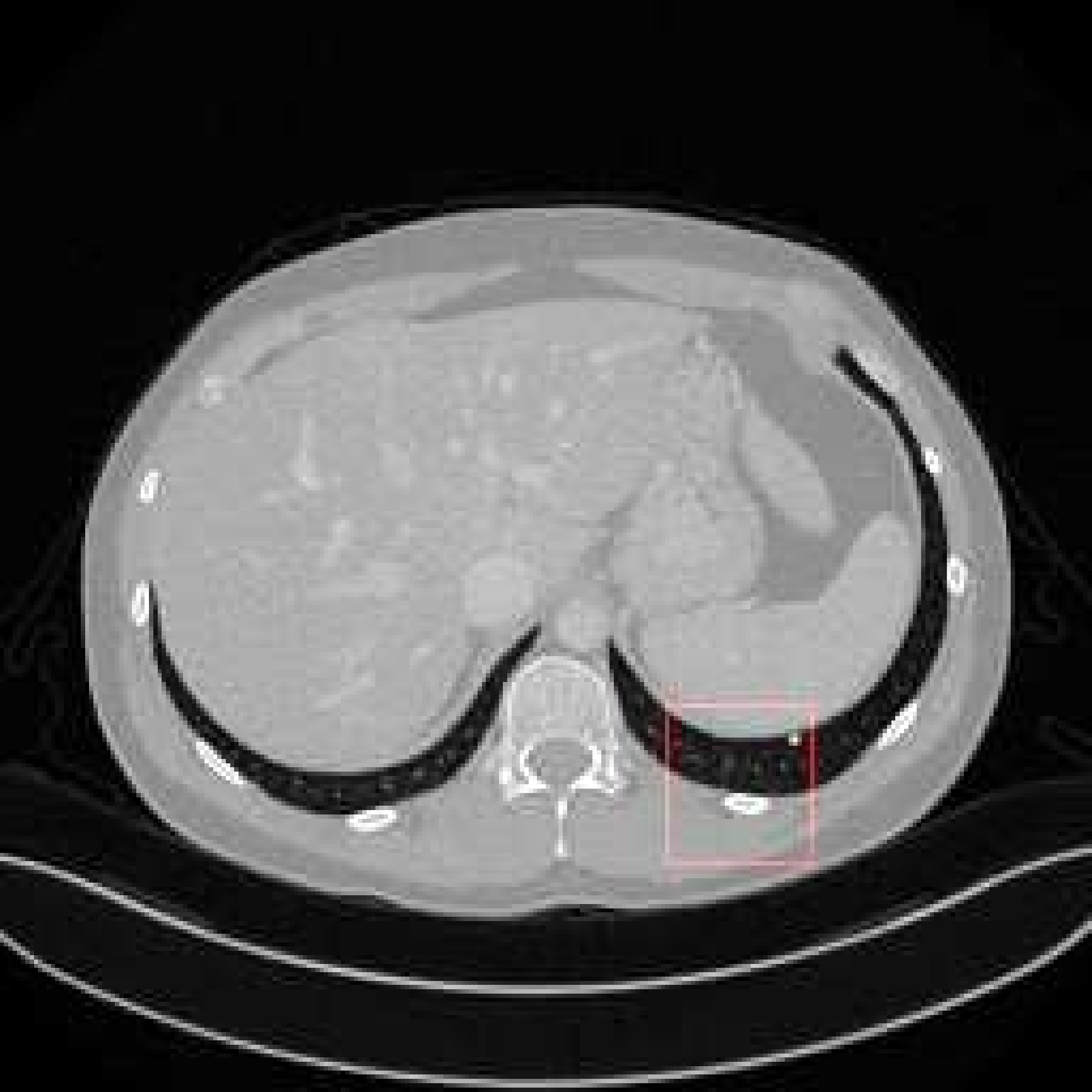}
		\includegraphics[width=0.28\columnwidth]{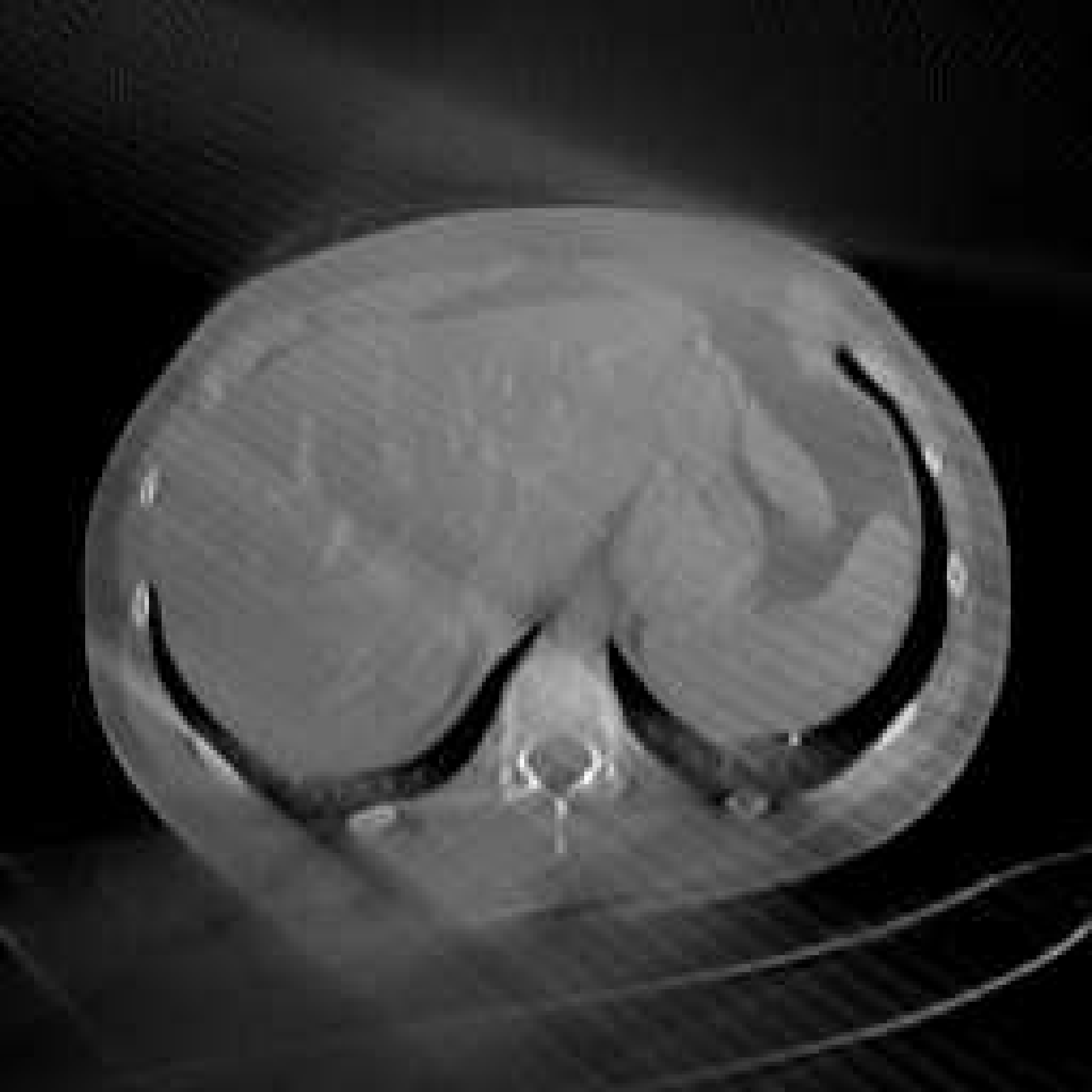}
		\includegraphics[width=0.28\columnwidth]{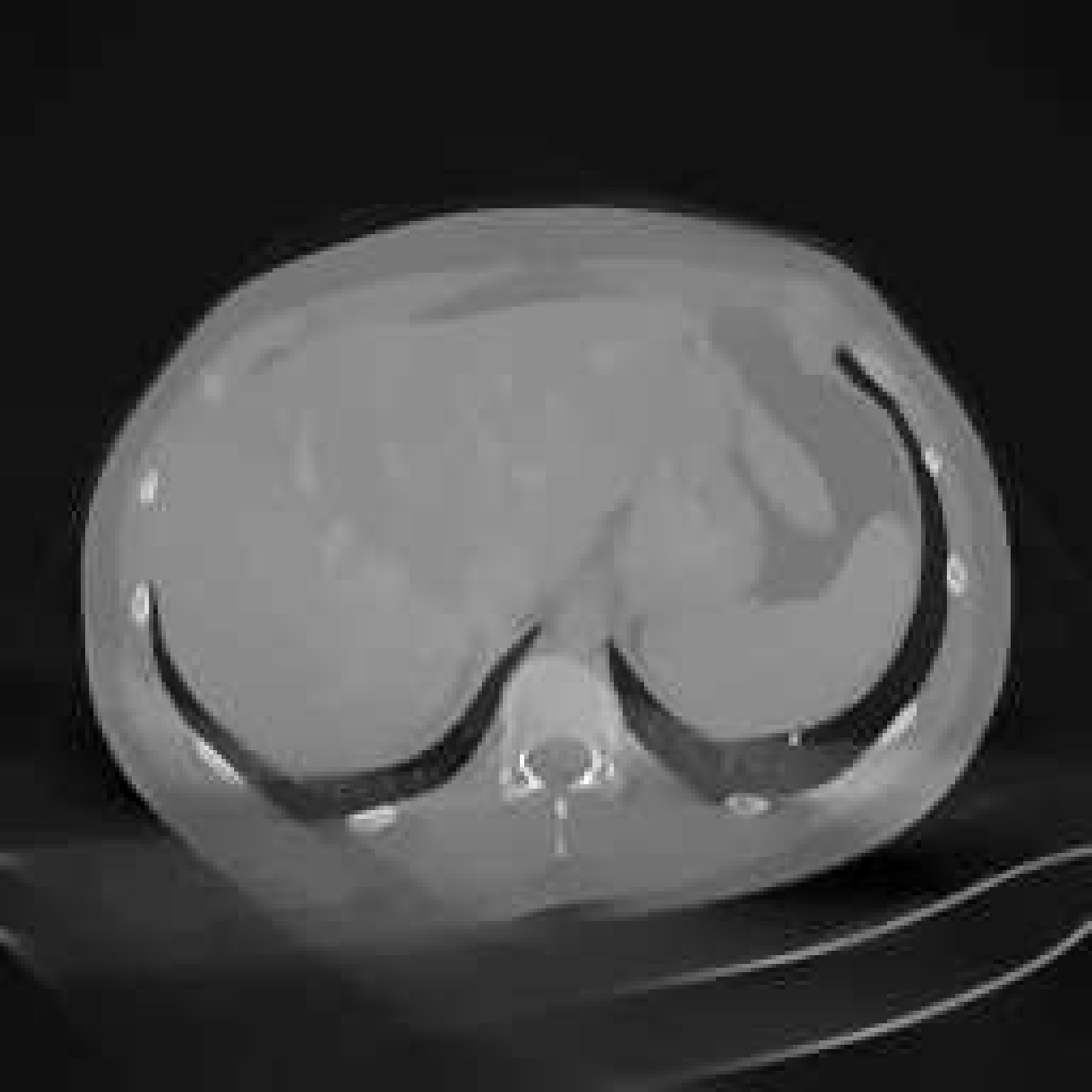}
		\includegraphics[width=0.28\columnwidth]{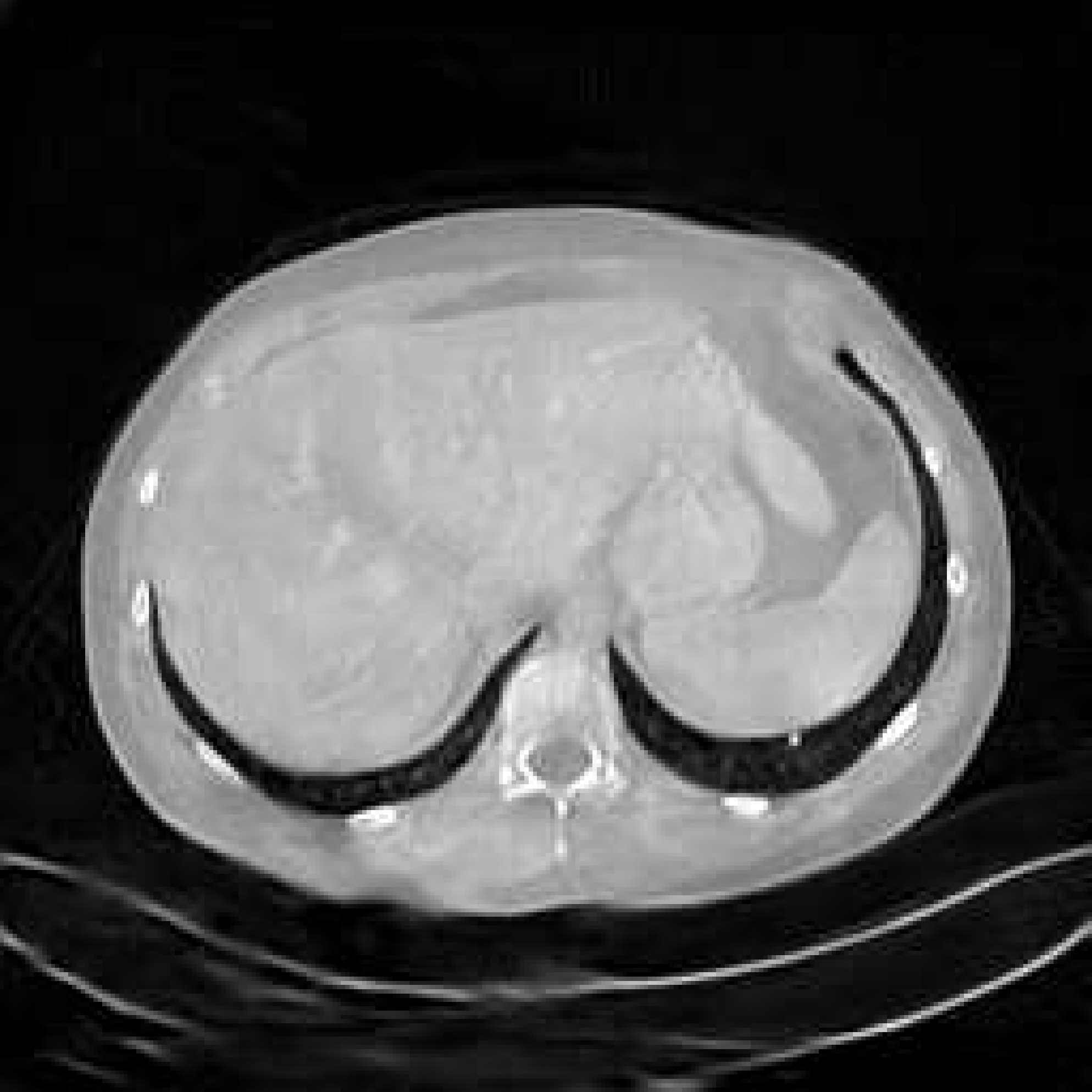}
		\includegraphics[width=0.28\columnwidth]{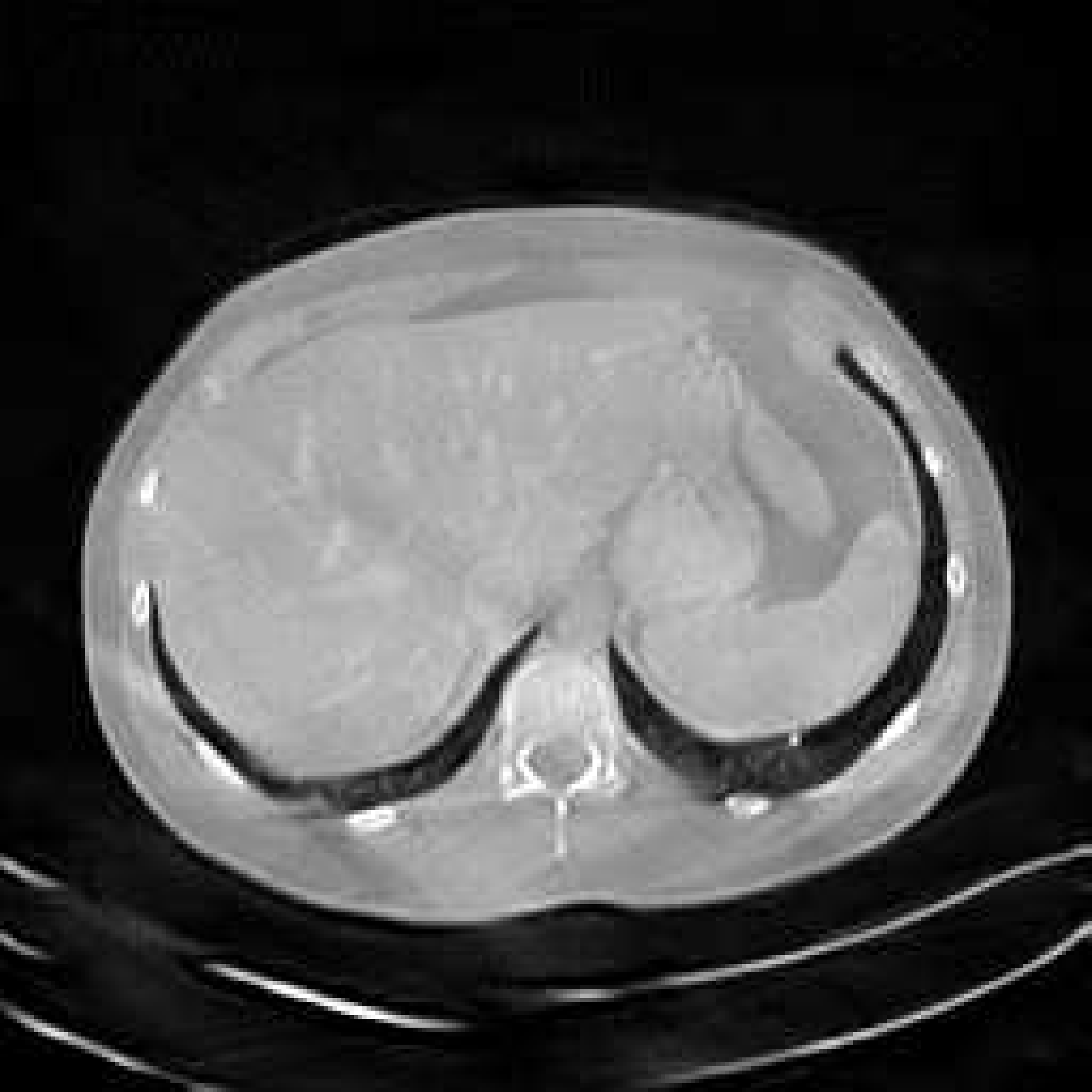}
		\includegraphics[width=0.28\columnwidth]{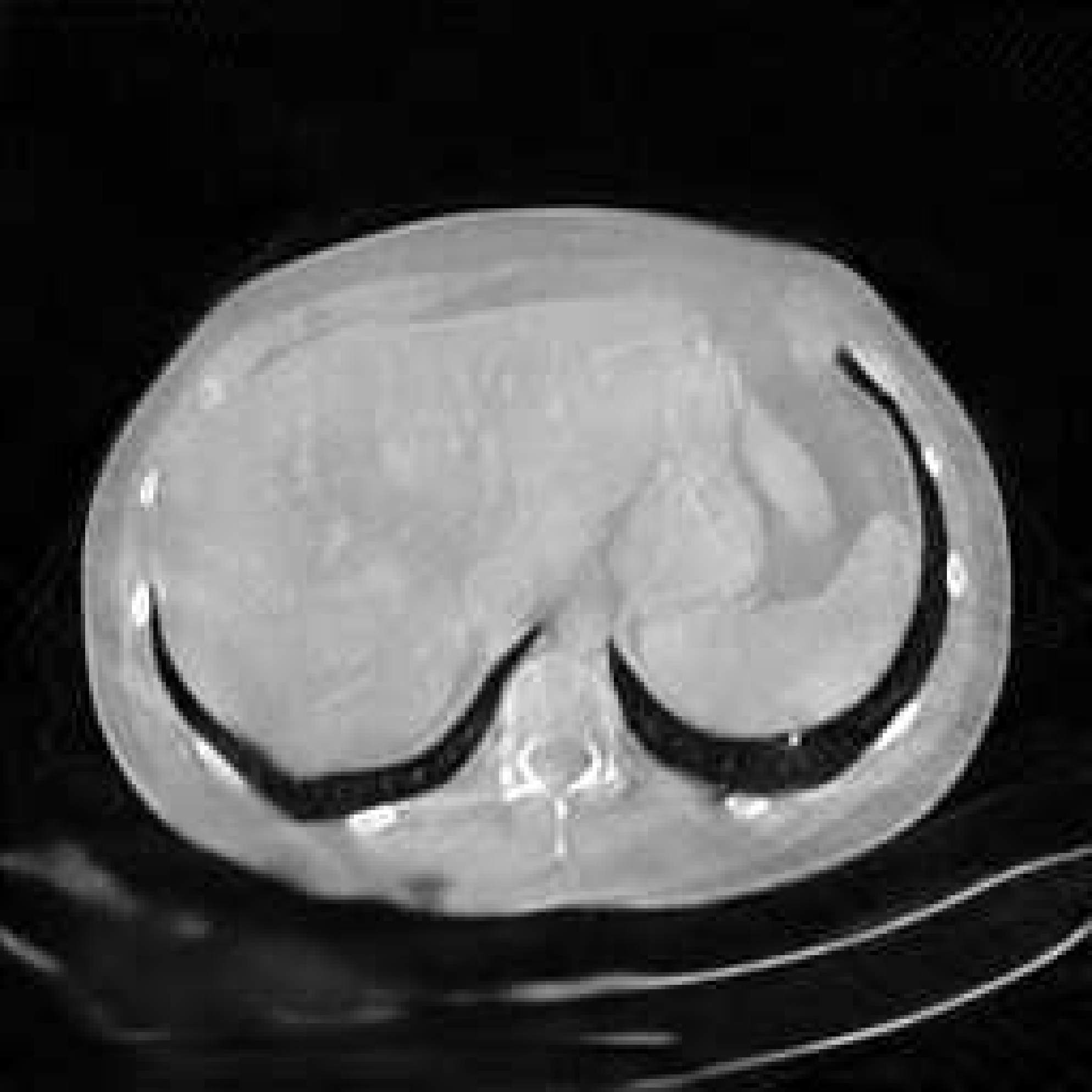}
		\includegraphics[width=0.28\columnwidth]{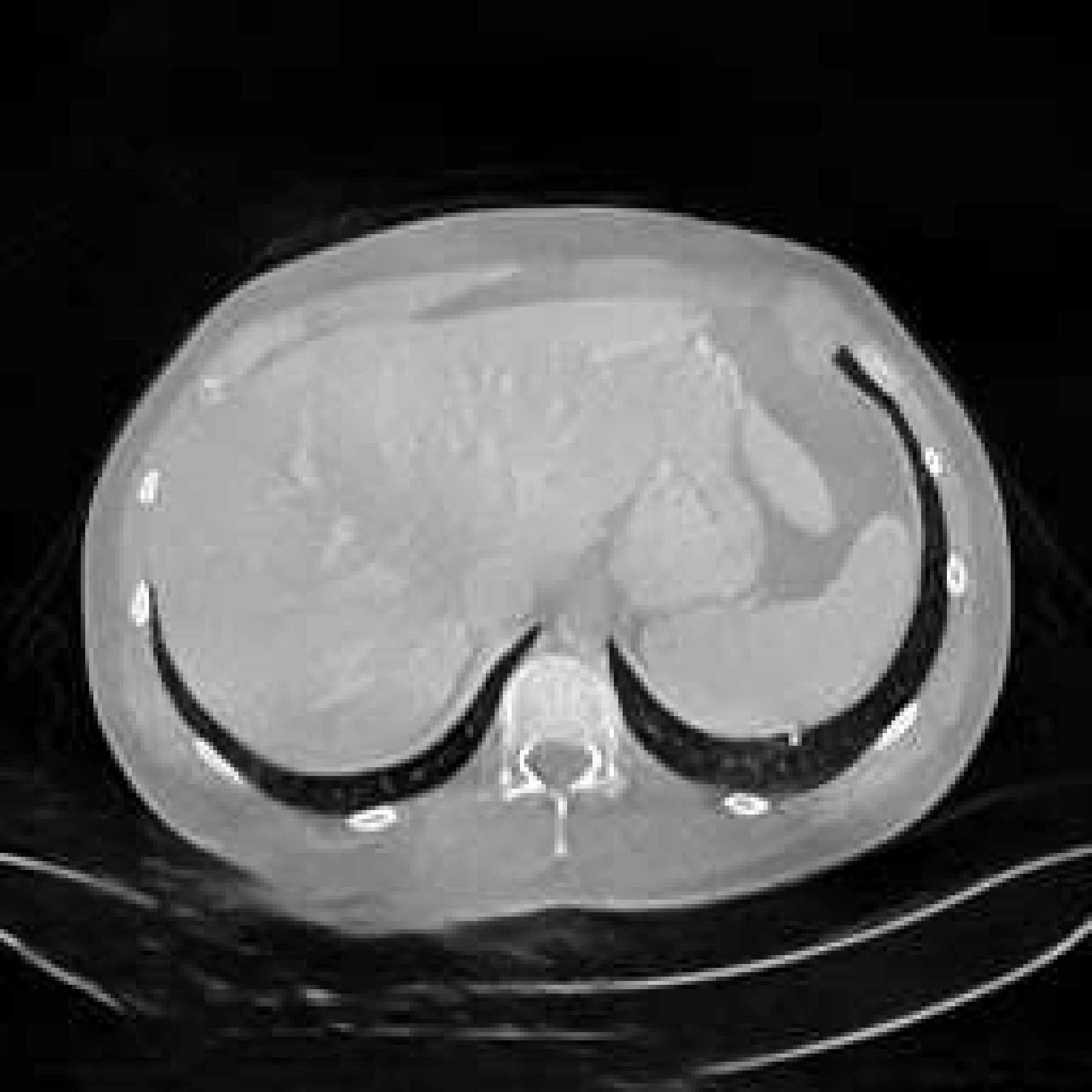}
	}
	\centerline{
		\includegraphics[width=0.28\columnwidth]{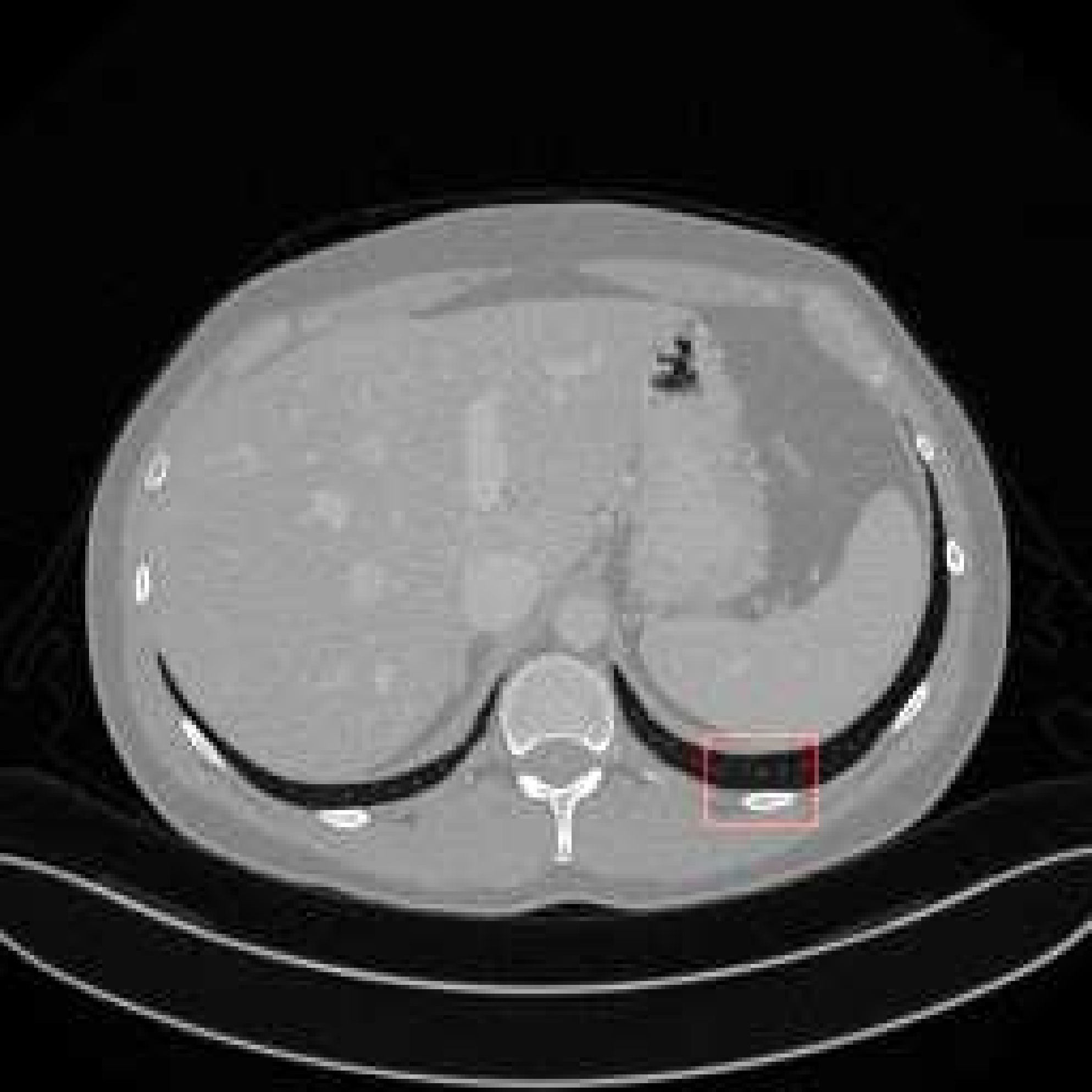}
		\includegraphics[width=0.28\columnwidth]{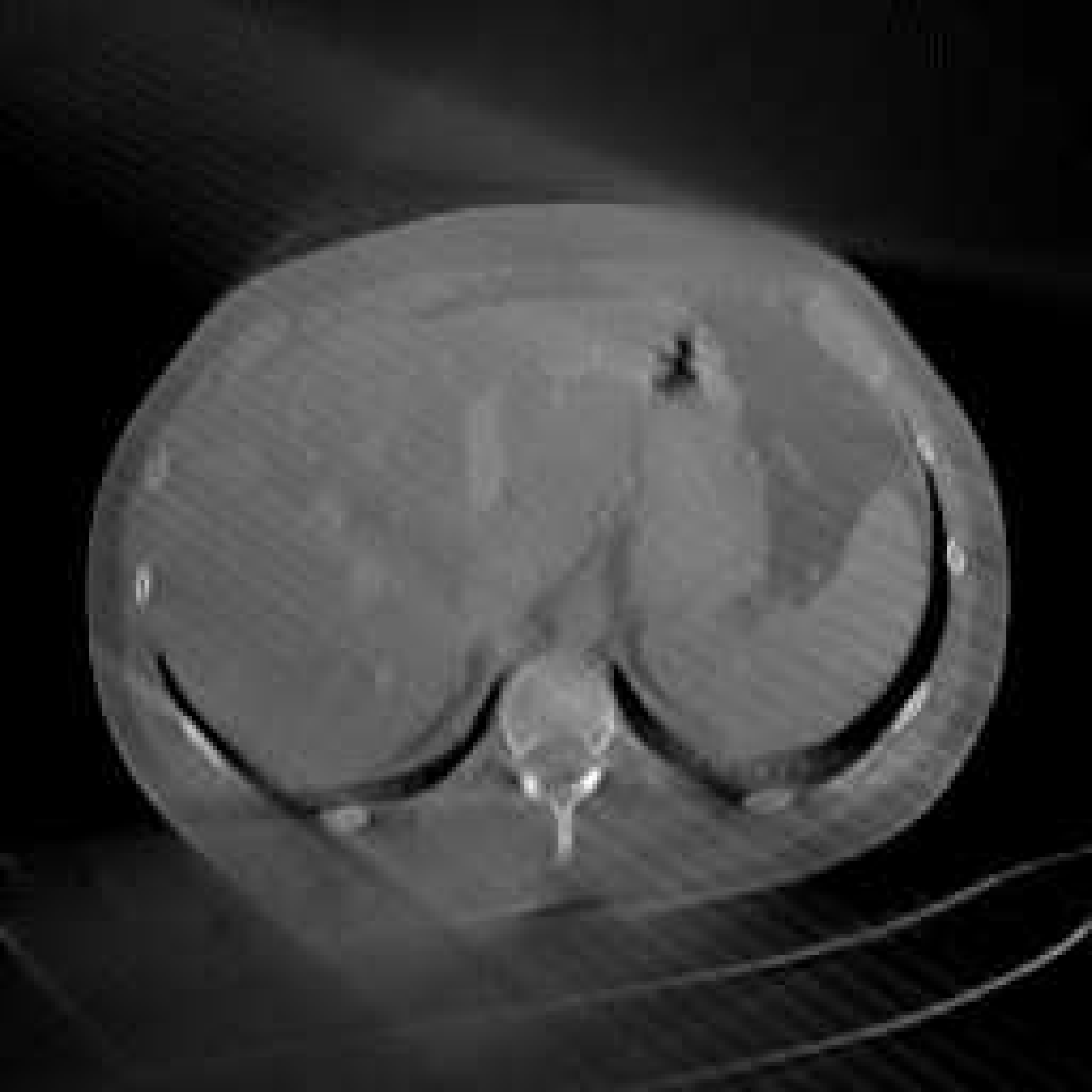}
		\includegraphics[width=0.28\columnwidth]{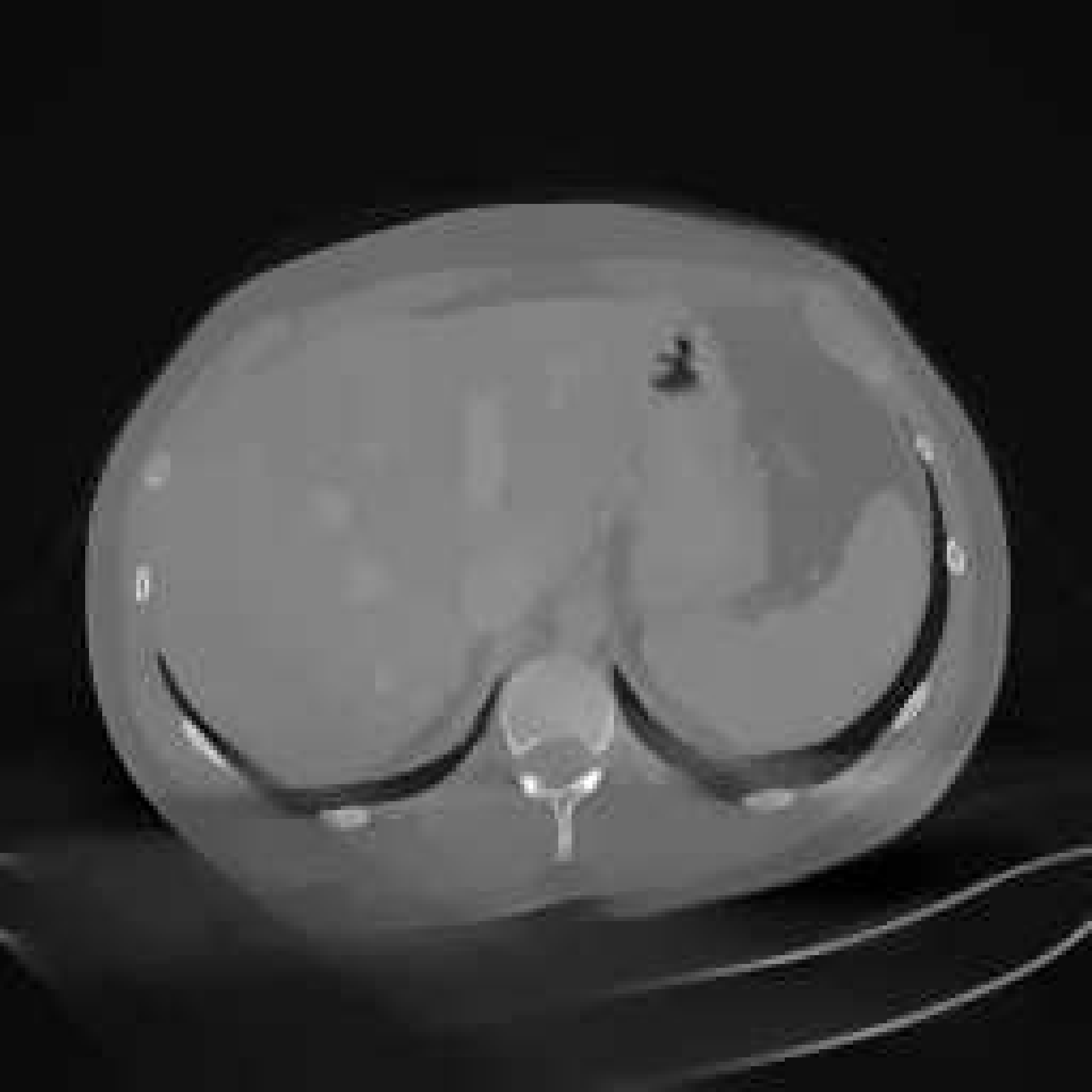}
		\includegraphics[width=0.28\columnwidth]{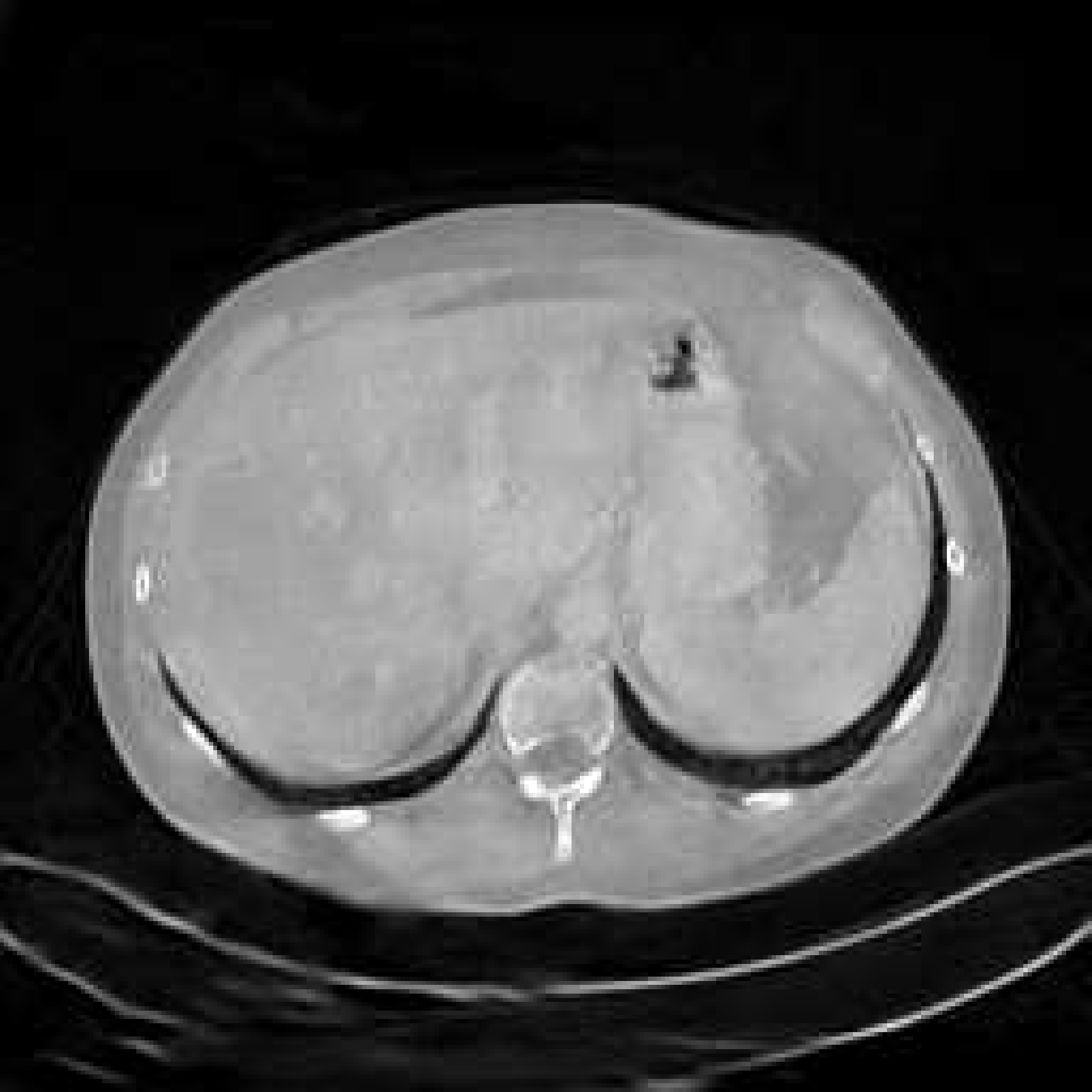}
		\includegraphics[width=0.28\columnwidth]{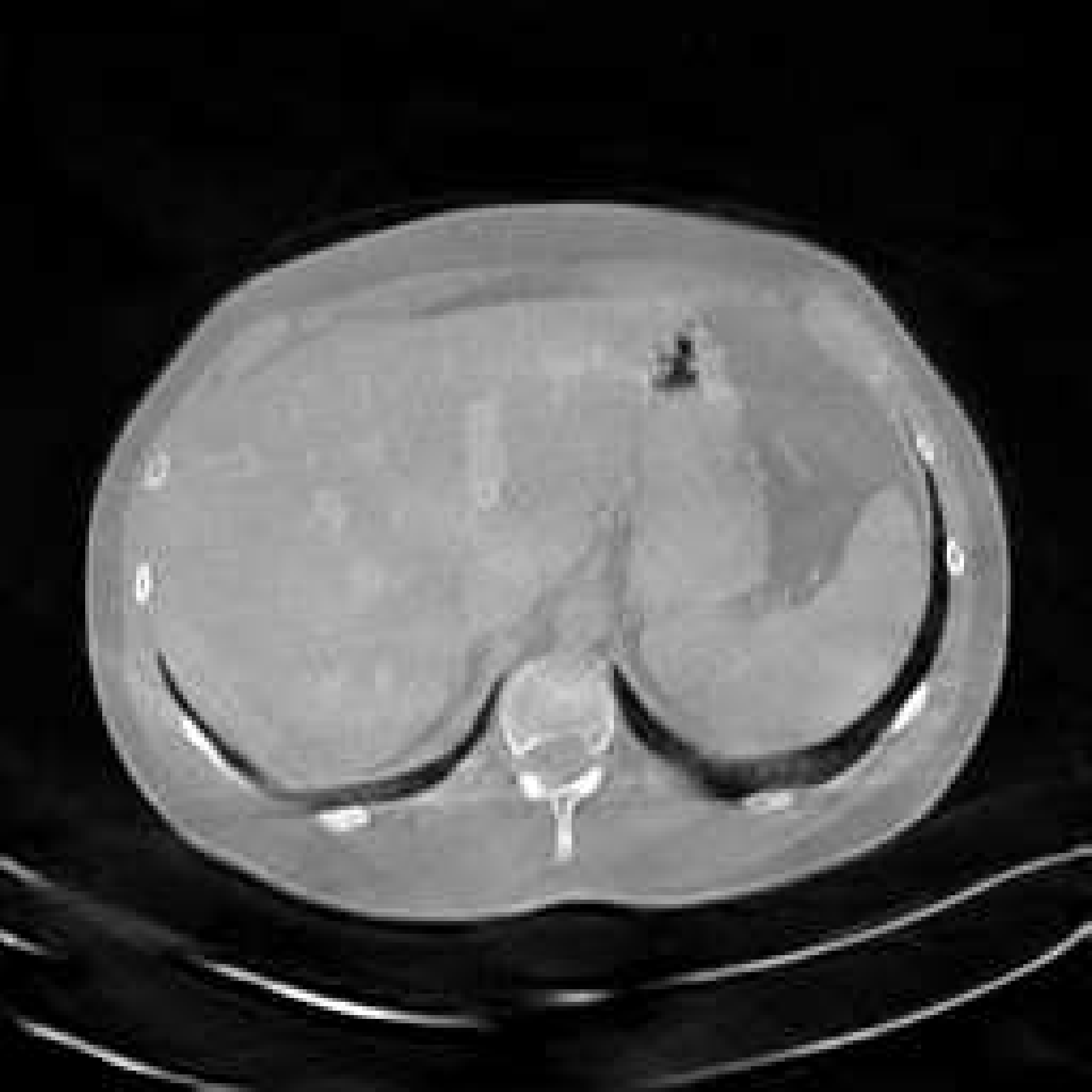}
		\includegraphics[width=0.28\columnwidth]{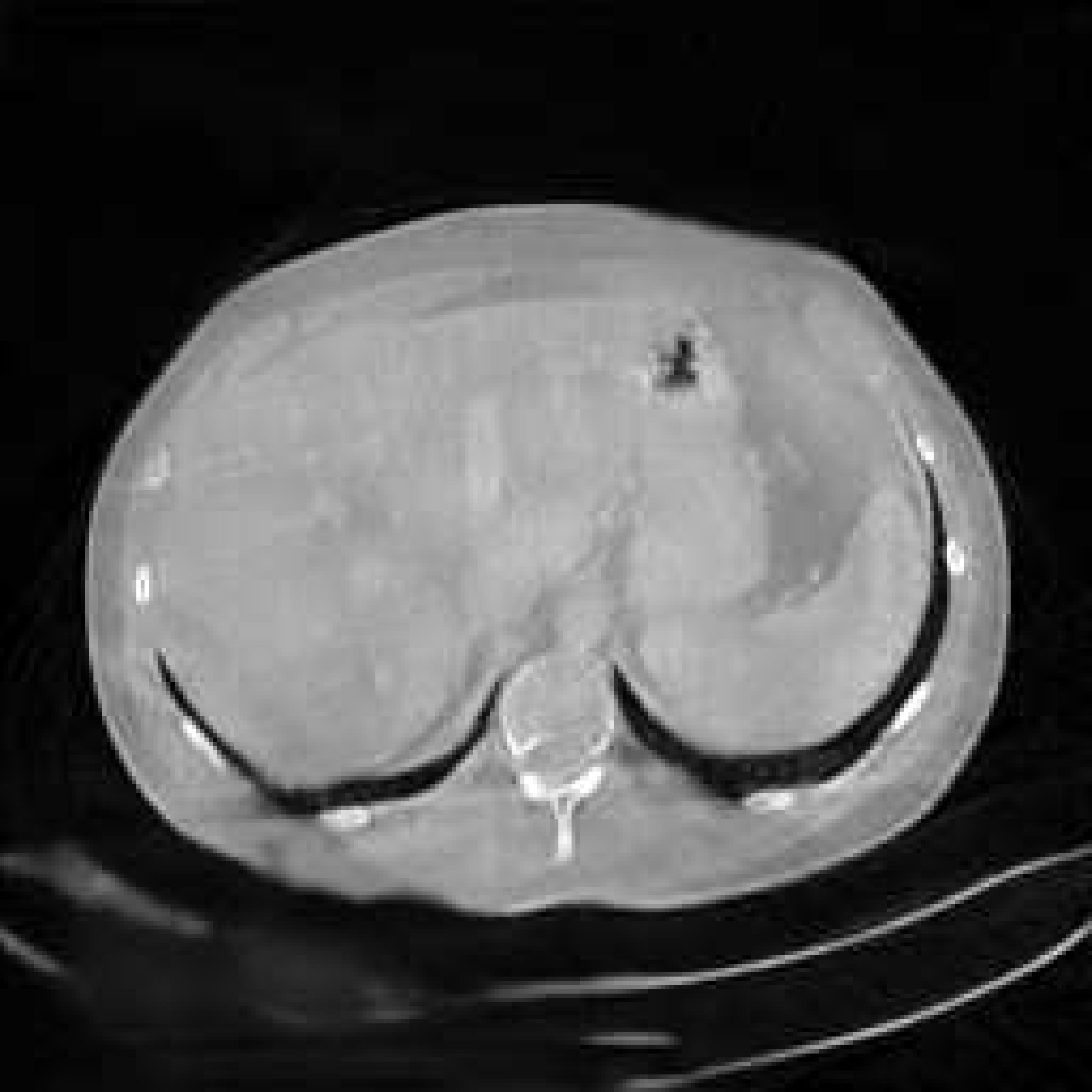}
		\includegraphics[width=0.28\columnwidth]{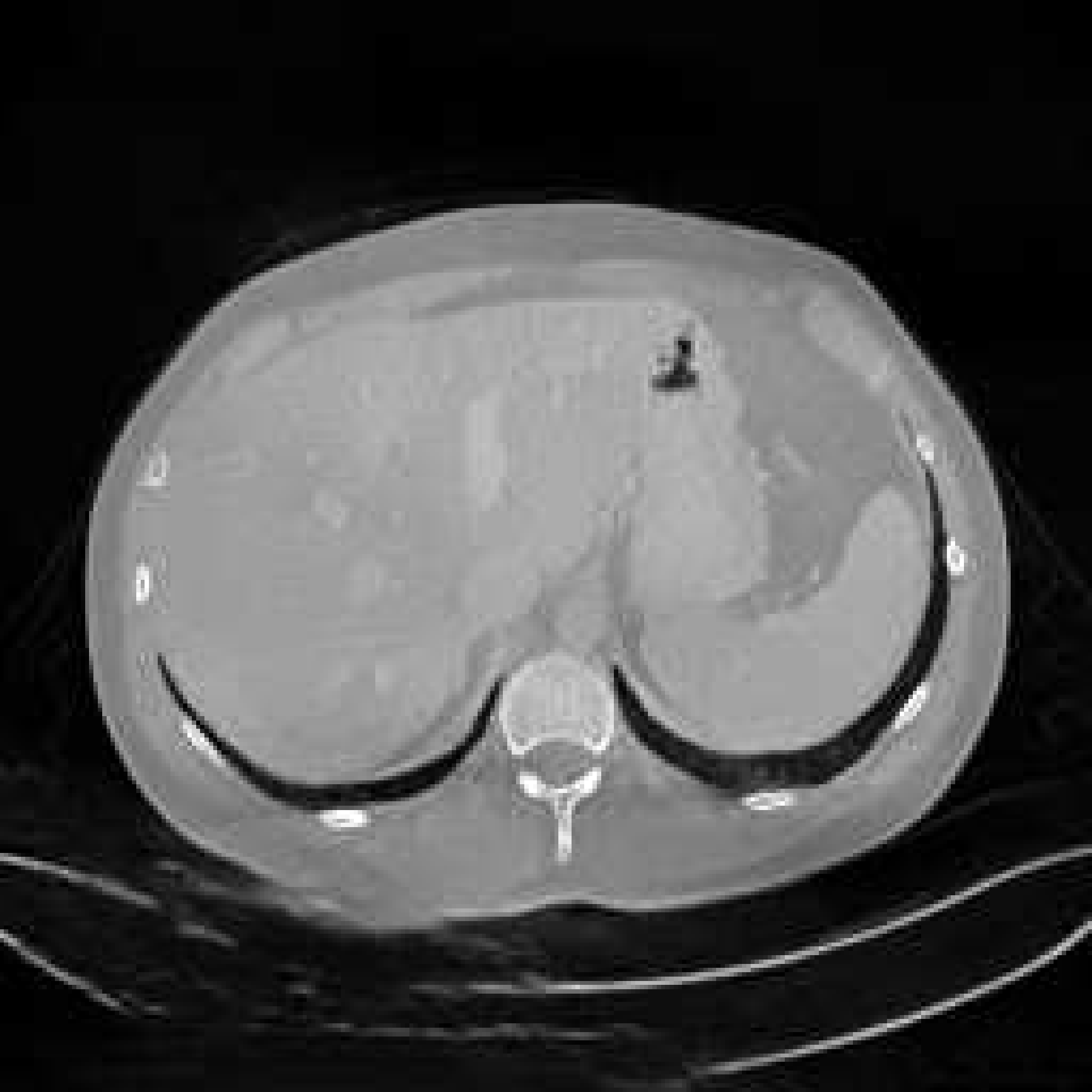}
	}
	\centerline{
		\subfigure[Label]{\includegraphics[width=0.28\columnwidth]{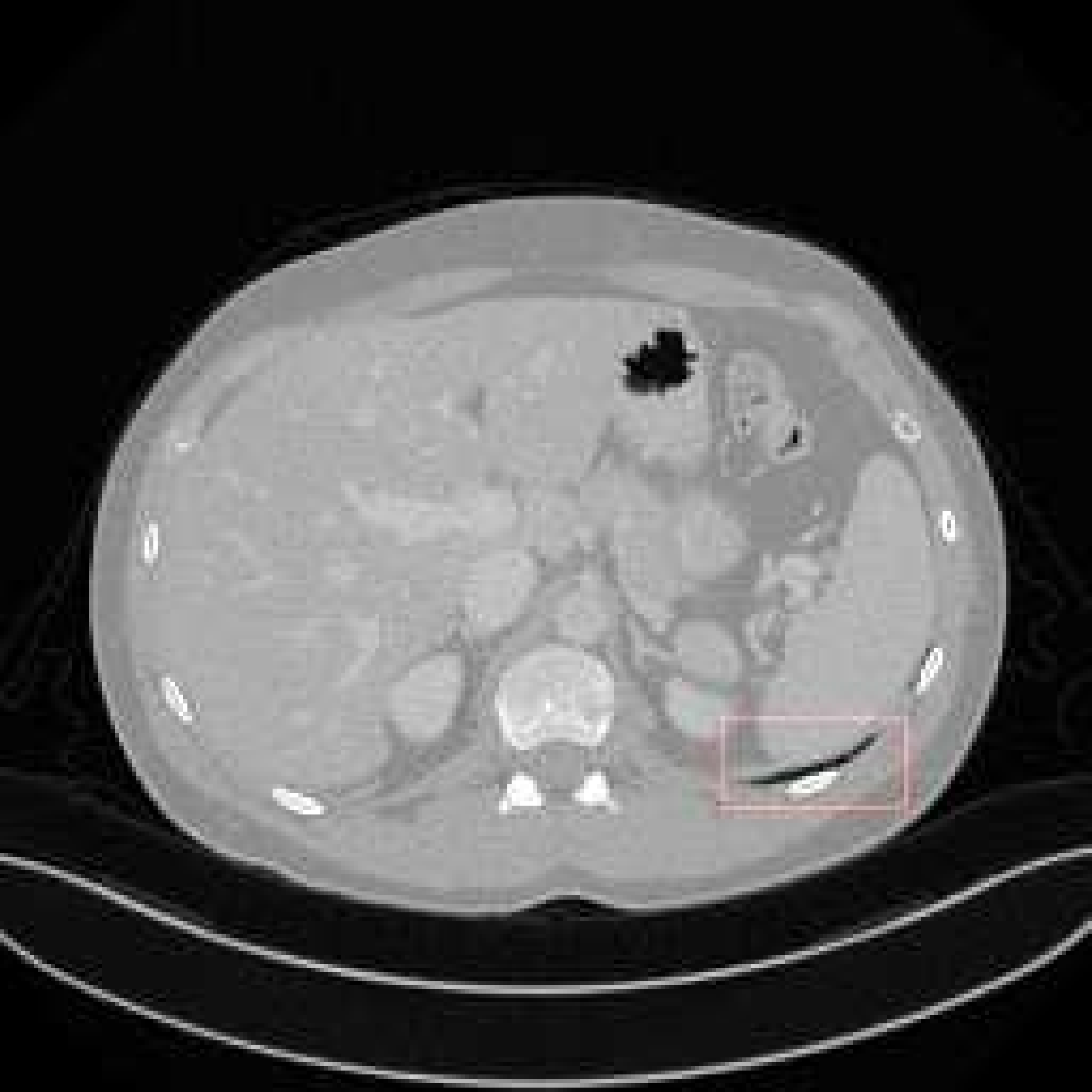}}
		\subfigure[FBP]{\includegraphics[width=0.28\columnwidth]{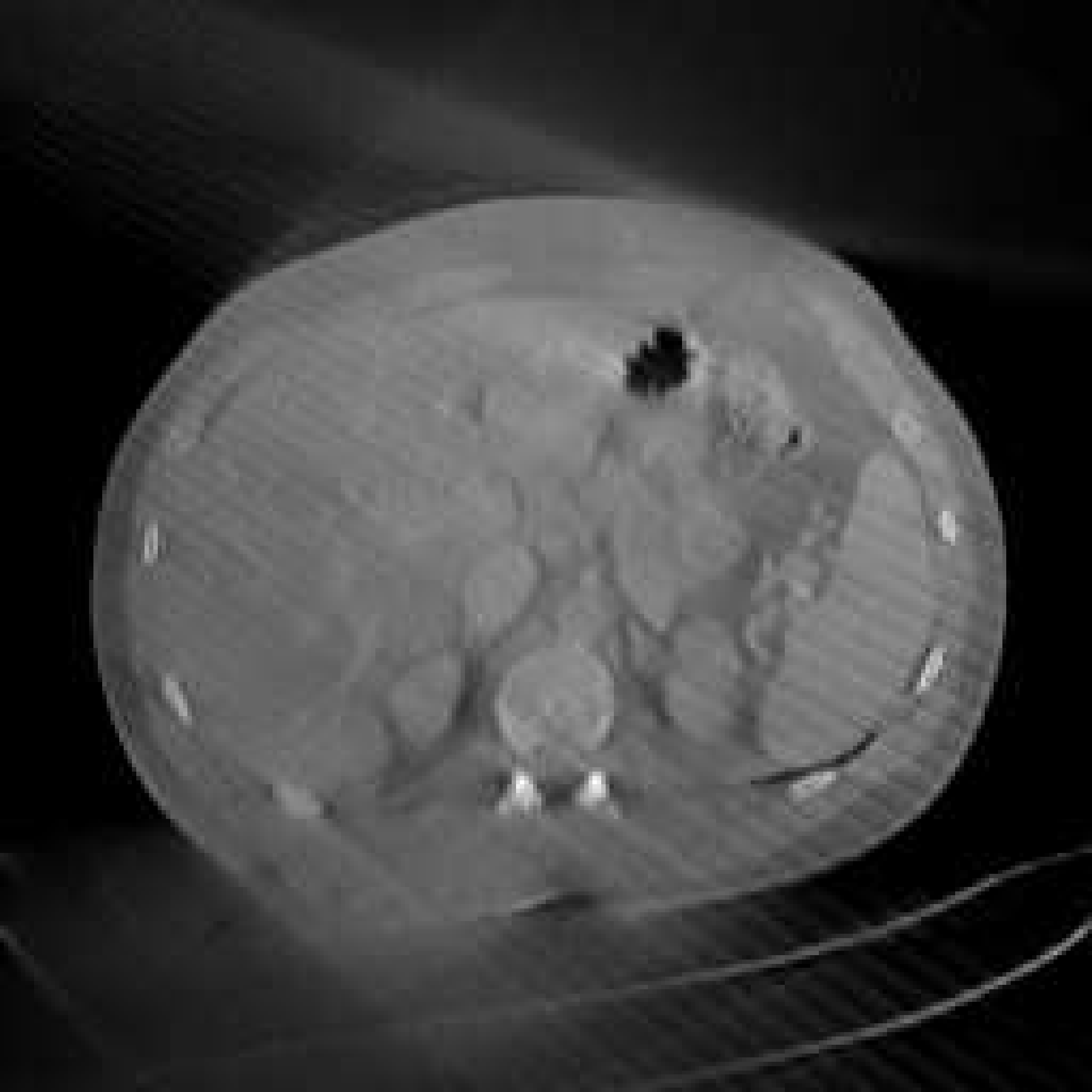}}
		\subfigure[TV-regularization]{\includegraphics[width=0.28\columnwidth]{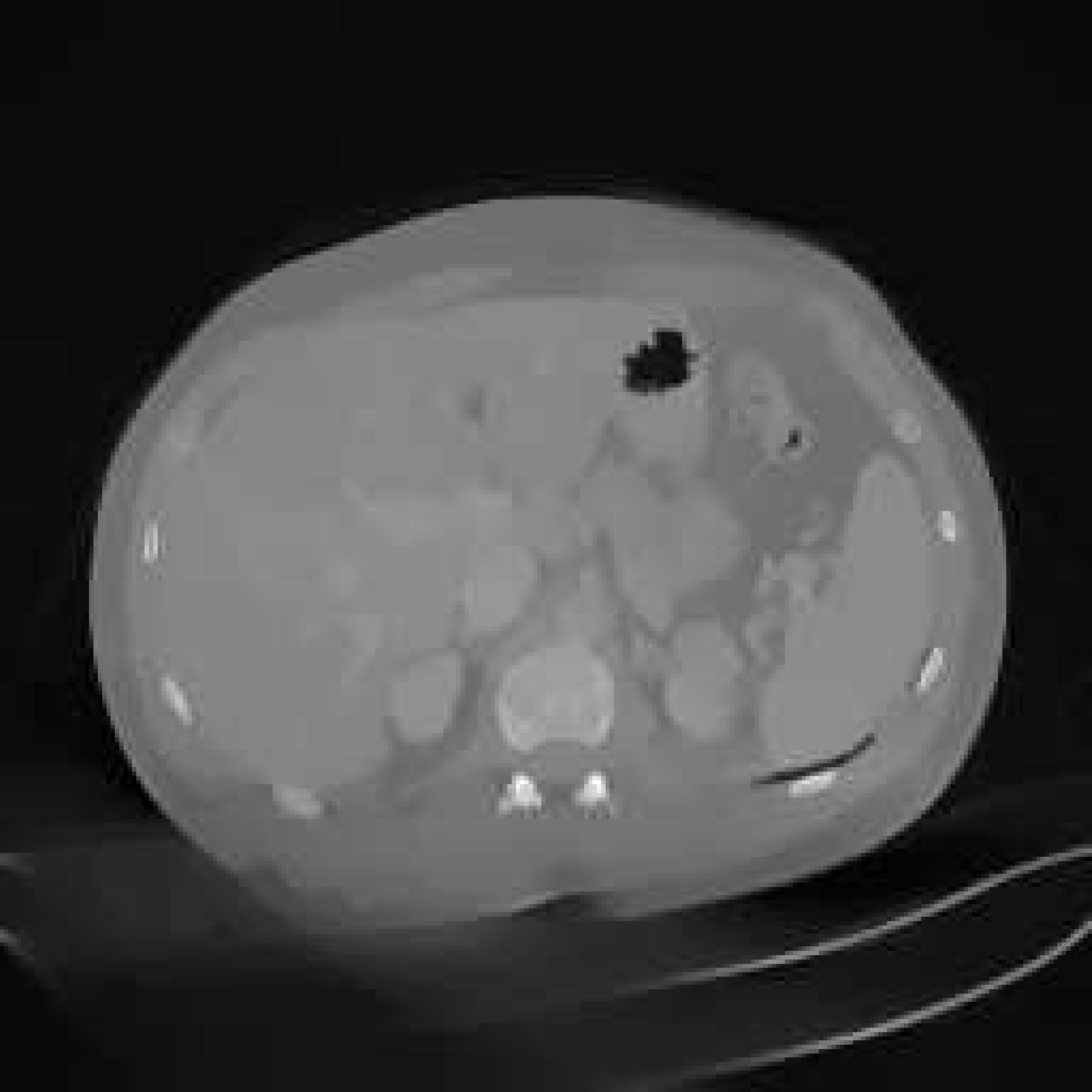}}
		\subfigure[Red-CNN]{\includegraphics[width=0.28\columnwidth]{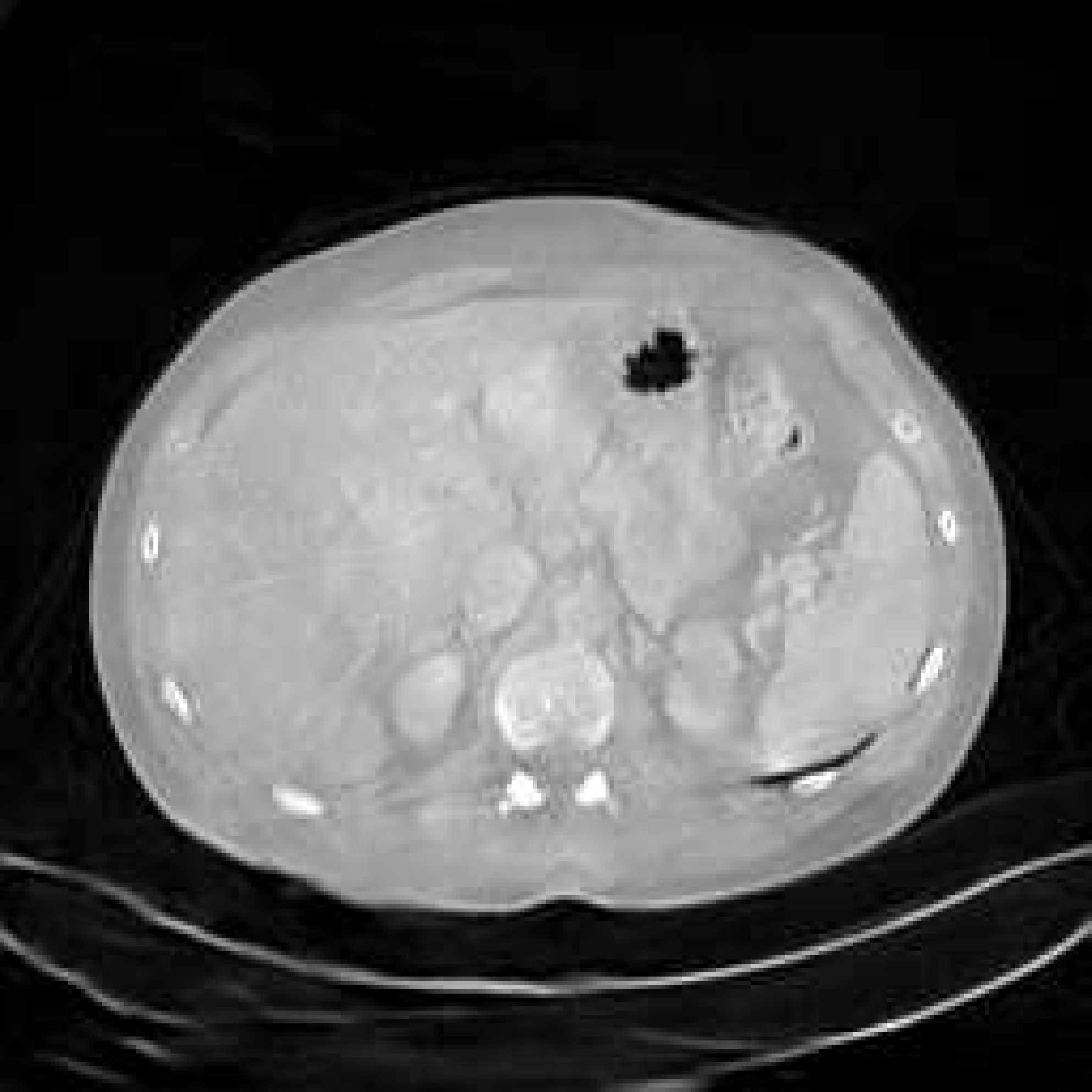}}
		\subfigure[FBP-Conv]{\includegraphics[width=0.28\columnwidth]{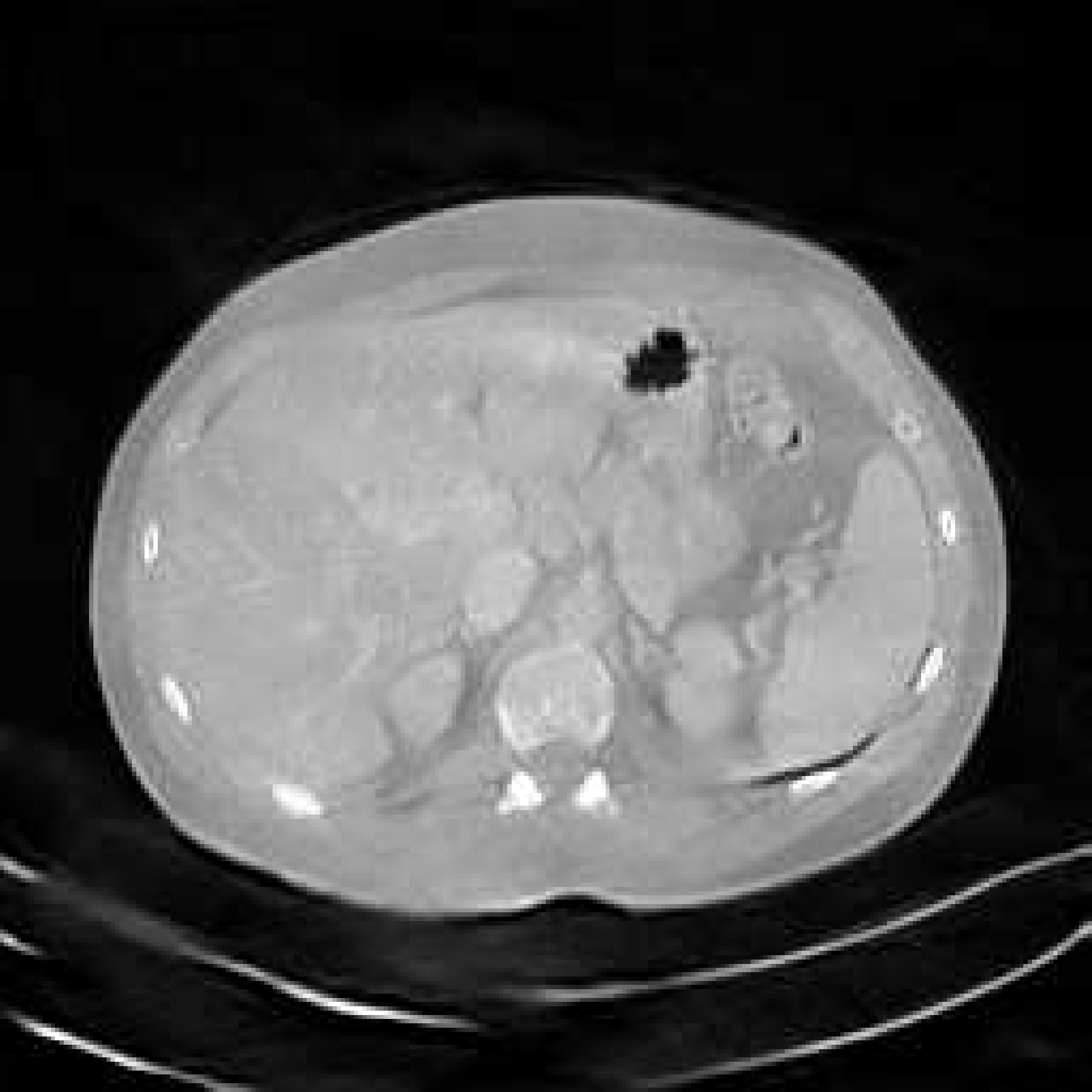}}
		\subfigure[DD-Net]{\includegraphics[width=0.28\columnwidth]{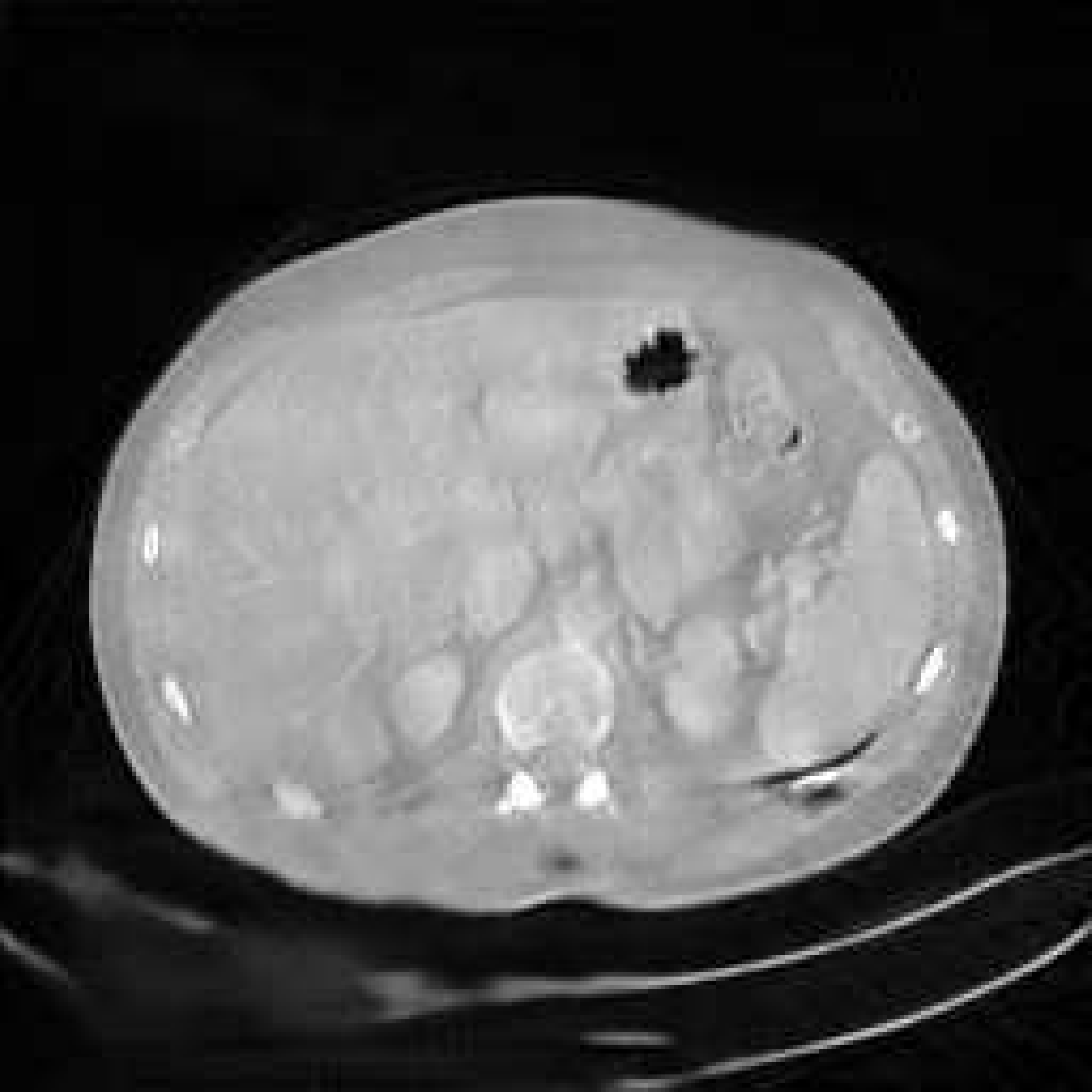}}
		\subfigure[Ours]{\includegraphics[width=0.28\columnwidth]{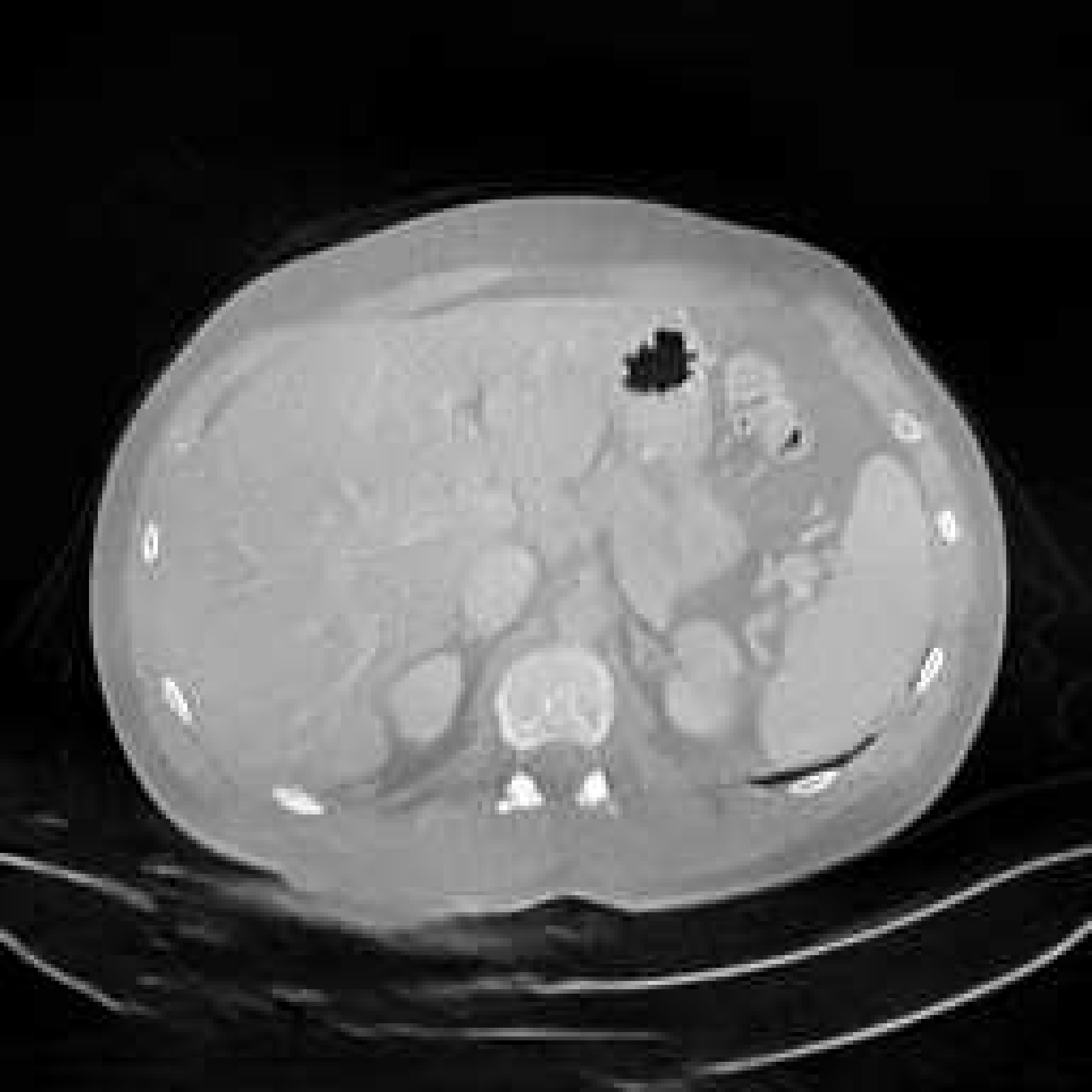}}
	}
	\caption{The reconstructed CT images by the six methods for fan-beam geometry.}
	\label{F8}
\end{figure*}

\begin{figure*}[!t]

	\centerline{
		\includegraphics[width=0.28\columnwidth]{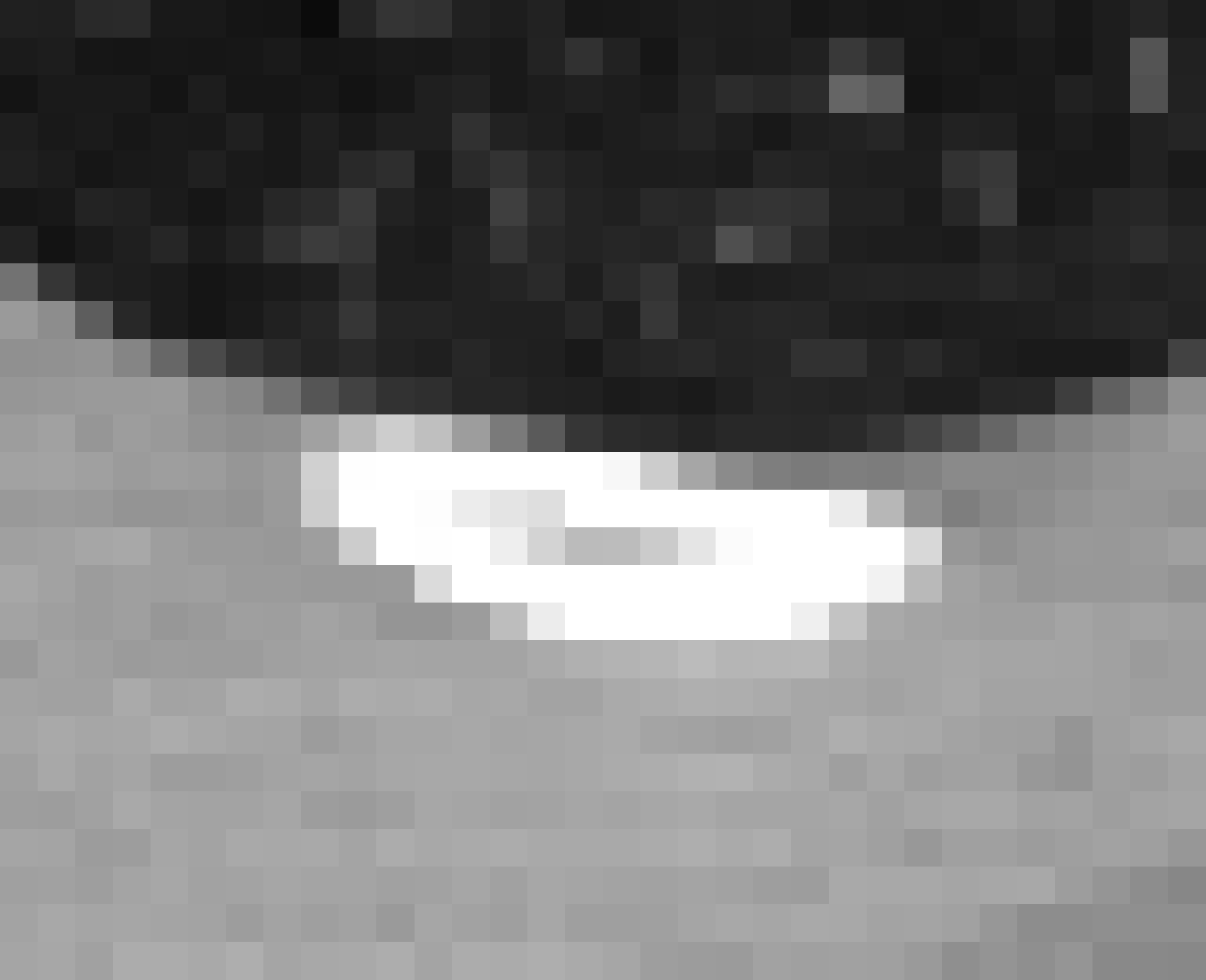}
		\includegraphics[width=0.28\columnwidth]{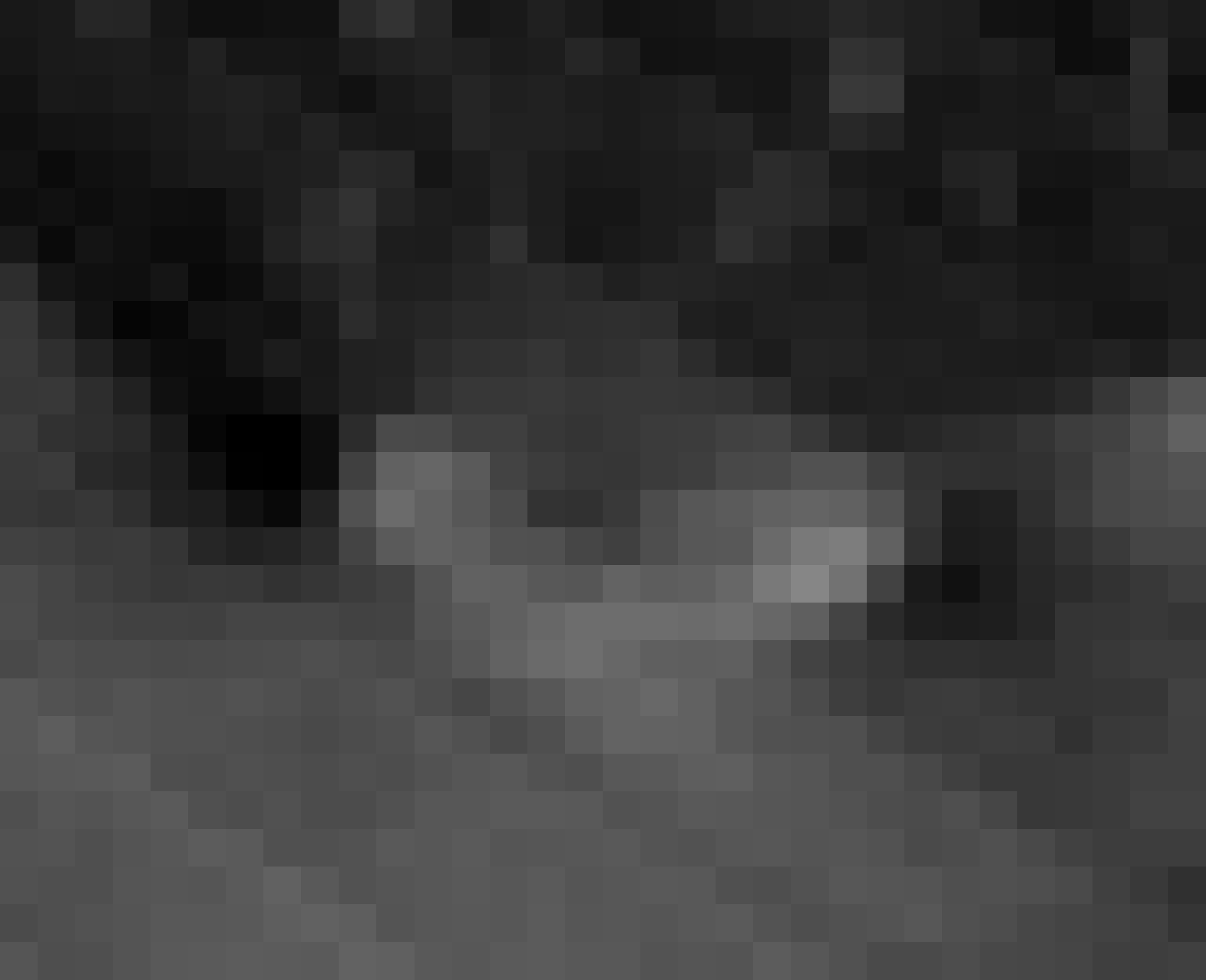}
		\includegraphics[width=0.28\columnwidth]{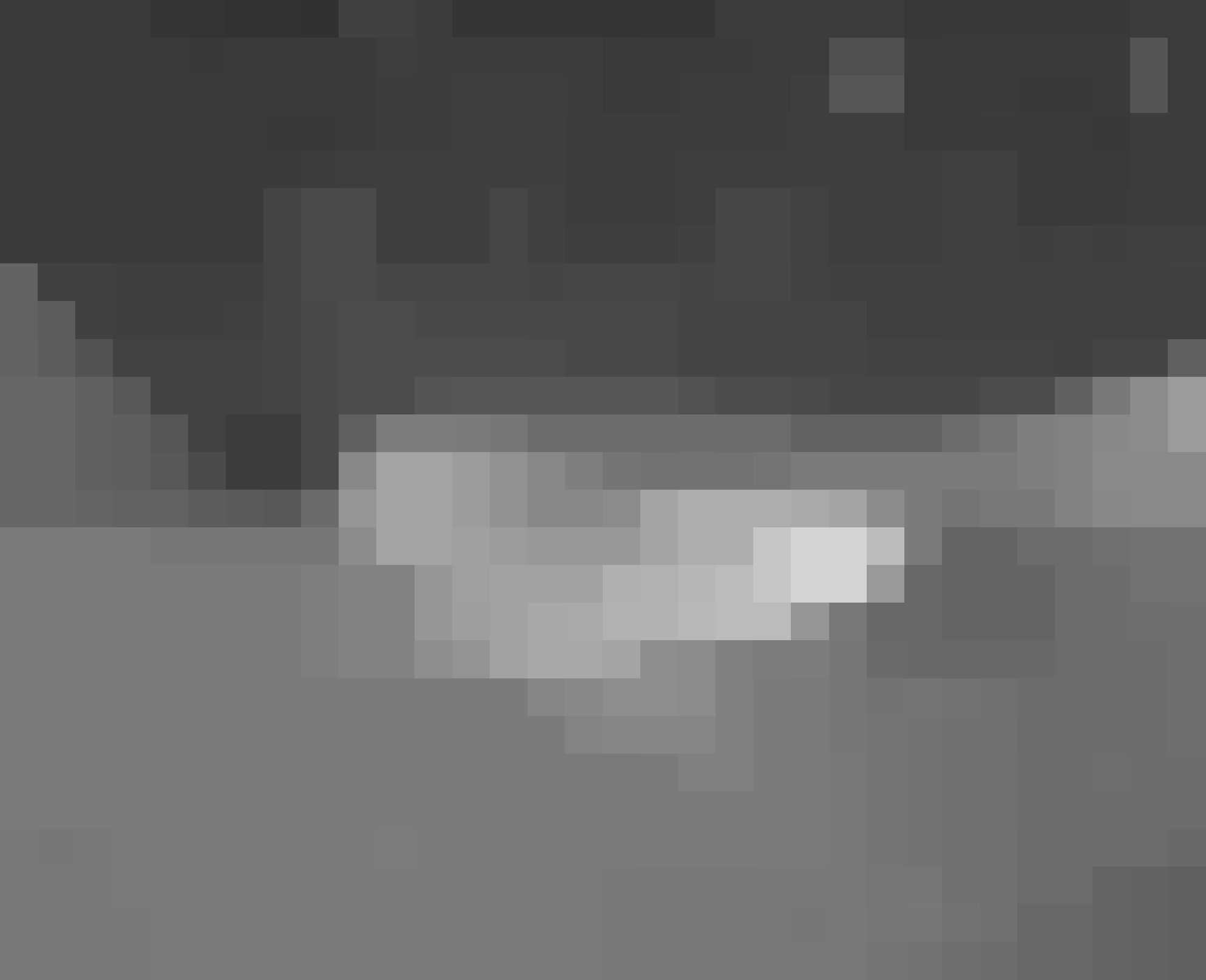}
		\includegraphics[width=0.28\columnwidth]{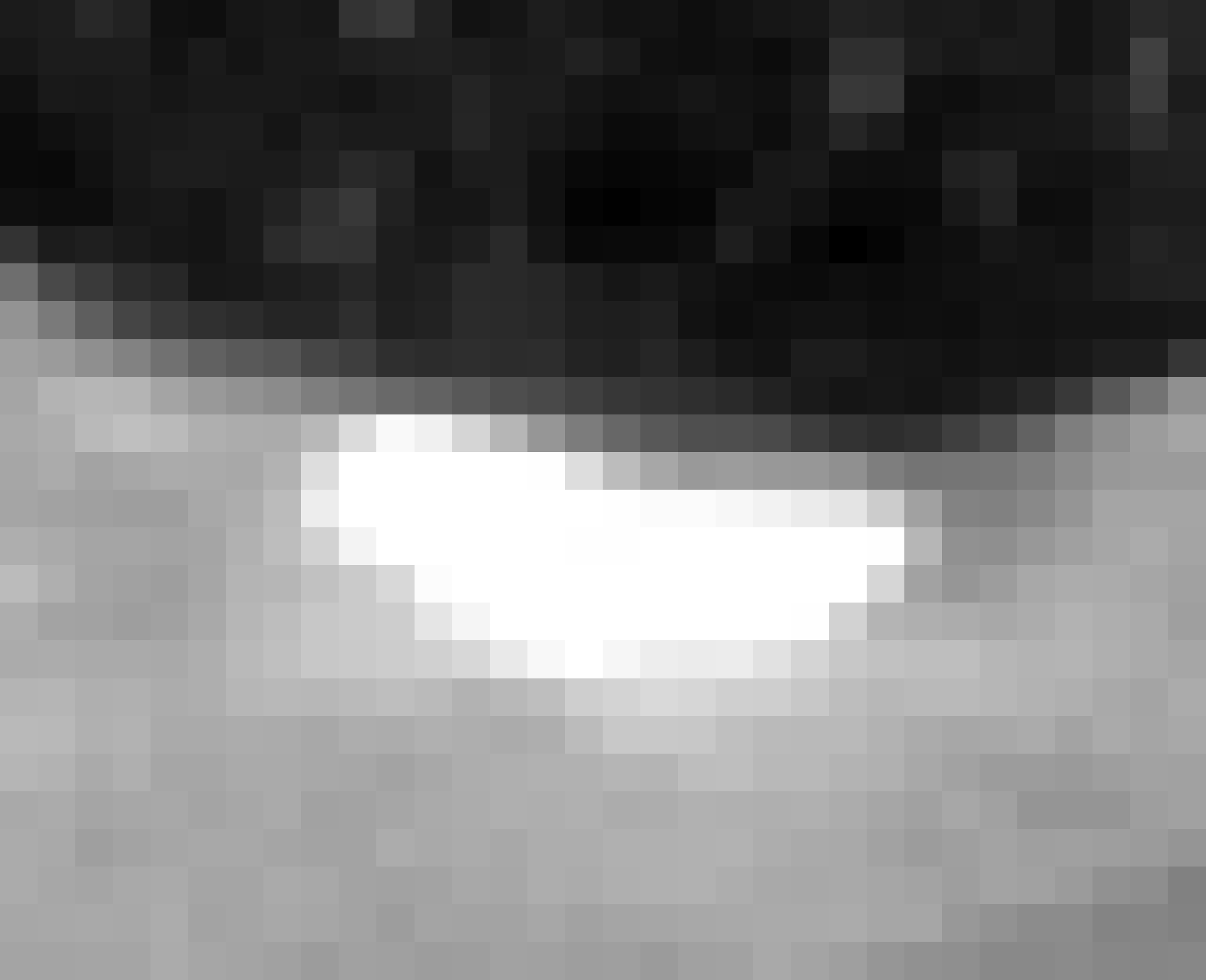}
		\includegraphics[width=0.28\columnwidth]{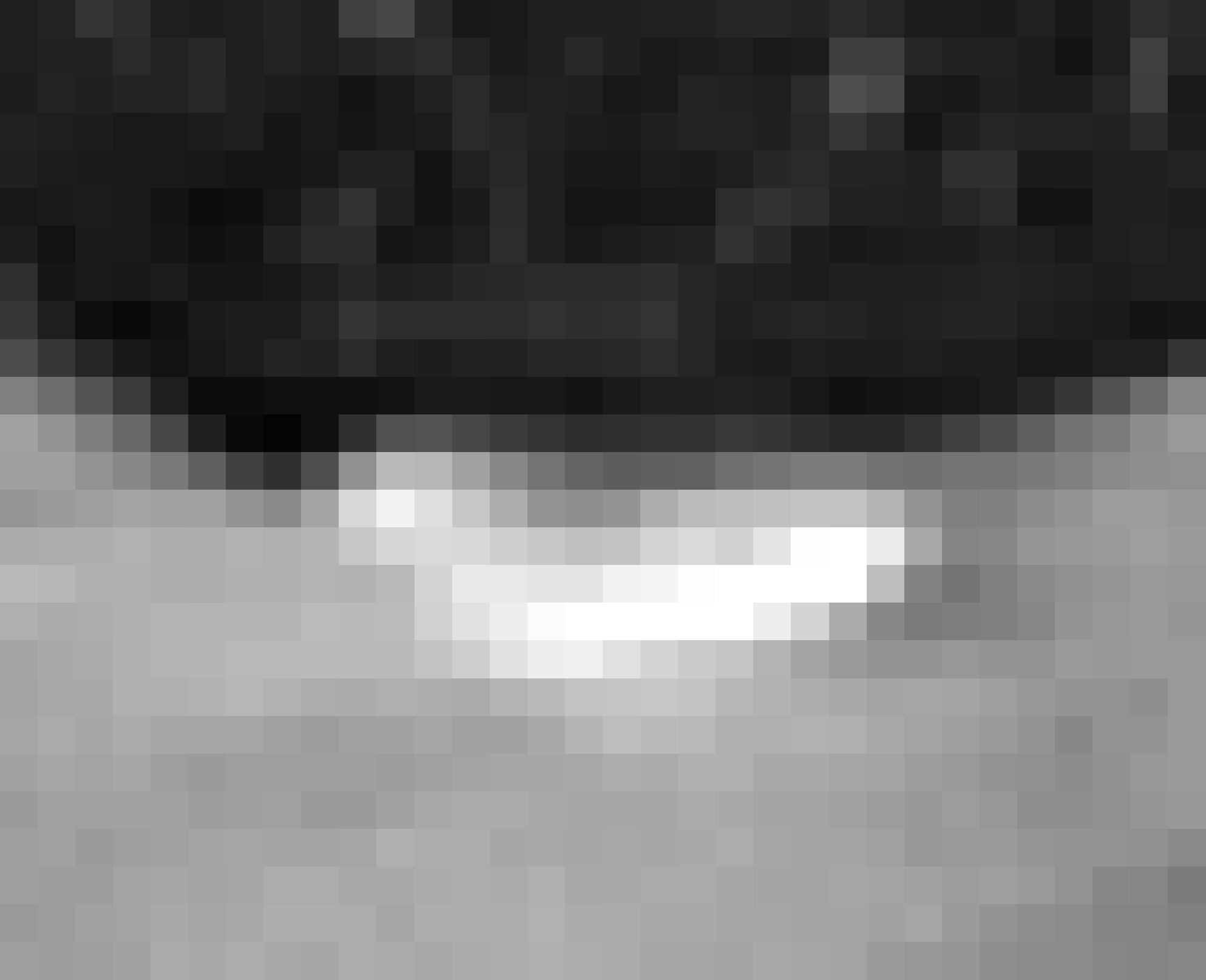}
		\includegraphics[width=0.28\columnwidth]{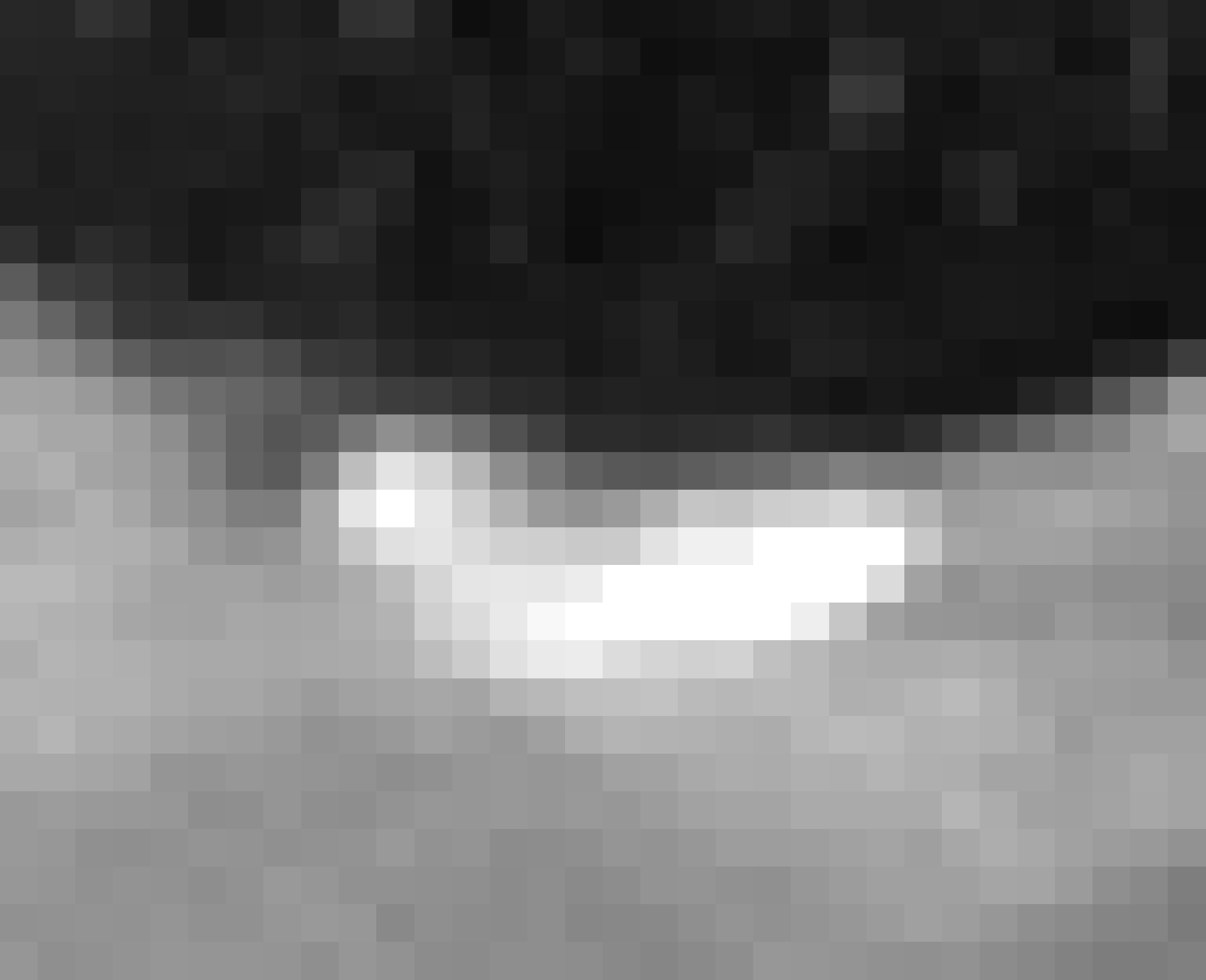}
		\includegraphics[width=0.28\columnwidth]{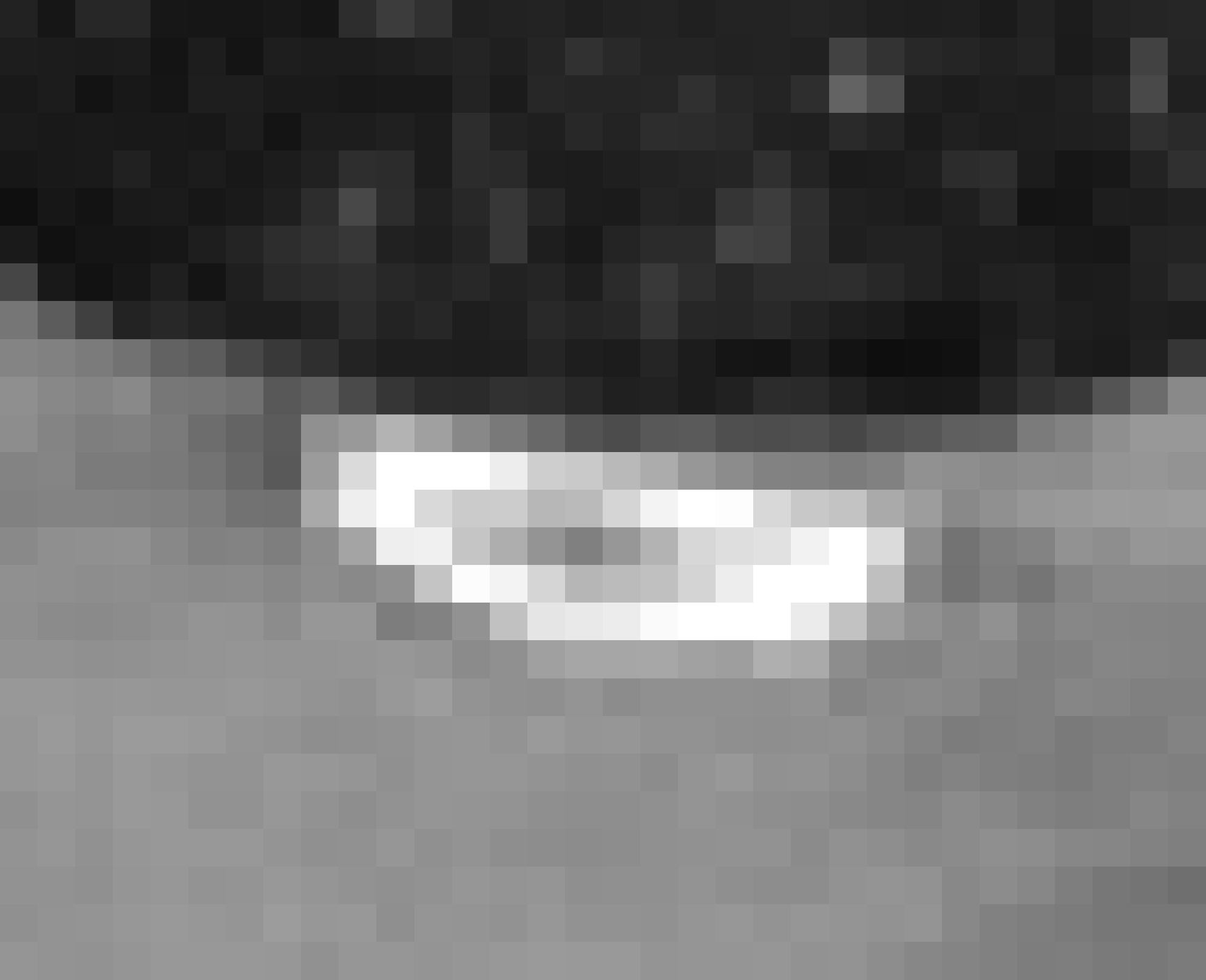}
	}
	\vspace{0.5mm}
	\centerline{
		\includegraphics[width=0.28\columnwidth]{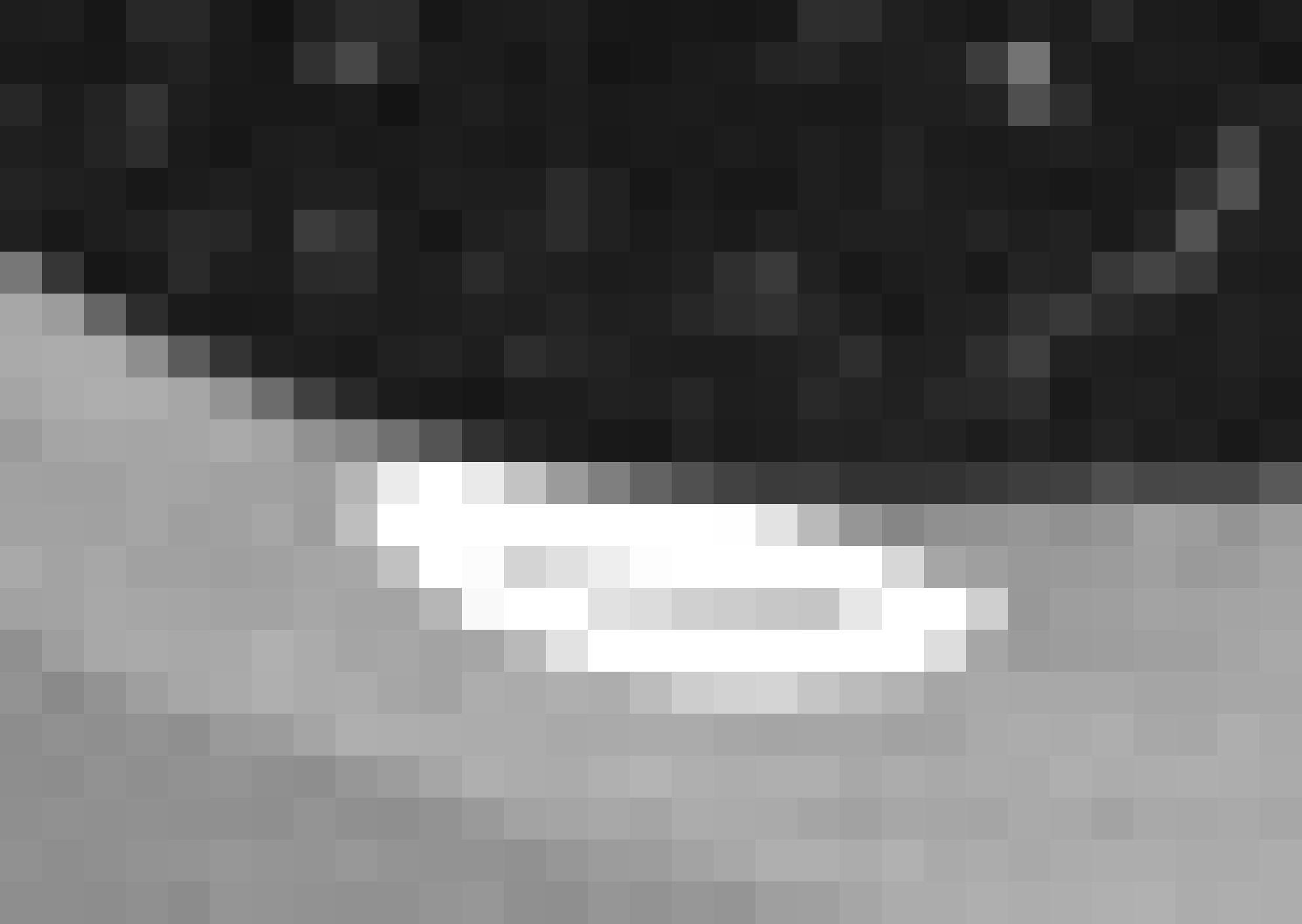}
		\includegraphics[width=0.28\columnwidth]{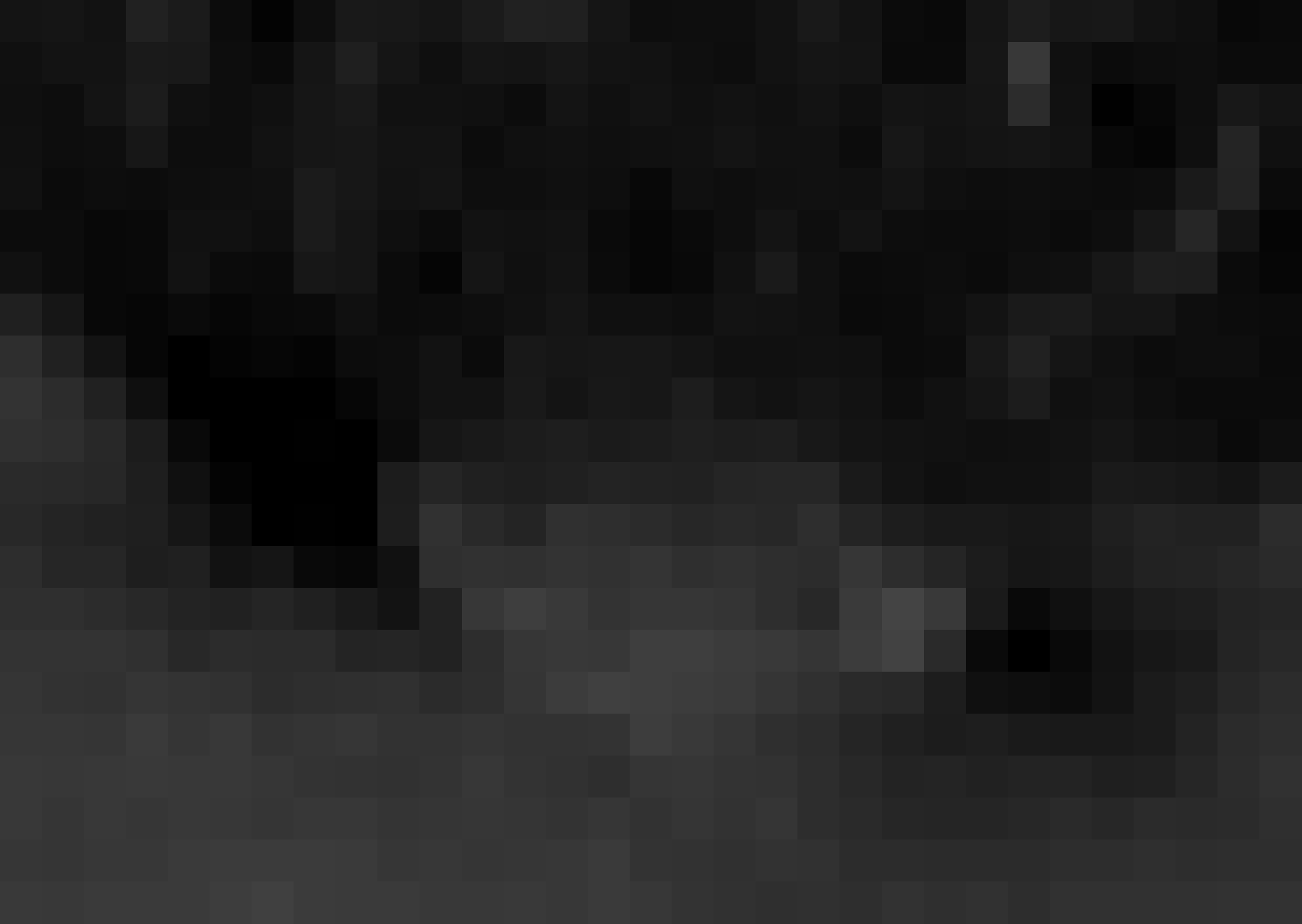}
		\includegraphics[width=0.28\columnwidth]{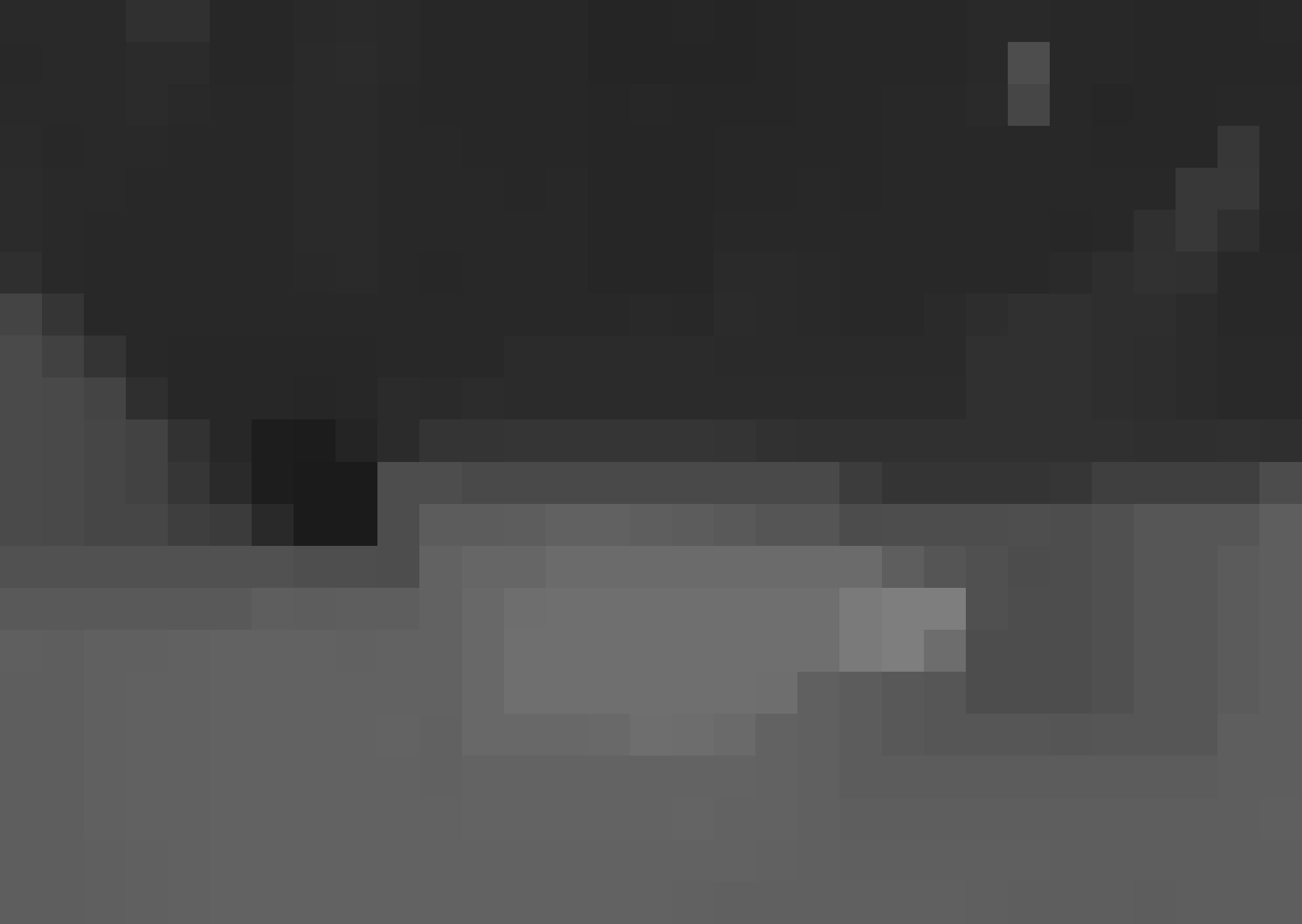}
		\includegraphics[width=0.28\columnwidth]{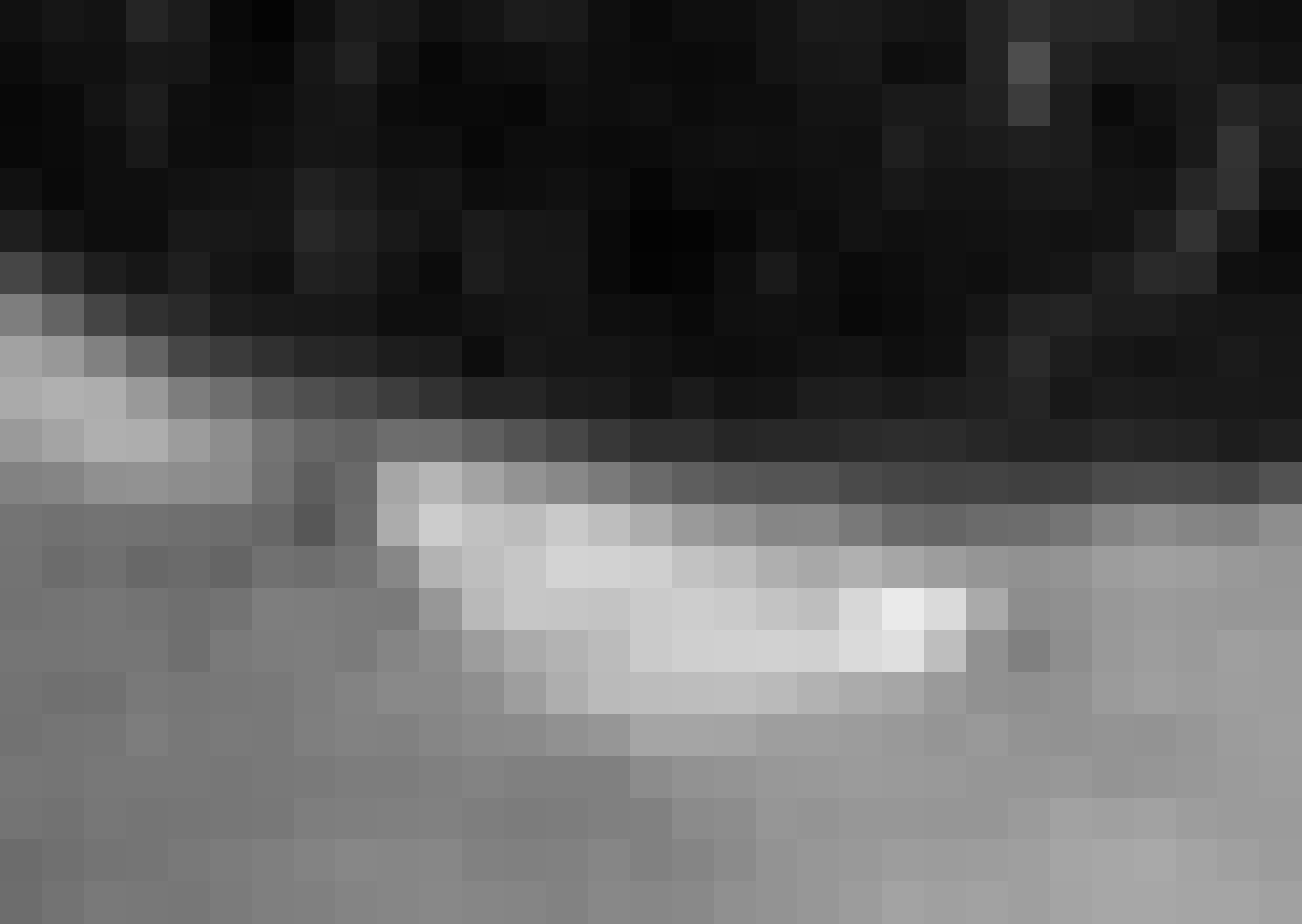}
		\includegraphics[width=0.28\columnwidth]{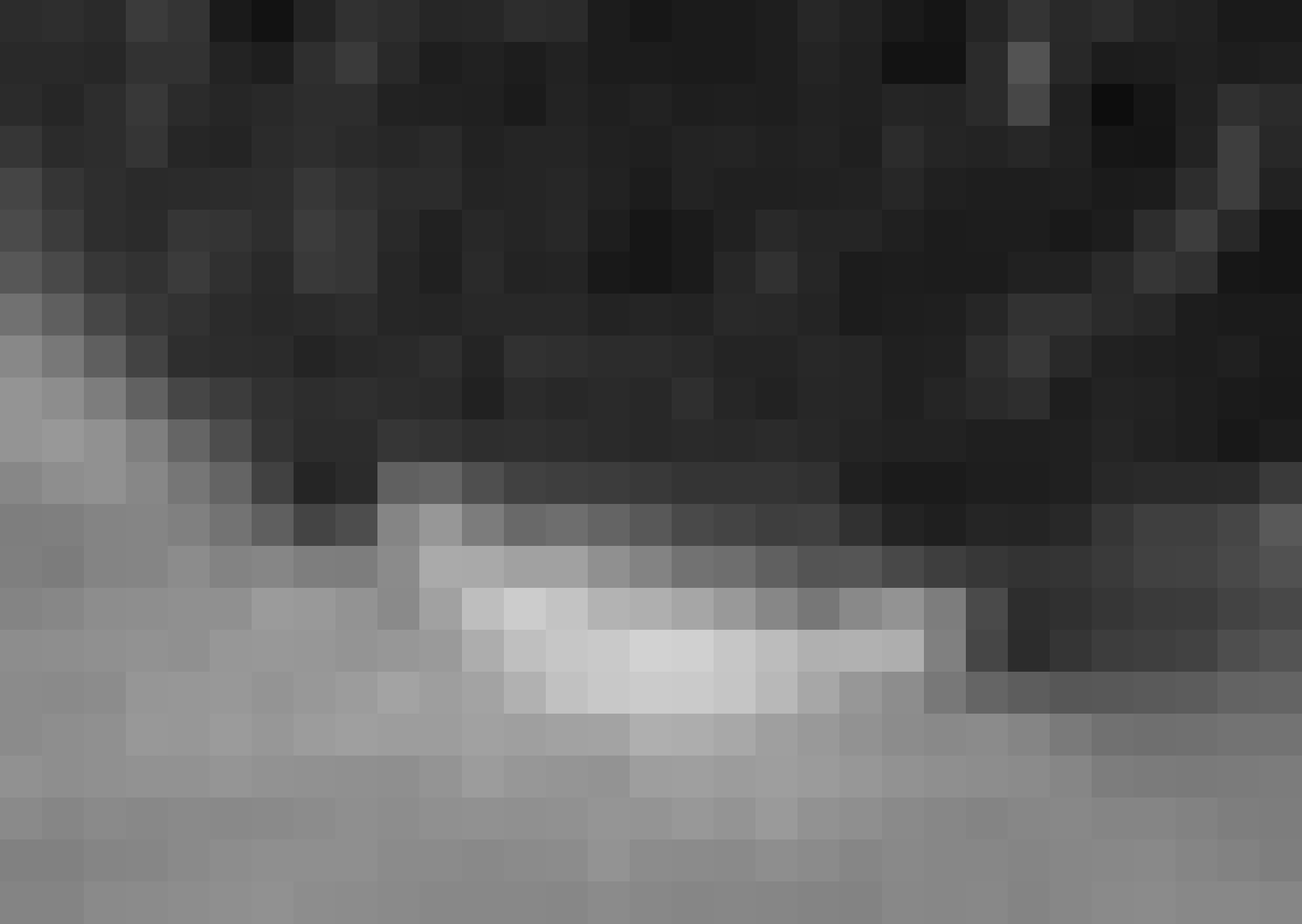}
		\includegraphics[width=0.28\columnwidth]{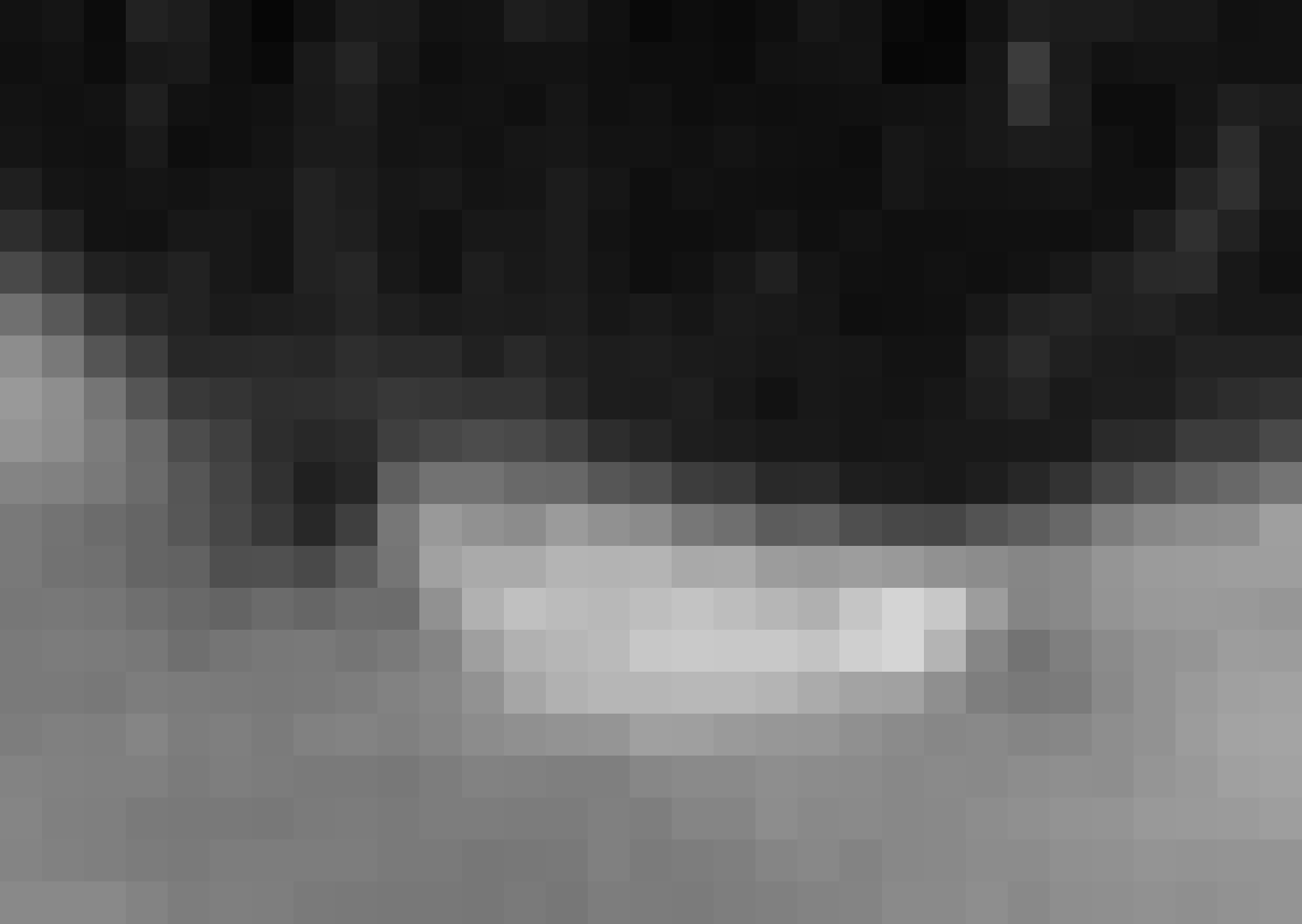}
		\includegraphics[width=0.28\columnwidth]{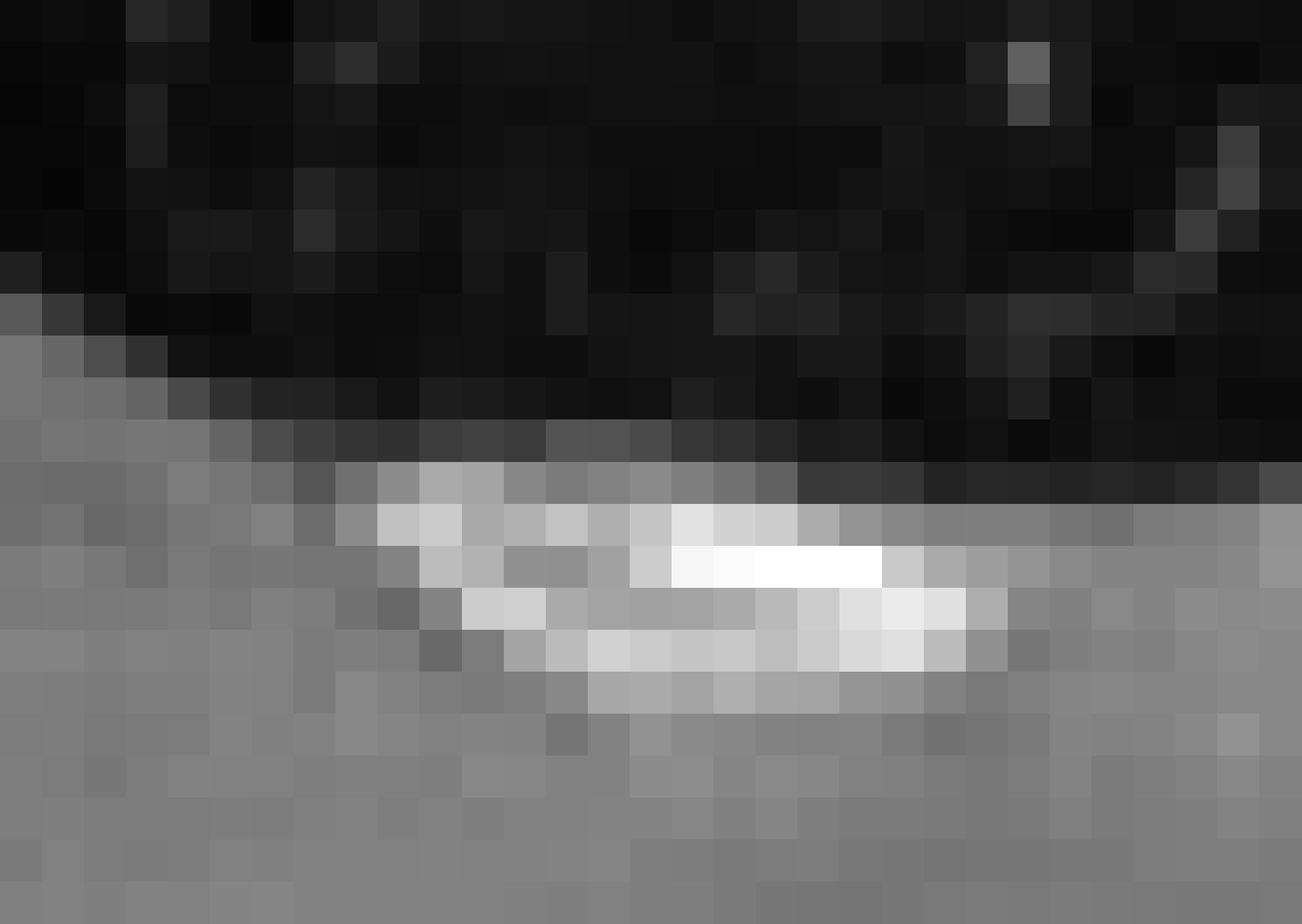}
	}
	\vspace{0.5mm}
	\centerline{
		\includegraphics[width=0.28\columnwidth]{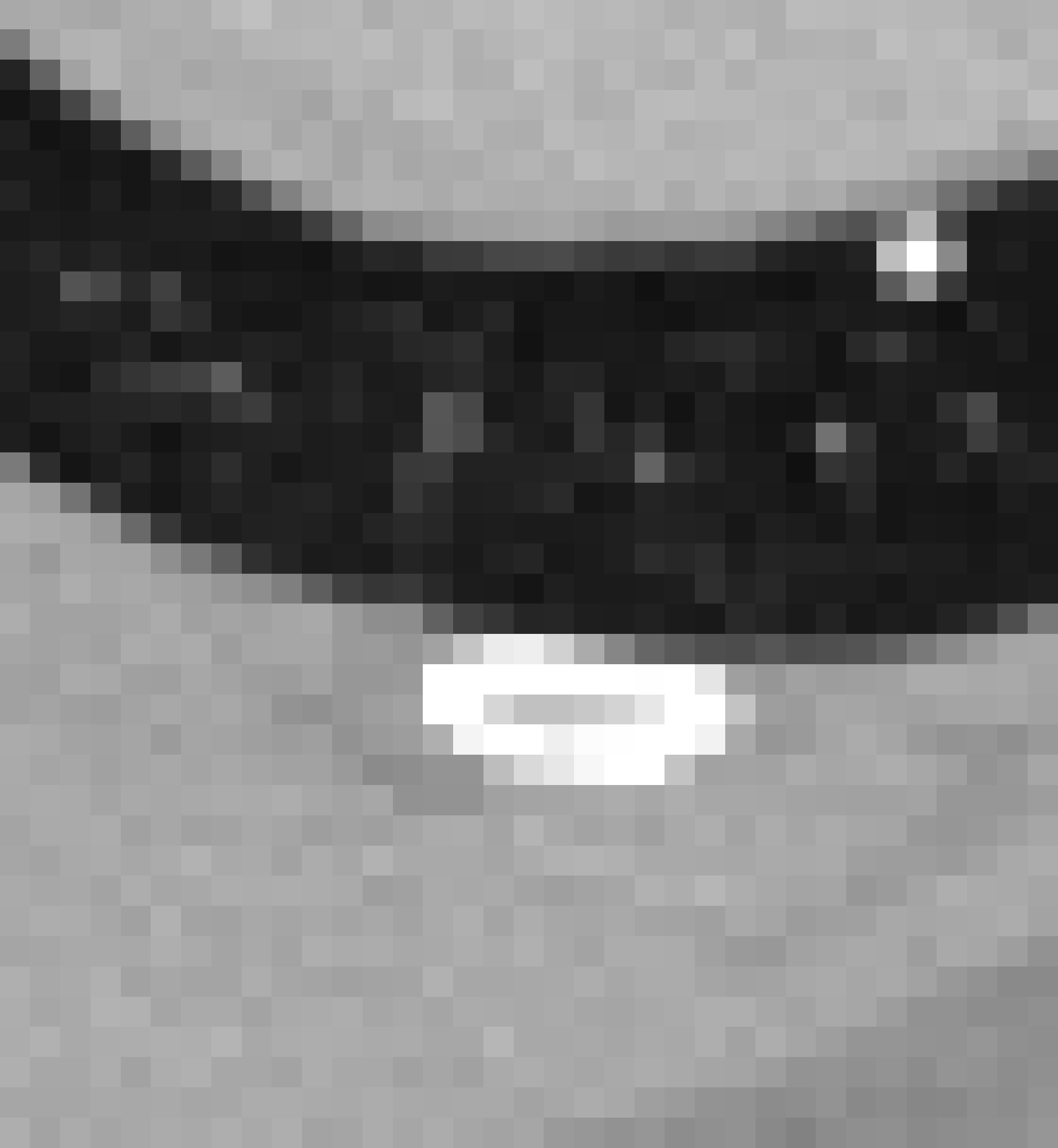}
		\includegraphics[width=0.28\columnwidth]{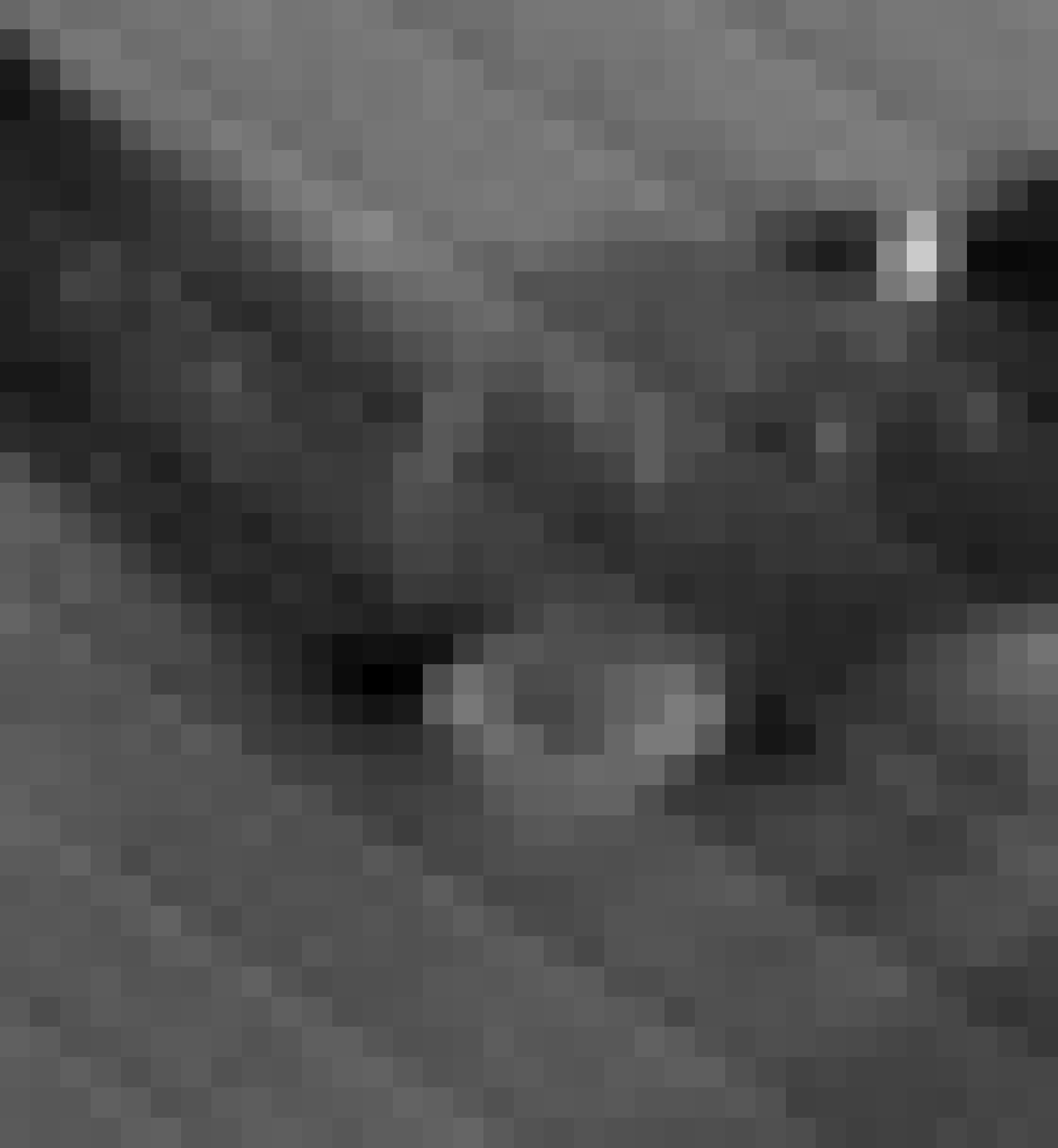}
		\includegraphics[width=0.28\columnwidth]{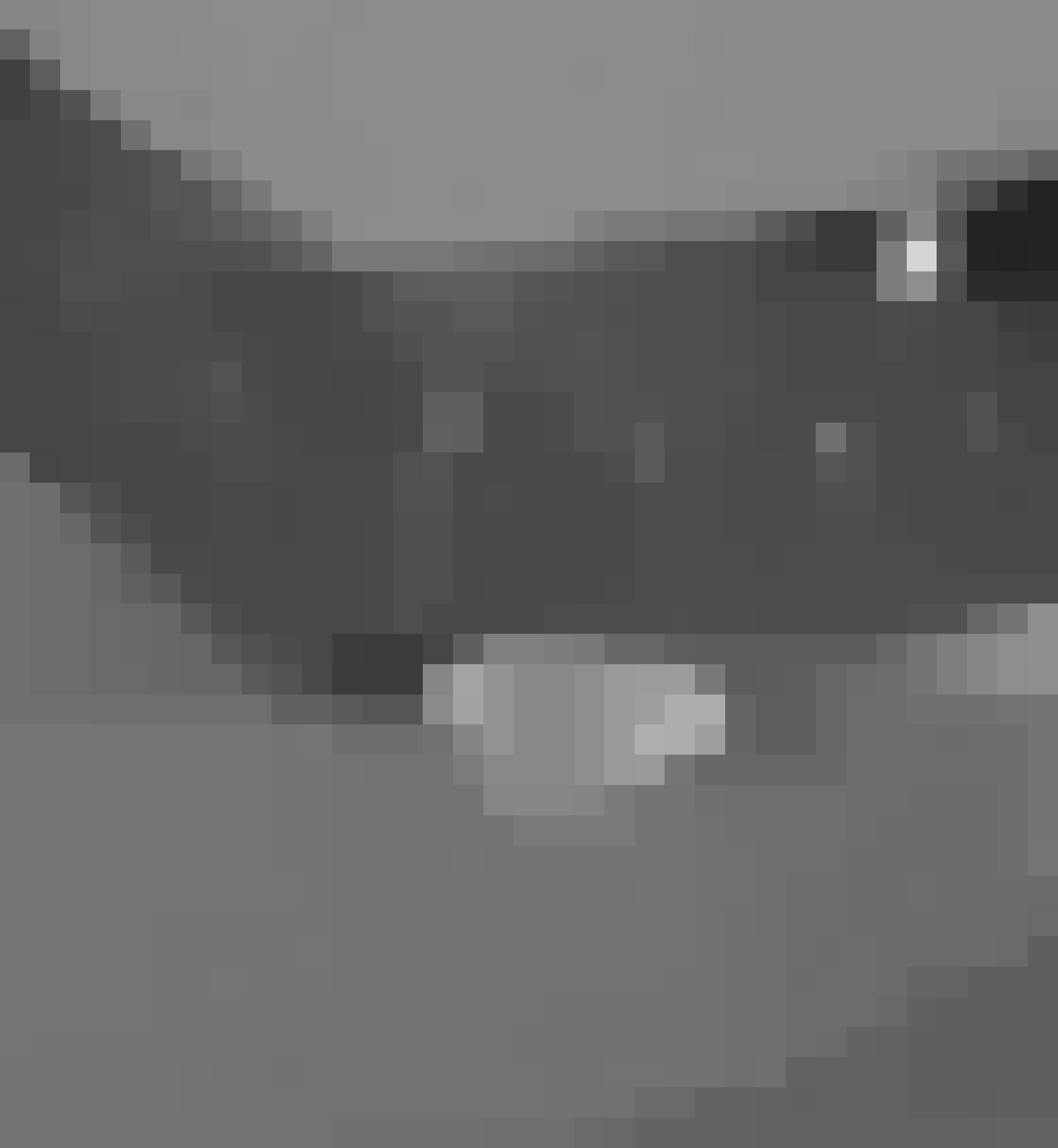}
		\includegraphics[width=0.28\columnwidth]{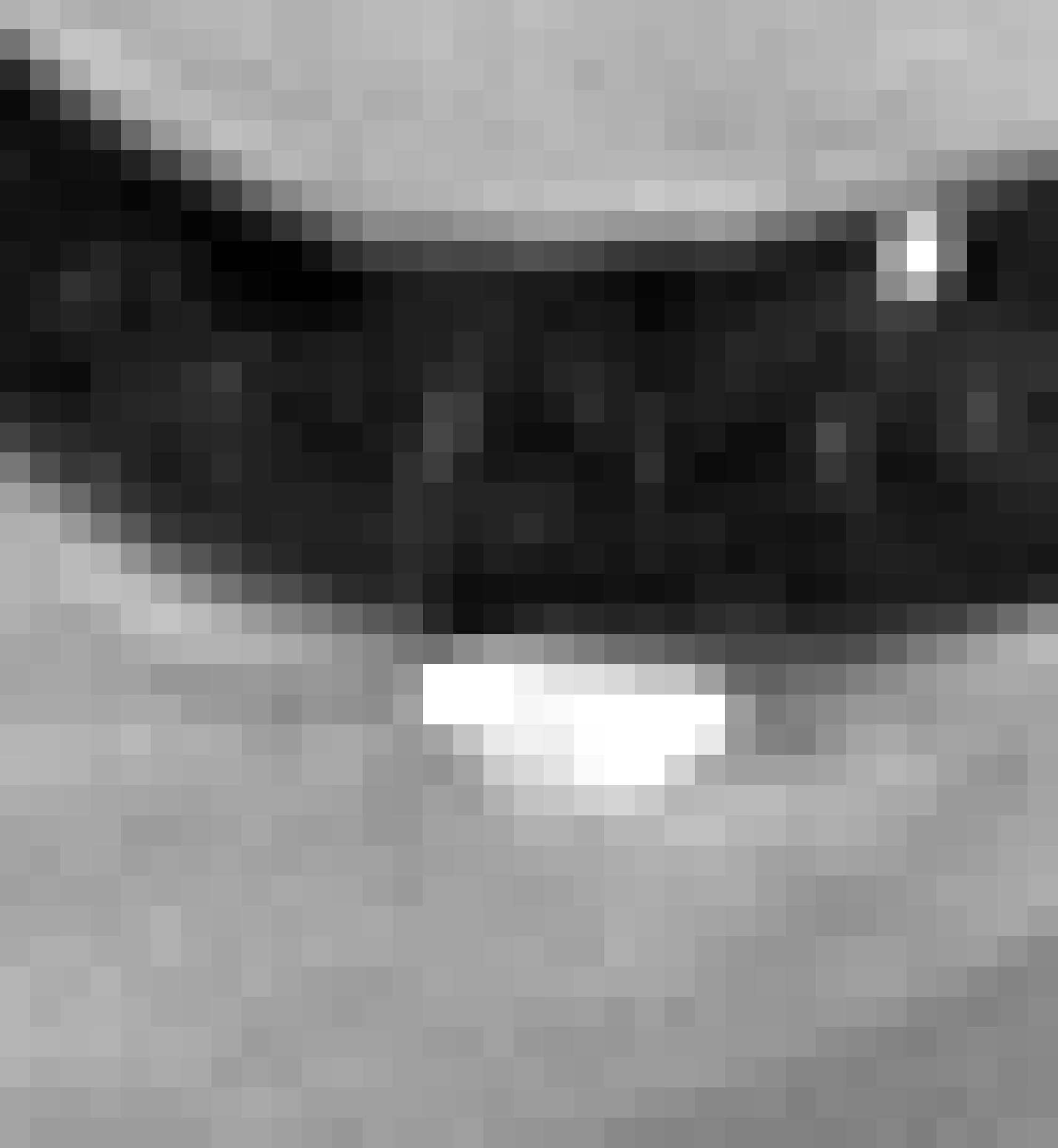}
		\includegraphics[width=0.28\columnwidth]{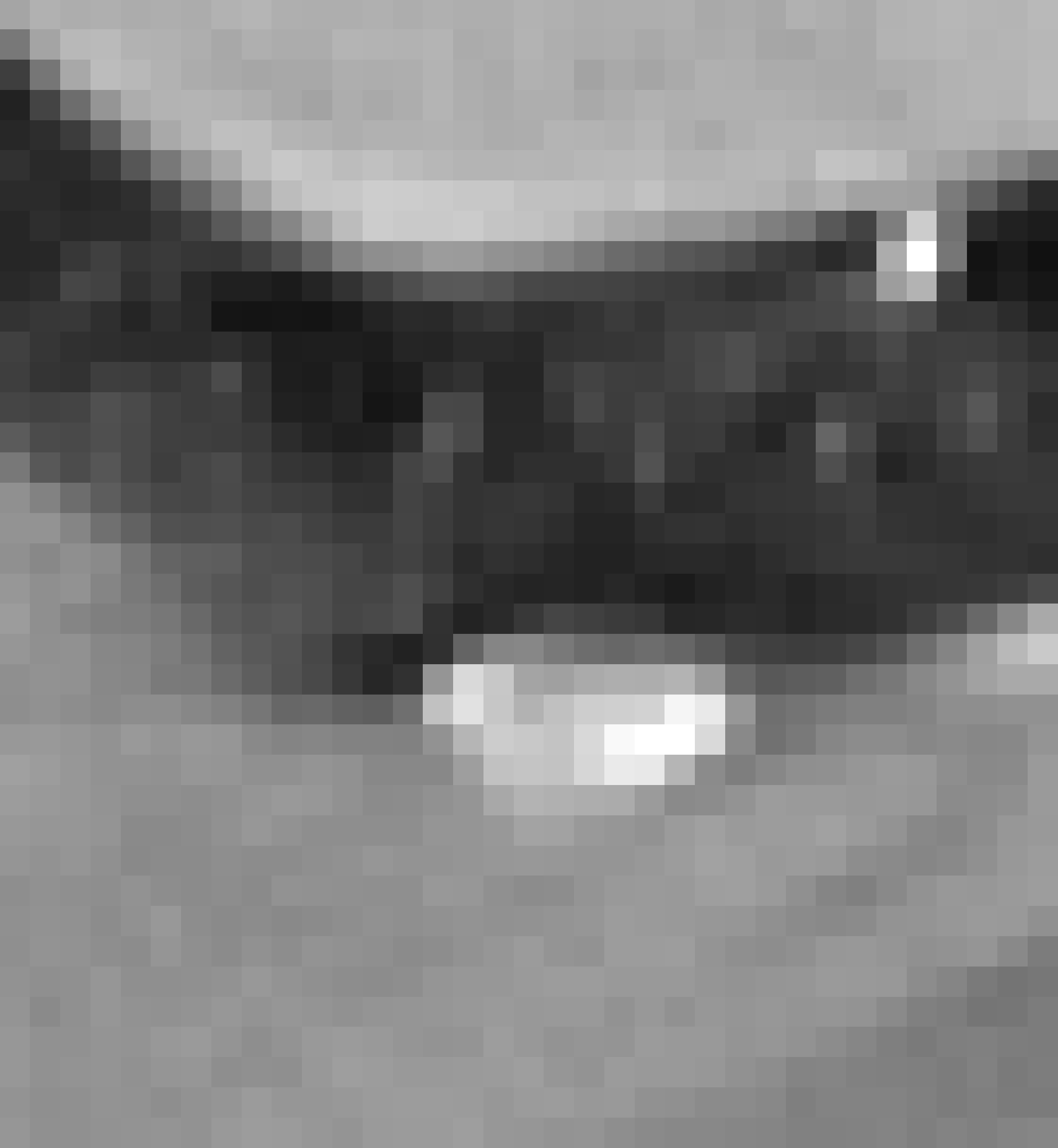}
		\includegraphics[width=0.28\columnwidth]{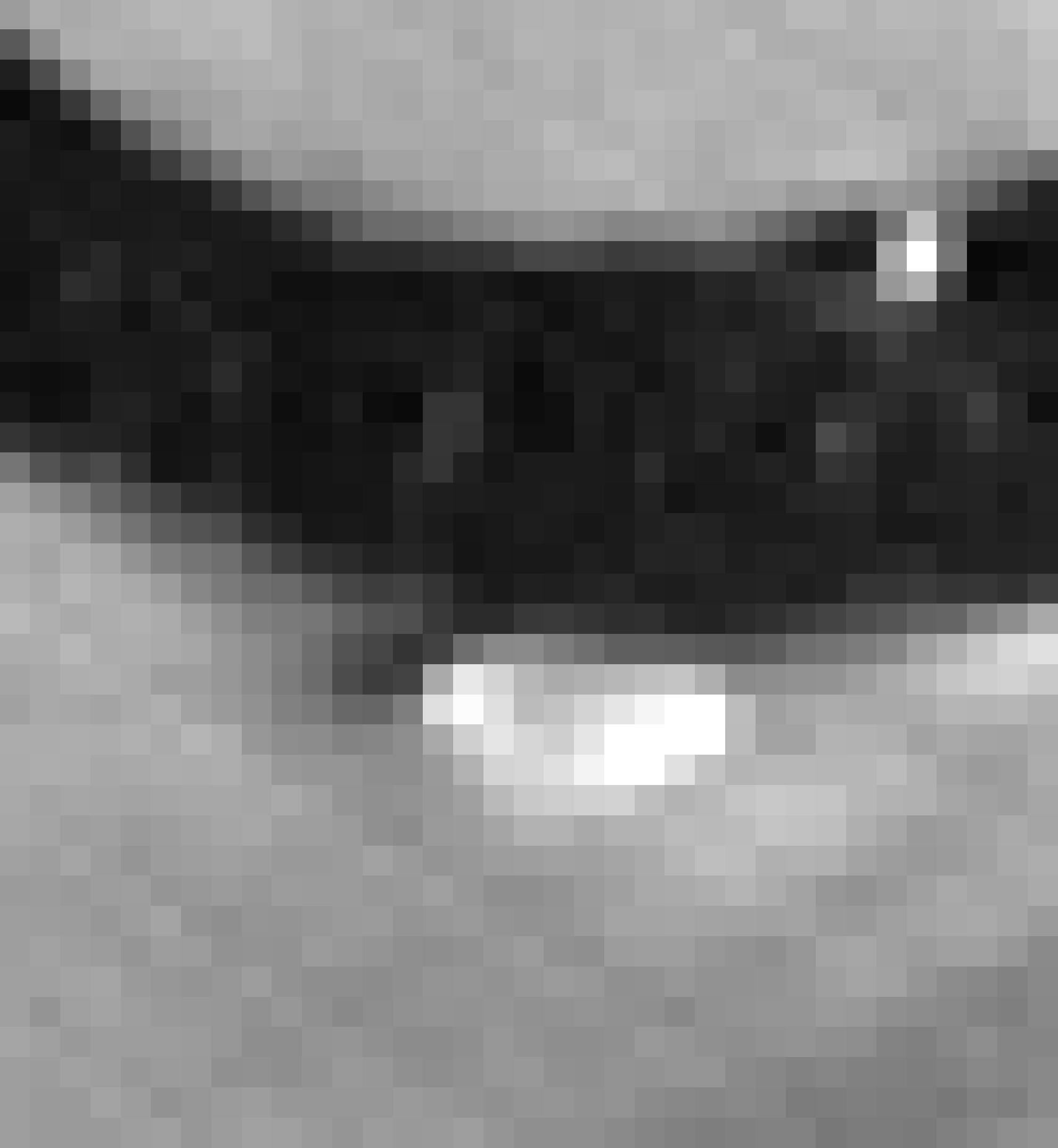}
		\includegraphics[width=0.28\columnwidth]{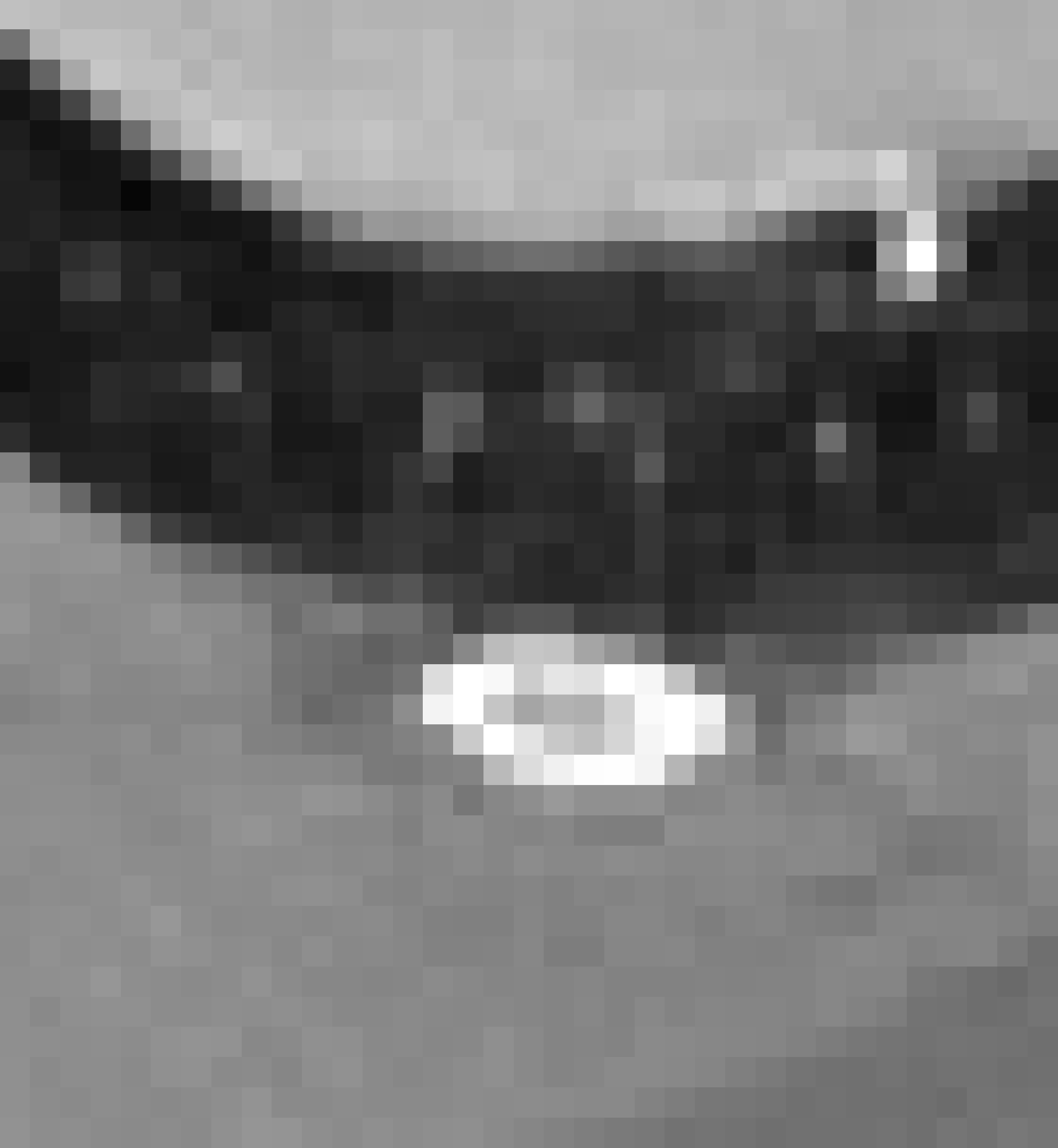}
	}
	\vspace{0.5mm}
	\centerline{
		\includegraphics[width=0.28\columnwidth]{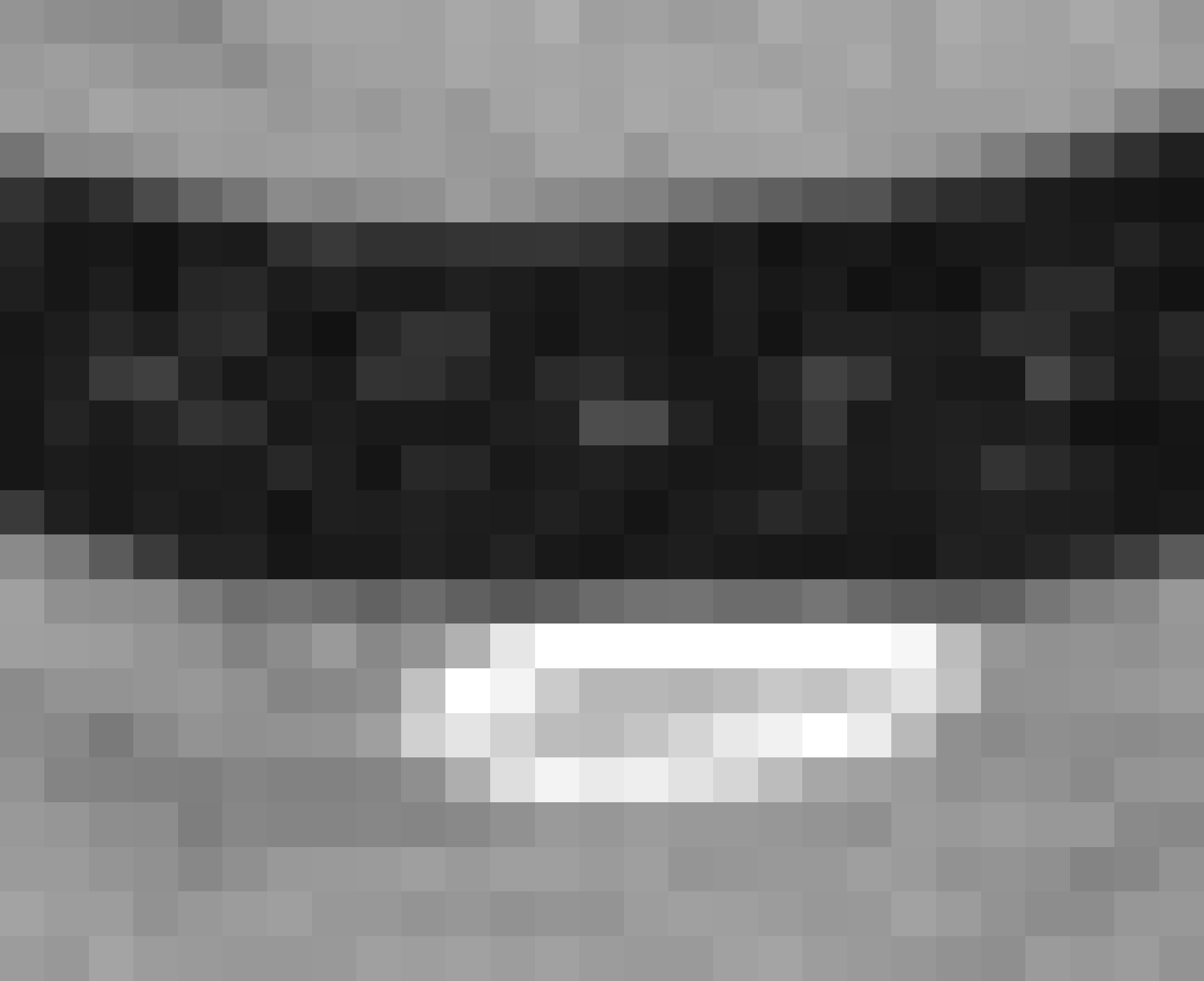}
		\includegraphics[width=0.28\columnwidth]{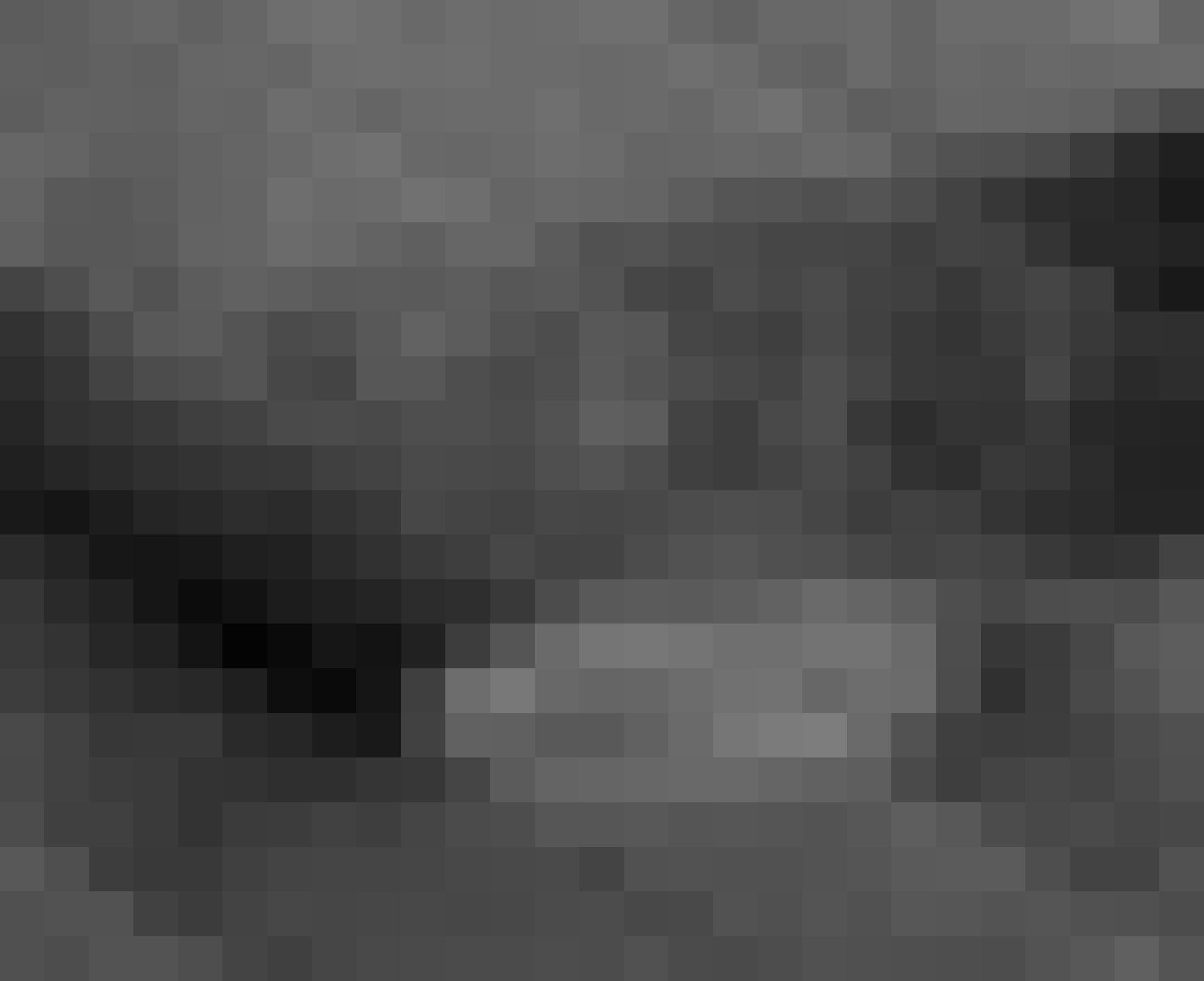}
		\includegraphics[width=0.28\columnwidth]{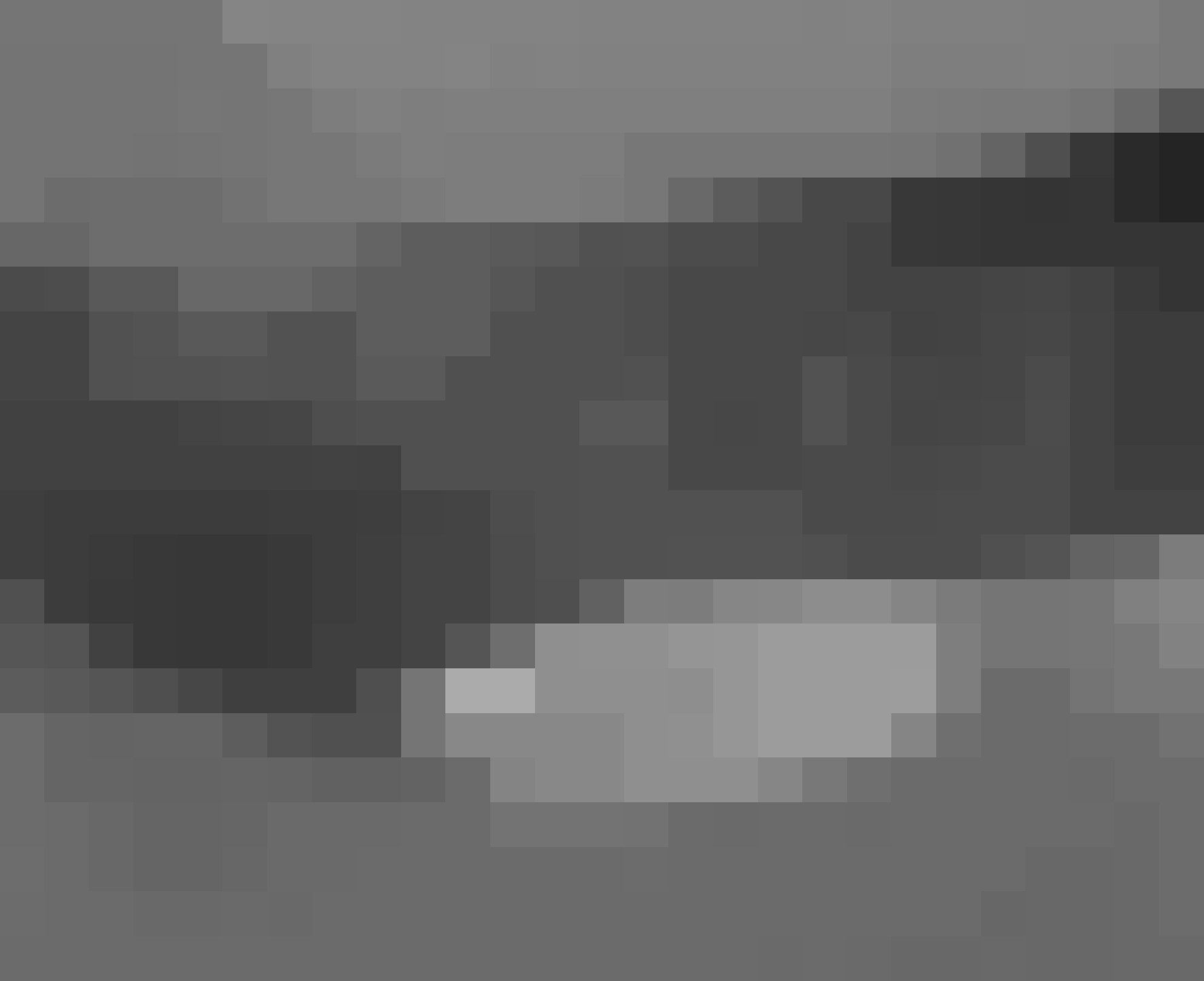}
		\includegraphics[width=0.28\columnwidth]{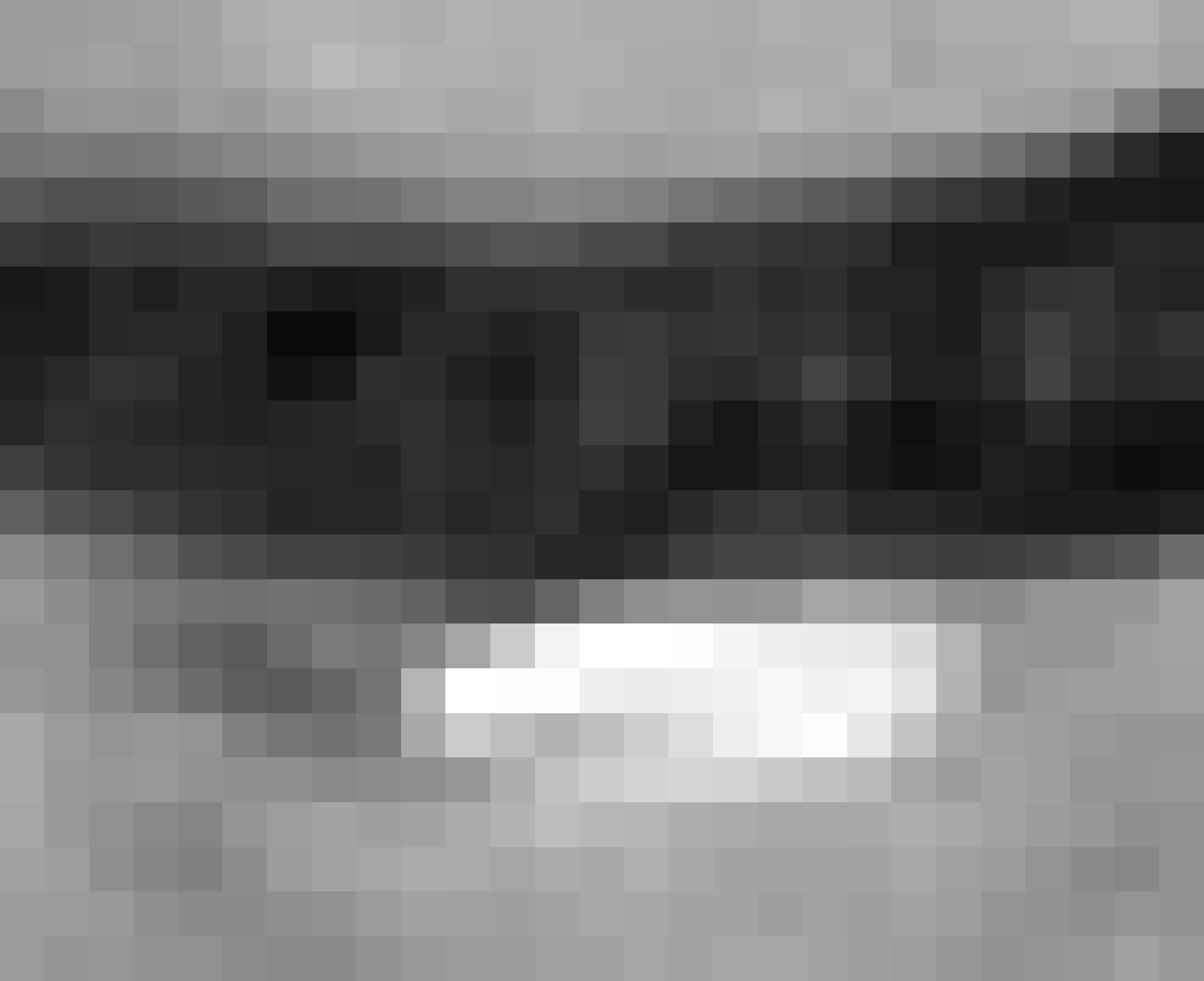}
		\includegraphics[width=0.28\columnwidth]{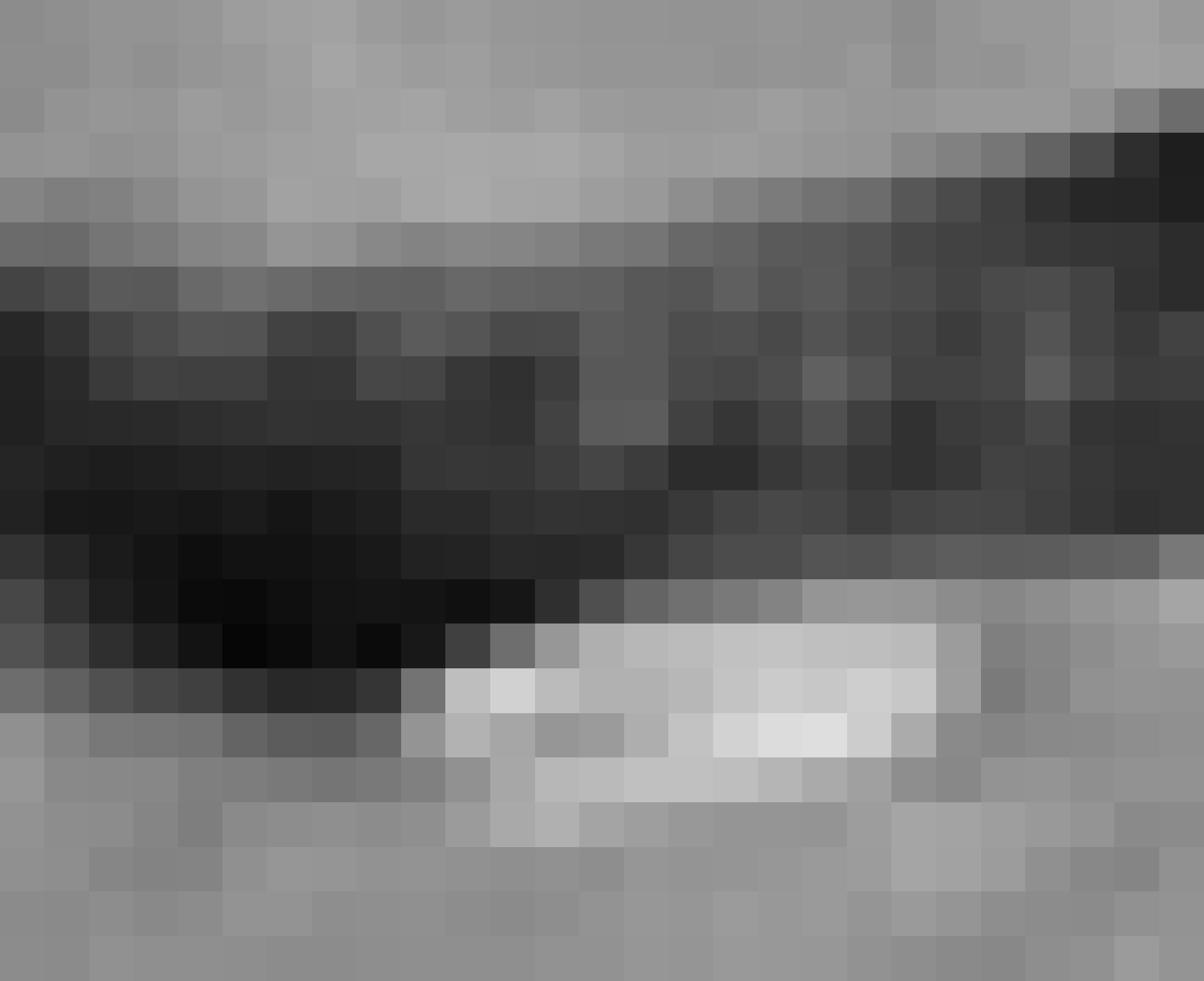}
		\includegraphics[width=0.28\columnwidth]{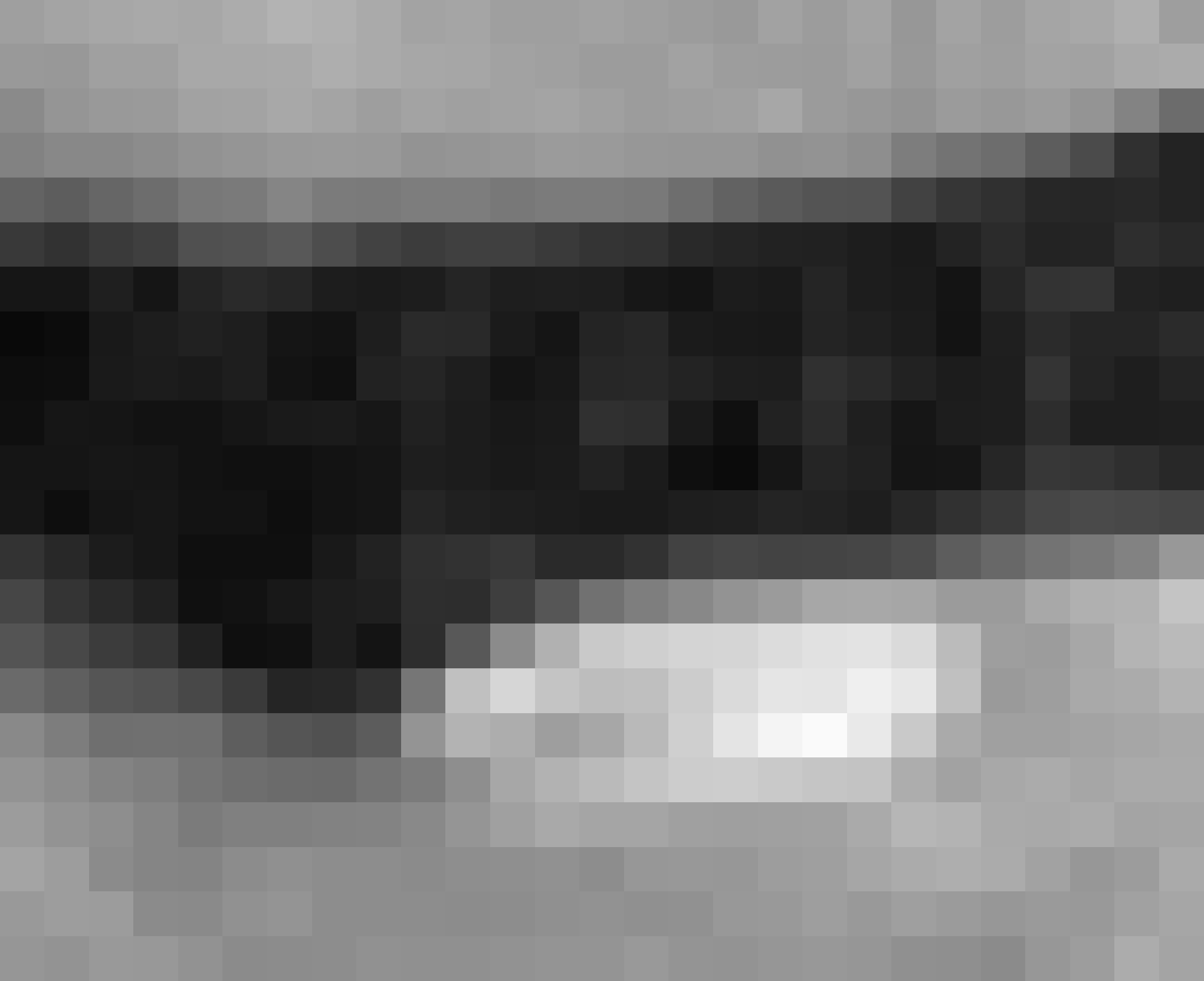}
		\includegraphics[width=0.28\columnwidth]{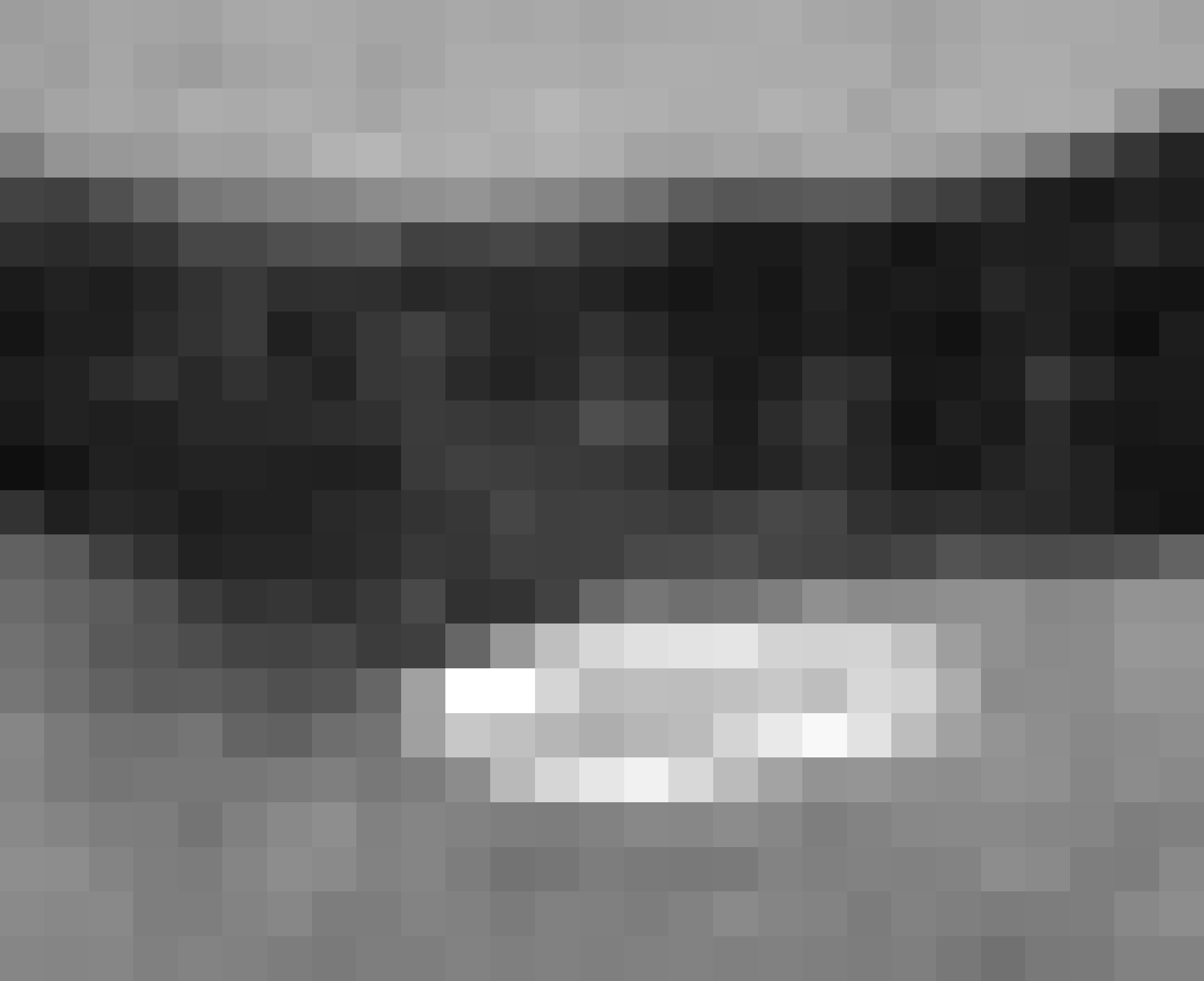}
	}
	\vspace{0.5mm}
	\centerline{
		\subfigure[Label]{\includegraphics[width=0.28\columnwidth]{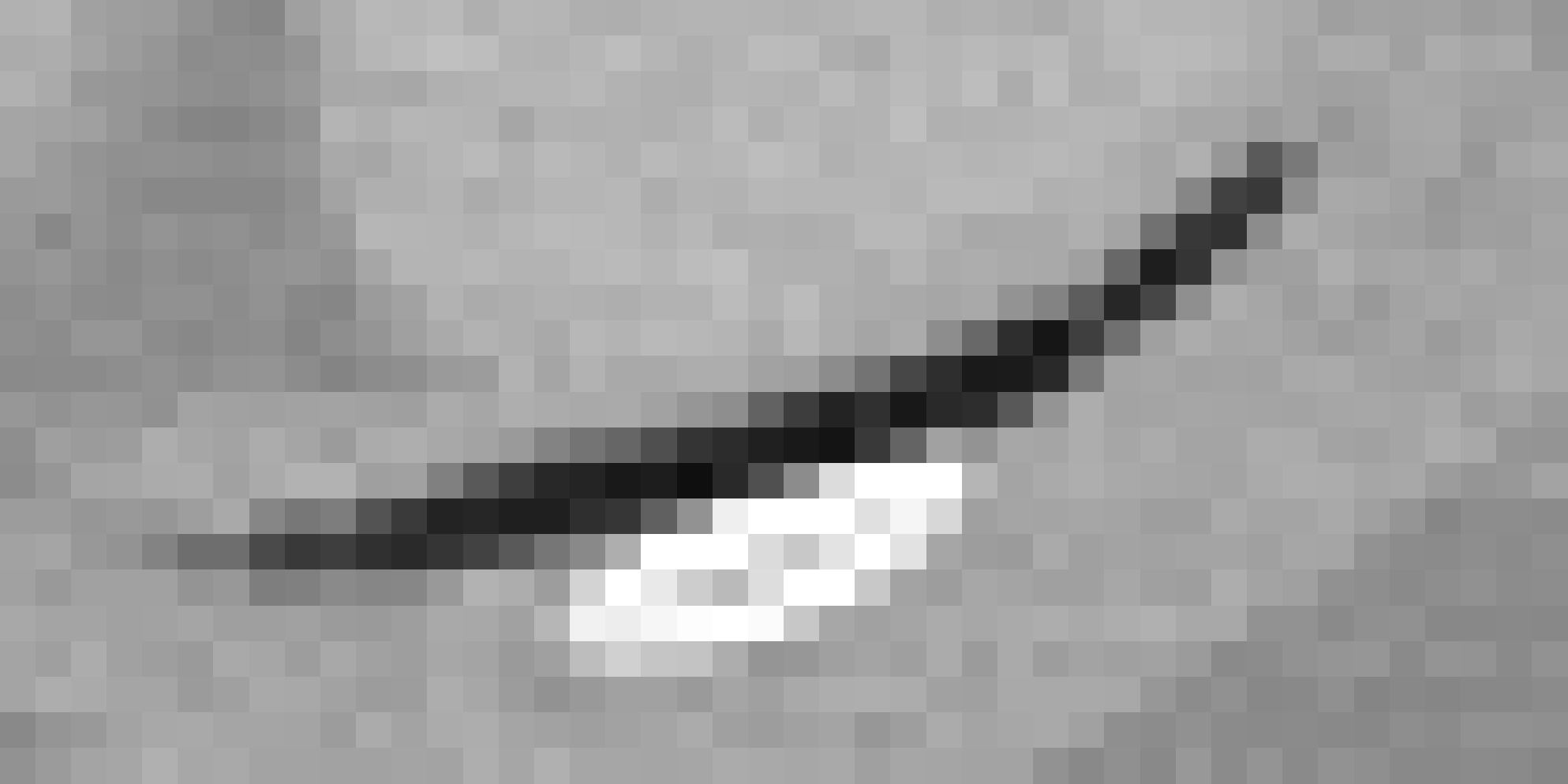}}
		\subfigure[FBP]{\includegraphics[width=0.28\columnwidth]{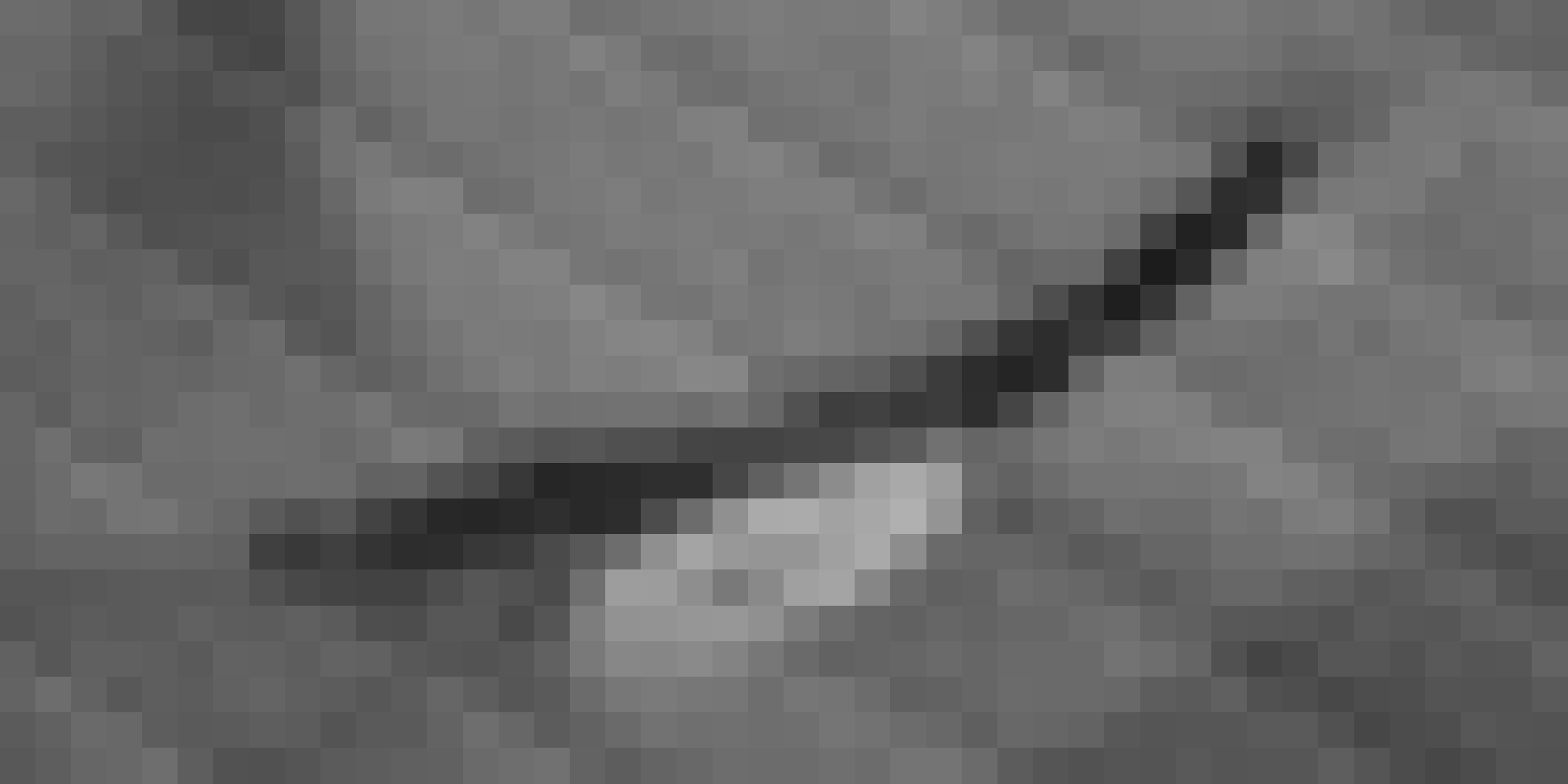}}
		\subfigure[TV-regularization]{\includegraphics[width=0.28\columnwidth]{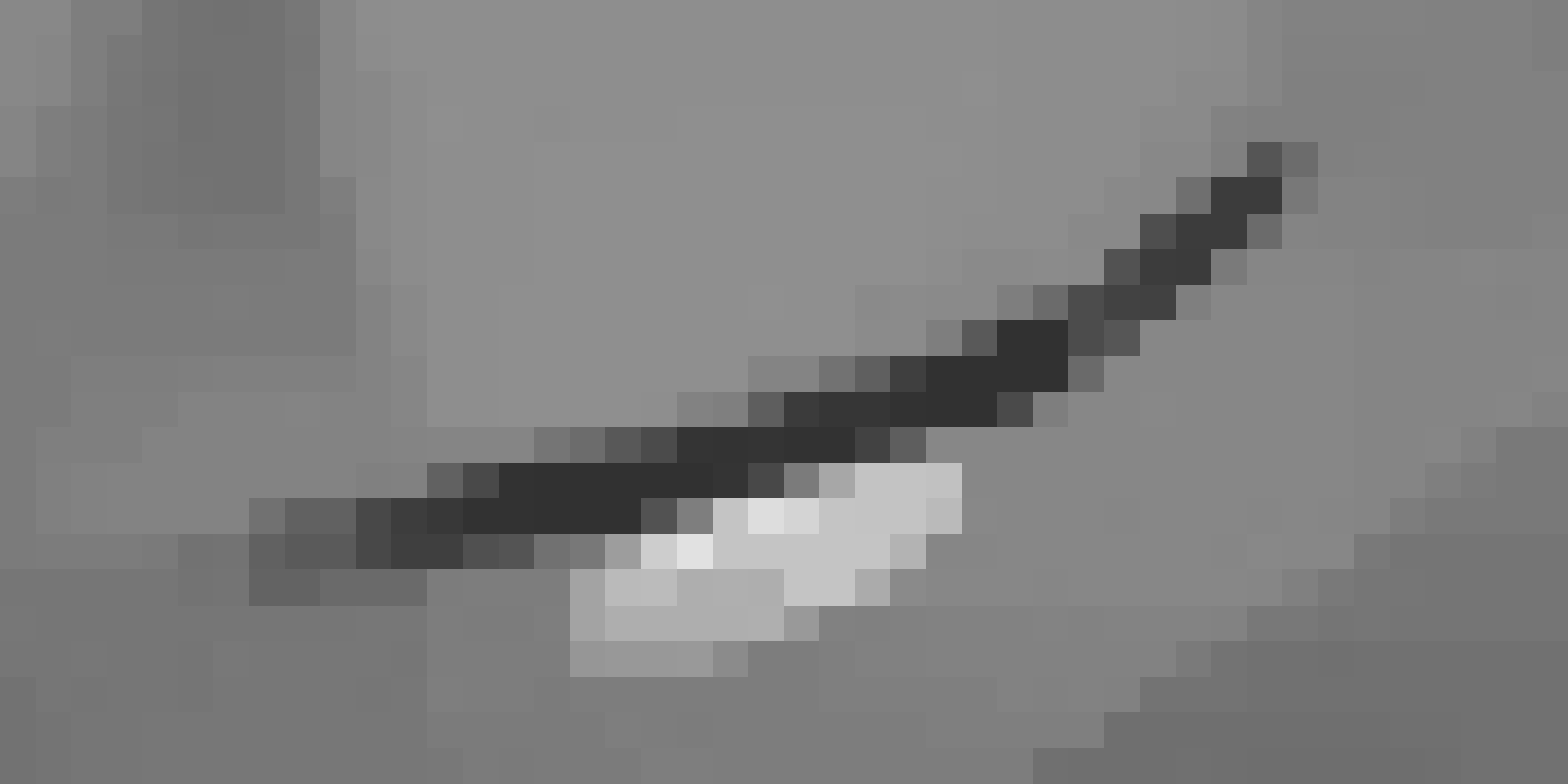}}
		\subfigure[Red-CNN]{\includegraphics[width=0.28\columnwidth]{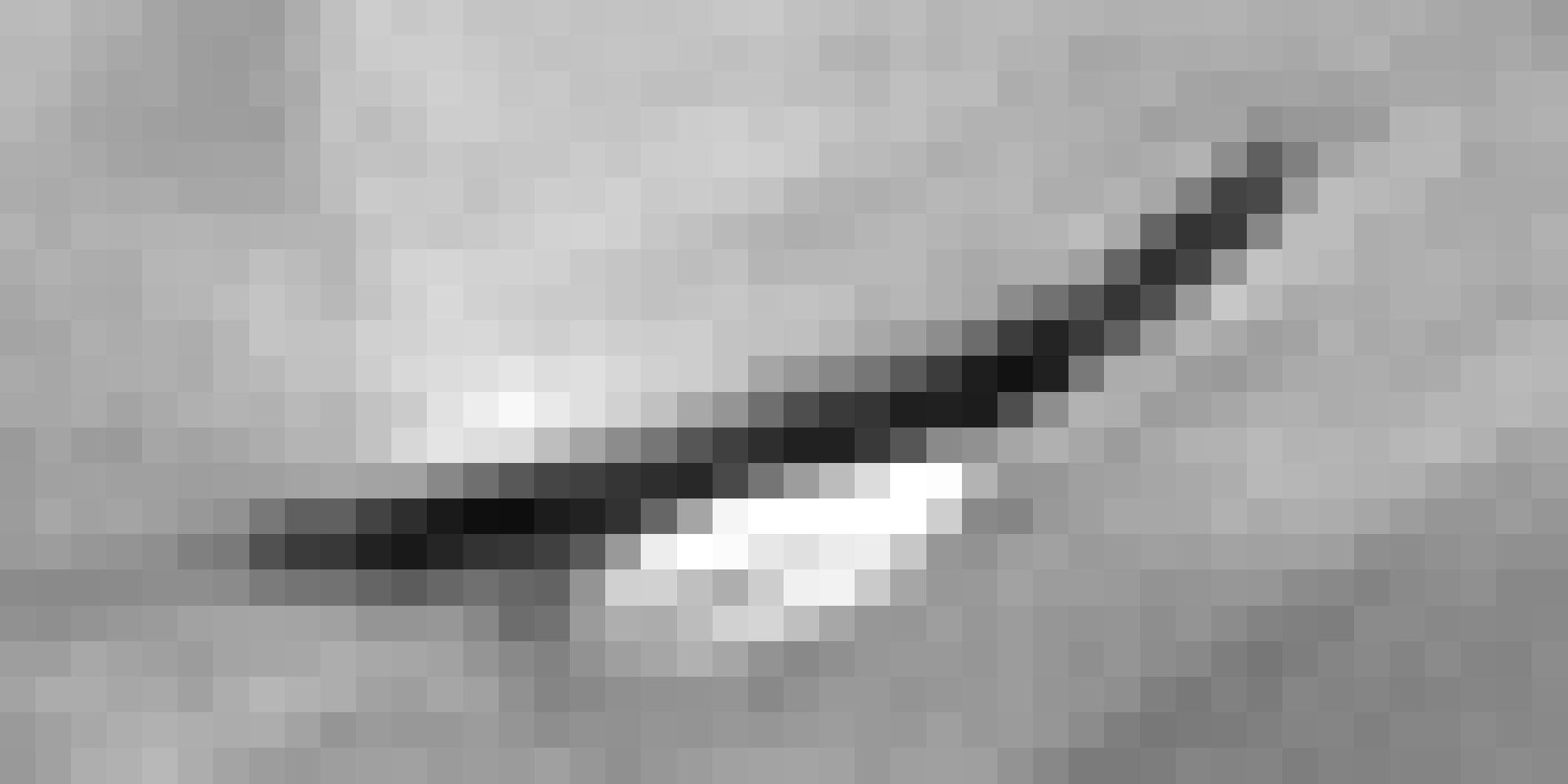}}
		\subfigure[FBP-Conv]{\includegraphics[width=0.28\columnwidth]{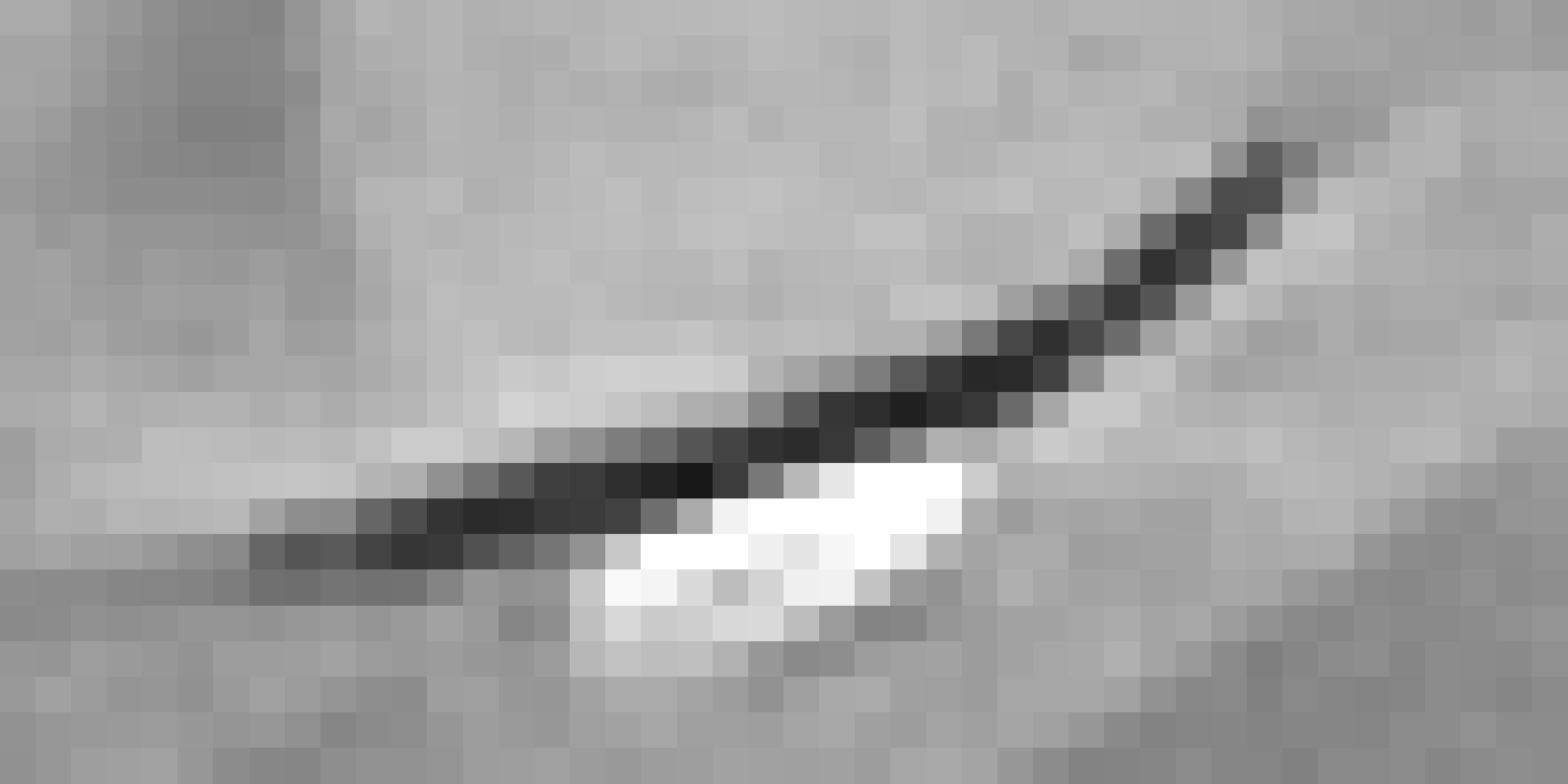}}
		\subfigure[DD-Net]{\includegraphics[width=0.28\columnwidth]{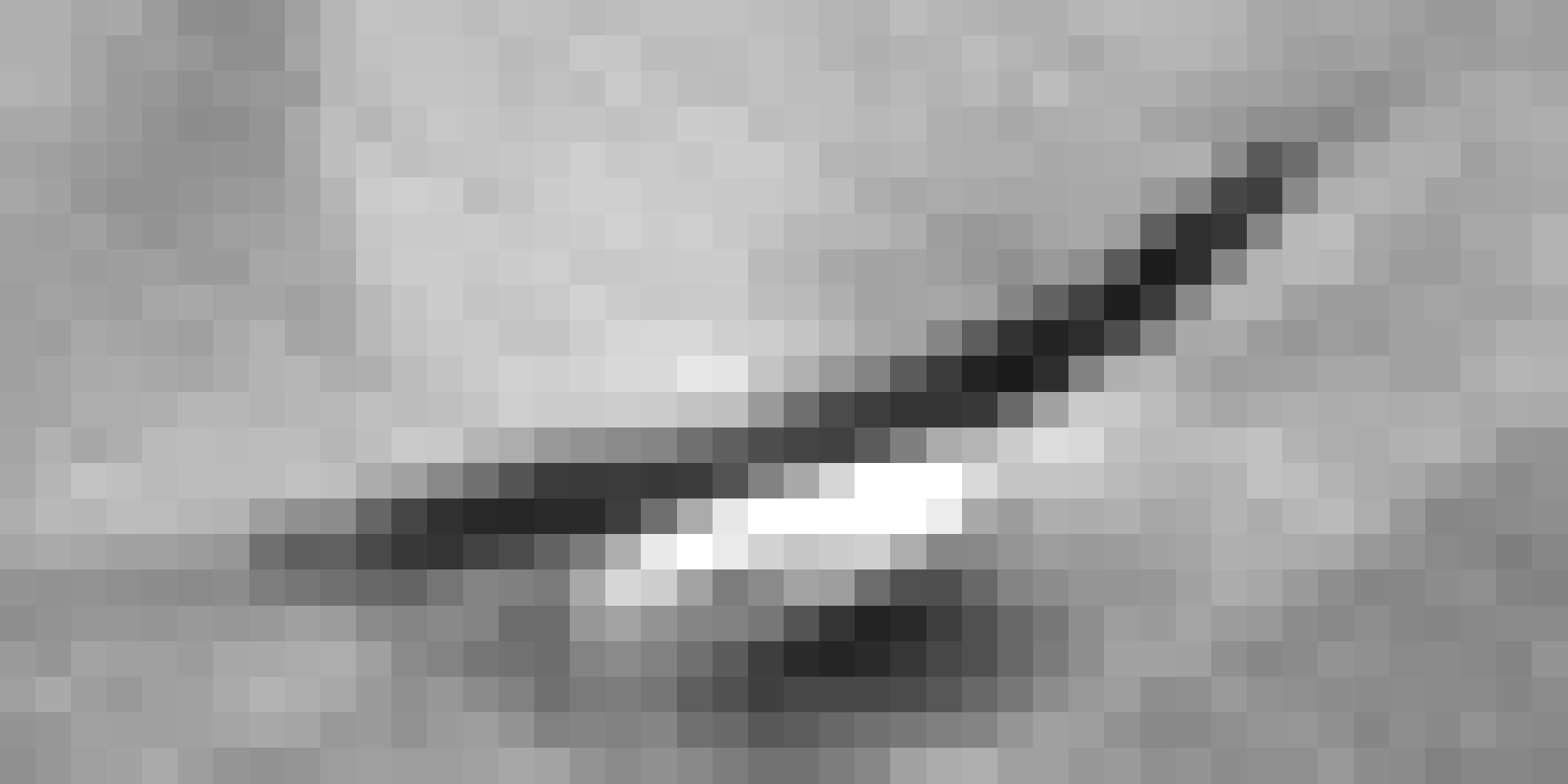}}
		\subfigure[Ours]{\includegraphics[width=0.28\columnwidth]{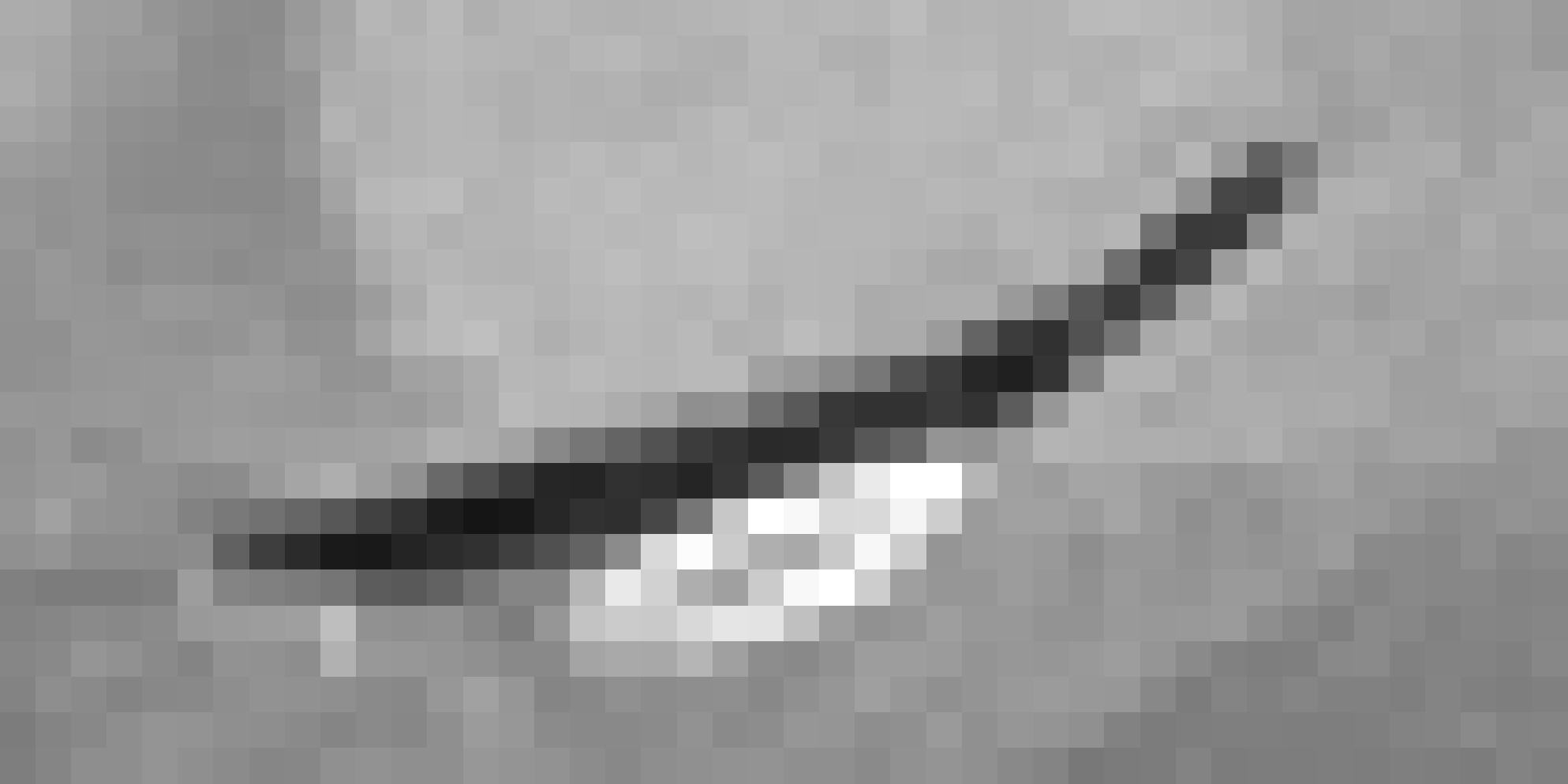}}
	}
	
	\caption{The zoomed regions marked by the red box in Fig. \ref{F8}a.}
	\label{F9}
\end{figure*}
 To reconstruct the CT images, we need to scan the object at least  $180^\circ$ for parallel-beam geometry and $180^\circ$+fan angles
for fan-beam geometry. So, when the number of scanning angles is limited and the same, 150, the qualities of the reconstructed CT images from the parallel-beam CT are higher than those from the fan-beam CT. Therefore, for the 2D limited-angle CT image reconstruction, the parallel-beam geometry is superior over the fan-beam geometry. However, to verify the ability of our network to reconstruct CT images from fan-beam limited-angle sinograms, we still perform the fan-beam CT reconstruction experiments.  
\subsubsection{Data Preparation}
The same 1500 full dose CT images (of size 512$\times$512) from ``the 2016 NIH-AAPM-Mayo
Clinic Low Dose CT Grand Challenge"
are used as the CT image labels $\varsigma^{label}$, where 995 of them
are used as the training set, another 5 of them are used as the
validation set and the other 500 as the test set.

Since it needs more GPU memories to implement the Randon transform for the fan-beam geometry CT, we downsample the CT image labels to reduce their size to 256$\times$256, i.e., we perform the fan-beam CT image reconstruction experiments on CT images of size  256$\times$256.

The radius of the fan-beam geometry is set as $R=600$. We use the Matlab function ``fanbeam" to  generate the sinogram labels $z^{label}$ of size $721\times360$, where 360 is the number of scanning angles corresponding to $\frac{\pi}{180}\times[0:1:359]$ and $721$ is the number of detectors corresponding to $\frac{\pi}{180}\times[-18:0.05:18]$ and $18^\circ$ is half of the fan angles.  We downsample  $z^{label}$ at scanning angles $\frac{\pi}{180}\times[0:1:149]$ to get the simulated limited-angle sinogram $g$ of size $150\times721$.  The limited-angle sinogram $g$ is the input of our network and  the reconstructed CT image from $g$ by the FBP algorithm is used as the initial guesses of CT images $u_0$. For the  compared networks, Red-CNN, FBP-Conv and DD-Net, the reconstructed CT image from $g$ by the FBP algorithm is also used as their inputs.

\subsubsection{Parameter Setup}
The parameters $\Theta_\varsigma^n$ and $\Theta_z^n$ in $Res_\varsigma^n(\cdot)$ and $Res_z^n(\cdot)$ of our network are automatically initialized by Tensorflow using  the default values. The initial values of parameters $\Theta_u$ in $Merge_u^n(\cdot)$ are set as $t_1=1$ and $t_2=t_3=t_4=0.1$. The number of iterations is set as $N_{iter}=3$. The batch size is set as 2  and the number of training epochs is $100$.

The parameters for Red-CNN, FBP-Conv, and DD-NeT are set as
described in their corresponding papers and the initial values in their networks are initialized by Tensorflow automatically. The batch size is 5 and
the training epochs
for Red-CNN, FBP-Conv and DD-Net are all 500. 

The parameters of the TV regularization algorithm for fan-beam CT are set as
$\lambda_3=20$, $\rho=0.1$, $t_5=0.1$, the number of maximal iterations $M_{iter}=300$ and the convergent criteria $\frac{\|u^{n+1}-u^{n}\|^2_2}{\|u^{n+1}\|^2_2}\le\epsilon=10^{-4}$.
 
\subsubsection{Subjective Evaluation}
Fig. \ref{F8} shows some reconstructed results from the test set by the six methods. It can be observed that the reconstructed images in Fig. \ref{F8}b by the FBP algorithm have severe  artifacts caused by the incomplete data. 
The results of the TV regularization algorithm  still have some  artifacts and some of the edges and structures in Fig. \ref{F8}c are blurred.
From Fig. \ref{F8}d to Fig. \ref{F8}g, we can observe that
the artifacts are suppressed to different degrees by
the different networks. However, the results of Red-CNN in Fig. \ref{F8}d and DD-Net in Fig. \ref{F8}f  have some dark areas which are caused by the intensity inhomogeneous. The boundaries of the reconstructed images by the FBP-conv in Fig. \ref{F8}e are somewhat blurred. Compared to these methods, the results of our method can preserve more fine edges and structures and have no inhomogeneous intensities.

To better demonstrate that our method can preserve more edges and boundaries, Fig. \ref{F9} shows the zoomed regions marked by the  red box in Fig. \ref{F8}a. It can be easily observed that  the boundaries and edges of the reconstructed images  by our method are the most consistent with the labels compared to those by the other methods. 

\begin{figure*}[!t]
	\centerline{
		\includegraphics[width=0.28\columnwidth]{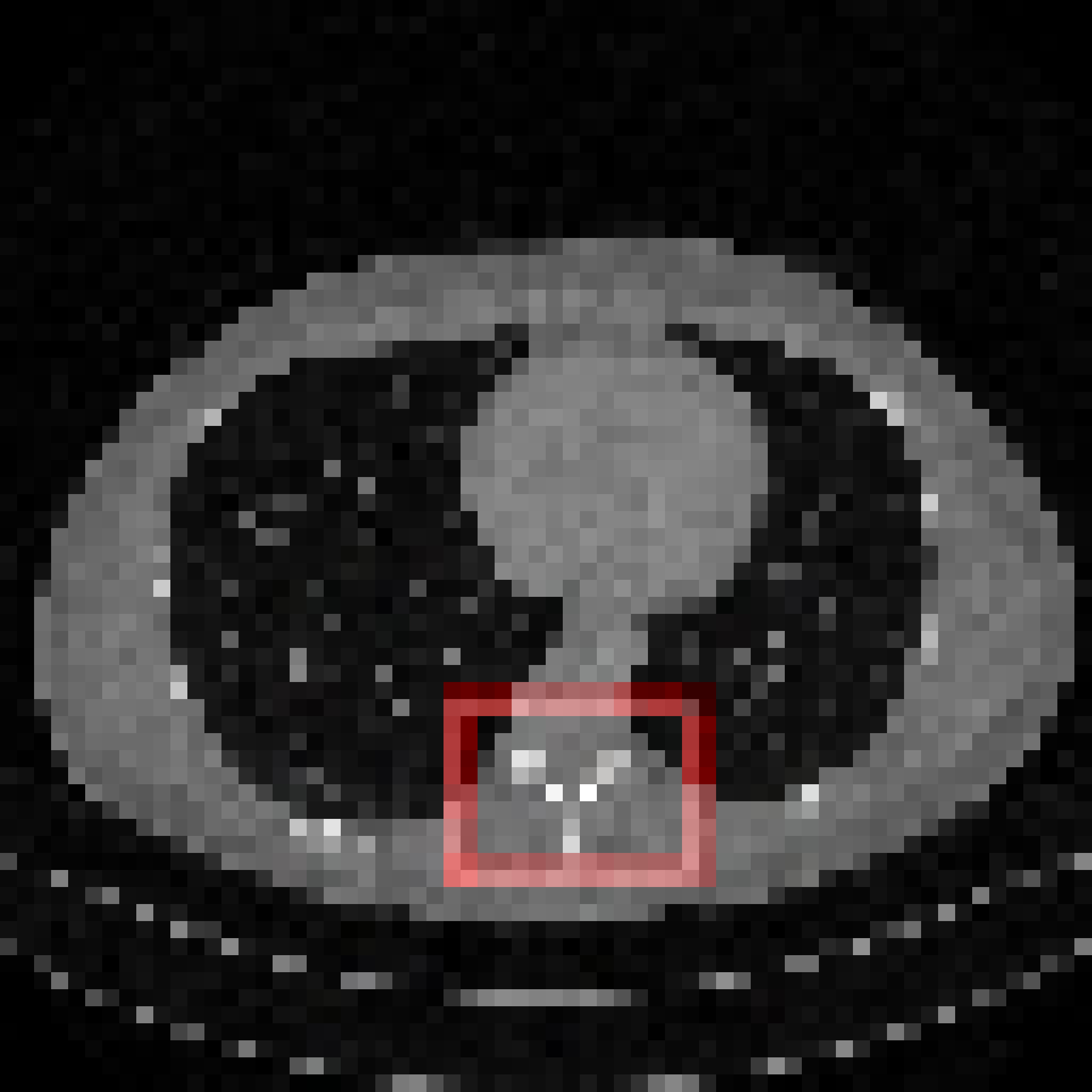}
		\includegraphics[width=0.28\columnwidth]{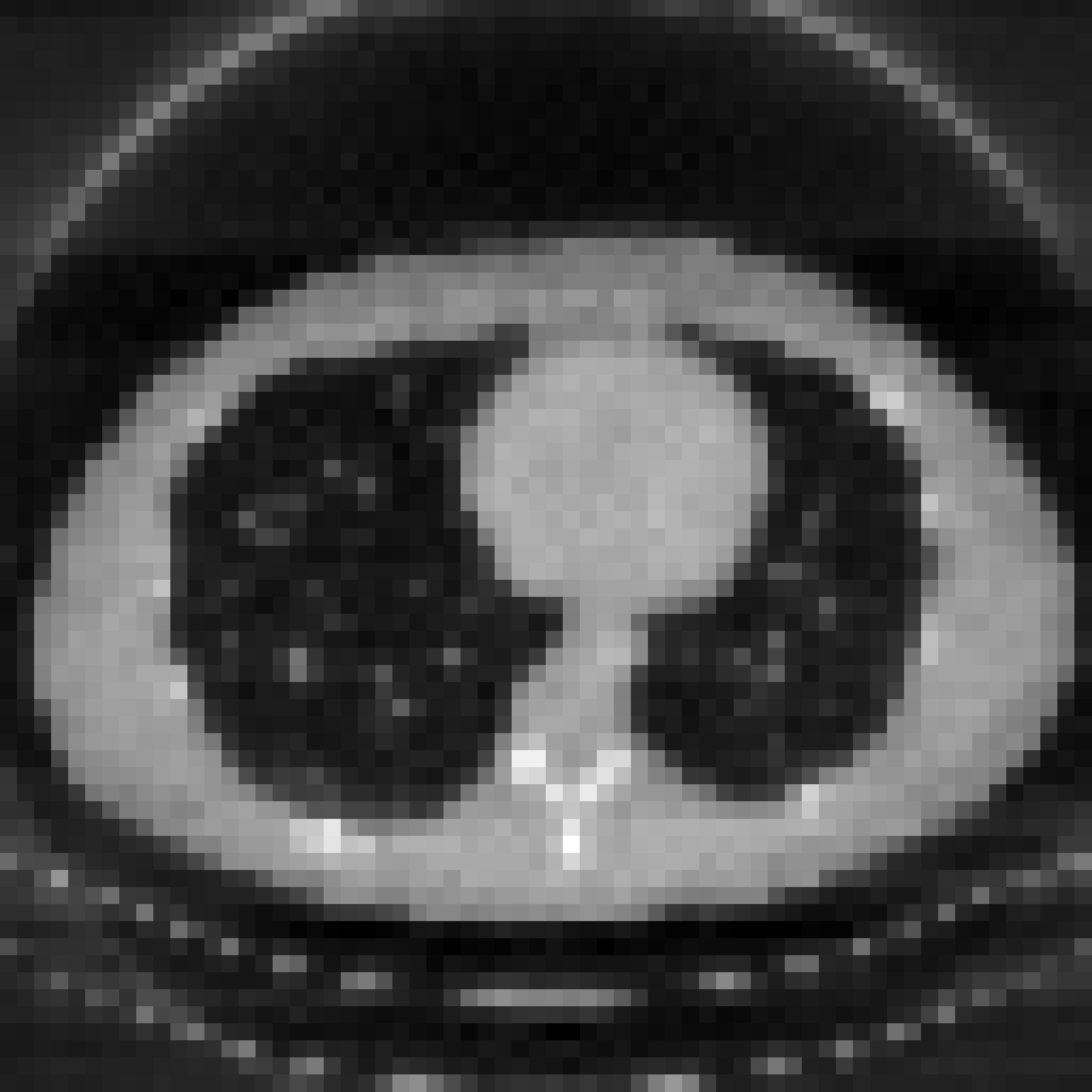}
		\includegraphics[width=0.28\columnwidth]{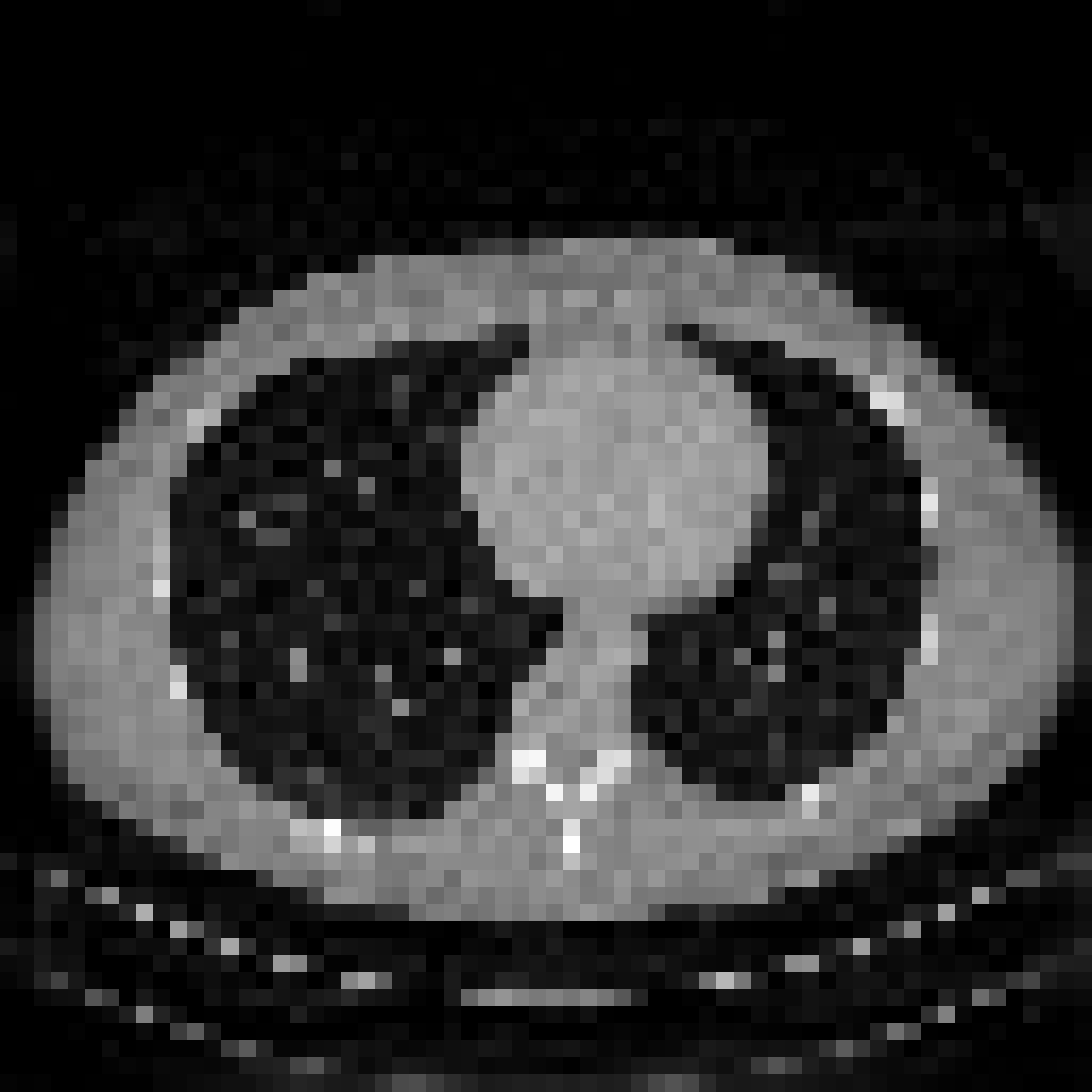}
		\includegraphics[width=0.28\columnwidth]{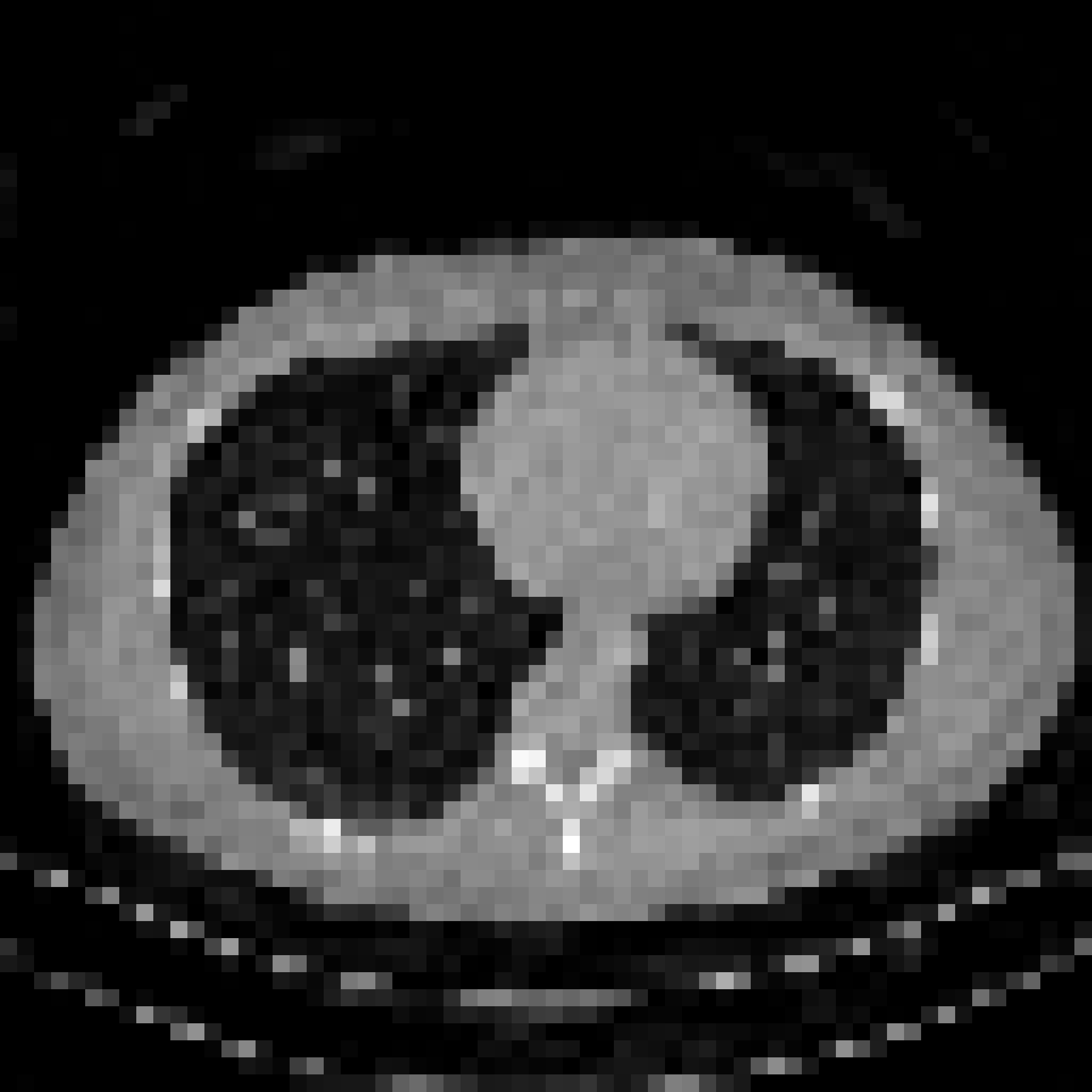}
		\includegraphics[width=0.28\columnwidth]{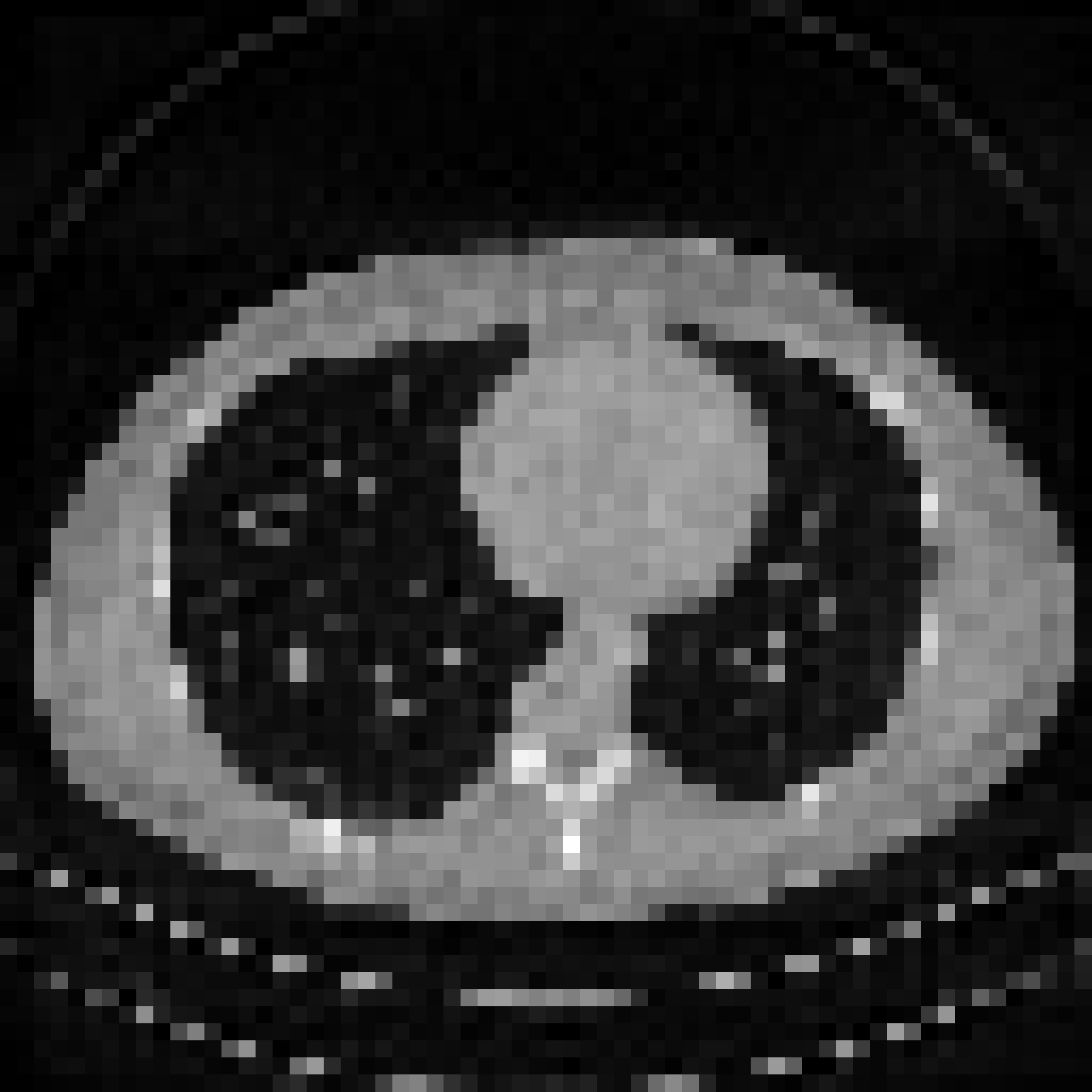}
		\includegraphics[width=0.28\columnwidth]{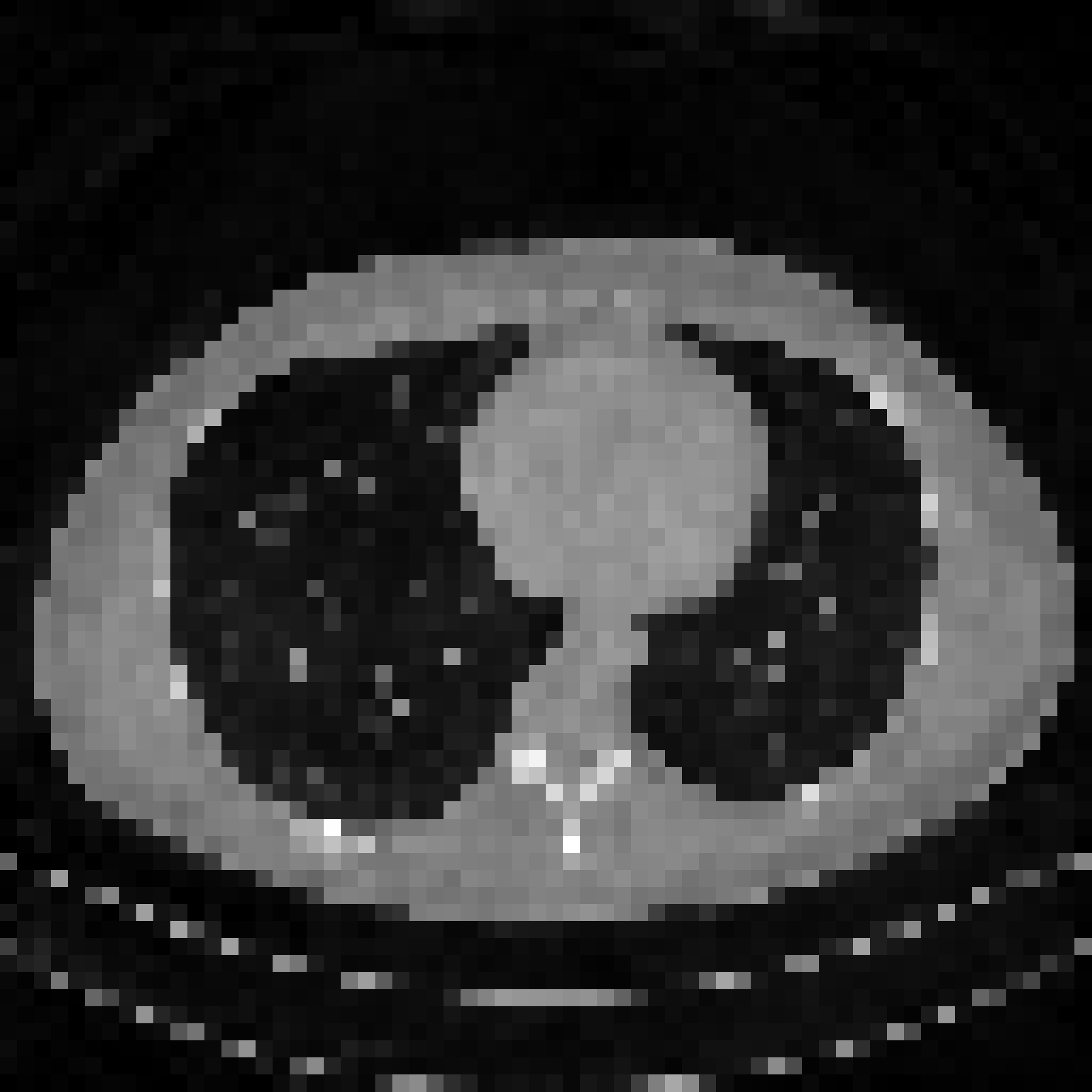}
	}
	
	\centerline{
		\includegraphics[width=0.28\columnwidth]{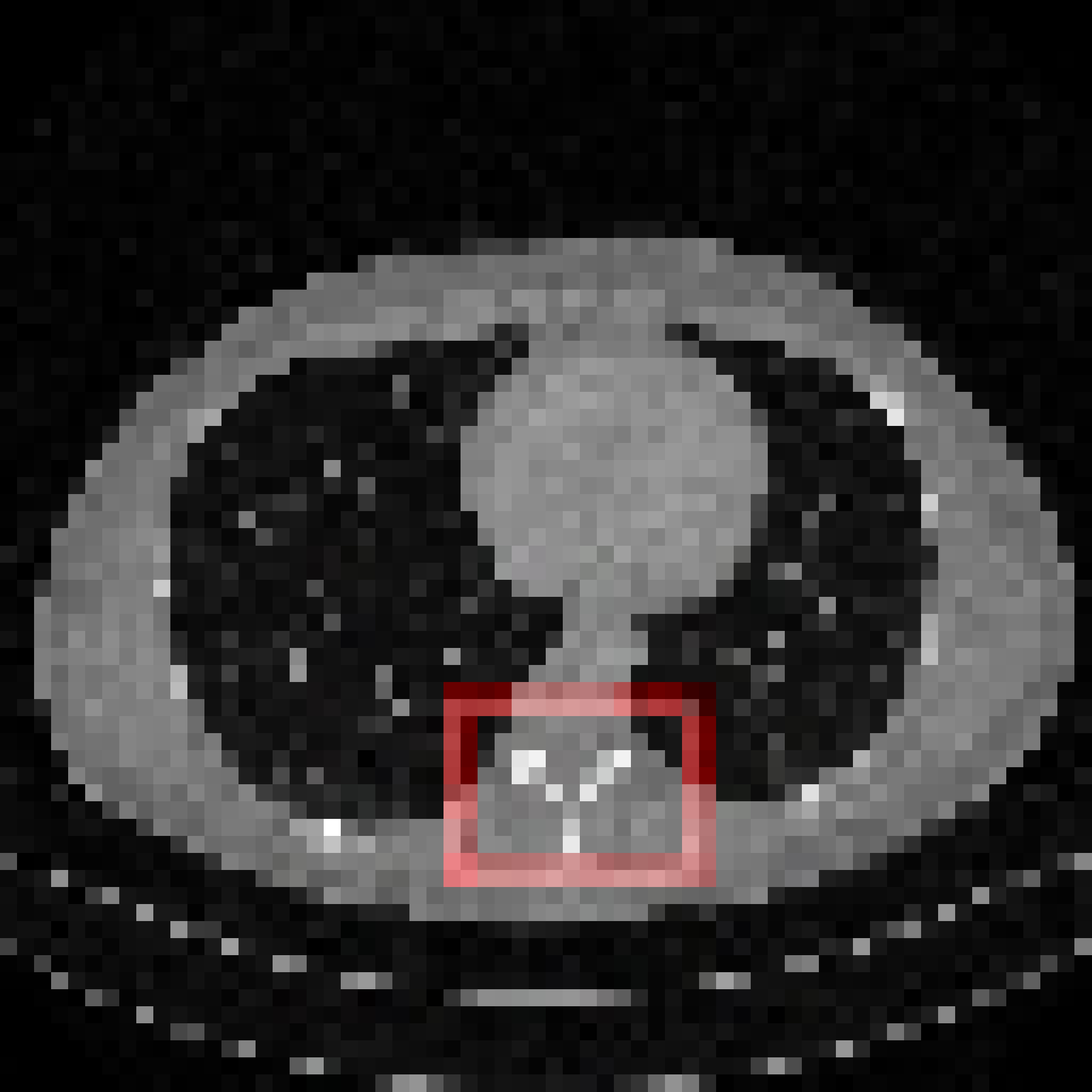}
		\includegraphics[width=0.28\columnwidth]{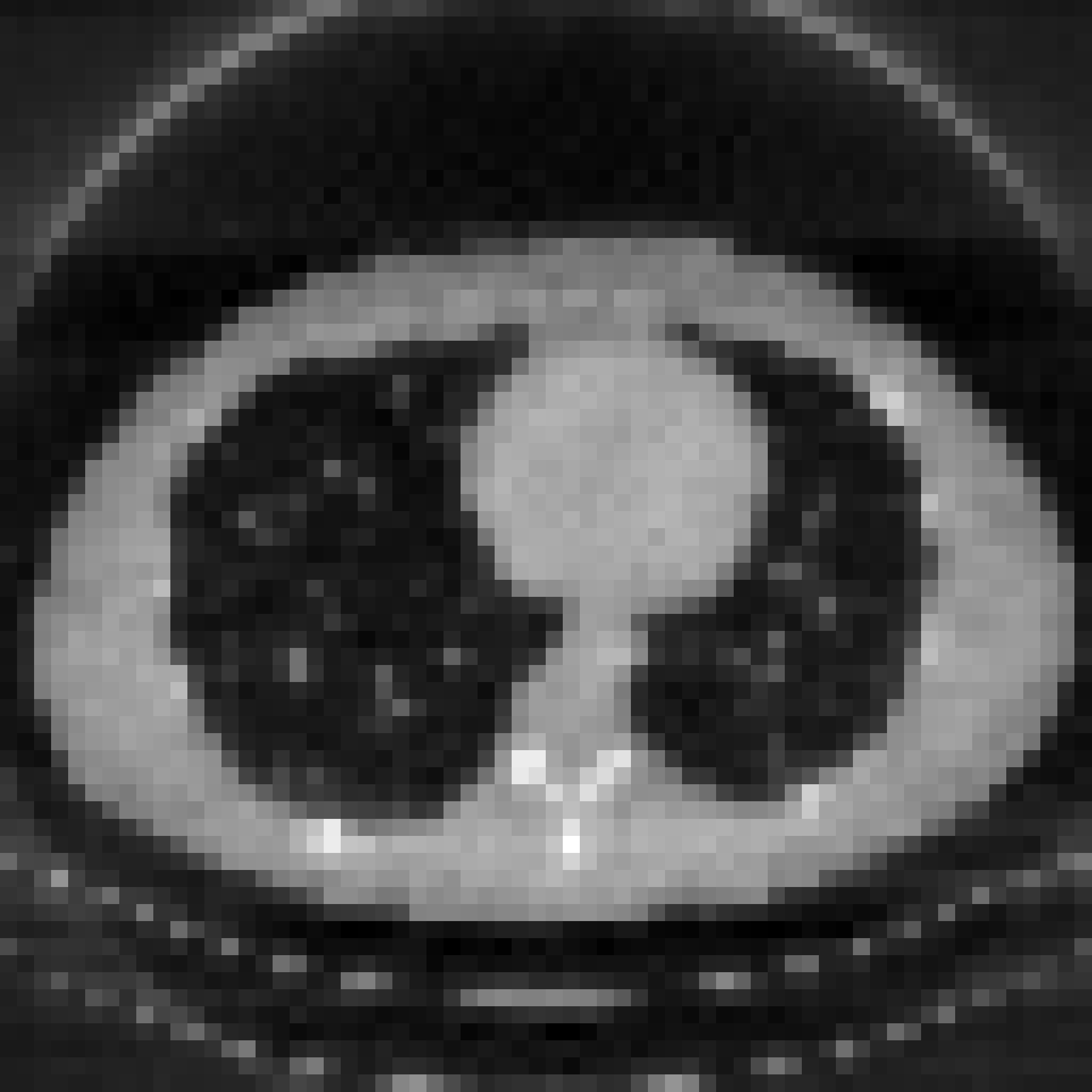}
		\includegraphics[width=0.28\columnwidth]{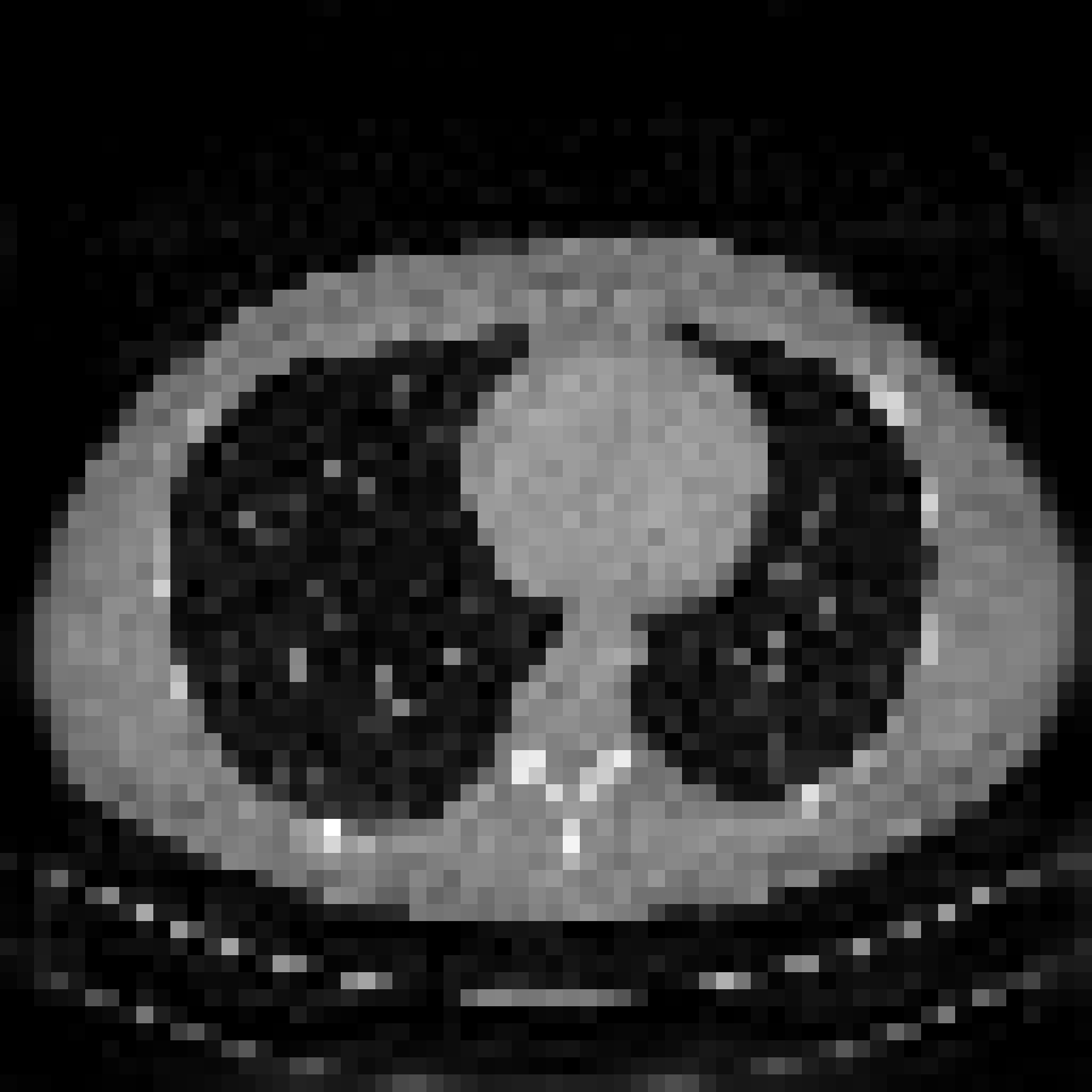}
		\includegraphics[width=0.28\columnwidth]{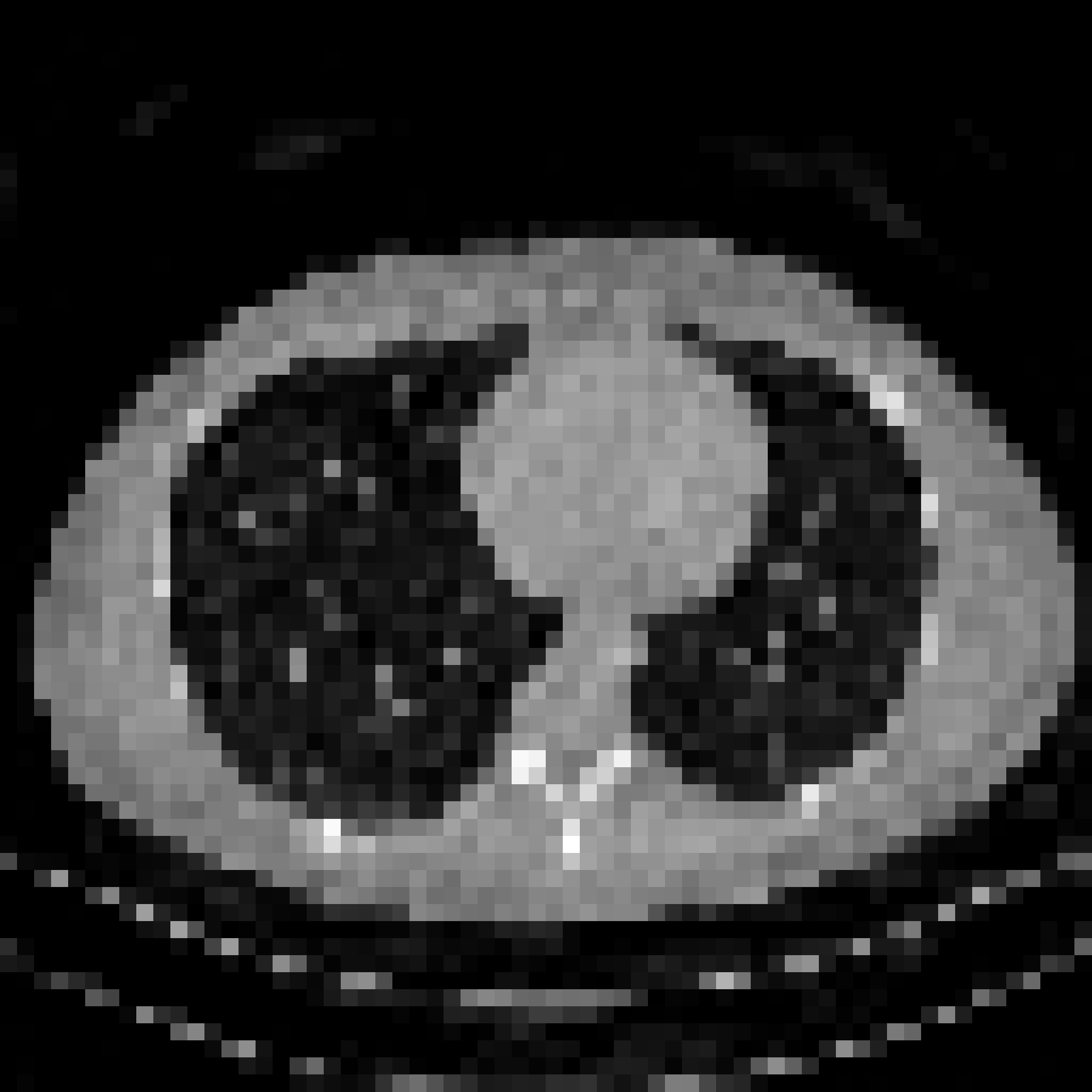}
		\includegraphics[width=0.28\columnwidth]{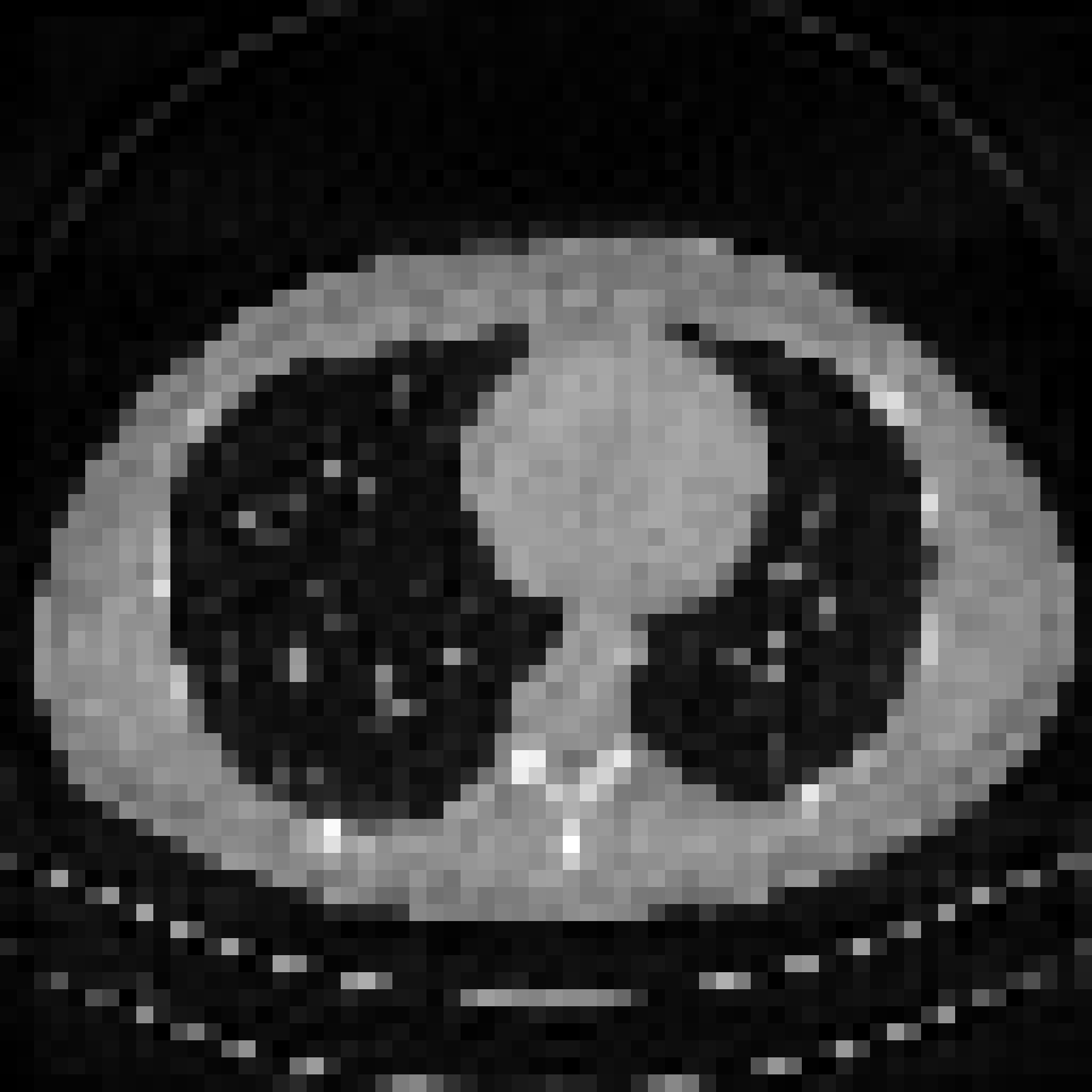}
		\includegraphics[width=0.28\columnwidth]{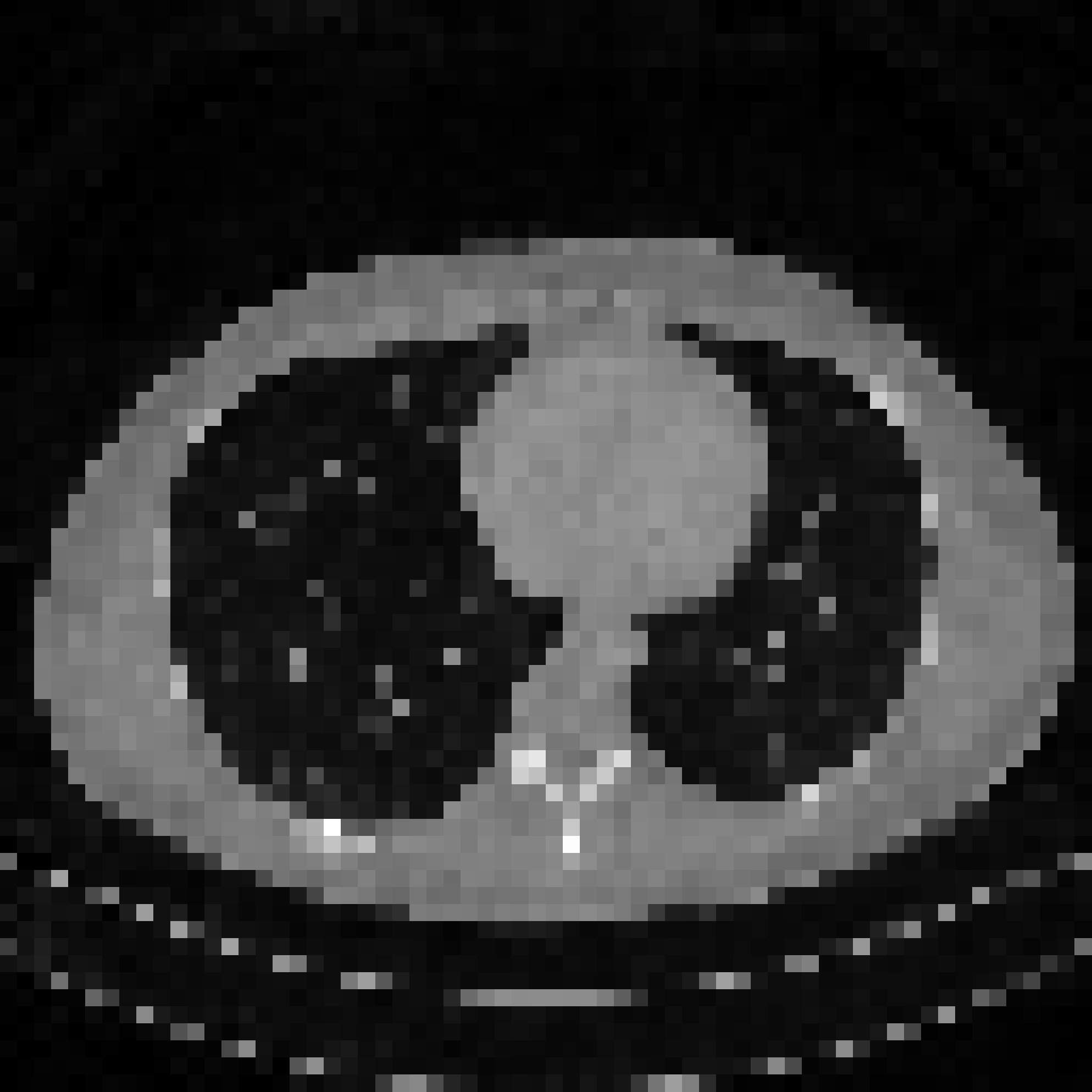}
	}
	\centerline{
		\includegraphics[width=0.28\columnwidth]{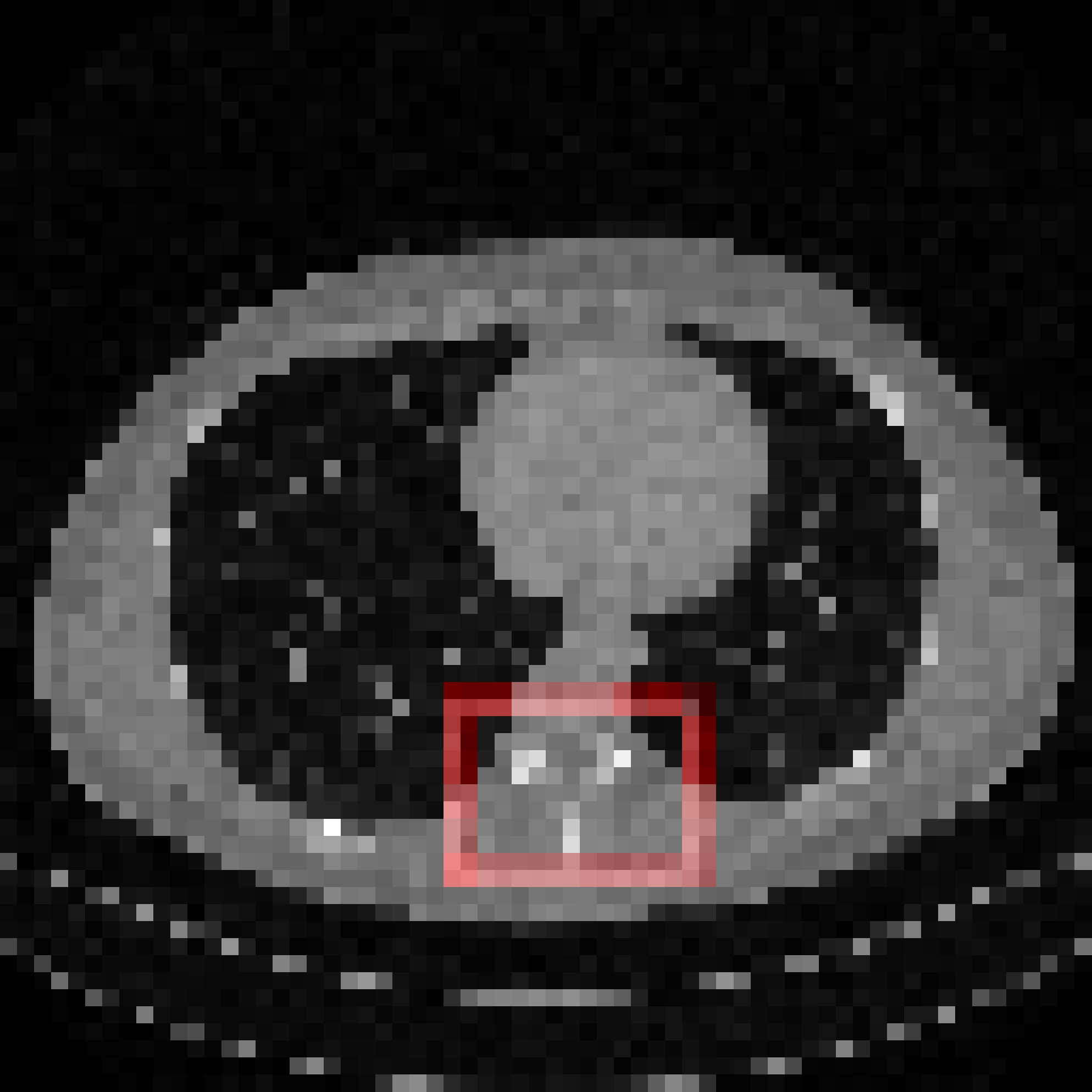}
		\includegraphics[width=0.28\columnwidth]{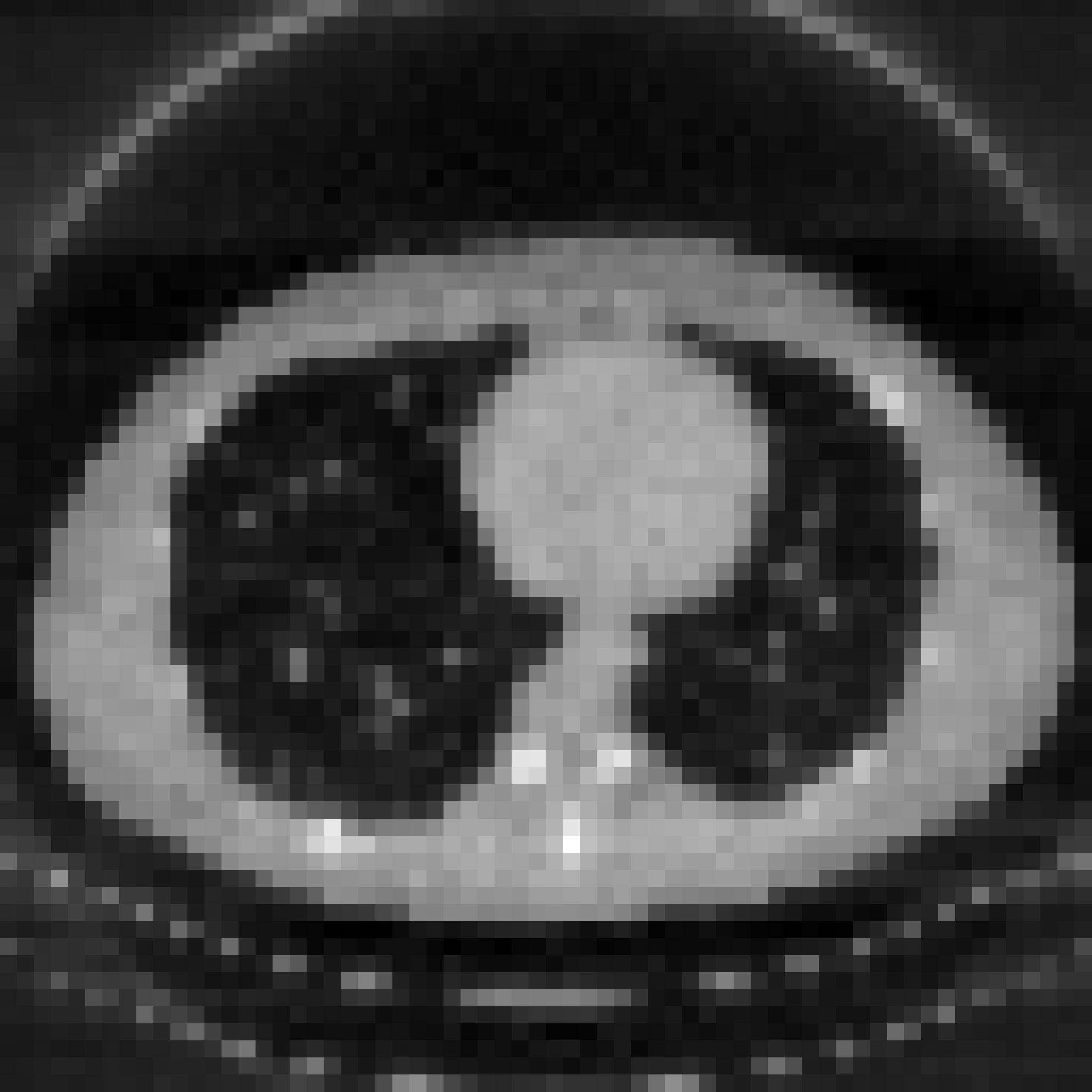}
		\includegraphics[width=0.28\columnwidth]{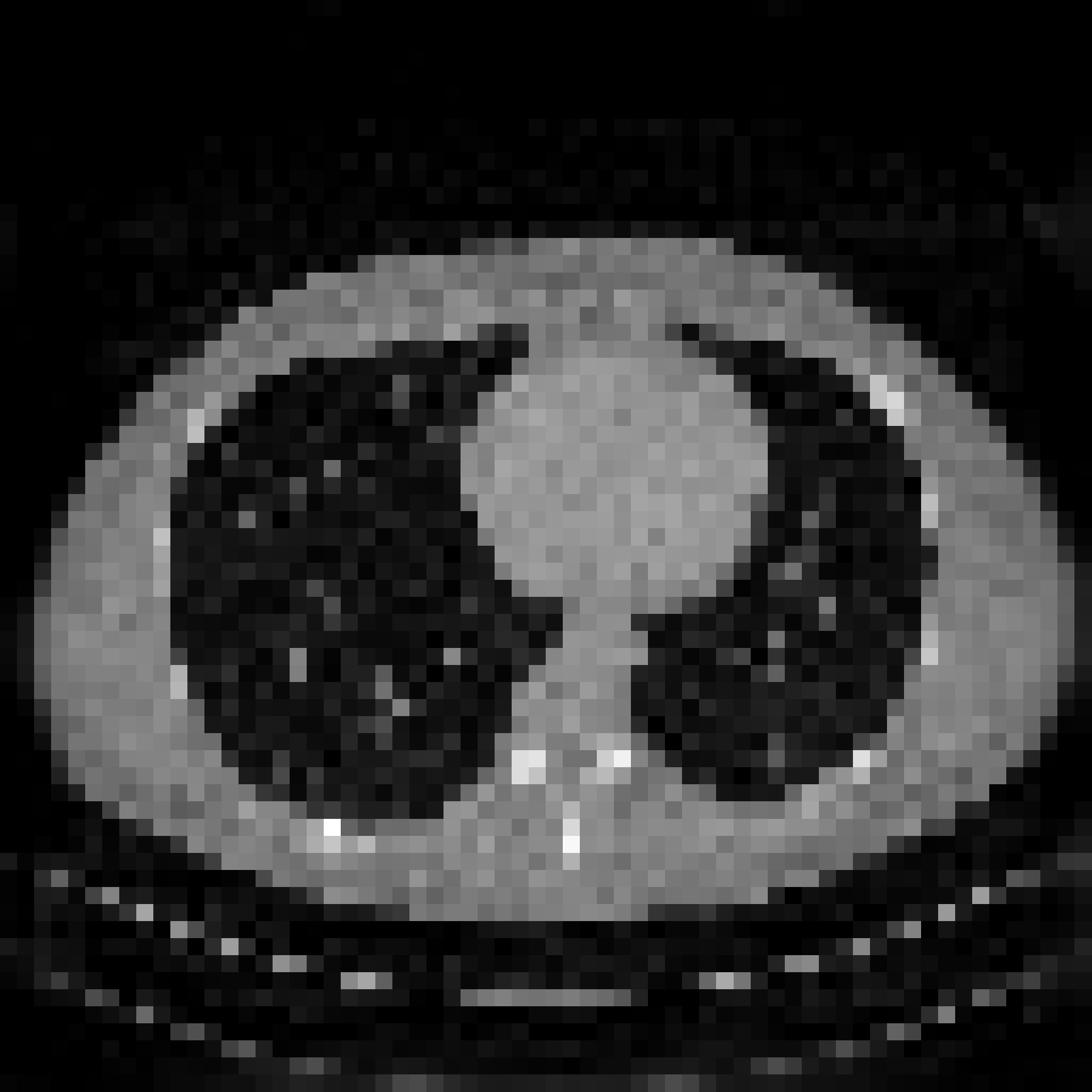}
		\includegraphics[width=0.28\columnwidth]{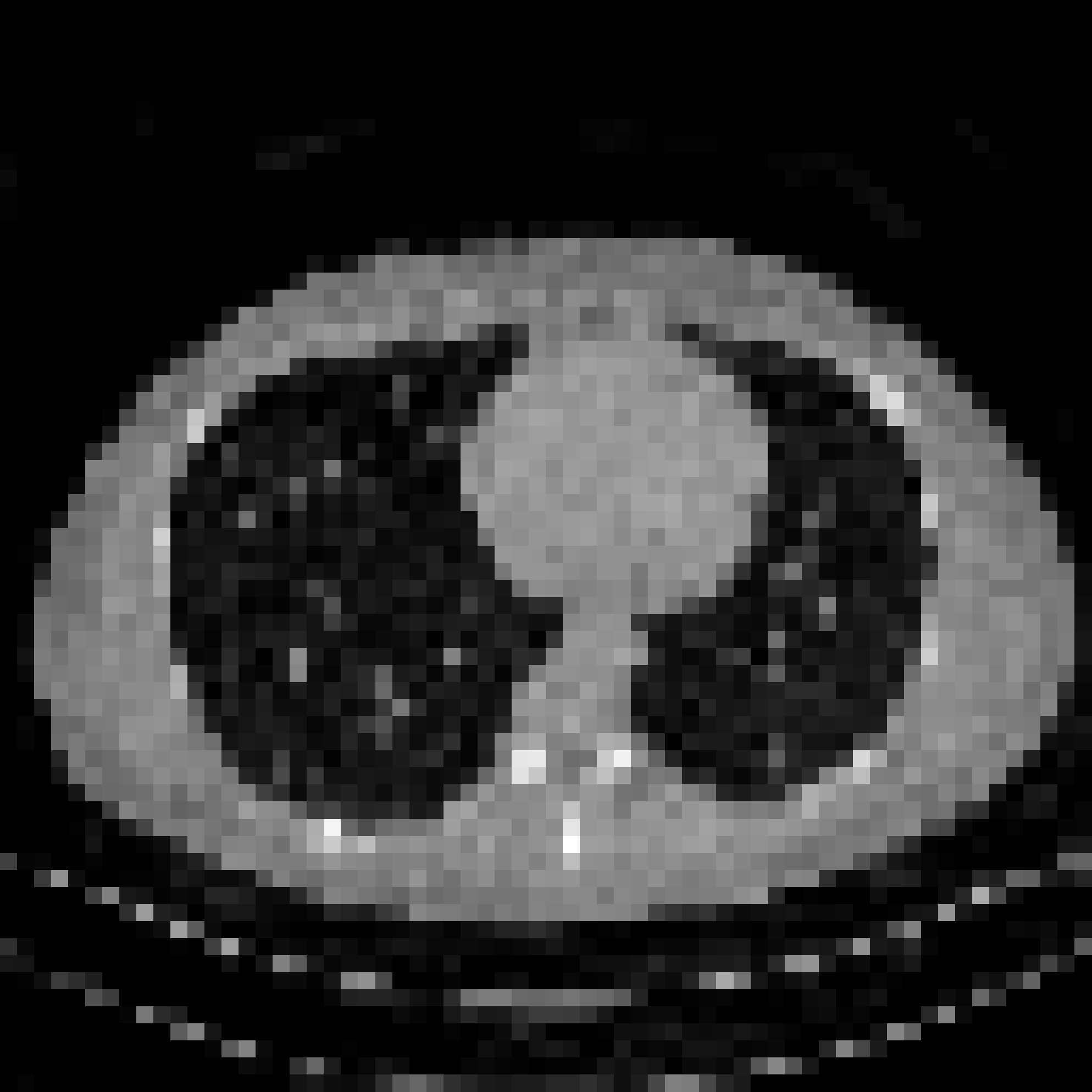}
		\includegraphics[width=0.28\columnwidth]{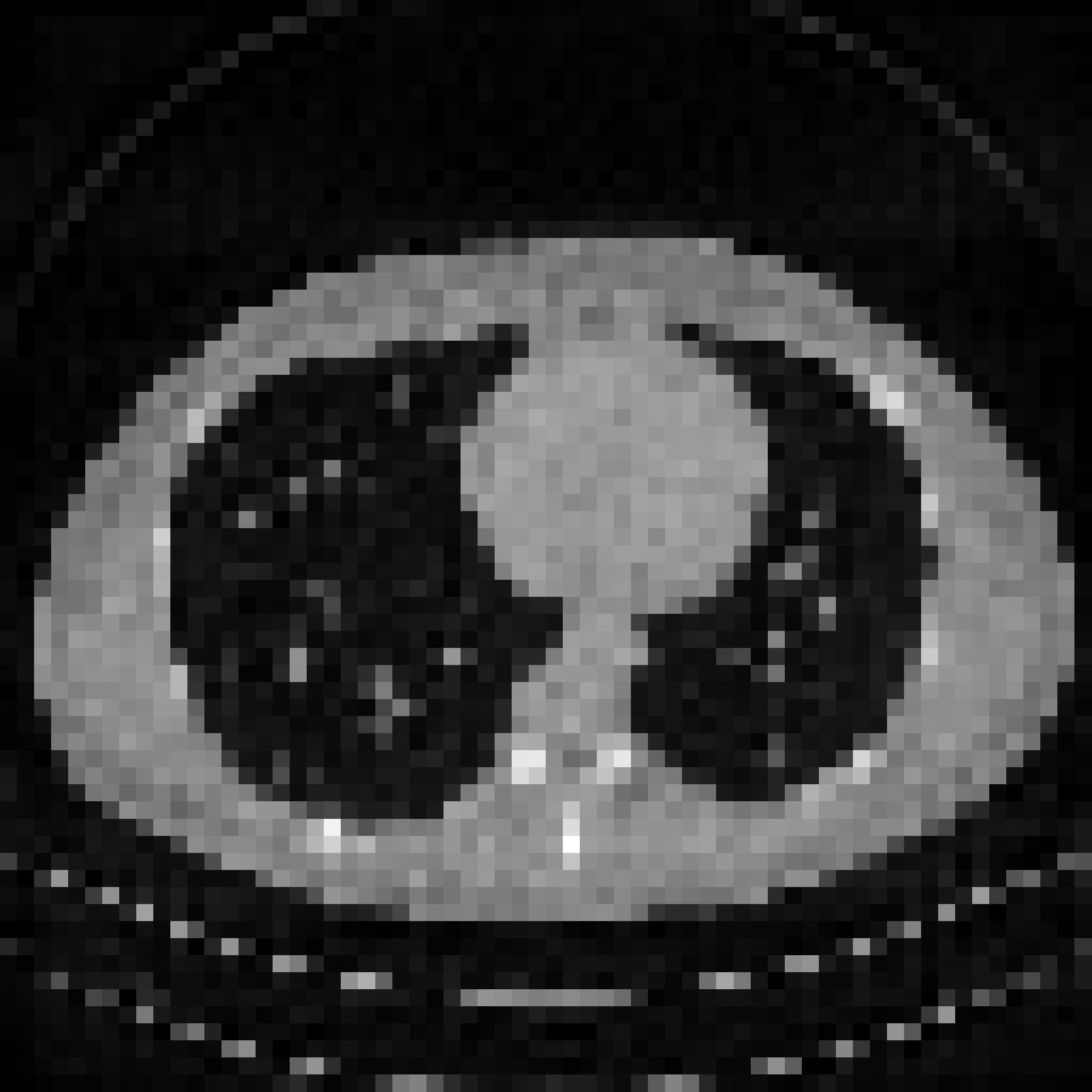}
		\includegraphics[width=0.28\columnwidth]{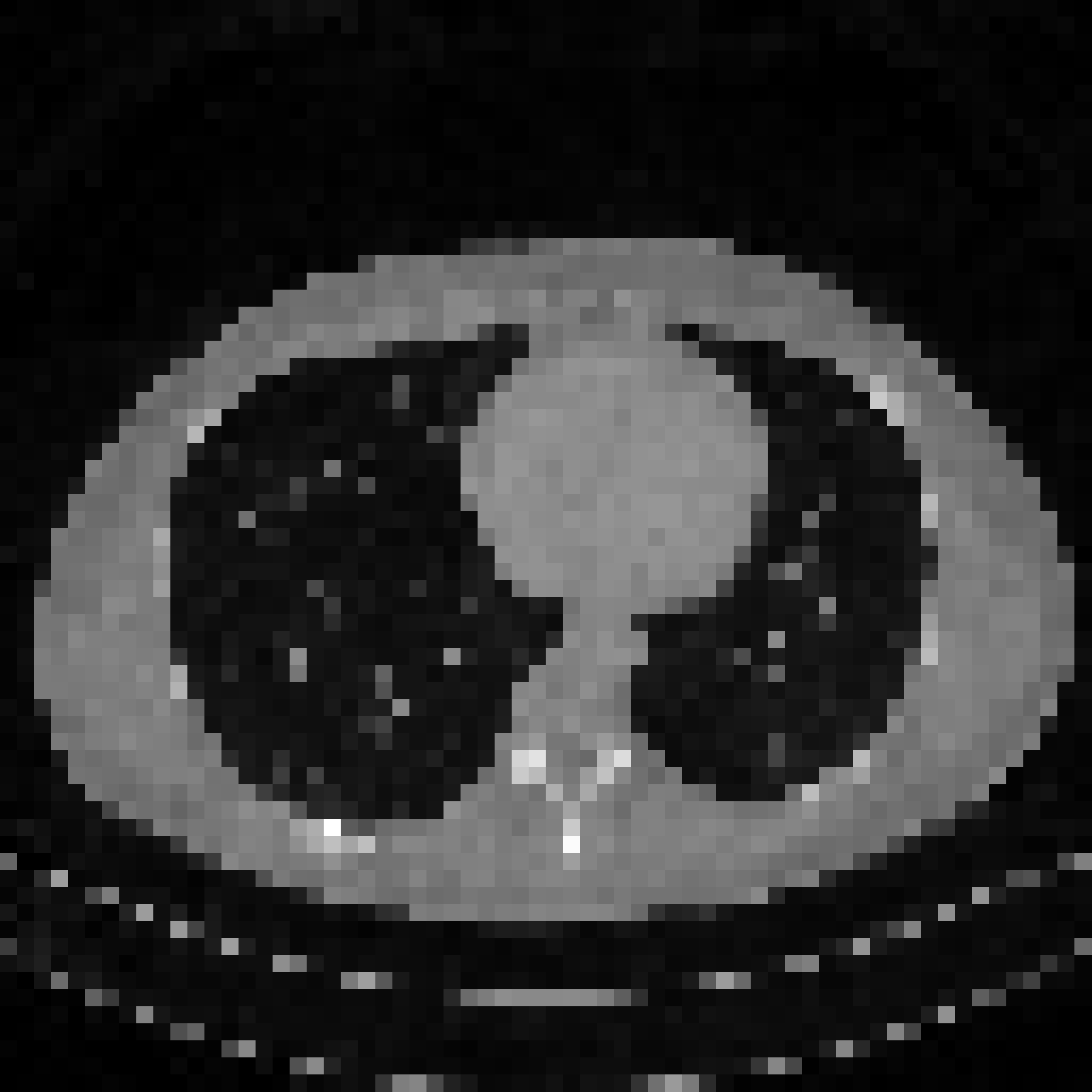}
	}
	\centerline{
		\subfigure[Label]{\includegraphics[width=0.28\columnwidth]{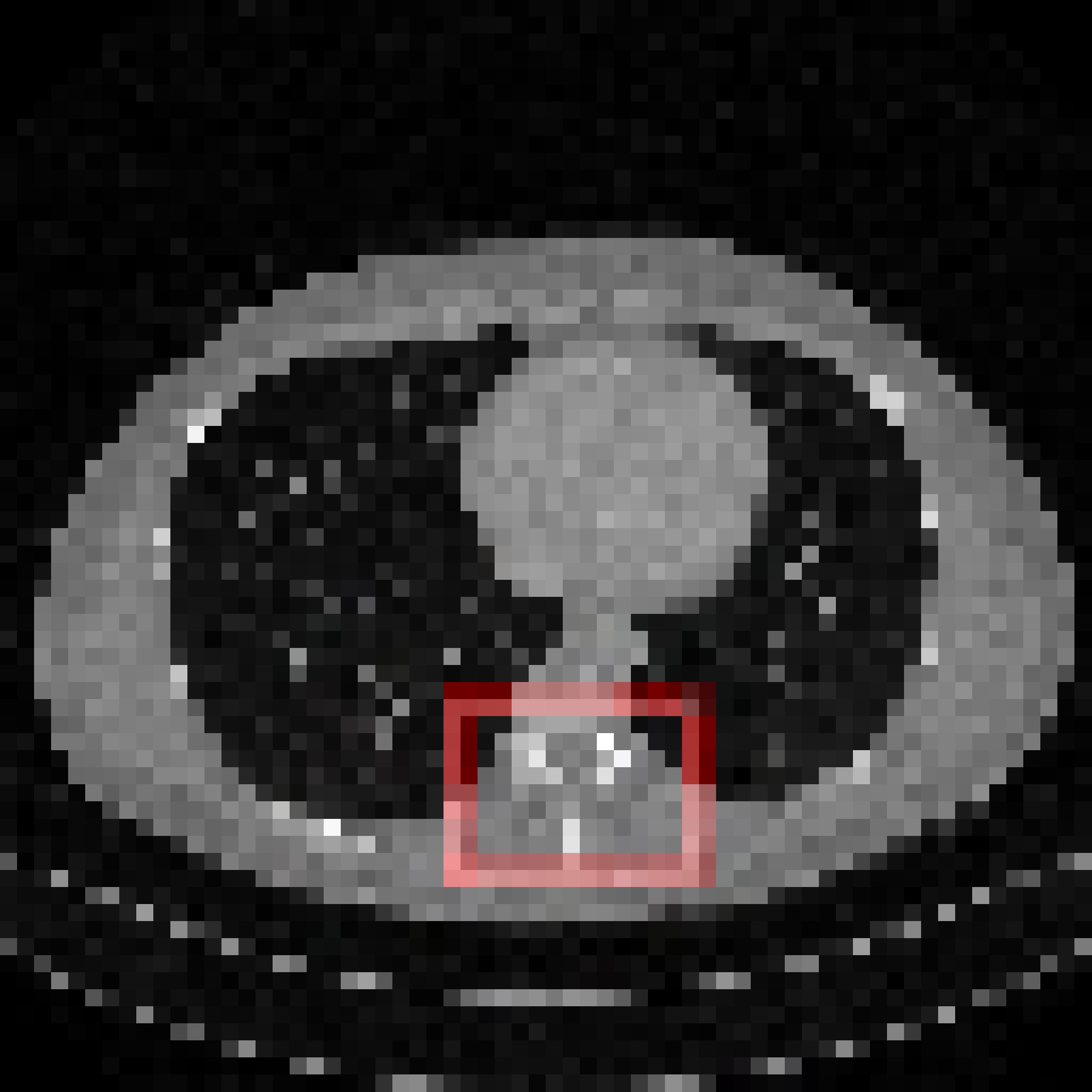}}
		\subfigure[FDK]{\includegraphics[width=0.28\columnwidth]{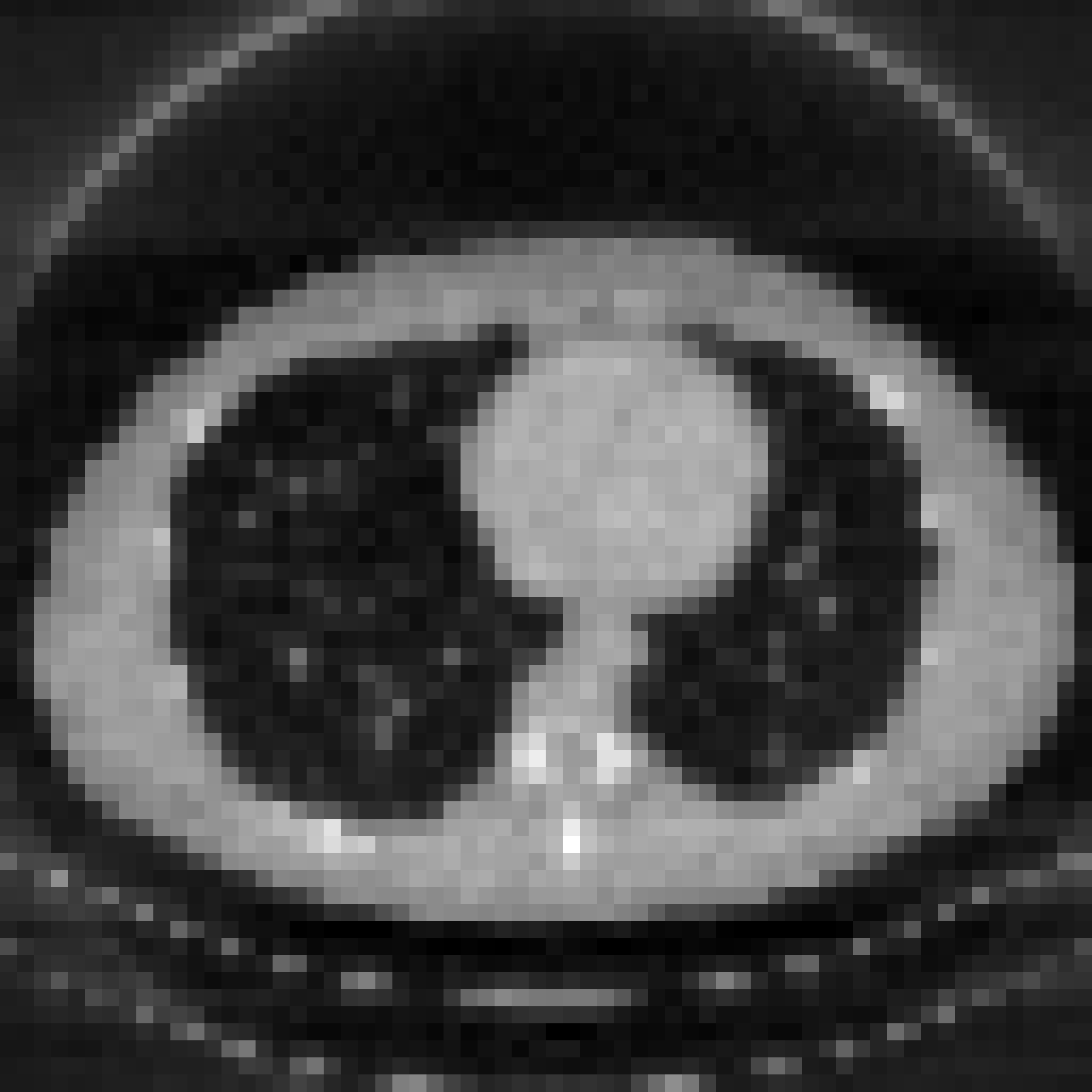}}
		\subfigure[Red-CNN]{\includegraphics[width=0.28\columnwidth]{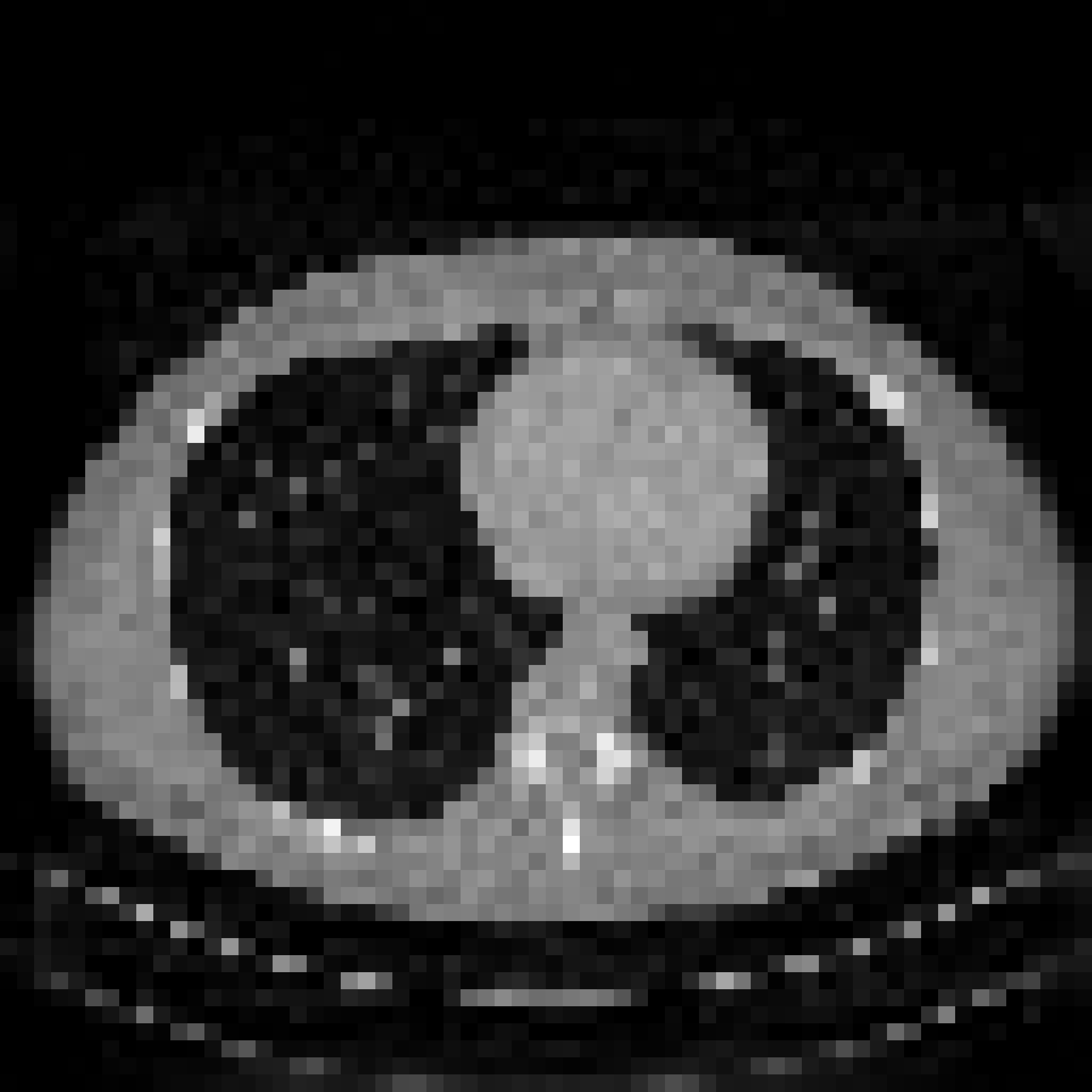}}
		\subfigure[FBP-Conv]{\includegraphics[width=0.28\columnwidth]{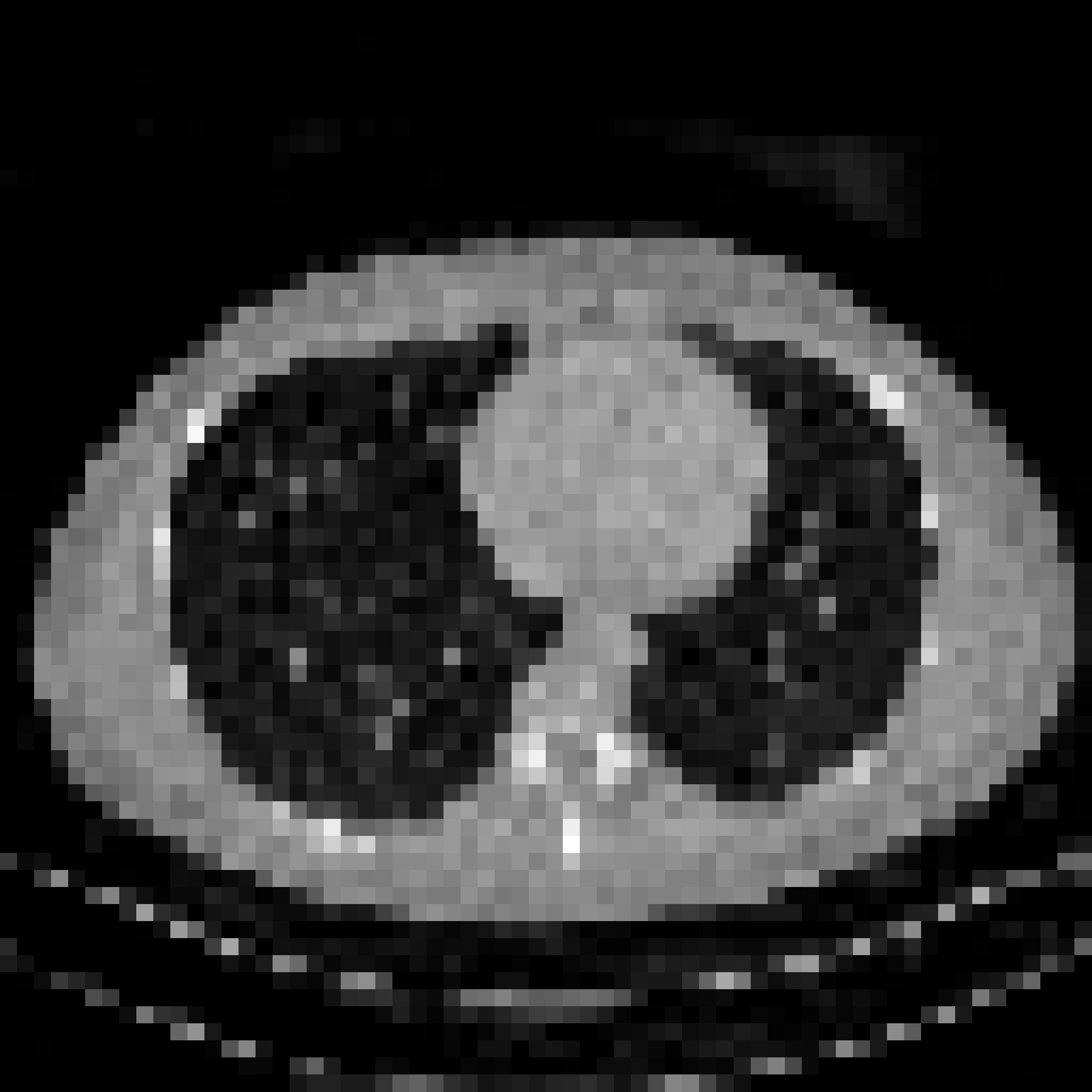}}
		\subfigure[DD-Net]{\includegraphics[width=0.28\columnwidth]{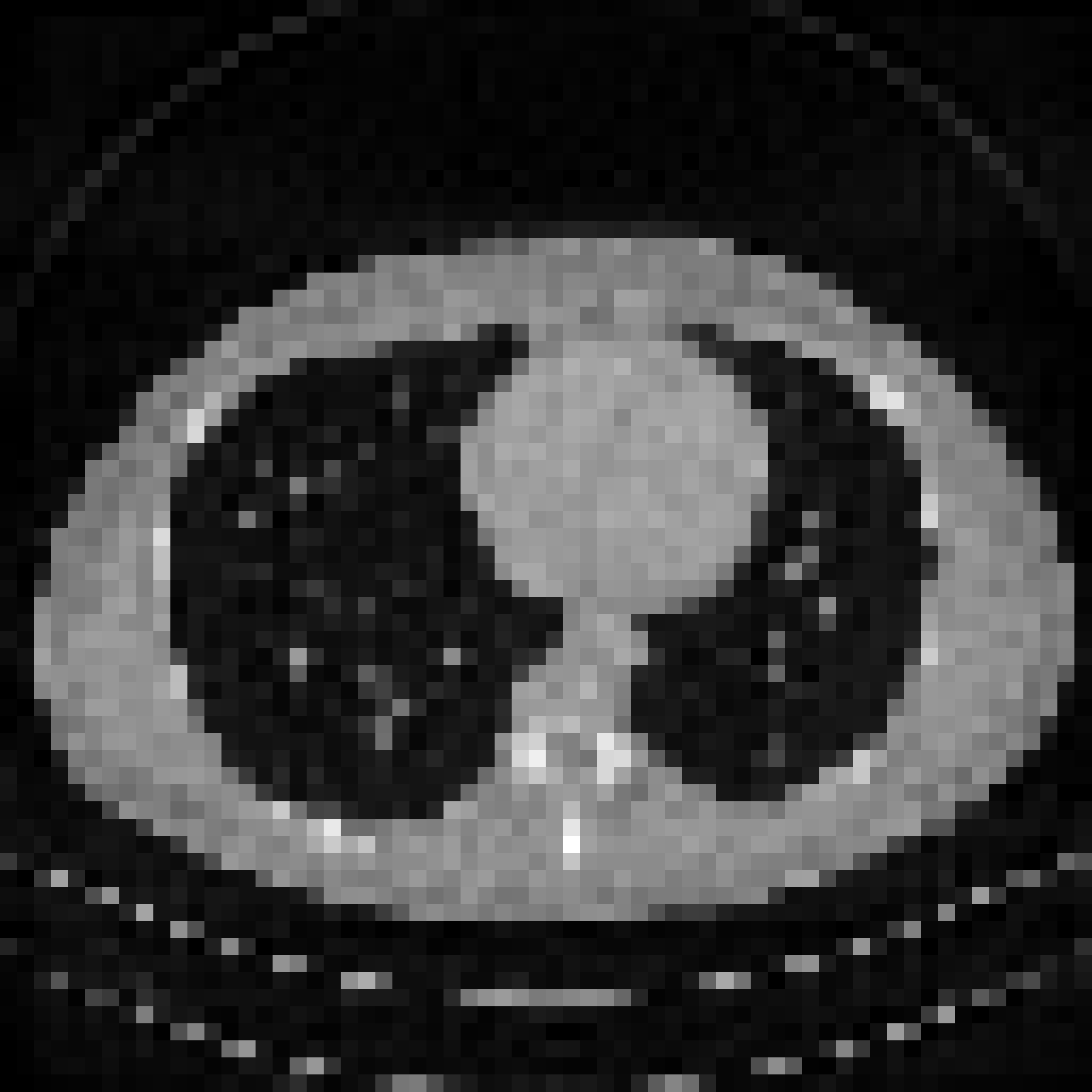}}
		\subfigure[Ours]{\includegraphics[width=0.28\columnwidth]{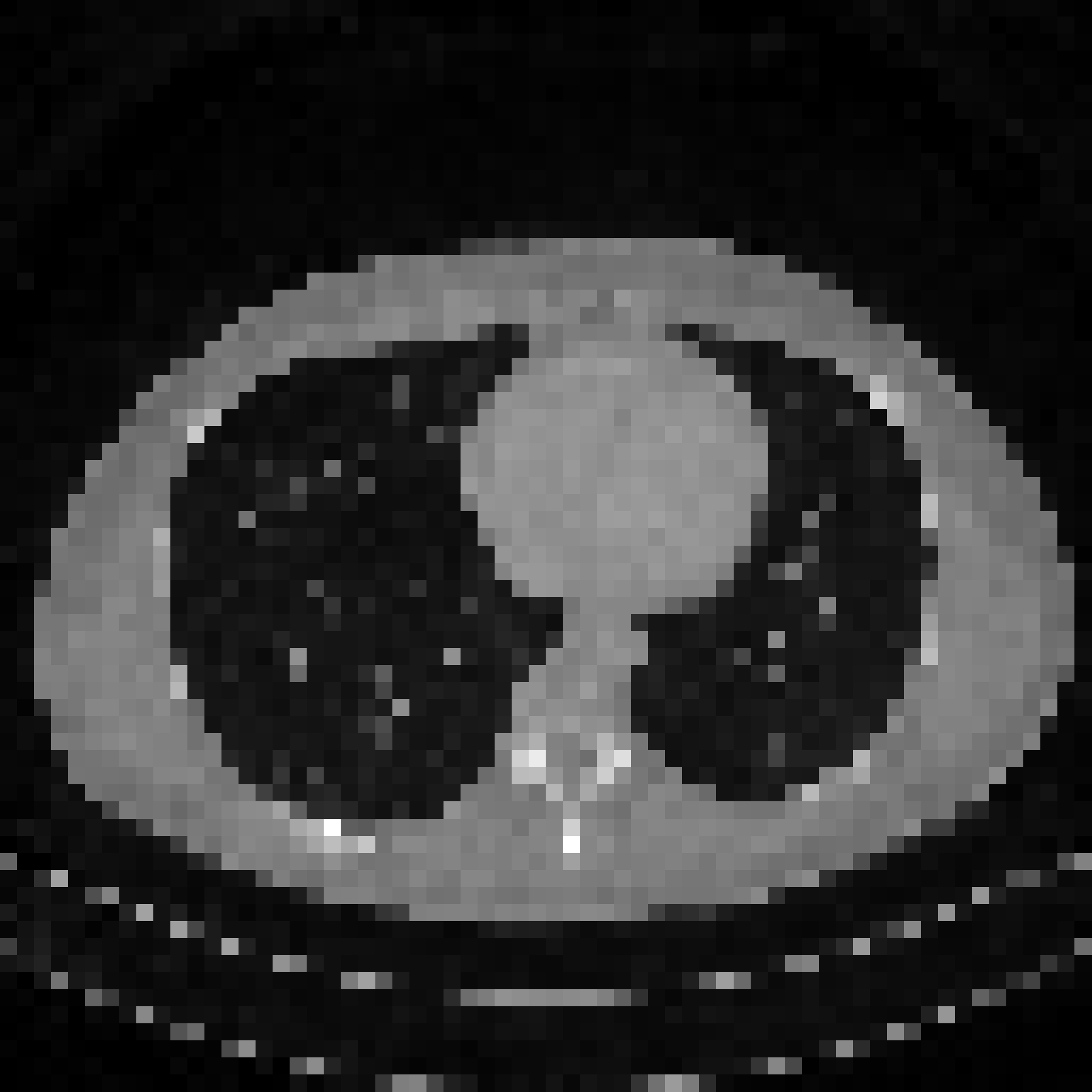}}
	}
	
	\caption{The reconstructed CT images by the five methods for circle cone-beam geometry.}
	\label{F10}
\end{figure*}

\begin{figure*}[!t]

	\centerline{
		\includegraphics[width=0.28\columnwidth]{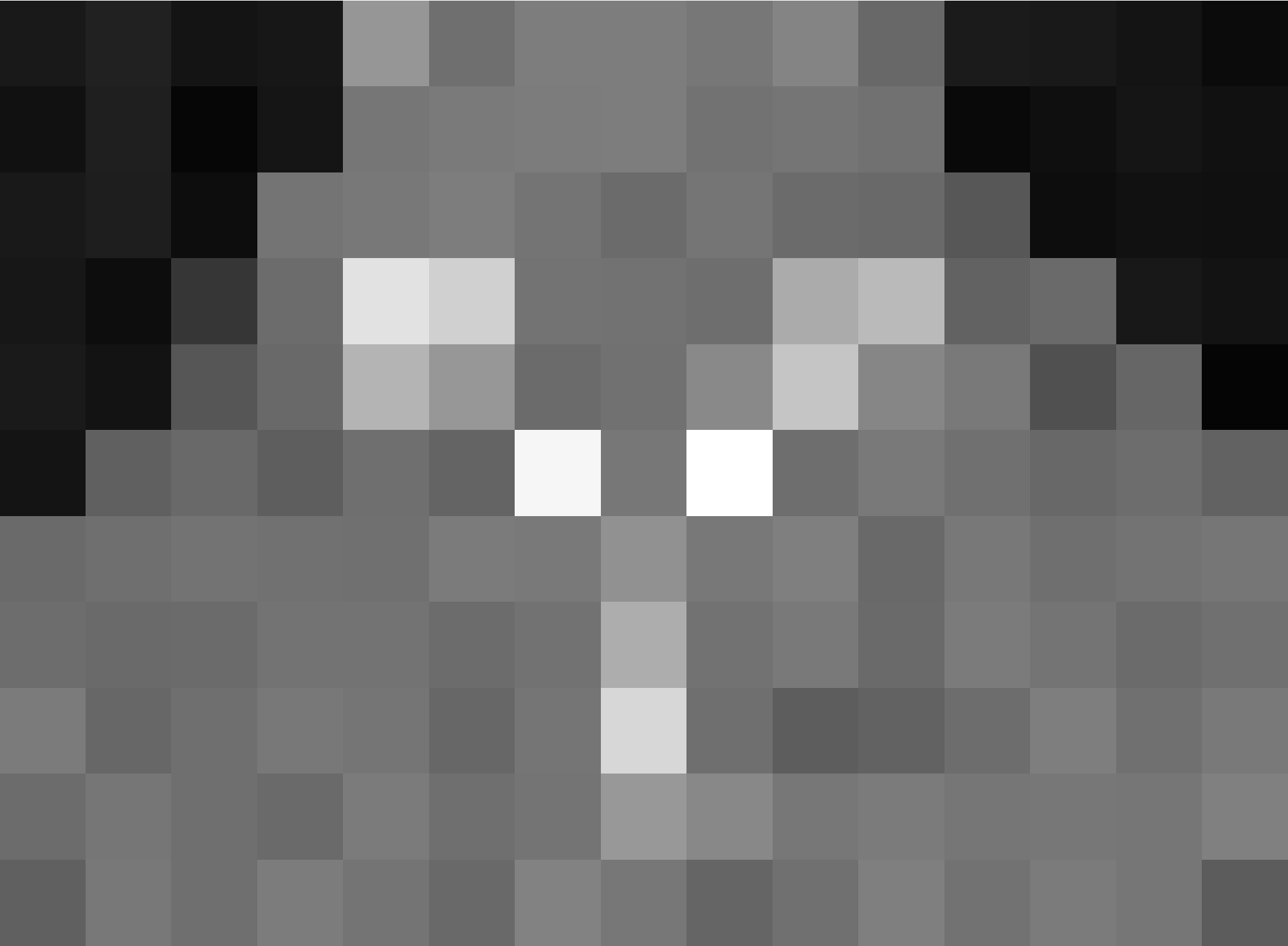}
		\includegraphics[width=0.28\columnwidth]{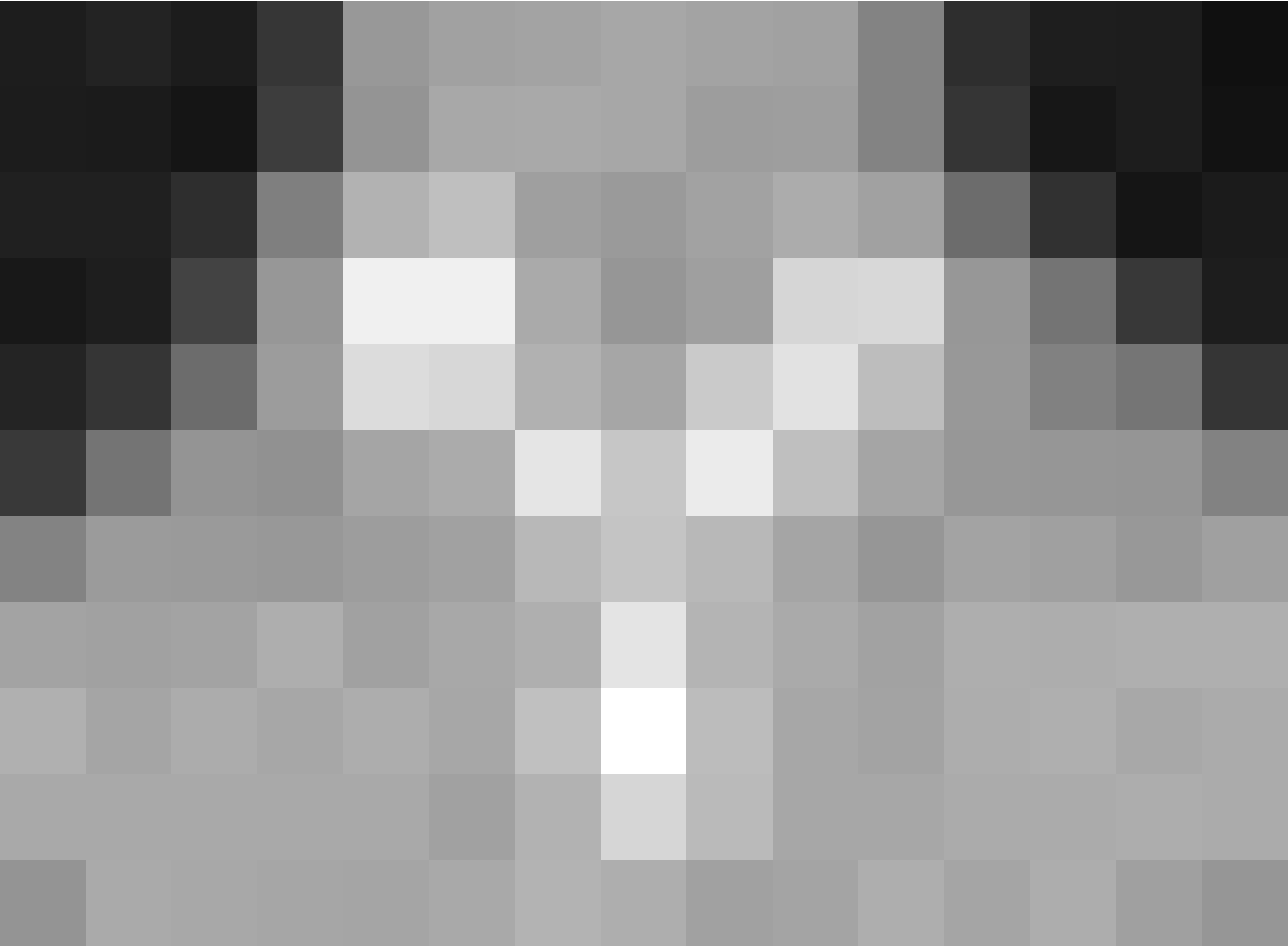}
		\includegraphics[width=0.28\columnwidth]{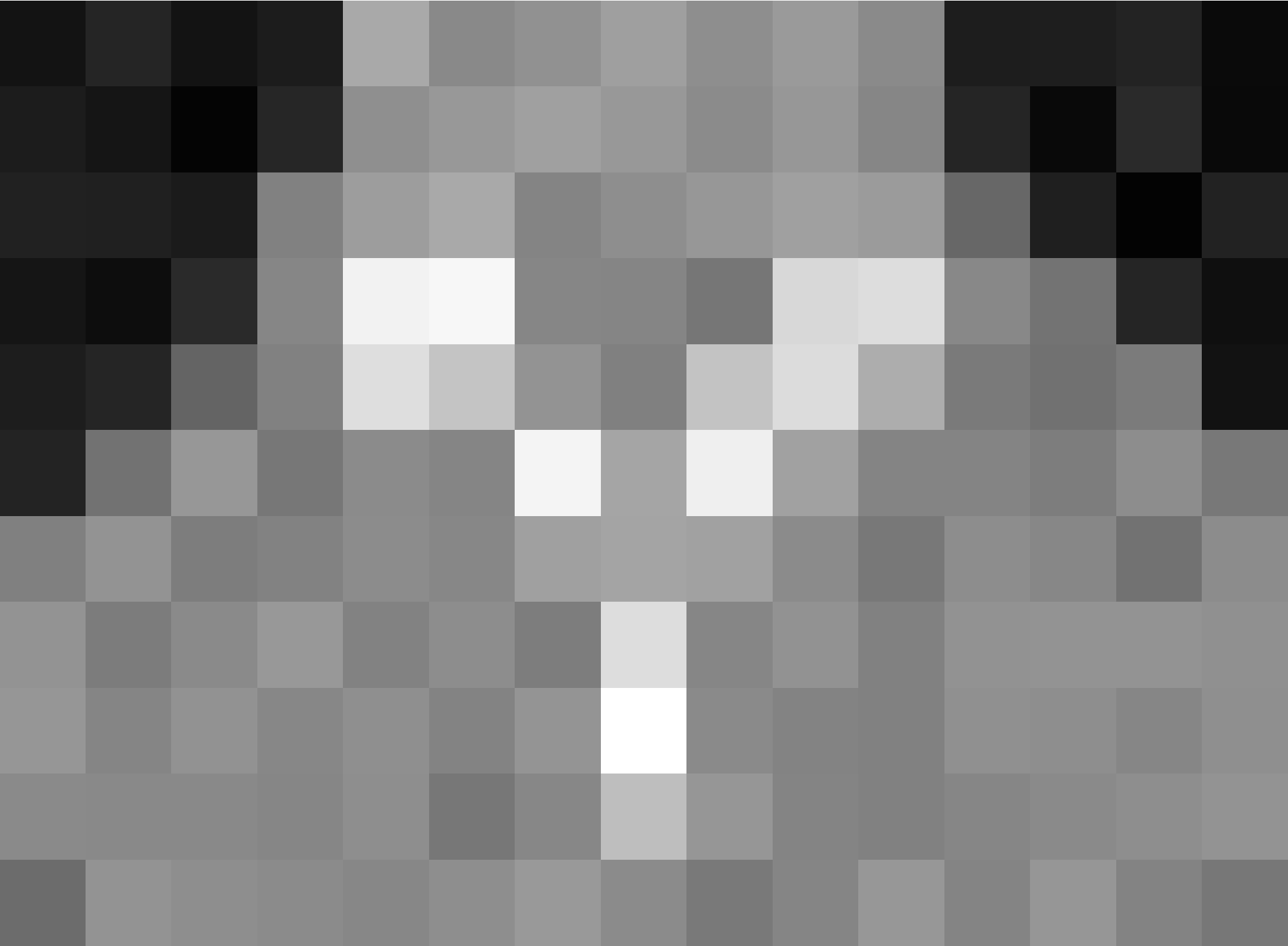}
		\includegraphics[width=0.28\columnwidth]{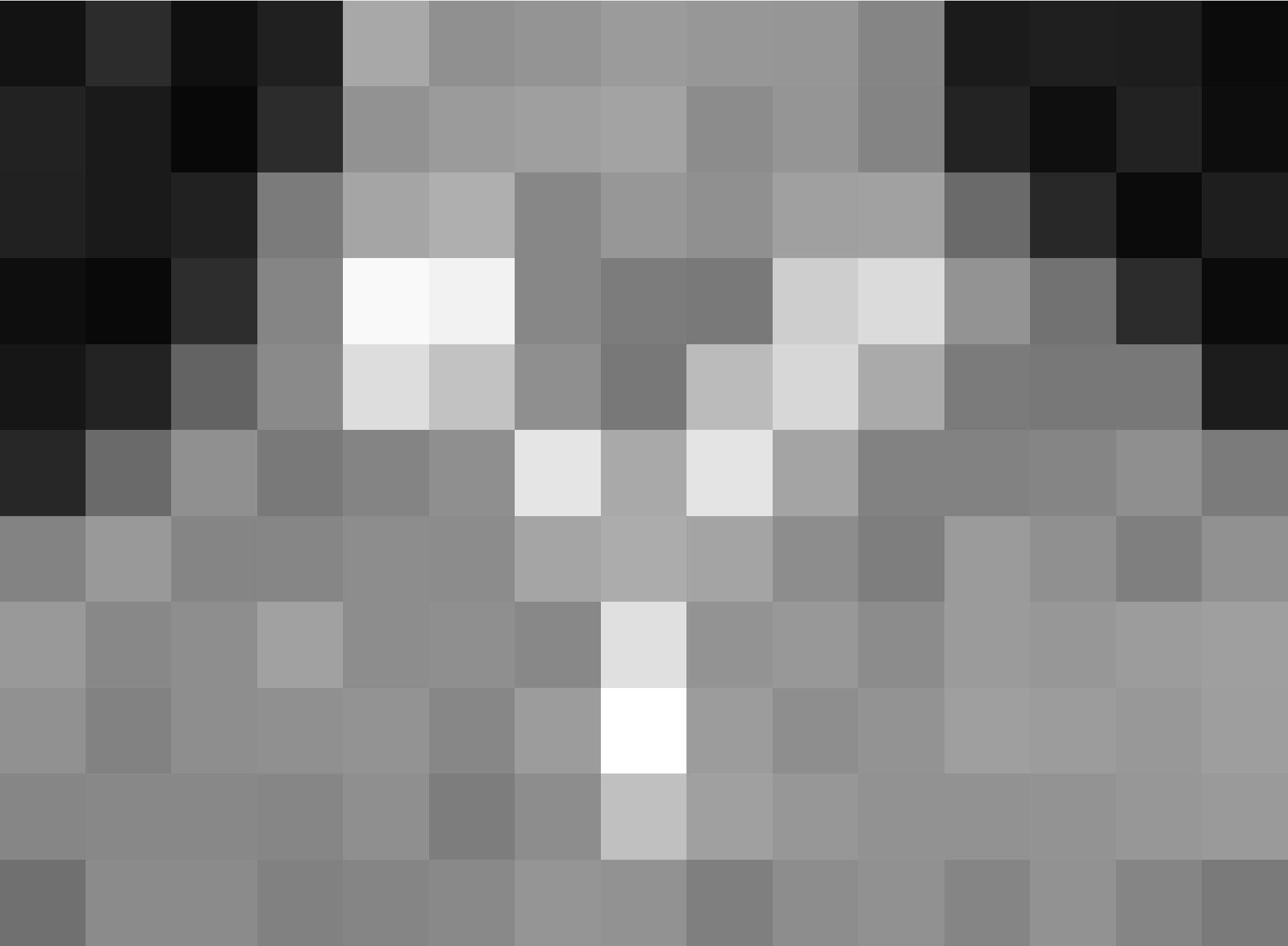}
		\includegraphics[width=0.28\columnwidth]{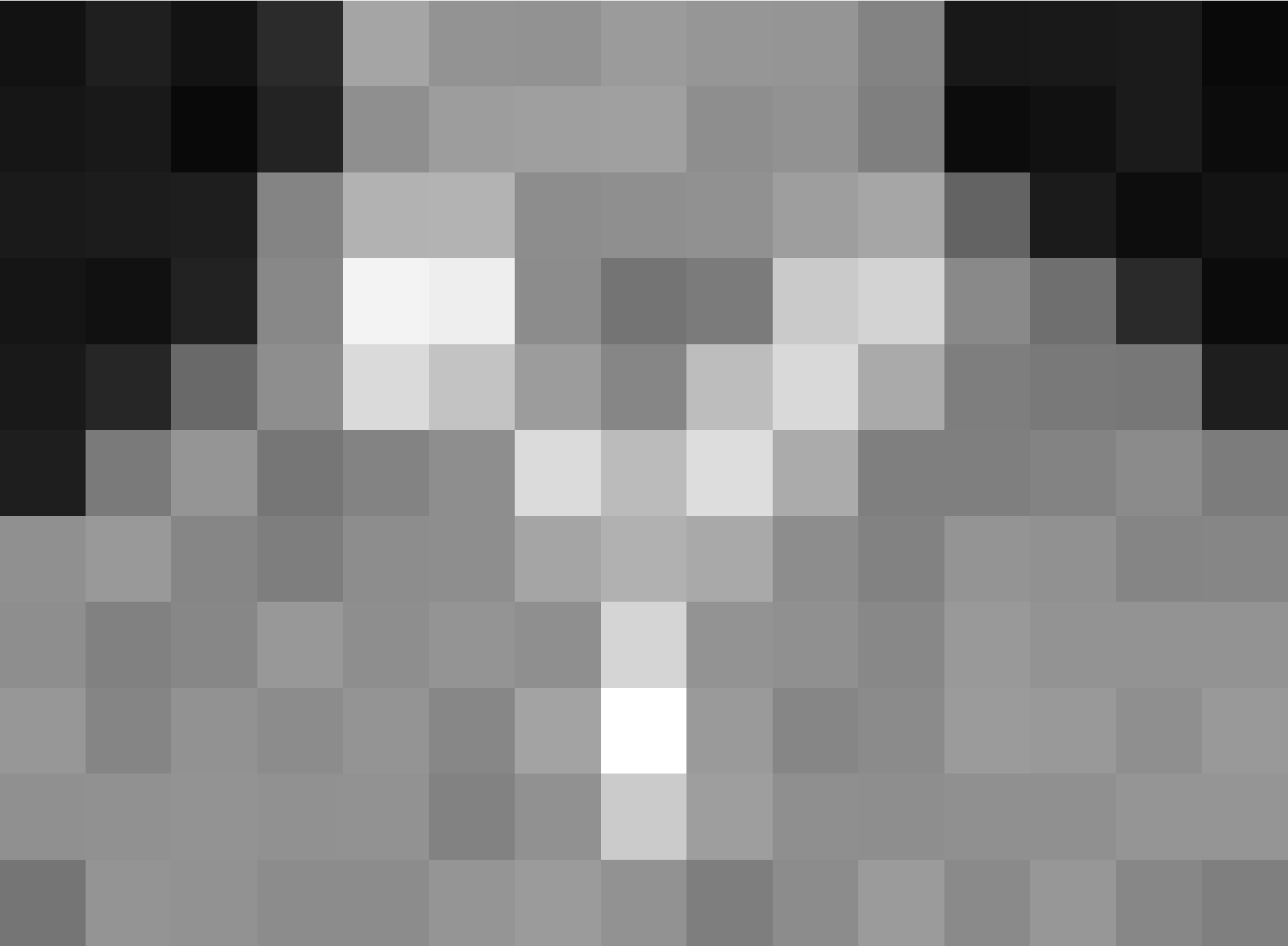}
		\includegraphics[width=0.28\columnwidth]{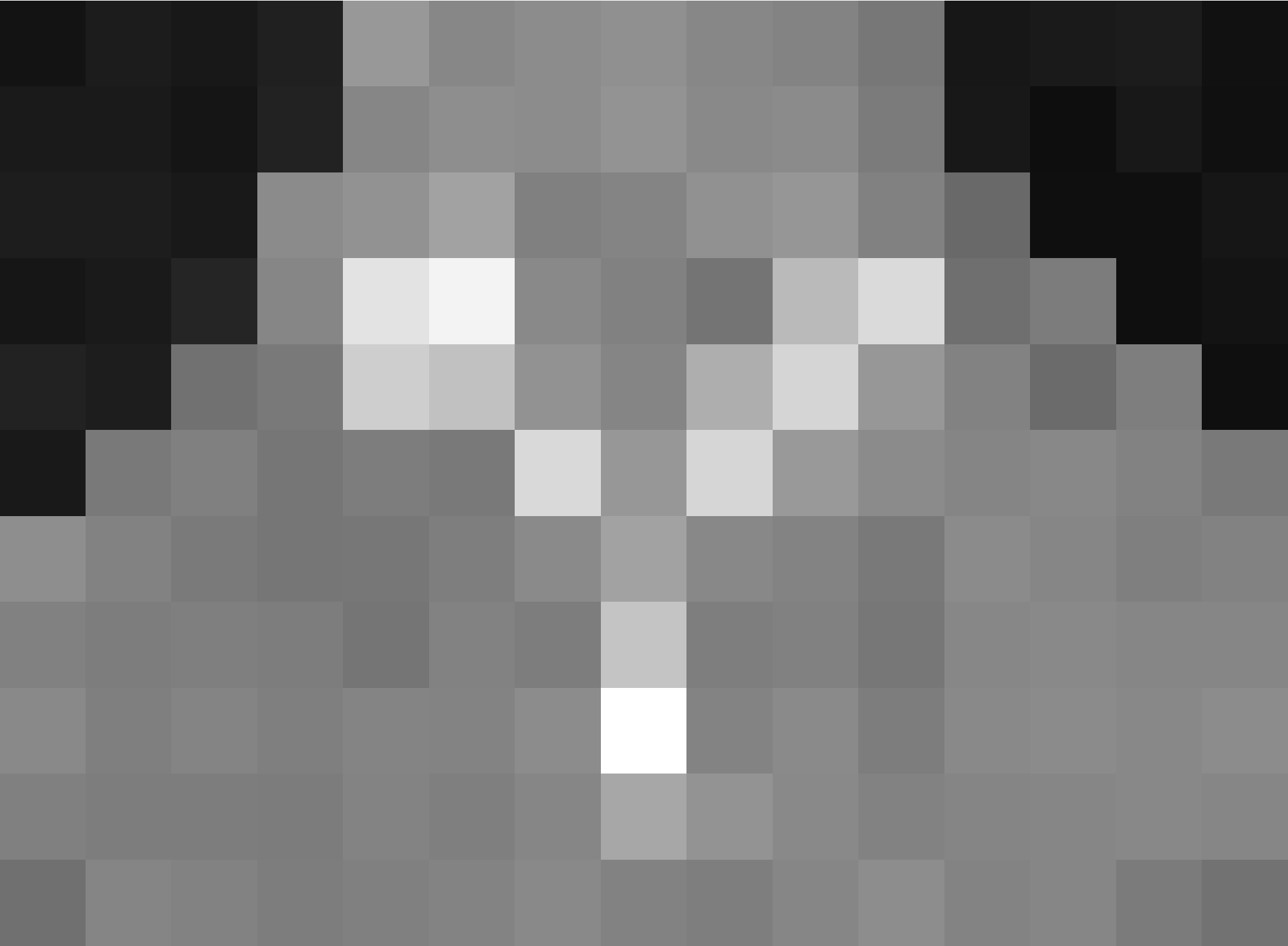}
	}
	\vspace{0.5mm}
	\centerline{
		\includegraphics[width=0.28\columnwidth]{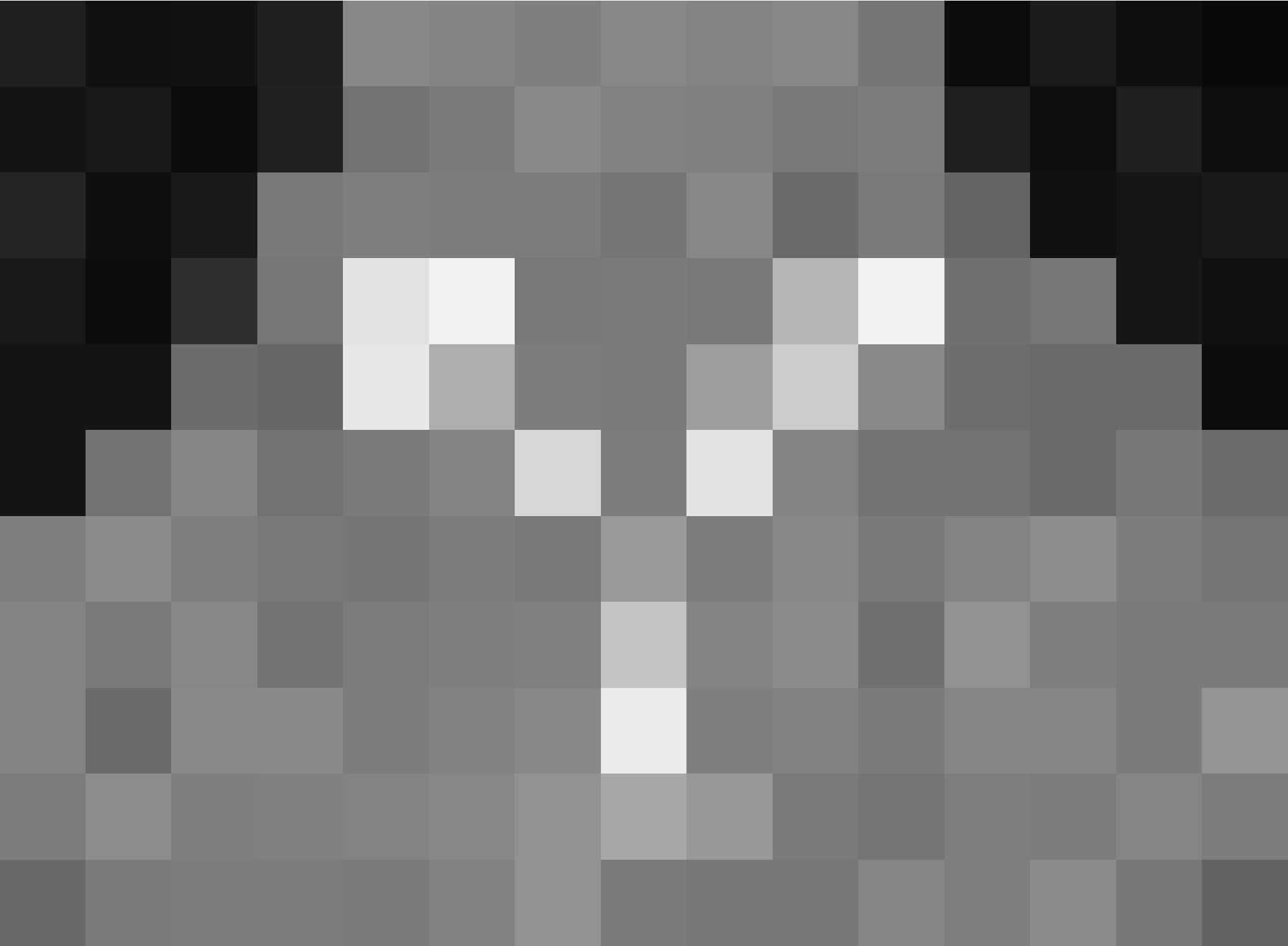}
		\includegraphics[width=0.28\columnwidth]{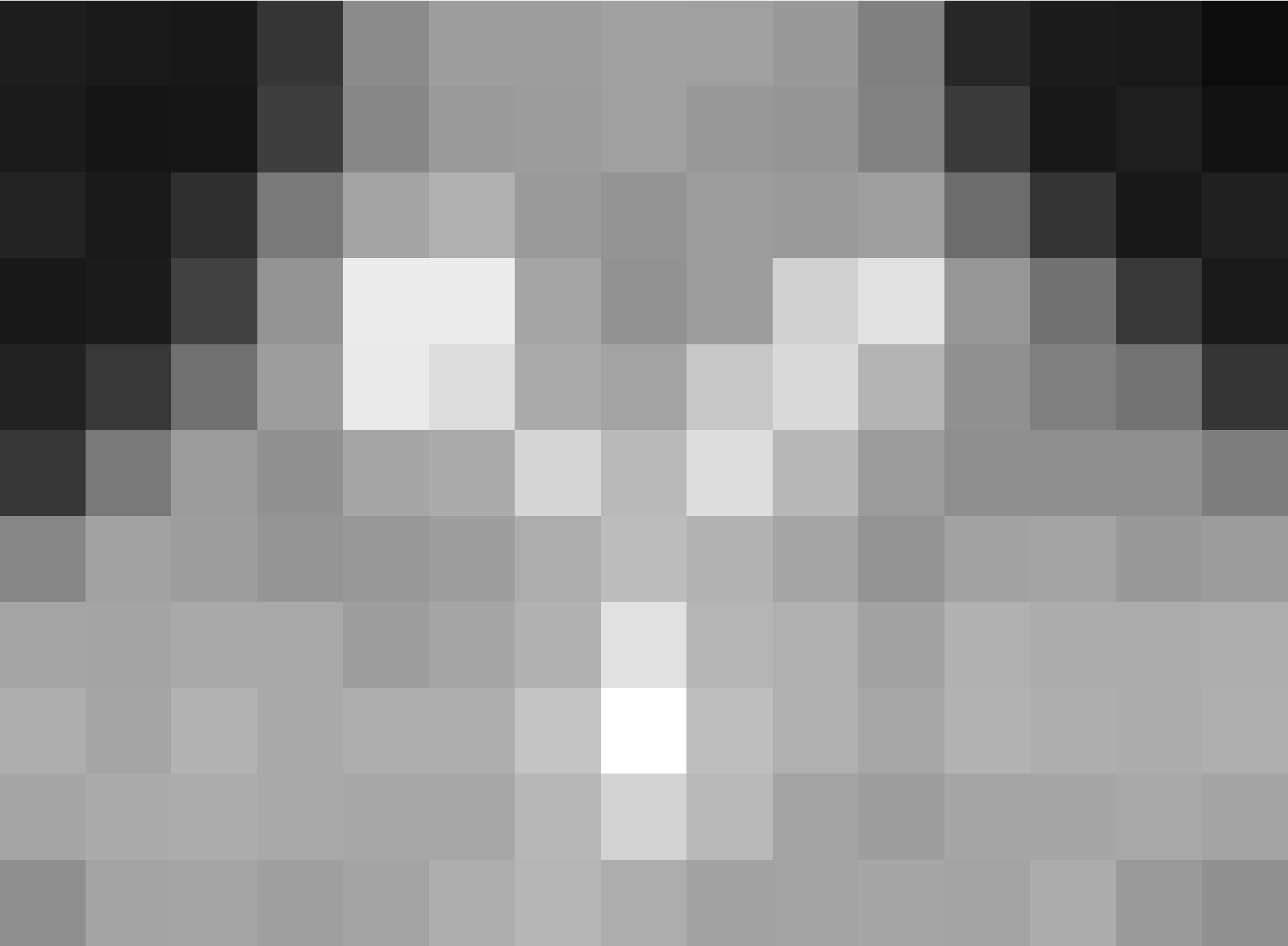}
		\includegraphics[width=0.28\columnwidth]{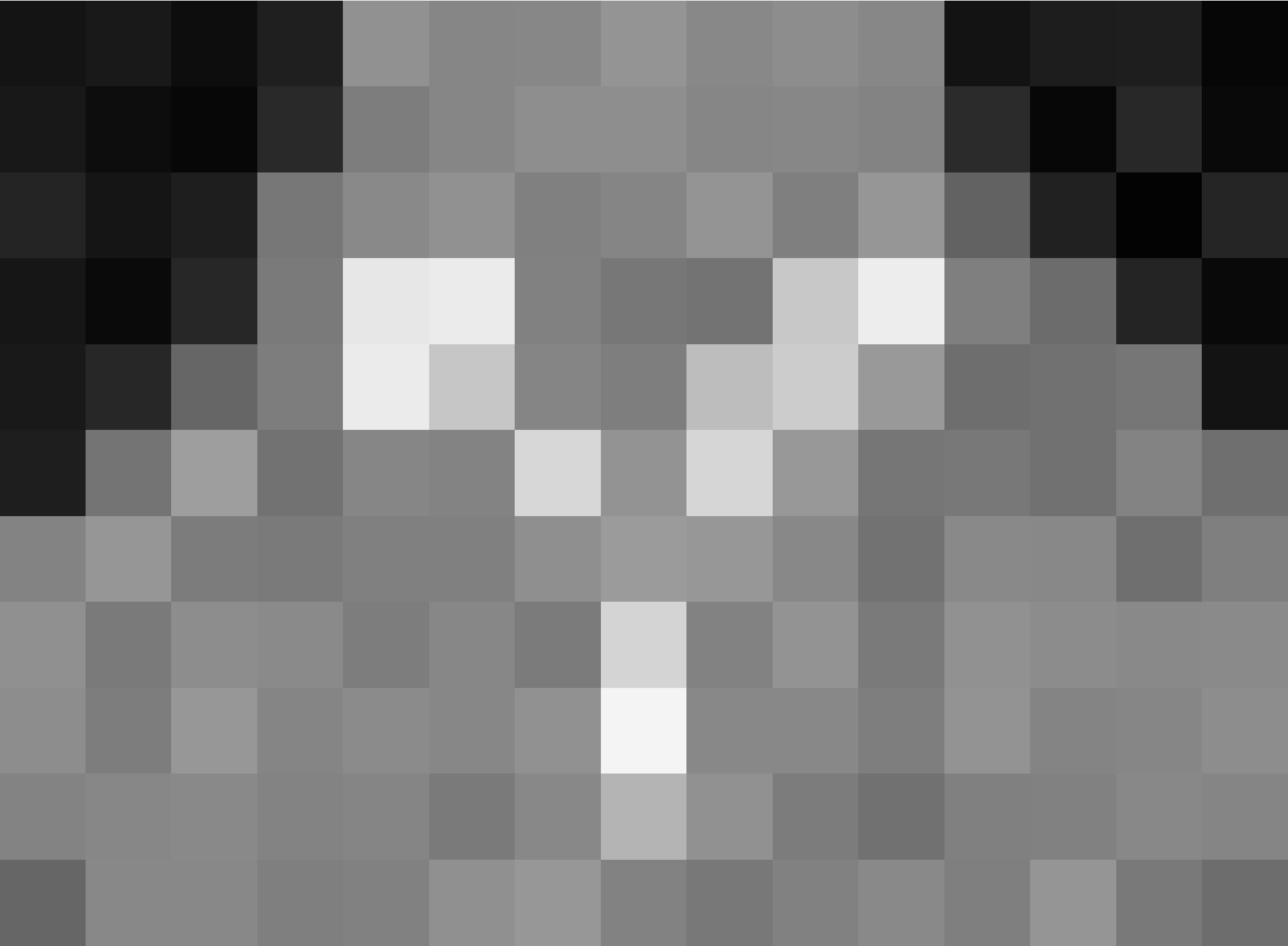}
		\includegraphics[width=0.28\columnwidth]{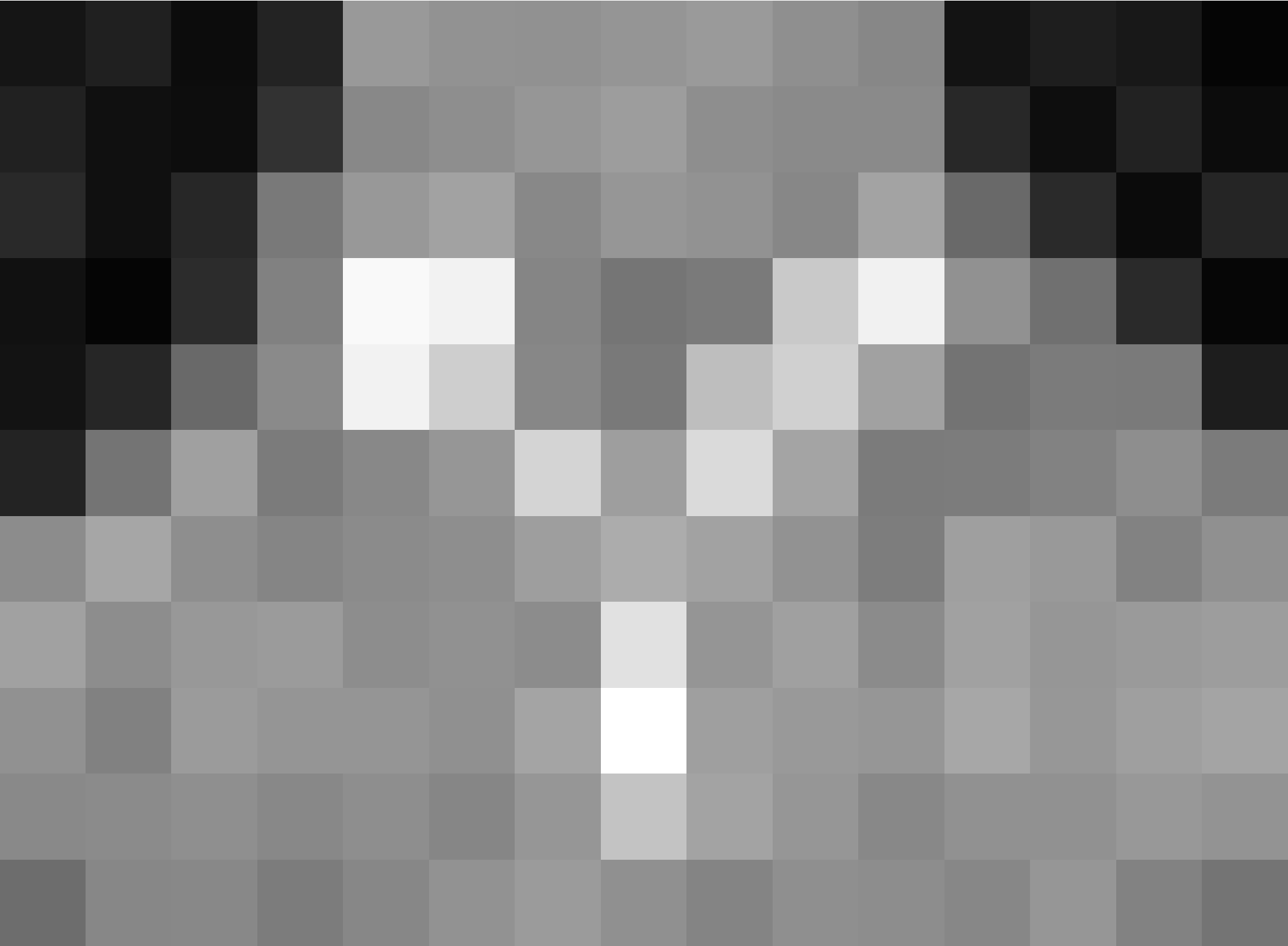}
		\includegraphics[width=0.28\columnwidth]{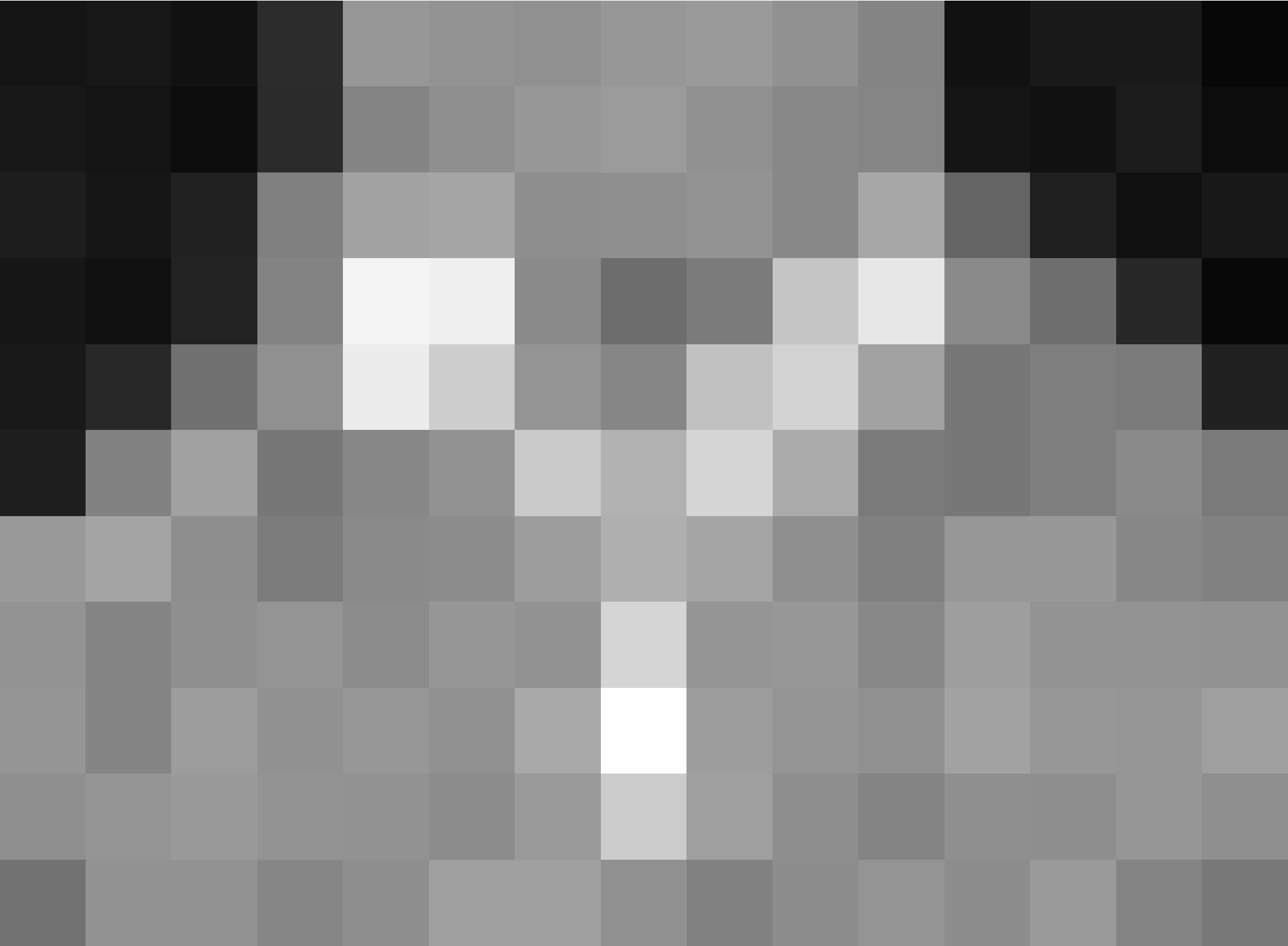}
		\includegraphics[width=0.28\columnwidth]{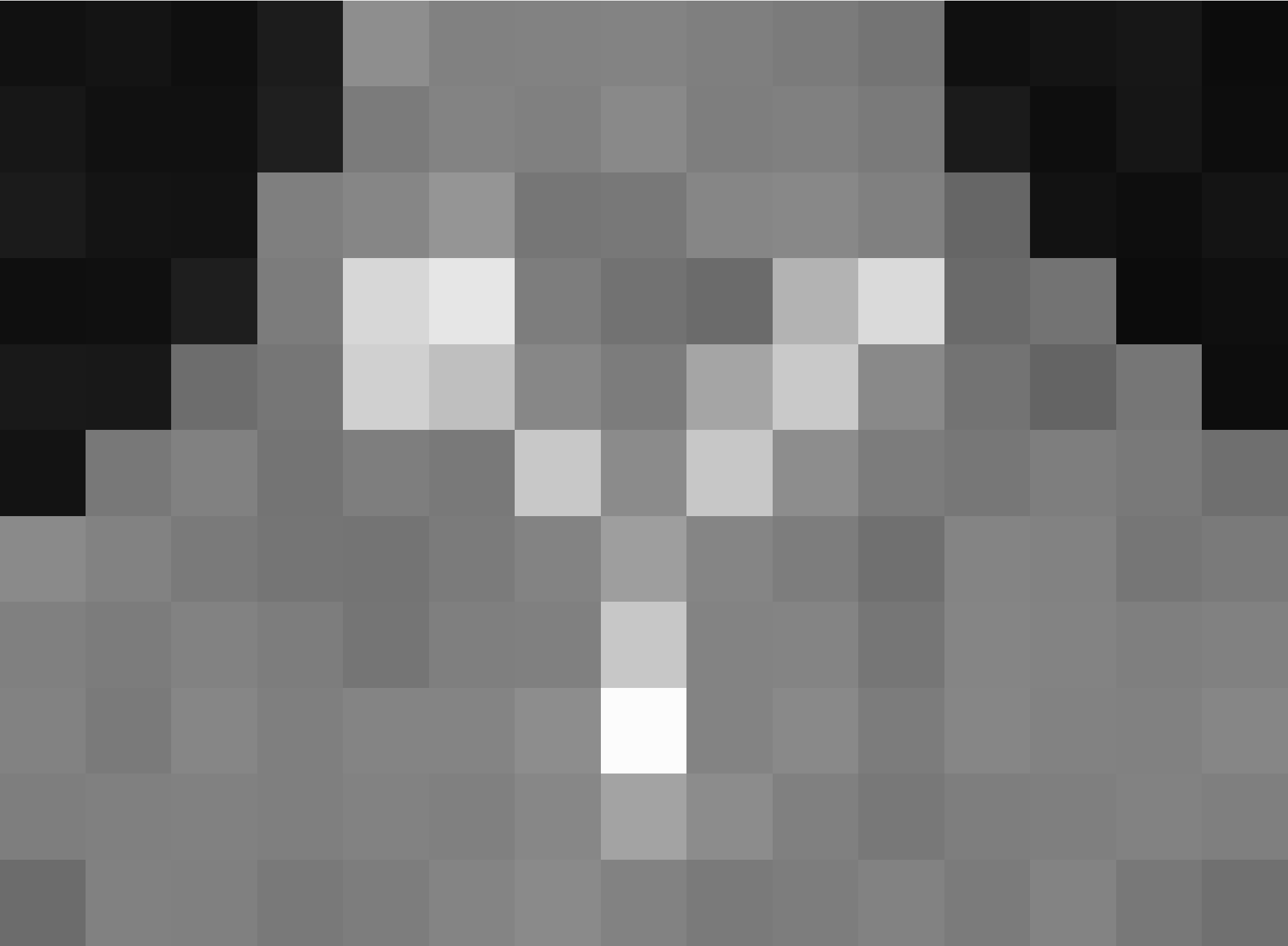}
	}
	\vspace{0.5mm}
	\centerline{
		\includegraphics[width=0.28\columnwidth]{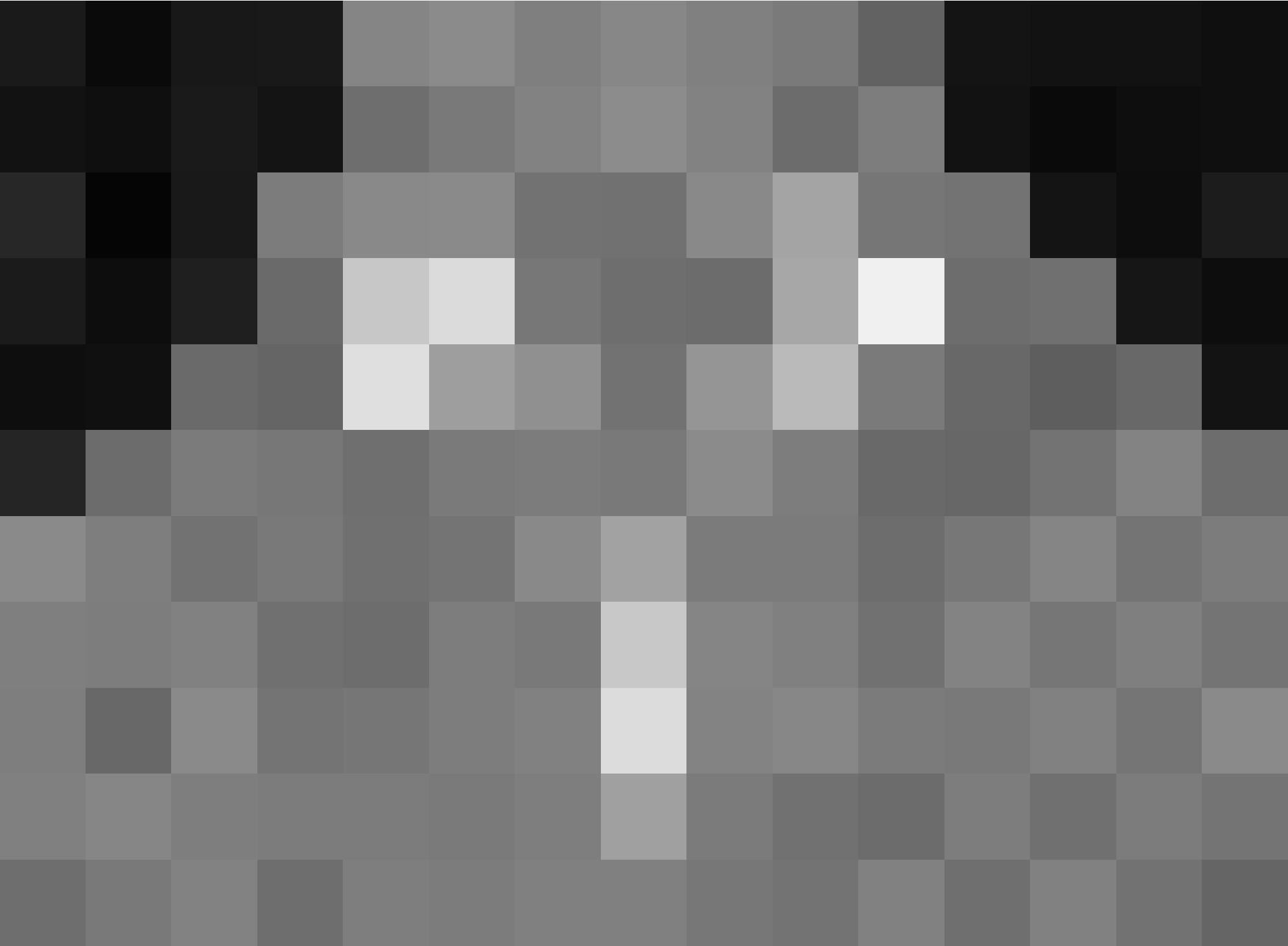}
		\includegraphics[width=0.28\columnwidth]{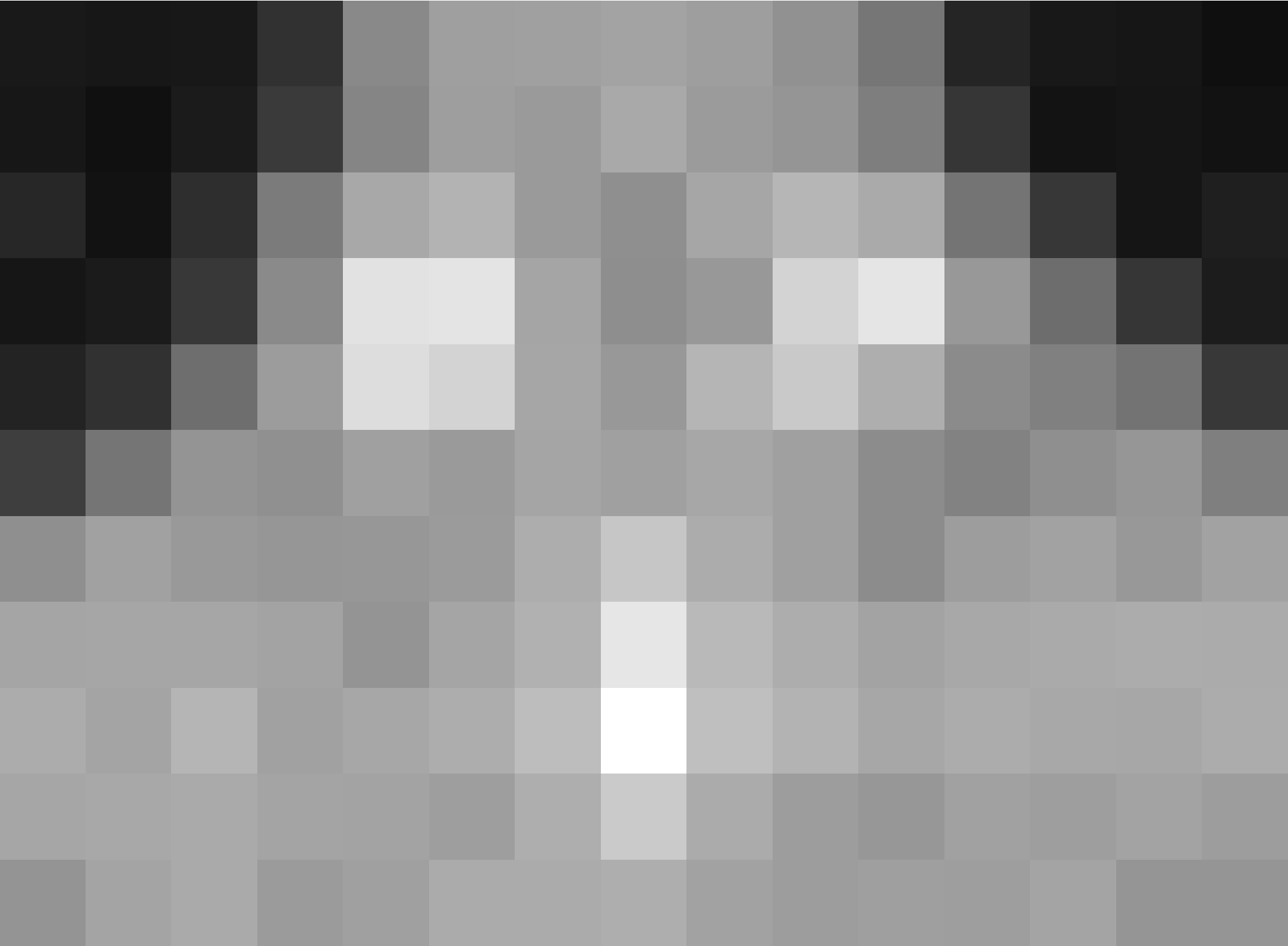}
		\includegraphics[width=0.28\columnwidth]{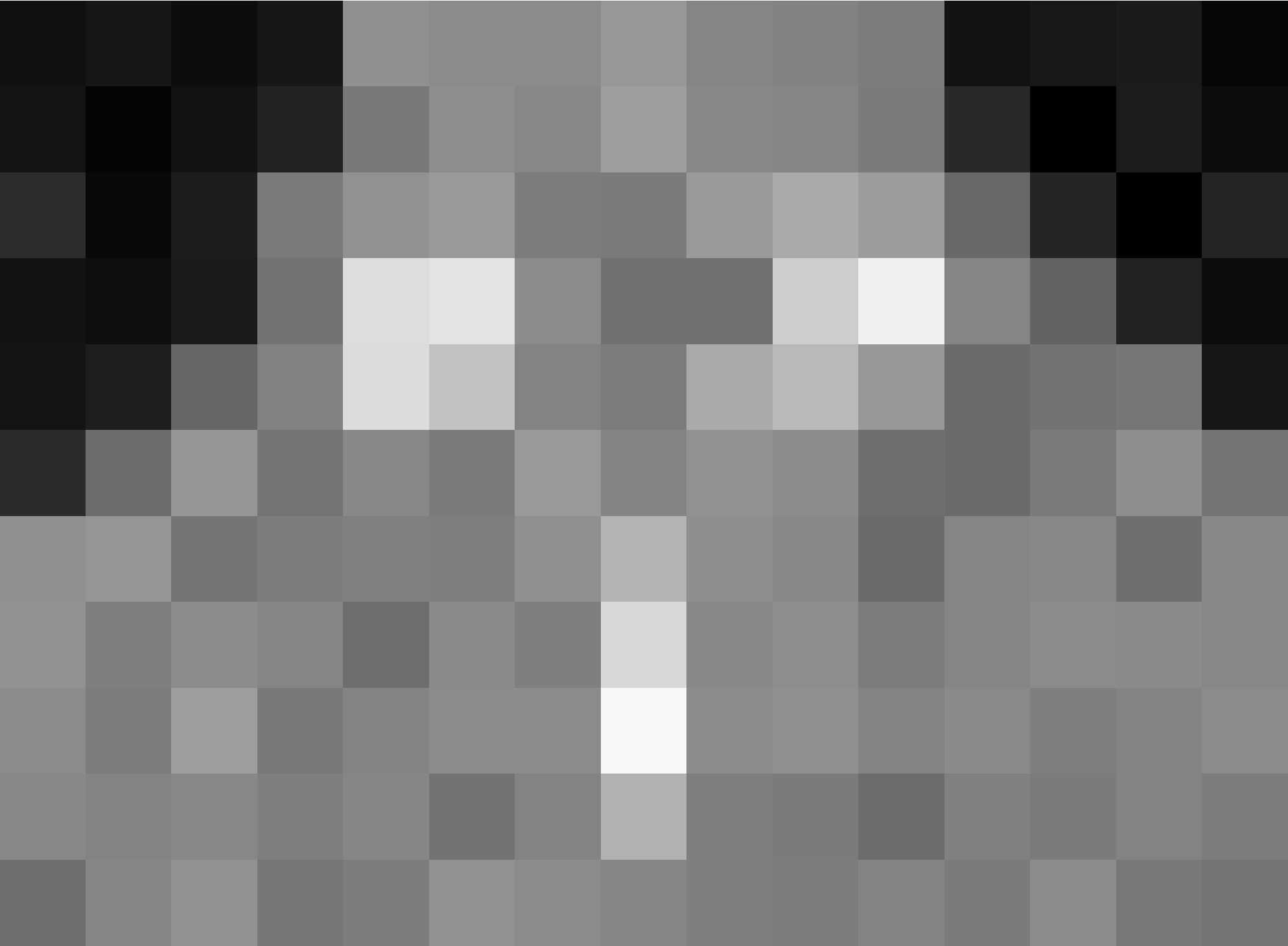}
		\includegraphics[width=0.28\columnwidth]{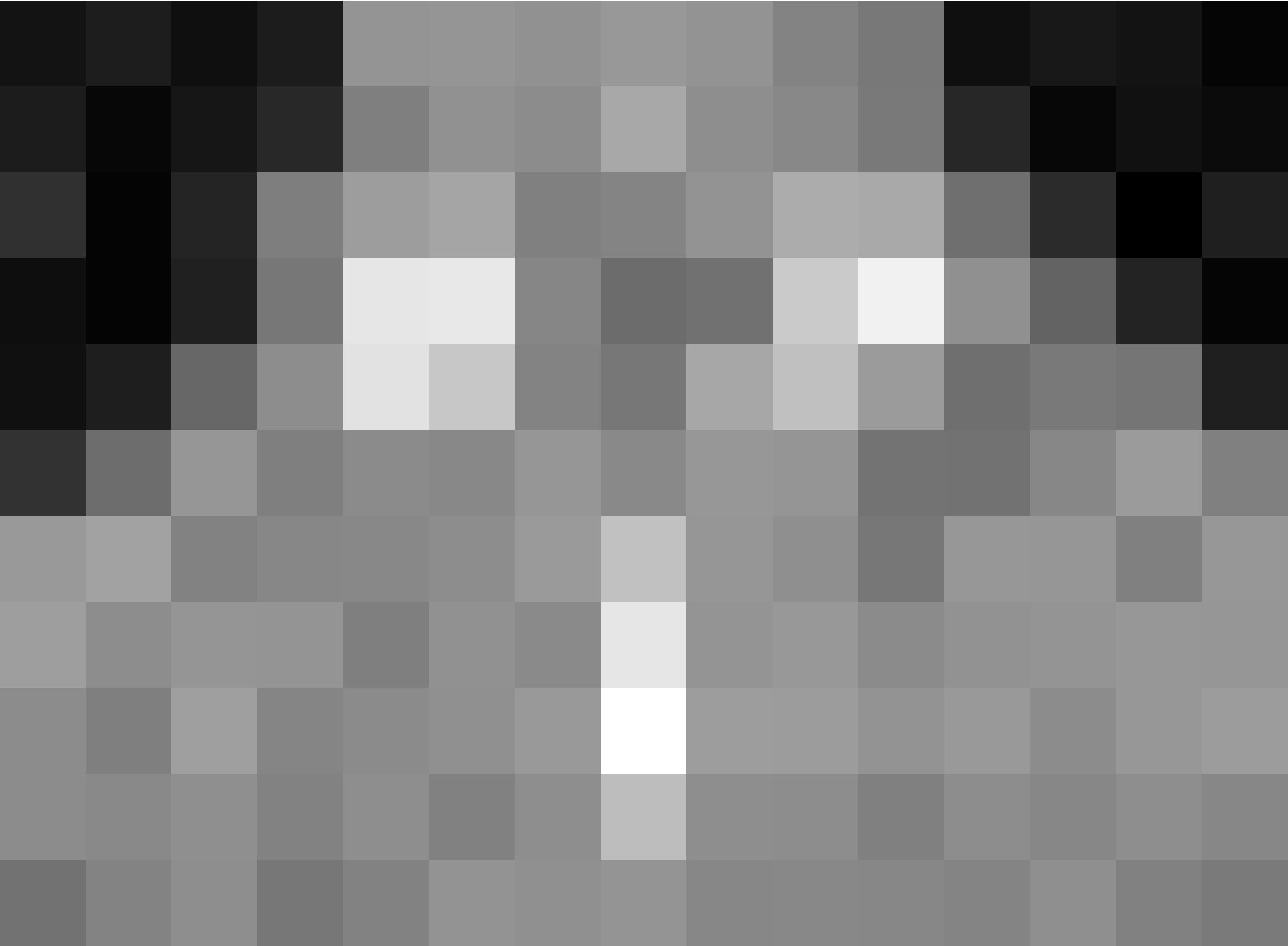}
		\includegraphics[width=0.28\columnwidth]{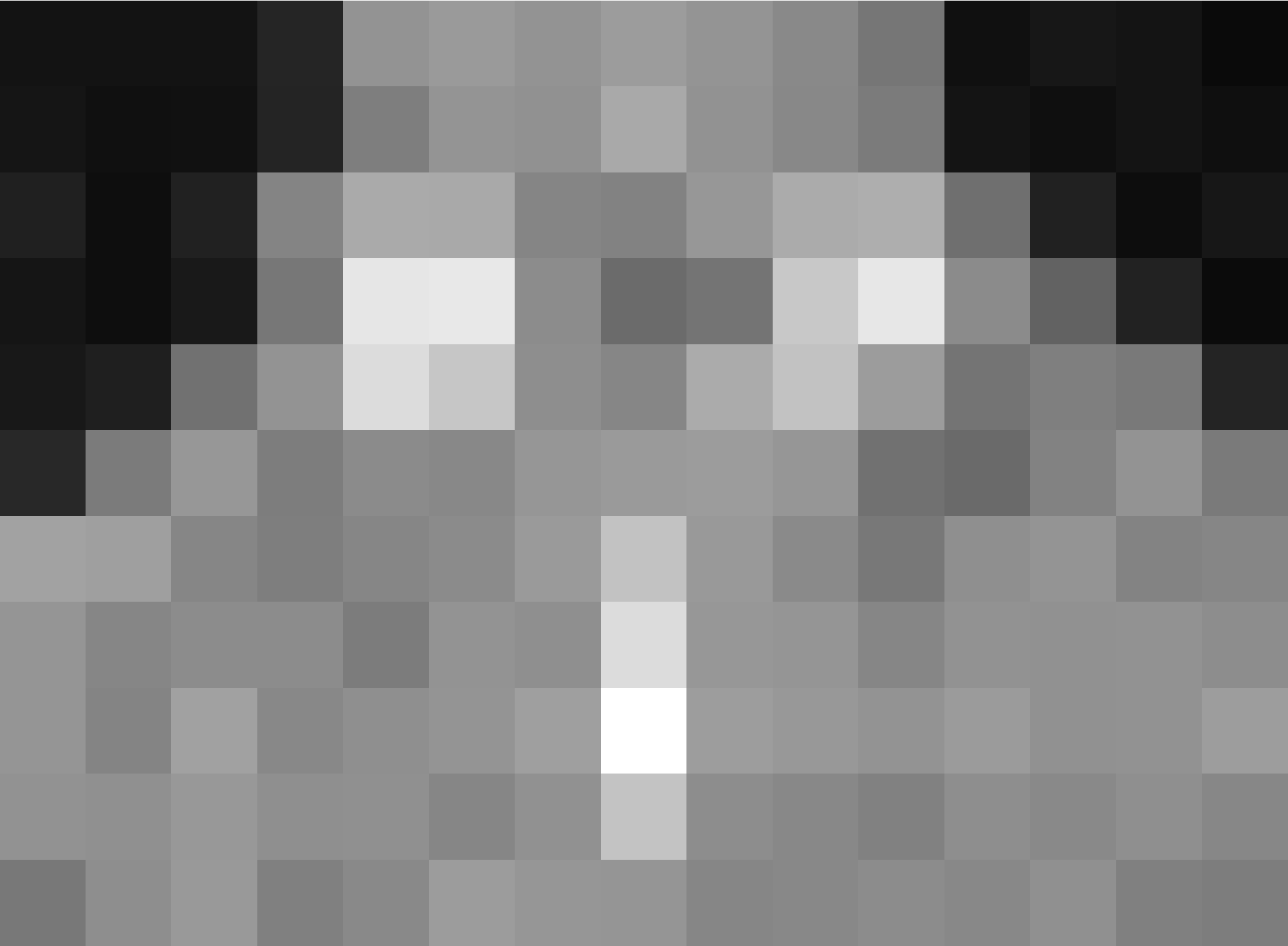}
		\includegraphics[width=0.28\columnwidth]{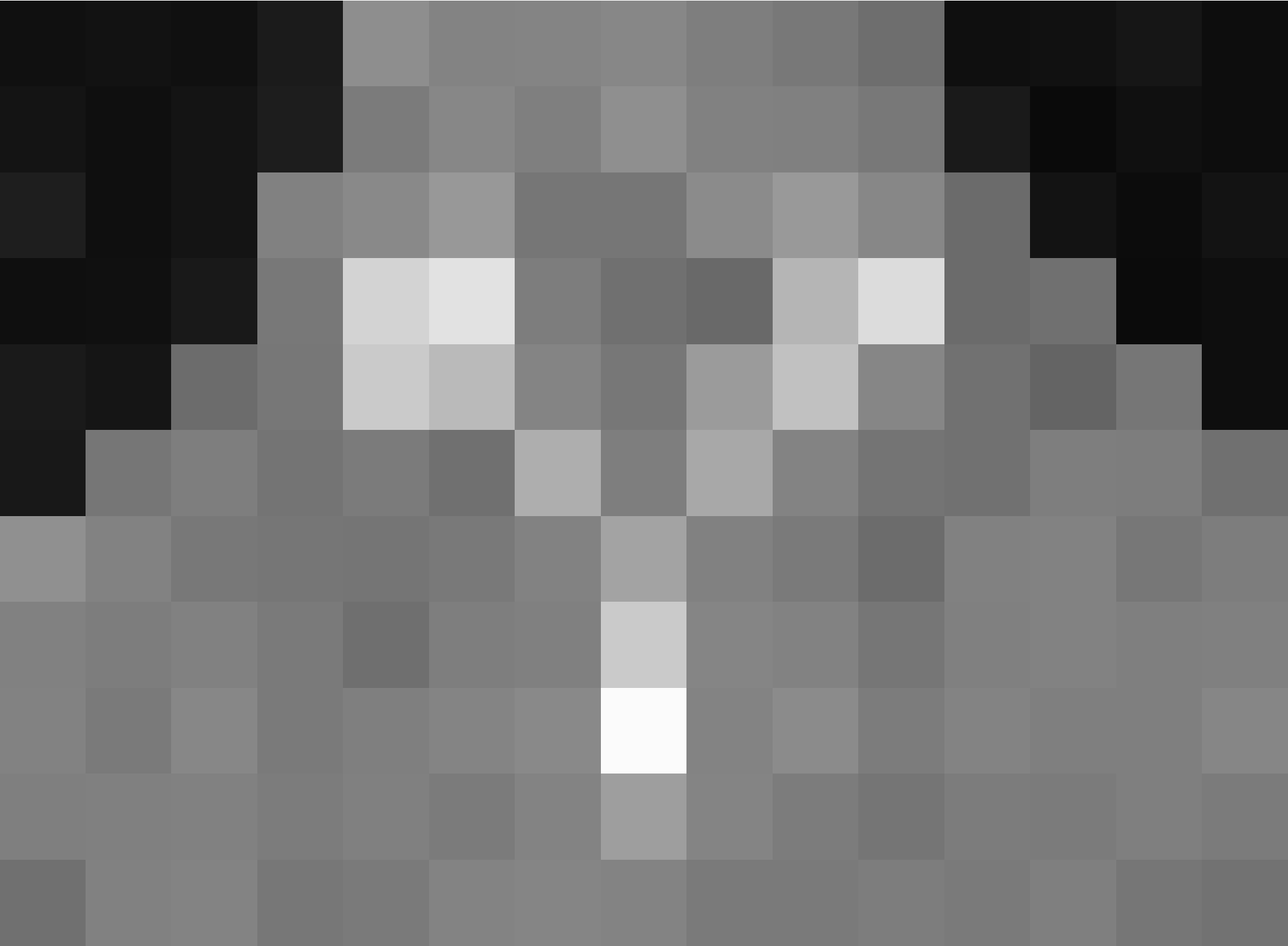}
	}
	\vspace{0.5mm}
	\centerline{
		\subfigure[Label]{\includegraphics[width=0.28\columnwidth]{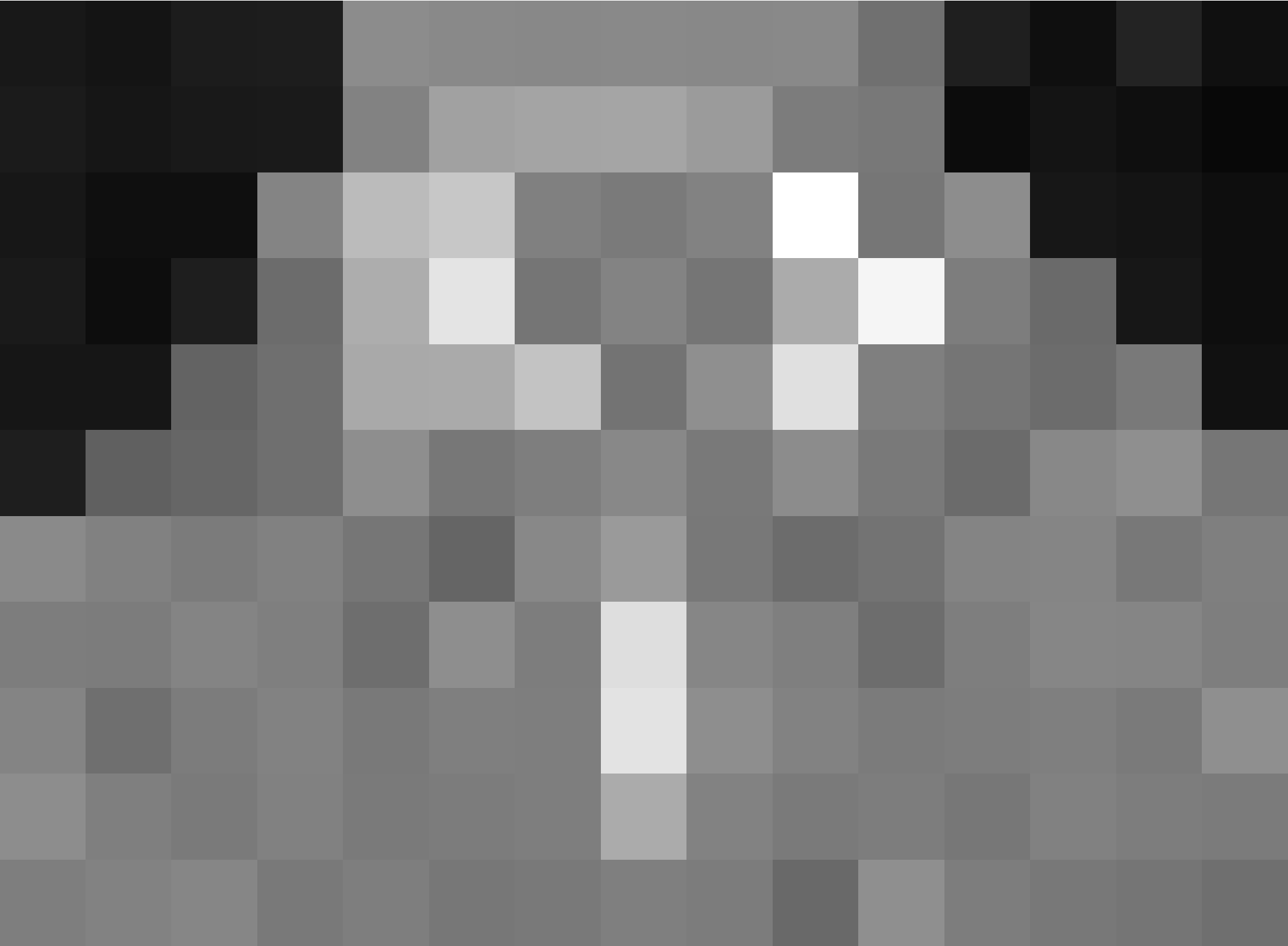}}
		\subfigure[FDK]{\includegraphics[width=0.28\columnwidth]{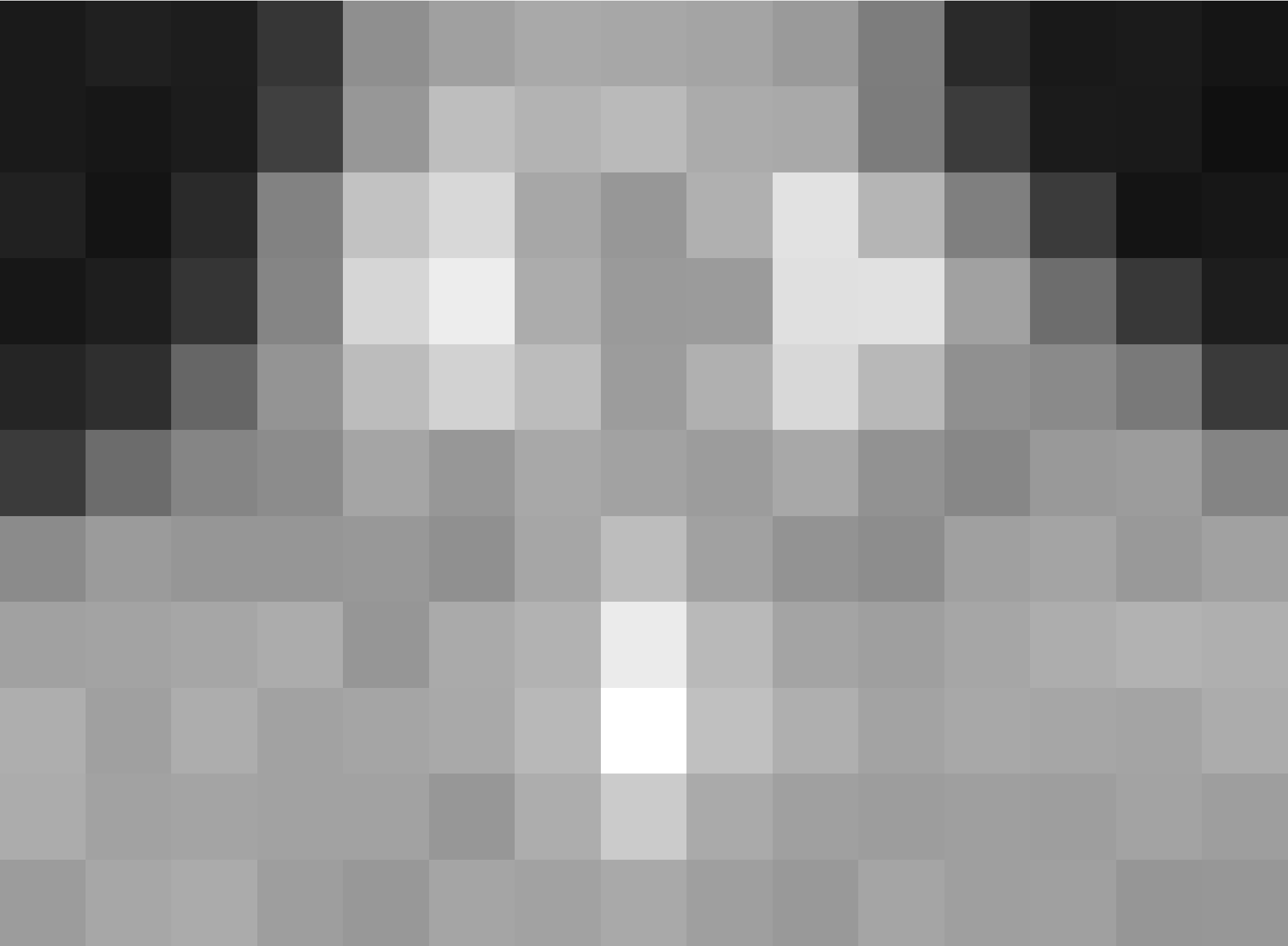}}
		\subfigure[Red-CNN]{\includegraphics[width=0.28\columnwidth]{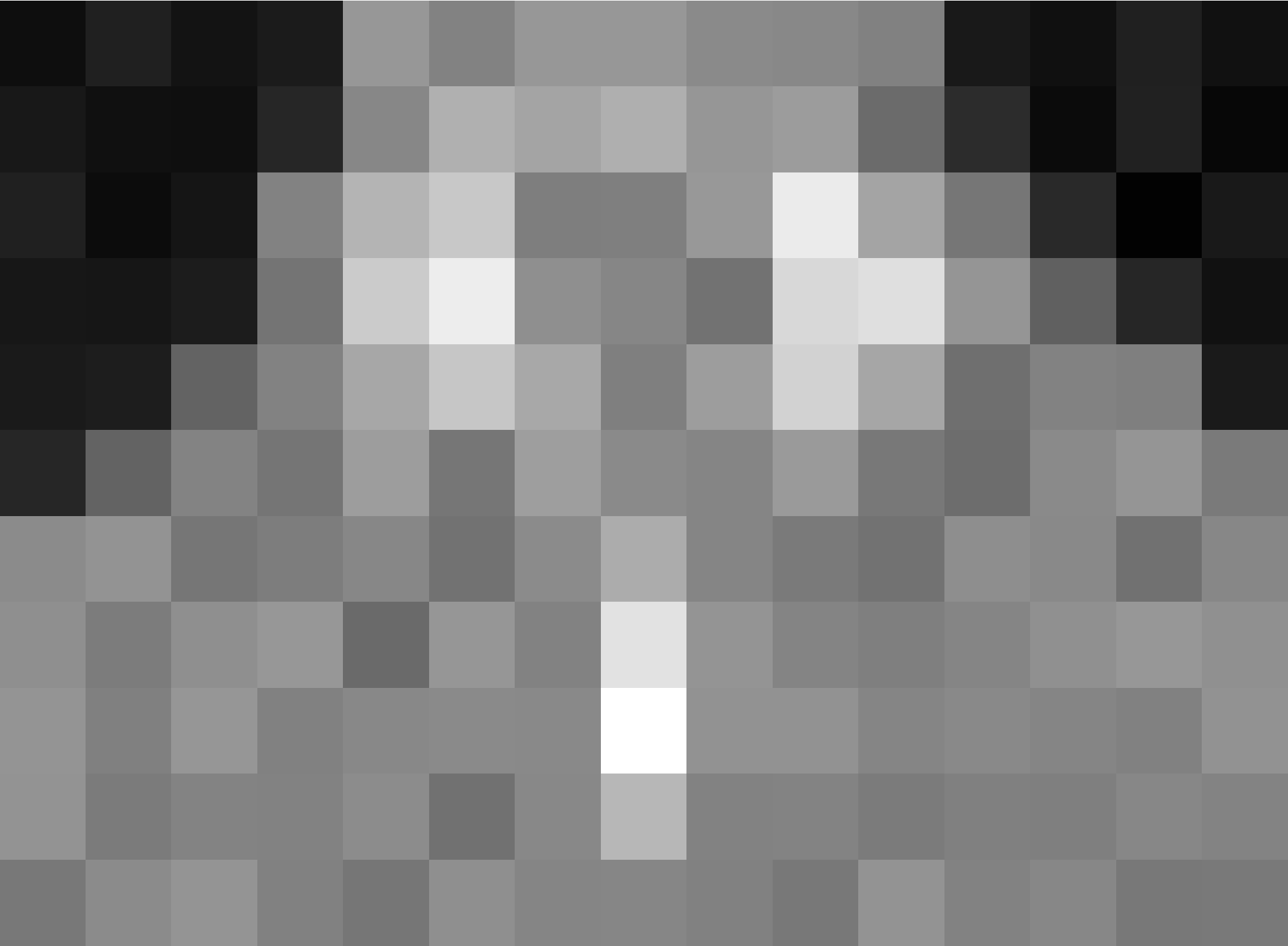}}
		\subfigure[FBP-Conv]{\includegraphics[width=0.28\columnwidth]{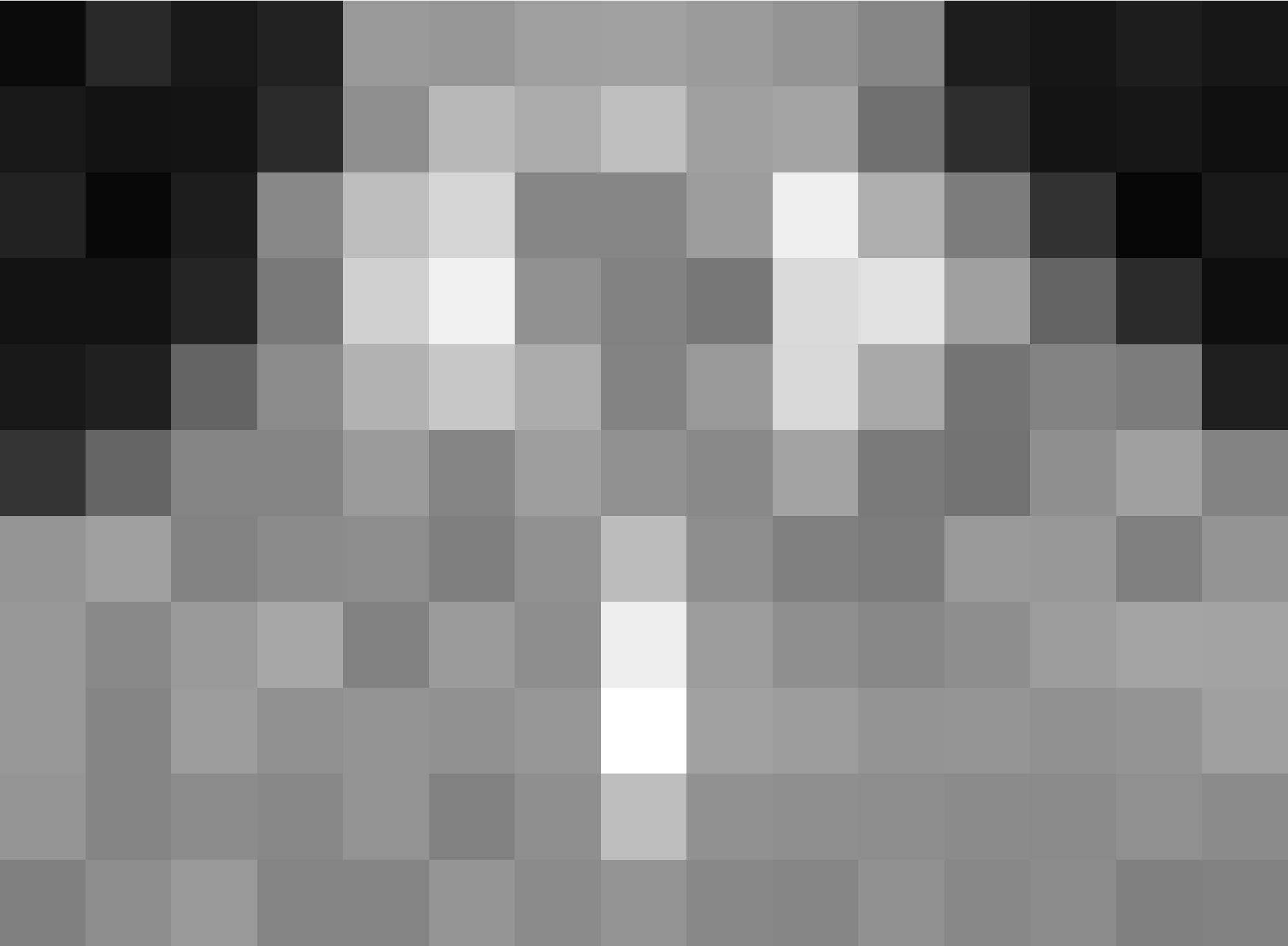}}
		\subfigure[DD-Net]{\includegraphics[width=0.28\columnwidth]{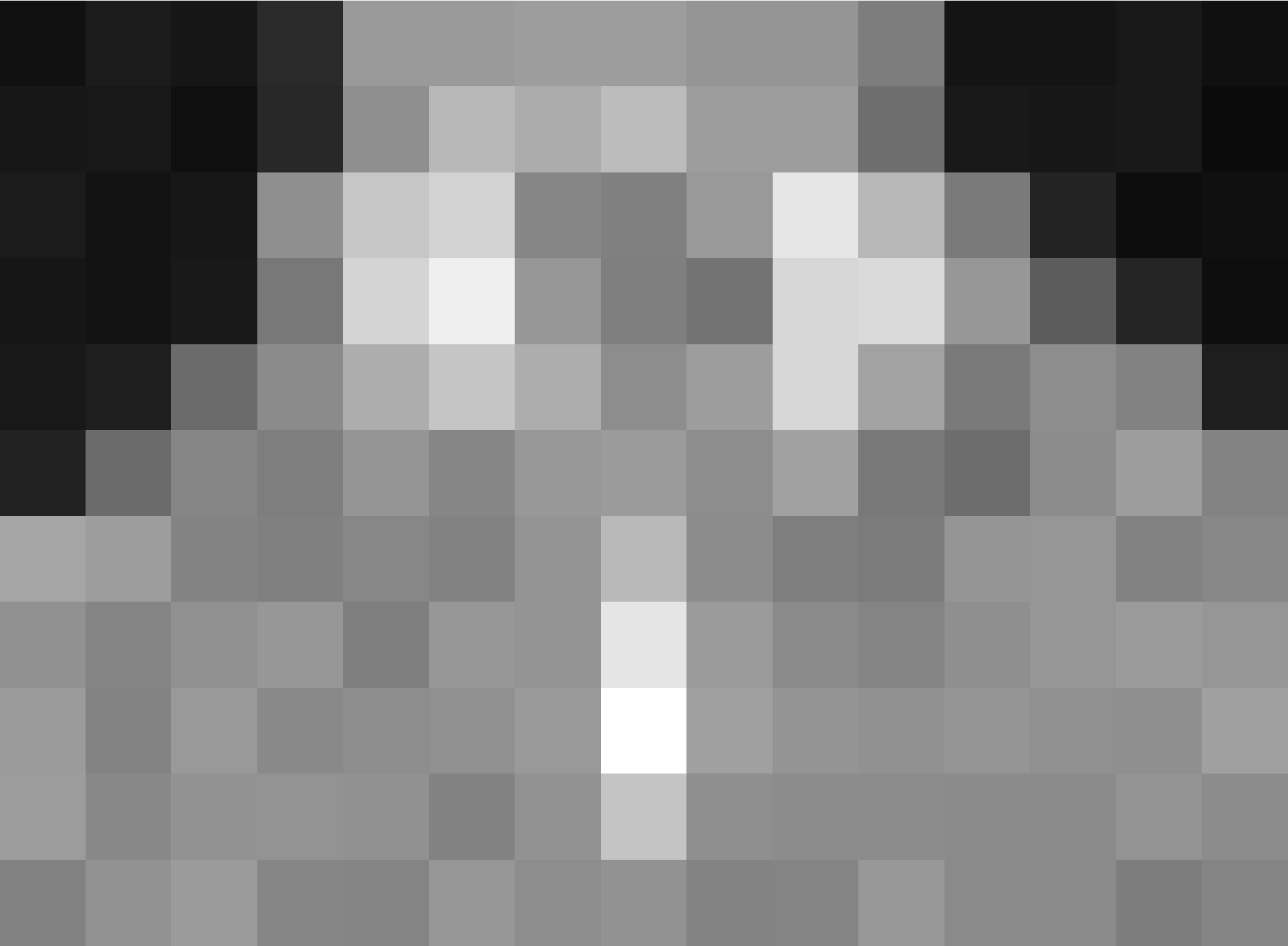}}
		\subfigure[Ours]{\includegraphics[width=0.28\columnwidth]{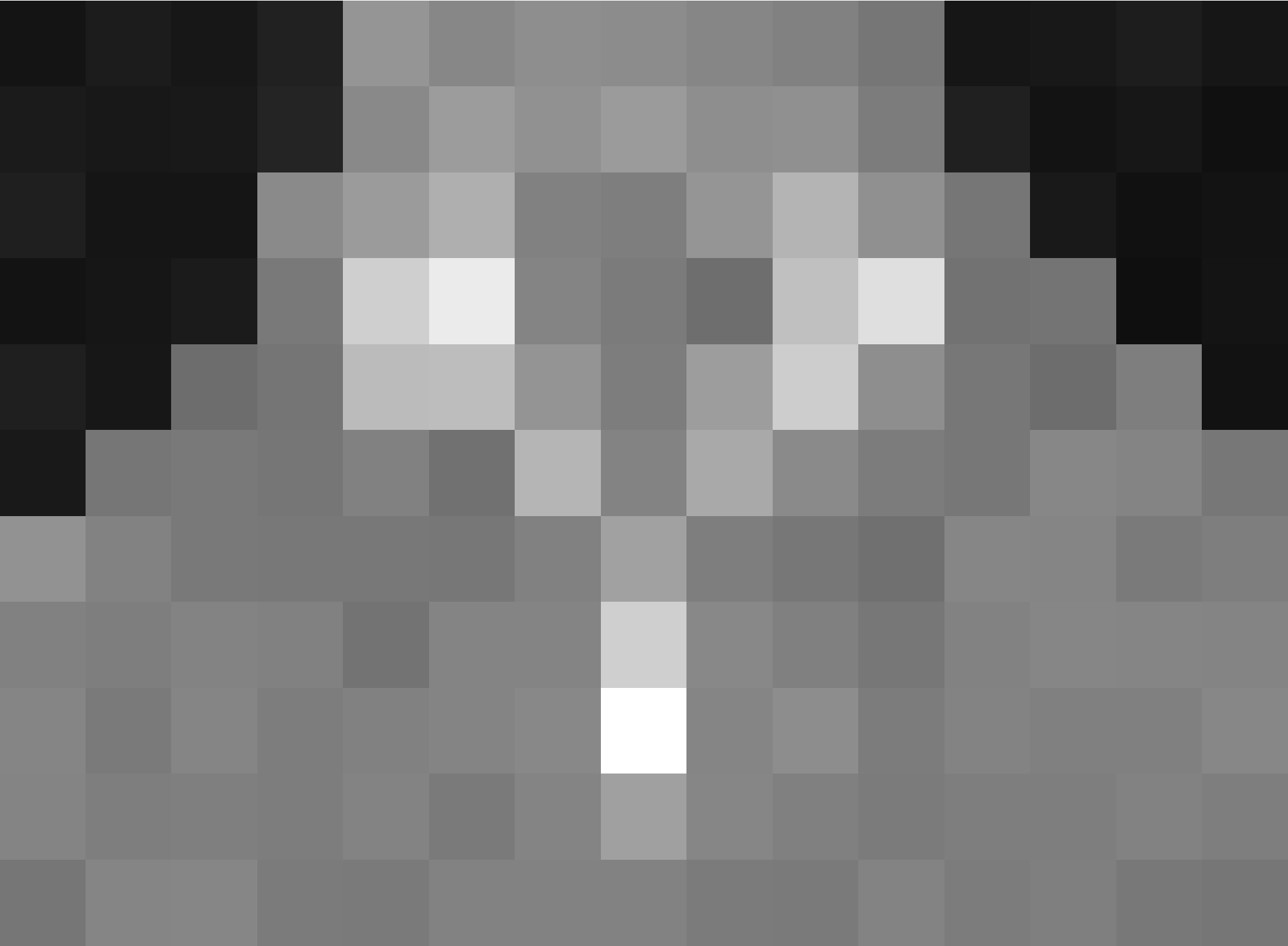}}
	}
	\vspace{0.5mm}
	
	\caption{The zoomed regions marked by the red box in Fig. \ref{F10}a.}
	\label{F11}
\end{figure*}

\subsubsection{Objective Evaluation}
Quantitative analysis for the  CT images reconstructed from the test set
by the six methods has also been performed and the
results are shown in Table \ref{T2}. It can be observed that our method has the highest SSIM and PSNR in average.  Compared to  the second highest, Red-CNN, the PSNR of our method is about 0.7db higher than it. 

\begin{table}[htbp]
	\caption{{The Averaged PSNR and SSIM of  CT Images Reconstructed by the Six Methods for Parallel-Beam CT}}
	\label{T2}
	\setlength{\tabcolsep}{3pt}
	\centering
	\begin{tabular}{c|c|c}
		\hline
		& PSNR  & SSIM   \\ \hline
		FBP	     &15.43	&0.51 \\
		TV regularization & 26.95  &0.84\\
		Red-CNN	 &28.69	&0.86 \\
		FBP-Conv &27.97	&0.84  \\
		DD-Net	 &27.15	&0.82  \\
		Ours     &\textbf{29.41 }    &\textbf{0.87}   \\
		\hline 
	\end{tabular}%
\end{table}%

\subsection{Circle Cone-beam}
\subsubsection{Data Preparation}
The  full dose CT images  from ``the 2016 NIH-AAPM-Mayo
Clinic Low Dose CT Grand Challenge"
are also used as the CT image labels $\varsigma^{label}$. 
Due to the limitation of GPU memory, we choose only 4 continuous CT images and downsample them to form the  object labels $\varsigma^{label}$ of size $64\times64\times4$. We totally assemble 700 objects, where 596 of them are used as the training set, another 4 of them are used as the validation set and the other 100 as the test set. 

The radius of the circle cone-beam geometry is set as $DSO=500$, the distance between the X-ray source and the flat detector panel is set as $DSD=565$. The size of the detector array is $79\times7$ and the distances between each detector are $1$.  The objects are centrally located at the origin, where $[-32:31]\times[-32:31]$ corresponds to the $x-y$ coordinates and $[-2:1]$ corresponds to the z-coordinate. We obtain the sinogram labels $z^{label}$ by rotating the X-ray source in the  $x-y$ plane along the $z$ coordinate to sample the object at angels $\frac{\pi}{180}\times[0:1:219]$. Thus the size of the sinogram labels $z^{label}$ is $79\times7\times220$.  We downsample  $z^{label}$ at scanning angles $\frac{\pi}{180}\times[0:1:179]$ to get the simulated limited-angle sinogram $g$ of size $79\times7\times180$. The  limited-angle sinogram $g$ is the input of our network and  the reconstructed CT image from $g$ by the short-scan FDK \cite{ISI:A1994NZ27700005} algorithm is used as the initial guesses of CT images $u_0$. For the  compared networks, Red-CNN, FBP-Conv and DD-Net, all slices of the reconstructed CT image from $g$ by the short-scan FDK  algorithm are used as their inputs.
\subsubsection{Parameter Setup}
The parameters $\Theta_\varsigma^n$ and $\Theta_z^n$ in $Res_\varsigma^n(\cdot)$ and $Res_z^n(\cdot)$ of our network are automatically initialized by Tensorflow using  the default values. The initial values of parameters $\Theta_u$ in $Merge_u^n(\cdot)$ are set as $t_1=1$ and $t_2=t_3=t_4=0.1$. The number of iterations is set as $N_{iter}=5$. The batch size is set as 4  and the number of training epochs is $50$.

The parameters for Red-CNN, FBP-Conv, and DD-NeT are set as
described in their corresponding papers and the initial values in their networks are initialized by Tensorflow automatically. The batch size is 4  and
the training epochs
for Red-CNN, FBP-Conv and DD-Net are all 500. 

\subsubsection{Subjective Evaluation}
Fig. \ref{F10} shows four slices of one object from the test set by the five methods, FDK, Red-CNN, FBP-Conv, DD-Net and ours and Fig. \ref{F11} shows the zoomed regions marked by the  red box in Fig. \ref{F10}a. From  Fig. \ref{F11}, it can be observed that  the contrast of the reconstructed images  by our method is the most consistent with the labels  compared to those by the other methods, which demonstrates that our method can preserve more edge information.

\subsubsection{Objective Evaluation}
The average PSNR and SSIM of the reconstructed CT image slices from the test set are listed in Table \ref{T3}. It can be observed that our method has the highest SSIM and PSNR in average.  Compared to  the second highest, DD-Net, the PSNR of our method is about 3.8db higher.

\begin{table}[htbp]
	\caption{{The Averaged PSNR and SSIM of  CT Images Reconstructed by the five Methods for Circle Cone-Beam CT}}
	\label{T3}
	\setlength{\tabcolsep}{3pt}
	\centering
	\begin{tabular}{c|c|c}
		\hline
		& PSNR  & SSIM   \\ \hline
		FDK	     &8.19	&0.55 \\
		Red-CNN	 &30.10	&0.89 \\
		FBP-Conv &29.15	&0.89  \\
		DD-Net	 &30.22	&0.82  \\
		Ours     &\textbf{34.04 }    &\textbf{0.91}   \\
		\hline 
	\end{tabular}%
\end{table}%

\section{Conclusion}

In this paper, we first proposed a variational model with two regularizations for the limited-angle CT image reconstruction, where one regularization utilizes the prior information of  sinograms in the frequency domain and the other utilizes the prior information of  CT images in the spatial domain. Then we used the penalty method to convert the variational model into three iterative subproblems, where the first subproblem completes  the sinograms and the second refines the CT images,  and the last merges the outputs of the first two subproblems. Instead of giving any explicit form of the regularizations and using the optimal algorithms to solve the first two subproblems, we used the CNNs of four layers to approximate the solutions of the first two subproblems. Therefore, by unrolling the iterative algorithm, we obtained an end-to-end deep network for the limited-angle CT image reconstruction. Experimental results showed that our deep network outperformed the existing algorithms  under parallel-beam, fan-beam and circle cone-beam scanning  geometries for the limited-angle CT image reconstruction.

%Due to the limitation of GPU memory, we only performed the 2D CT image reconstruction experiments for  parallel-beam and fan-beam scanning geometries. In the future, we will consider how to reduce the GPU memory usage of our network and use it to reconstructed  3D CT images. 

\balance
\bibliographystyle{IEEEtran}
\bibliography{11}
\end{document}